\documentclass{jfm}
\usepackage{graphicx}
\usepackage{stackengine}
\usepackage{booktabs}
\usepackage{subcaption}
\usepackage{float}
\usepackage{amssymb}
\usepackage{amsmath}
\usepackage{color}

\graphicspath{{images/}}

\shorttitle{Linear instability of viscoelastic pipe flow}
\shortauthor{I.~Chaudhary, P.~Garg, G.~Subramanian, and V.~Shankar}

\title{Linear instability of viscoelastic pipe flow}

\author{Indresh Chaudhary\aff{1}, Piyush Garg \aff{2},
Ganesh Subramanian\aff{2}\corresp{\email{sganesh@jncasr.ac.in}}
\and  V.~Shankar\aff{1}\corresp{\email{vshankar@iitk.ac.in}}
}
\affiliation{\aff{1}Department of Chemical Engineering, Indian Institute of Technology, Kanpur 208016, India
\aff{2}Engineering Mechanics Unit, Jawaharlal Nehru Centre for Advanced Scientific Research, Bangalore 560064, India}
\begin{document}

\maketitle

\begin{abstract}

  A modal stability analysis shows that pressure-driven pipe flow of
  an Oldroyd-B fluid is linearly unstable to axisymmetric
  perturbations, in stark contrast to its Newtonian counterpart which
  is linearly stable at all Reynolds numbers. The dimensionless groups
  that govern stability are the Reynolds number
  $\Rey = \rho U_{max} R /\eta$, the elasticity number
  $E = \lambda \eta/(R^2 \rho)$ and the ratio of solvent to solution
  viscosity $\beta = \eta_s/\eta$; here, $R$ is the pipe radius,
  $U_{max}$ is the maximum velocity of the base flow, $\rho$
  is the fluid density, and $\lambda$ is the microstructural
  relaxation time. The unstable mode has a phase speed close to
  $U_{max}$ over the entire unstable region in ($\Rey$, $E$,
  $\beta$) space.  In the asymptotic limit $E (1-\beta) \ll 1$, the critical
  Reynolds number for instability diverges as
  $\Rey_c \sim (E (1-\beta))^{-3/2}$, the critical wavenumber
  increases as $k_c \sim (E (1-\beta))^{-1/2}$, and the unstable
  eigenfunction is localized near the centerline, implying that the
  unstable mode belongs to a class of viscoelastic center modes. In
  contrast, for $\beta \rightarrow 1$ and $E \sim 0.1$, $Re_c$ can be
  as low as $O(100)$, with the unstable eigenfunction no longer being
  localized near the centerline.

  Unlike the Newtonian transition which is dominated by nonlinear
  processes, the linear instability discussed in this study could be
  very relevant to the onset of turbulence in viscoelastic pipe
  flows. The prediction of a linear instability is, in fact,
  consistent with several experimental studies on pipe flow of polymer
  solutions, ranging from reports of `early turbulence' in the 1970's
  to the more recent discovery of `elasto-inertial turbulence'
  (Samanta {\em et al.}, {\em Proc. Natl. Acad. Sci.}, {\bf 110},
  10557--10562 (2013)). The instability identified in this study
  comprehensively dispels the prevailing notion of pipe flow of
  viscoelastic fluids being linearly stable in the $Re-W$ plane
  ($W = Re \, E$ being the Weissenberg number), marking a possible
  paradigm shift in our understanding of transition in rectilinear
  viscoelastic shearing flows. The predicted unstable eigenfunction
  should form a template in the search for novel non-linear
  elasto-inertial states, and could provide an alternate route to the
  maximal drag-reduced state in polymer solutions. The latter has thus
  far been explained in terms of a viscoelastic modification of the
  nonlinear Newtonian coherent structures.

\end{abstract}

\section{Introduction}

Laminar pipe flow of a Newtonian fluid is well known to be linearly
stable at all Reynolds numbers
\citep{Drazinreid,SchmidHenningsonBook,meseguer_trefethen_2003}, and a
rigorous theoretical description of the onset of turbulence in this
flow has therefore remained an outstanding challenge in fluid dynamics
research for more than a century \citep{eckhardt_etal_2007}.
Experiments since the classic work of \cite{reynold1883}
have shown that the transition to turbulence occurs at a
Reynolds number $\Rey \approx 2000$ \citep{avila2011,mullin_2011}, in
stark contrast to the aforementioned prediction of linear stability
theory.  As shown originally by Reynolds himself, the transition can
be delayed considerably, even up to $\Rey \sim 10^5$
\citep{pfenniger1961boundary}, by carefully minimizing external
perturbations, thus pointing to the importance of nonlinear effects.
The relatively recent discovery of nonlinear three-dimensional
solutions (termed `exact coherent states') of the Navier-Stokes
equations for pipe flow has considerably advanced our understanding in
this regard by providing the framework for a nonlinear, subcritical route to
transition. Such solutions are disconnected from the laminar state,
appearing via saddle-node bifurcations with increasing $Re$, and
closely resemble coherent structures in the turbulent buffer layer
\citep{waleffe_1998,kerswell_2005,eckhardt_etal_2007}.  
The existence of such solutions has led to a new dynamical systems perspective, wherein transitional turbulence in a pipe is interpreted as a wandering trajectory in an appropriate 
phase space which visits the neighbourhood of multiple invariant sets (including the aforementioned solutions) in a seemingly unpredictable manner \citep{budanur_etal_2017}.

The  onset of turbulence in pipe (and channel) flow of 
  viscoelastic polymer solutions, however, remains largely unexplored
\citep{larson1992}.  Polymer solutions are known to be susceptible to
purely elastic linear instabilities even in the absence of inertia,
but only in flows with curved streamlines as in the Taylor-Couette or
Dean geometries \citep{shaqfeh1996}; the instability eventually leads
to a disorderly flow state \citep[termed `elastic
turbulence';][]{Groisman2000}, and the transition manifests as an
enhanced drag above a threshold Weissenberg number,
$W$, defined as the product of the shear rate and the longest polymer
relaxation time.  In contrast, addition of small
amounts of polymers to turbulent pipe flow leads to a drastic
reduction in the frictional drag \citep{virk1975}, a phenomenon called
turbulent drag reduction that has been extensively investigated
\citep{white_mungal_2008,graham_2014,Xi2019DRreview}. There is relatively little
discussion in the drag reduction literature, however, of the role of the added
polymers on turbulence onset.  Nevertheless, there have been some
reports of `early turbulence' in pipe flow of polymer solutions,
beginning in the 1960s
\citep{ram_etal_1964,goldstein_etal_1969,forame_etal_1972,hansen_etal_1973,hansen_little_1974,jones_etal_1976,
  hoyt_1977,zakin_etal_1977}, wherein transition was observed to occur
at $Re$'s much lower than 2000.  Recent experiments
\citep{samanta_etal2013,srinivas_kumaran_2017,choueiri_etal_2018,chandra_etal_2018,chandra_shankar_das_2020}
have convincingly demonstrated that at sufficiently high polymer
concentrations ($> 300$ppm for pipes and $>80$ppm for channels), flow
of polymer solutions in pipes and channels does indeed become unstable
at Reynolds numbers much lower ($\sim 800$ for pipes and $\sim 200$
for micro-channels) than those corresponding to the Newtonian
transition.  To differentiate it from conventional Newtonian
turbulence, the ensuing flow state has been referred to as
`elasto-inertial turbulence' \citep[abbreviated
`EIT'; see][]{samanta_etal2013} pointing to the importance of both elastic
and inertial forces in the underlying dynamics.

While the possibility of a linear instability in viscoelastic plane
shear flows has occasionally been speculated upon \citep{graham_2014},
most of the literature has extrapolated the Newtonian scenario to the
viscoelastic case, assuming viscoelastic pipe flows to also be
linearly stable.  This viewpoint has been explicitly stated in several
earlier studies \citep[see, for example][in
particular]{bertola_etal_2003,morozov_saarloos2005,pan_etal_2013,sid_etal_2018} despite
the absence of a systematic exploration of the larger parameter space
in the viscoelastic case where, in addition to the Reynolds number
$Re$, the elasticity number $E$ (which is a ratio of the polymer relaxation
to the momentum diffusion timescales; $E = W/Re$) and the
ratio of solvent to total solution viscosity $\beta$ are also expected
to influence stability.  Indeed, the presumed stability of
viscoelastic pipe flow to infinitesimal disturbances is so ingrained
in the field that, prior to the present effort, there has not been a
linear stability analysis using a realistic constitutive model for
viscoelastic pipe flow! The only reported stability analysis for the
pipe geometry \citep{hansen1973,hansen_etal_1973} neglects the crucial
convected nonlinearities in the Oldroyd-B constitutive relation, and
hence does not account for an essential feature of
polymer rheology. The lack of emphasis on a viscoelastic transition
triggered by a linear instability is particularly perplexing in the
light of the unambiguous experimental evidence of the critical
Reynolds numbers being same for the unperturbed and externally
perturbed transition scenarios for sufficiently concentrated ($\sim 300$ppm onwards) polymer solutions \citep[see figure~3a of][]{samanta_etal2013}.

In a recent Letter \citep{Garg2018}, we demonstrated, for the first
time, that elastic, viscous and inertial effects in polymer solutions
(modelled as Oldroyd-B fluids) can combine to render viscoelastic pipe
flow {\em linearly unstable} at Reynolds numbers much lower than
2000. In this paper, we build on this discovery by (i)
providing a detailed picture on the origin of the instability, (ii)
augmenting the original results by exploring a larger parameter space,
and (iii) comparing our theoretical predictions to existing experimental
observations and direct numerical simulations. We also provide a perspective on how the presence of a
linear instability in viscoelastic pipe flow can potentially alter the
prevailing paradigm for laminar-turbulent transition and turbulent
drag reduction in polymer solutions.  In the remainder of this
Introduction, we review relevant earlier work on this subject under the
following headings: (i) Newtonian transition, (ii) turbulent drag
reduction, (iii) experimental studies on the onset of turbulence in
viscoelastic flows, (iv) computational bifurcation studies and direct
numerical simulations, and (v) stability analyses of viscoelastic shearing
flows. Finally, the specific objectives for the present work are laid out in
the context of the existing paradigm vis-a-vis the viscoelastic transition.

\subsection{Newtonian pipe-flow transition}
\label{subsec:Newtoniantransition}

Classical modal stability analyses
\citep{corcos_sellars,gill_1965a,gill_1965b,salwen_grosch_1972,garg_rouleau_1972}
have found fully-developed pipe flow to be linearly stable even up to
$\Rey \sim 10^7$ \citep{meseguer_trefethen_2003}.  The Newtonian
eigenspectrum for pipe flow, for sufficiently high $\Rey$, conforms to
the characteristic `Y-shaped' locus  known for canonical shearing flows
\citep[plane Couette and Poiseuille flows;
see][]{SchmidHenningsonBook}, with three distinct branches: the `A
branch' corresponding to `wall modes' with phase speeds approaching
zero, the `P branch' corresponding to `center modes' with phase speeds
tending to the maximum base flow velocity, and the `S branch' with
modes having a phase speed intermediate between those for wall and
center modes.  While a wall mode belonging to the A branch becomes
unstable in plane channel flow of a Newtonian fluid at $\Rey > 5772$
\citep[the Tollmien-Schlichting instability, see][]{Drazinreid}, all
three branches remain stable for Newtonian pipe flow regardless of
$\Rey$, with the phase speed of the modes belonging to the S-branch
equalling two-thirds of the base-state maximum.  The prediction of stability
to infinitesimal disturbances at any Reynolds
number is broadly consistent with experiments, wherein, as
mentioned before, the transition can be delayed 
upto $\Rey \sim 10^5$ \citep{pfenniger1961boundary}, by carefully controlling
the inlet conditions. Henceforth, we will refer to this transition
scenario, which is highly sensitive to inlet conditions, as
``natural'' transition, while the transition which occurs at the
oft-quoted Reynolds number of around 2000 will be referred to as
``forced'' transition. 
While the natural transition for the Newtonian case is a sensitive function of experimental conditions, the forced transition is quite robust. The difference between 
the associated threshold $Re$'s arises, of course, due to the subcritical nature of the Newtonian transition.

The predictions from a modal analysis are only concerned
with asymptotic behaviour at long times.  More than a century after Reynolds' experiments, a series
of studies in the early 1990s
\citep{butler_farrell,Trefethen1993,reddy_henningson} demonstrated the
possibility of short-time growth of the disturbances, even when all
eigenmodes are stable.  This early-time growth was
attributed to the non-normal nature of the linearized operator
underlying Newtonian stability, leading to the
eigenfunctions corresponding to different eigenvalues not being
orthogonal \citep{grossman,schmid2007}. The (non-exponential) growth,
variously referred to as non-modal, transient or algebraic growth, was
regarded as the reason for amplification of initial disturbances to a
sufficiently large magnitude such that non-linearities can become
important, in turn leading to a subcritical transition.  It is worth
mentioning, however, that the aforementioned non-modal analyses were
restricted to infinitesimal disturbances \citep[also
see][]{SchmidHenningsonBook}.  Thus, although the optimal disturbances
corresponding to maximum transient growth were identified in most
cases as counter-rotating stream-wise vortices aligned along the
span-wise direction giving rise to  growing streaks,
the detailed manner in which this growth would eventually be modified by nonlinear effects was not
addressed.  While recent developments
\citep{Pringle_Kerswell_2010,kerswell2018} have obtained
three-dimensional spatially localized structures, by accounting for
the effects of nonlinearity within a more general optimization
framework, it was \cite{waleffe_1997}'s effort which first accounted
for the back-coupling of the growing streaks to the original
stream-wise vortices via a wiggling instability, thereby leading to a
self-sustaining process.

The effort of \cite{waleffe_1997}  helped highlight the 
physical mechanism underlying finite-amplitude travelling-wave 
solutions that had recently been discovered for plane Couette flow
\citep{nagata_1990,clever_busse_1992}, and their role in the transition process.
 A more complete understanding of pipe flow transition has since been
achieved via the characterization of an increasing number of such
solutions \citep[both steady, time-periodic,
see][]{wedin_kerswell_2004}, dubbed `exact coherent states', all of
which are disconnected from the laminar state (on account of its
linear stability), and emerge via saddle-node bifurcations at $Re$'s
lower than that corresponding to the experimentally observed
transition. All of the ECS's have a common underlying structure
consisting of a mean shear with superimposed wavy stream-wise vortices
and stream-wise velocity streaks. The ECS's thus provide explicit
constructs of the aforementioned self-sustaining process proposed by
\cite{waleffe_1997}.  The discovery of ECS solutions has paved the way
for a dynamical-systems-based interpretation of the Newtonian
transition. This picture posits that pipe flow may be viewed as a dynamical
system in an appropriate phase space which includes the fixed point
corresponding to the steady laminar state, 
and the invariant sets corresponding to the various ECS solutions 
(fixed points, periodic, relative periodic orbits, etc.), with their stable and unstable manifolds.
 Close
to onset, the transitional flow may be interpreted as a phase-space
trajectory sampling 
neighbourhoods of these multiple sets in an unpredictable manner \citep[see][and references
therein]{budanur_etal_2017}. Transition is effected when  a
(finite-amplitude) perturbation takes the flow away from the (shrinking) basin of attraction of the steady laminar state.

\subsection{Turbulent drag reduction}
\label{subsec:dragreduction}
Addition of polymers to a Newtonian solvent renders the solution
viscoelastic, leading to phenomena such as die swell, rod-climbing
etc., in the laminar regime \citep{birdvol1}.  One of the most
dramatic consequences of polymer addition is the phenomenon of
`turbulent drag reduction'
\citep{virk1975,toms_1977,virk1997,white_mungal_2008} wherein addition
of small quantities ($10$ppm onwards) of polymer to a fully
turbulent pipe flow of a Newtonian fluid results in a 70-80\% reduction in
the pressure drop.  Experimental data is often represented on a
`Prandtl-Karman' plot of $1/\sqrt{f}$ vs. $\log(\Rey \sqrt{f})$, $f$ being
the friction factor, where data in the turbulent regime (corresponding to high
$Re\sqrt{f}$) appears as a straight line of slope $4$ reflecting
the log-law for Newtonian turbulence \citep{schlictingbook}.
Upon addition of polymer, the data follows the
Newtonian turbulent asymptote until the onset of drag reduction at an $\Rey \sqrt{f}$ 
independent of the concentration \citep[see, for example, Fig.~1a of][]{virk1997}.
In the drag-reduced regime, the slope increases with increasing polymer concentration, corresponding to a progressively lower pressure drop. At sufficiently high $Re\sqrt{f}$, however, the data 
for different concentrations collapse
onto a single curve termed the `maximum drag-reduction' (MDR)
asymptote \citep[Fig.~7 of][]{virk1975}, which appears to be universal for flexible polymers.  This scenario, where the initial transition to turbulence is unaffected by added polymer, is referred to as `Type A' drag reduction.  Importantly, experiments also exhibit another approach to MDR \citep[Fig.~1b
of][]{VrikNature1975}, dubbed `Type-B drag reduction', wherein onset of drag
reduction occurs immediately after transition without 
 an intermediate Newtonian turbulent regime. In the Type-B scenario, at sufficiently
high concentrations, the MDR asymptote is approached right after the
transition, implying that MDR is not necessarily a high-$\Rey$
phenomenon. Most experimental efforts have, however, focussed on larger $Re\sqrt{f}$'s of $O(10^3)$, and not much attention has therefore been paid to the $Re$ corresponding to onset.

%
%
%
%

\subsection{Early transition and Elasto-inertial turbulence}
\label{subsec:expts}
While the pioneering work by \cite{virk1975} found transition in
pipe flow of dilute polymer solutions to occur roughly at the same
$\Rey$ as the Newtonian one, there have been reports of a delayed
transition \citep{Giles1967,Castro1968,White1970}.  
Significantly,  there have also been several reports of `early
turbulence', wherein transition is reported at an $\Rey$ as low as $500$
(Goldstein, Adrian \& Kreid 1969; Forame, Hansen \& Little 1972;
Hansen, Little \& Forame 1973; Hansen \& Little 1974; Little et
al. 1975; Hoyt 1977; Zakin et al. 1977; Draad, Kuiken \& Nieuwstadt
1998), although these early experimental efforts were not corroborated
and followed up in a systematic manner.  The conflicting conclusions of
delayed or early transition could perhaps be attributed to poor
characterization of the polymer solutions used.  In a
recent important paper, \cite{samanta_etal2013} examined the flow of
polyacrylamide solutions of varying concentrations in pipes of
diameter 4 and 10 mm.  Two experimental protocols were followed: one
in which the transition was `forced' by fluid injection
to the flow near the inlet, and the other corresponding to a natural transition (at $Re \sim 8000$ for the Newtonian case). With increasing polymer concentration, the natural transition threshold decreased while that for the forced transition increased, and for concentrations greater than $300$ ppm, the two threshold $Re$'s were found to coincide and decrease with further increase in concentration, with $Re \sim 800$ for the $500$ ppm solution. Further, structural signatures such as puffs, characteristic of sub-critical Newtonian dynamics, were absent for such concentrated solutions.
%
%
%
%

The independence of the transition $\Rey$ 
 with respect to perturbation amplitude is strongly suggestive of a linear instability mechanism underlying the transition process, although, rather surprisingly,
the authors both in the aforesaid paper and in later efforts
\citep{sid_etal_2018,choueiri_etal_2018} attribute their observations
to nonlinear processes regardless of polymer concentration.
 Due to the smaller pipe diameter and
 higher polymer concentrations, the elasticity numbers
probed in the experiments of \cite{samanta_etal2013} are significantly
higher than those in the earlier experiments discussed above \citep{draad_etal_1998}. The flow
state that results after this non-hysteretic transition (for
sufficiently high polymer concentrations) has been referred to as
`elasto-inertial turbulence' \citep{samanta_etal2013},
 to contrast it with both purely elastic instabilities \citep[discussed above;
see][]{shaqfeh1996} in viscoelastic flows with curved streamlines even in the absence of inertia, and purely inertial Newtonian turbulence. 
%
%
The lack of a hysteretic signature in the
transition served as a primary motivation in our search
\citep{Garg2018} for a linear instability in viscoelastic pipe flow.
The recent experimental work of \cite{chandra_etal_2018,chandra_shankar_das_2020} further
corroborated the findings of \cite{samanta_etal2013}, and reported
a decrease in the transition $Re$ with increasing concentration in
the range $300 - 800$ppm.

In a significant departure from the prevailing
paradigm in drag reduction, a recent experimental study from Hof's group
\citep{choueiri_etal_2018} has demonstrated the non-universal nature
of the MDR asymptote.
The authors showed that with increase in polymer
concentration (at a fixed $Re < 3600$), it was possible to
exceed the MDR asymptote, with the flow relaminarizing completely, 
and the friction factor approaching its laminar value. 
As the polymer concentration is further increased, the laminar state 
becomes unstable and the drag increases further,
again reaching MDR at sufficiently high polymer concentration.
It follows from the sequence described above, as also 
alluded to in our earlier work \citep{Garg2018}, that the MDR regime could
also be be viewed as a `drag-enhanced' state arising from an
instability of the laminar state, rather than as a
drag-reduced state accessible only from Newtonian turbulence.  Both the
\cite{samanta_etal2013} and \cite{choueiri_etal_2018} studies show
 the EIT structures being oriented along the span-wise direction, in sharp contrast to the stream-wise vorticity known to  be dominant in Newtonian turbulent shearing flows.
Importantly, \cite{choueiri_etal_2018} also showed that the EIT state that
follows complete relaminarization is qualitatively similar to
the MDR state that occurs after Newtonian turbulence, 
 implying the relative robustness, with respect to the underlying parameters, of the span-wise-oriented coherent structures that characterize this state. These observations were, in fact, the original motivation for restricting the analysis in \cite{Garg2018}, and that presented here, to axisymmetric disturbances.

In summary, the experiments above suggest that the nature of viscoelastic
pipe flow transition, and the attainment of an MDR-like state, 
can be broadly classified  into weakly and
strongly elastic regimes, the underlying mechanisms being manifestly different in
the two cases.  At low polymer concentrations, the MDR regime is
accessed via the Newtonian-turbulent regime, with the transition from the
laminar state, in particular, being akin to the Newtonian case.  In contrast, for
sufficiently high polymer concentrations (moderately
elastic flows with $W \sim 1$, or strongly elastic flows with
$W \gg 1$), experiments are suggestive of an elasto-inertial linear instability, at an $\Rey$ substantially
lower than $2000$, that provides a direct
and continuous path to the MDR regime.

\subsection{DNS and computational bifurcation studies of viscoelastic flows}
\label{subsec:dns}

\subsubsection{Early DNS and computational bifurcation studies}
Several direct numerical simulation
(DNS) studies have been carried out, most often for the plane channel
geometry, to understand turbulence and drag reduction
\citep{sureshkumar_etal_1997,sibilla_baron2002,deangelies_etal,dubief_etal2004,xigrahamPRL2010}
in dilute polymer solutions 
 \citep[see][for a comprehensive review]{Xi2019DRreview} 
using the FENE-P model
\citep[][]{birdvol1} for the polymer.  These studies
showed that turbulence production in the buffer layer is altered by the
addition of polymers, and were able to successfully capture the
moderate drag reduction regime, i.e., at $\Rey$'s lower than those corresponding to the
MDR regime.  The DNS results  are broadly consistent with the experimental literature
on drag reduction which showed a thickening of the buffer layer on polymer addition
\citep{virk1975}. The viscoelastic modification of the buffer layer also served as a
motivation for a series of papers by
Graham and co-workers
\citep{stone_graham2002,stone_graham2004,li_etal_2006,Li_Graham_2007},  
which, based on the structural similarities shared by the ECS solutions and the turbulent buffer layer 
(see Section~\ref{subsec:Newtoniantransition}),
explored how viscoelasticity affects the ECS in channel flow.
They found that the $\Rey$ at which ECS
solutions emerge increases with increasing elasticity number $E$,
and appears to diverge at a critical $E$, suggesting that the ECS's are
absent in a sufficiently elastic polymer solution.  The disappearance
of the ECS's above a critical $E$ has been correlated to maximum drag
reduction, and was in fact proposed as an explanation for transition
delay by viscoelasticity, as reported in some of the experiments discussed
above, including those of \cite{samanta_etal2013} for concentrations
less than 200ppm.

Thus, the interpretation of turbulent drag reduction
(and, consequently, of laminar-turbulent transition) in viscoelastic
channel and pipe flows has been strongly influenced by the
aforementioned non-linear dynamical systems perspective developed in
the Newtonian context. Implicit in this picture is the assumption of
linear stability of viscoelastic pipe flow at all $\Rey$ and $E$ (or,
equivalently, $W$) and the existence of (disconnected) nonlinear ECS's over 
a subset of these parameters.
However, in the moderately and strongly elastic regimes referred to in
Sec.~\ref{subsec:expts}, the nonlinear ECS solutions are fully suppressed by
viscoelasticity, and hence there must be other qualitatively different
(linear or nonlinear) mechanisms that govern the transition.  The
experimental observations of
\citep{samanta_etal2013,choueiri_etal_2018}, in fact, clearly provide
evidence for a non-hysteretic transition in the strongly elastic
regime, which is strongly suggestive of a supercritical bifurcation
being triggered by a linear instability of the laminar state
\citep{Garg2018}.

\subsubsection{Recent DNS studies and the role of diffusion in the constitutive equation}
The pioneering DNS study of \cite{sureshkumar_etal_1997}, and the many
papers that followed it \citep{sibilla_baron2002,deangelies_etal,xigrahamPRL2010}, incorporated 
 an additional diffusive term in the constitutive  equation. While there must, strictly speaking, be such a diffusive term on account of the Brownian motion of the polymer molecules, the motivation for the introduction of diffusion in the aforementioned efforts was primarily numerical, with the aim of preserving the positive definiteness of the stress tensor. The magnitude of this stress diffusivity may be characterized by a Schmidt 
number, $Sc = \nu/D$, which is the ratio of the kinematic viscosity $\nu = \eta/\rho$ of the polymer solution 
to the stress diffusivity $D$. 
For dilute polymer
solutions involving high-molecular weight polymers, $Sc \sim 10^6$,
but earlier DNS studies have used a far smaller value of $Sc \approx 0.5$. The recent work of
\cite{sid_etal_2018}  showed that the two-dimensional structures characteristic of EIT are suppressed for $Sc < 9$, which might explain the reason the EIT state was not observed in the aforementioned simulation efforts.
A low $Sc$ is known to affect structures even outside of those
 pertaining specifically to drag reduction, for instance, those related to
low-$\Rey$ elastic turbulence
\citep{gupta2019}.
The recent DNS studies by
Dubief and co-workers \citep{dubief_etal2013,samanta_etal2013,sid_etal_2018}
in the absence of stress diffusion ($Sc \rightarrow \infty$) showed that the friction factor deviated from the laminar
value at $Re \sim 750$, while the Newtonian case remained laminar upto
$Re = 5000$ for identical initial forcing.
Further, the topological features of the structures in the unstable
region, as inferred from iso-surfaces of the second invariant of the
velocity gradient tensor, were span-wise oriented and stream-wise
varying, in stark contrast to span-wise varying and stream-wise
oriented vortices in Newtonian turbulence.   
While earlier simulations (for channel flow) by Graham and co-workers \citep{xigrahamPRL2010,Li_et_al2012,graham_2014} have shown the turbulence to exhibit long hibernating periods at large $W$, with the marginal state during these periods interpreted as that underlying the dynamics in the MDR regime, a recent study by \cite{lopez_choueiri_hof_2019} on viscoelastic pipe flow (at
$\Rey = 3500$)   showed that, on consideration of longer domains, the hibernating state above gives way to spatio-temporally intermittent turbulence, and for higher $W$, complete relaminarization.
At still higher $W$, the flow destabilizes again, and the resulting disorderly flow has been identified with EIT; the drag reduction in this regime approaches the MDR limit.  This study further
underscored the relevance of a new instability mechanism that directly
connects the laminar state to MDR, and
reinforced the importance of two-dimensional (or, axisymmetric, in the
case of pipe flow) effects in driving the elasto-inertial transition.
Most recently, the simulations of \cite{shekar_etal_2019}  have shown viscoelastic channel flow to destabilize via a non-linear mechanism triggered by finite-amplitude two-dimensional perturbations, and the resulting structures bore a strong 
resemblance to the Tollmien-Schlichting mode in Newtonian channel flow.
However, the conclusions of \cite{shekar_etal_2019} are only
applicable to channel flow; their relevance to transition
in viscoelastic channel flows will be discussed separately in a future
communication \citep{khalid_channel}.  We also argue below, in Sec.~\ref{sec:shekar}, that
the  axisymmetric
instability that is the subject of the present work bears no relation to
the Newtonian TS mode \citep[also see][]{Xi2019DRreview}.
Thus, barring the effort of \cite{shekar_etal_2019}, the aforementioned DNS studies suggest that the mechanism leading to EIT,  
which is also believed to underlie drag reduction (and MDR),
could be very different from the pathway that involves the elastically-modified ECS states, especially for pipe flow.
However, the work of
\cite{lopez_choueiri_hof_2019} again has 
$Sc = 0.5$ in their pipe flow simulations, and more work is required to
determine how the results of \cite{lopez_choueiri_hof_2019} would be
altered at higher $Sc$.
In Sec.~\ref{subsec:stressdiffusion}, we show that the unstable (axisymmetric) center mode analyzed in this work is suppressed  when the dimensionless diffusivity $E/Sc > 10^{-4}$, consistent with
the DNS results of \cite{sid_etal_2018} for channel flows (although, this does not rule out a sub-critical transition, again involving this mode, at lower $Sc$).

\subsection{Stability of viscoelastic shearing flows}
\label{subsec:vestability}
Prior to our Letter \citep{Garg2018} and the present work, there has
been no attempt \citep[barring that of][who neglected the convected
nonlinearities in the constitutive model]{hansen1973}
to examine the linear stability of viscoelastic pipe flow,
although  many studies \citep[for
e.g.,][]{gl67,lee_finlayson_1986,renardy1986linear,ho_denn_1977,sureshkumar1995linear}
have examined the stability of viscoelastic plane Couette and
Poiseuille flows.  A detailed survey of the literature 
on viscoelastic plane shearing flows has been presented in 
\cite{chaudhary_etal_2019}, and herein we restrict ourselves to summarizing the principal conclusions of 
\cite{Garg2018}.
\cite{Garg2018} showed that
viscoelastic pipe flow is indeed linearly unstable in parameter regimes where
experiments \citep{samanta_etal2013,chandra_etal_2018} observe an
instability. 
While the unstable mode has a finite radial spread for generic $Re$, $\beta$ and $E$, in the asymptotic limit $E(1-\beta) \ll 1$, when the critical Reynolds number required diverges as
$Re_c \sim [E(1-\beta)]^{-3/2}$, and the critical wavenumber increases as $ k_c \sim [E(1-\beta)]^{-1/2}$, the mode is confined to a thin region in the vicinity of the centerline. Regardless of localization, however, the phase speed of the unstable eigenfunction remains close to unity, indicating that the unstable mode belongs to a class of viscoelastic `center modes'.  The
linear, elasto-inertial wall-mode instability predicted for
viscoelastic channel flows in our earlier work
\citep{chaudhary_etal_2019}, along with the center-mode instability
reported in \cite{Garg2018} and expanded further in the present work, for pipe flow, show
that much remains to be understood with regard to (modal) stability
of viscoelastic shear flows.

\subsection{Objectives of the present study}
\label{subsec:motivation}

Thus, the above detailed survey of the existing literature serves as a clear motivation for the work reported here, which provides a comprehensive picture of the stability of viscoelastic pipe flow using the
Oldroyd-B model. The present work significantly differs
 from existing ones in that we analyze the linear
stability of flow of dilute polymer solutions in the $Re$-$W$-$\beta$ space,
rather than along the $W$ or $Re$ axis (which amounts to the neglect
of either inertia or viscoelasticity), and importantly, for the
canonical (and experimentally relevant) case of pressure-driven pipe
flow. It is well established in the literature that both plane Couette
and Poiseuille flows of a UCM fluid remain stable in the limit of zero
and small $\Rey$, and during the course of this study, we have verified
that pipe flow of UCM and Oldroyd-B fluids also remains stable at small $\Rey$,
reinforcing the consensus that elastic effects alone are not
sufficient to destabilize rectilinear viscoelastic flows.
\begin{figure}
  \centering
  \includegraphics[width = 6cm]{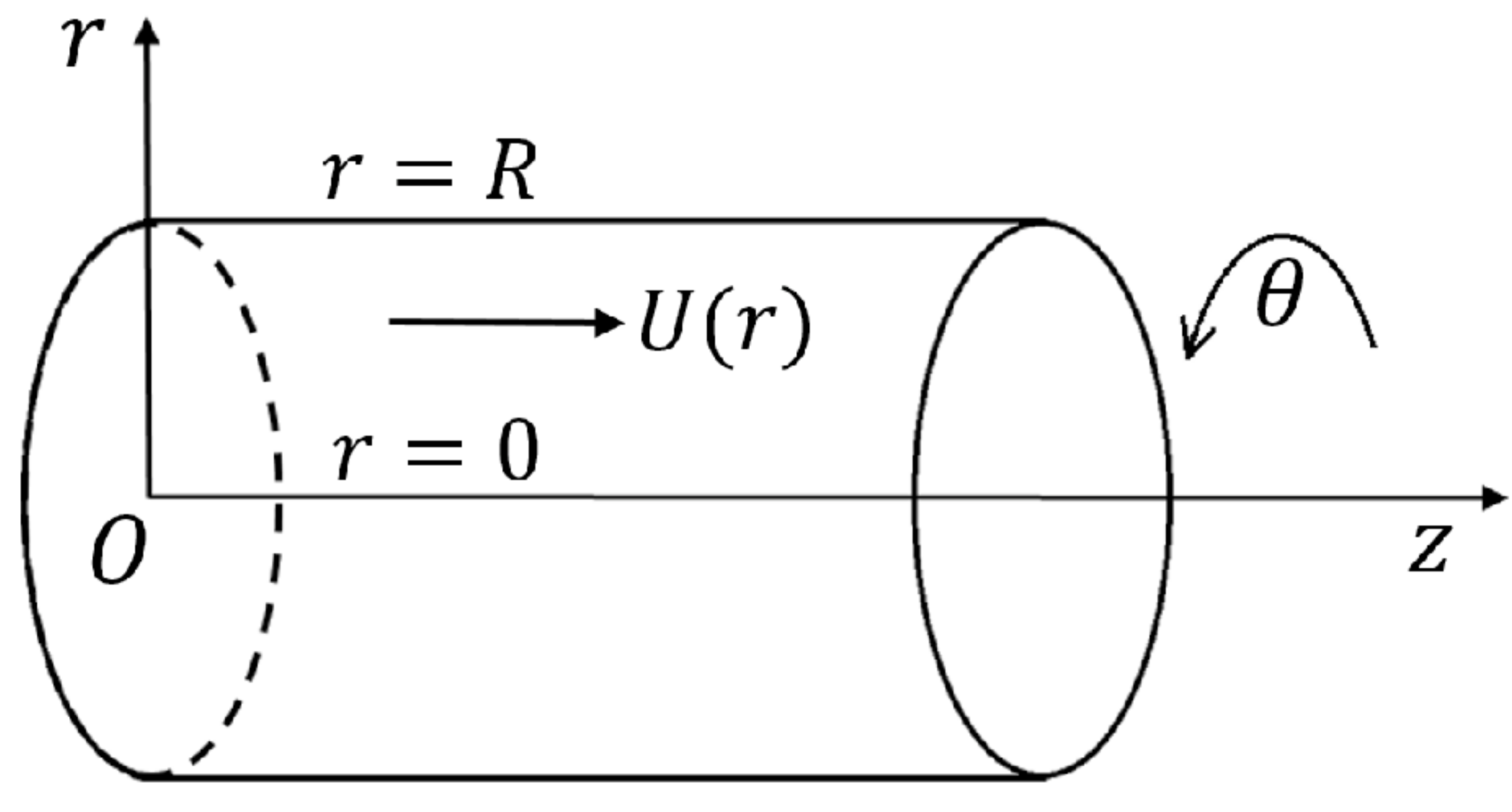}
  \caption{Schematic diagram showing the geometry and the
    coordinate system considered.}
  \label{fig:geometry}
\end{figure}

The rest of this paper is structured as follows: In
Section~\ref{sec:prob_num_method}, we 
outline the stability formulation for viscoelastic pipe Poiseuille flow subjected to infinitesimal  amplitude 
axisymmetric disturbances; the base state and governing
linearized differential equations are provided, followed by a brief
description of the numerical schemes employed. In
Section~\ref{sec:spectra_OB}, we first recapitulate  the key
features of the Newtonian pipe flow spectrum, which is followed
by a detailed discussion of the corresponding eigenspectra for an Oldroyd-B fluid,
as $E$ is varied for fixed $\beta$ (Sec.~\ref{subsec:varyingE}), wherein the centre mode instability is first
identified. 
The role of the continuous spectra (CS), in terms of their effect on the least
stable/unstable modes belonging to the Newtonian P-branch, is discussed in
Section~\ref{sec:centermode}. 
  In Section~\ref{subsec:varyingbeta}, we present 
the viscoelastic eigenspectra for fixed $E$ and varying $\beta$,
with the relation between the CS and the center mode being discussed in Sec.~\ref{subsec:originfixedEvaryingbeta}.
 The relative importance of the least
stable/unstable centre modes vis-a-vis wall modes in viscoelastic pipe flow is
highlighted in Section~\ref{sec:shekar},
where we  also contrast the pipe flow scenario with the recent DNS results for viscoelastic channel flow \citep{shekar_etal_2019} which point to the crucial role of the critical layer corresponding to the least stable wall mode (the elastically modified Tollmien-Schlichting mode). Neutral stability curves are presented in Section~\ref{sec:neutral_curves}, where
the behaviour of the neutral curves for $k\ll 1$ obtained
via a low-$k$ asymptotic analysis is shown to agree very well with those obtained from the full governing
equations for $k \ll 1$. 
For sufficiently small $E$, there is a remarkable
collapse of the neutral curves (Sec.~\ref{sec:collapses}) in the suitably rescaled $Re$--$k$
plane; a further collapse is obtained in the dual limit $E (1-\beta) \ll 1$ and $\beta \rightarrow 1$. In Section~\ref{sec:scalings}, we
demonstrate how the critical parameters $\Rey_c$, $k_c$ and $c_{r,c}$
scale with $E$ in the limit $E\ll 1$, and justify the numerical results via scaling
arguments in the limit of $Re \gg 1$,
$E \ll 1$, when the unstable mode is confined in the neighbourhood of the centerline.  In Sec.~\ref{subsec:stressdiffusion}, we examine the role of stress diffusion 
in the constitutive relation to show that
the unstable center mode persists for 
 physically realistic values of the diffusion coefficient.  Our theoretical predictions are compared (in a
parameter-free manner) with the experimental observations of
\cite{samanta_etal2013} and \cite{chandra_etal_2018} in
Section~\ref{sec:recent_studies}; herein, we also compare our 
predictions with the recent DNS results for viscoelastic pipe flow by
\cite{lopez_choueiri_hof_2019}.  Finally, in
Section~\ref{sec:conclusion}, we summarize the salient findings of
this study, and provide a discussion on how the discovery of a linear
instability in viscoelastic pipe flow can play a pivotal role in
clarifying the pathway to the MDR regime from
the laminar state.
  
\section{Problem formulation and numerical method}
\label{sec:prob_num_method}

\subsection{Governing Equations}
We consider the linear stability of steady fully-developed flow of a
viscoelastic fluid in a rigid circular pipe of radius $R$ as shown in
Fig.~\ref{fig:geometry}. A cylindrical polar coordinate system is used
with $r, \theta$ and $z$ denoting the radial, azimuthal and axial
directions. The following scales are used for nondimensionalizing the
governing equations: radius of the pipe $R$ for lengths, maximum
base-flow velocity $U_{max}$ for velocities, $R/U_{max}$ for time and
$\rho U_{max}^2$ for pressure and stresses, with $\rho$ being the
density of the fluid.

The governing (nondimensional) continuity and Cauchy momentum
equations are given by \refstepcounter{equation}
$$
\bnabla \bcdot \boldsymbol{v} = 0, \quad \frac{\p \boldsymbol{v}}{\p
  t} + (\boldsymbol{v} \bcdot \bnabla) \boldsymbol{v} = -\bnabla p +
\frac{\beta}{\Rey}\nabla^2\boldsymbol{v} + \bnabla\bcdot\mathsfbi{T}.
\eqno{(\theequation{\mathit{a},\mathit{b}})}
\label{eqn:conti_and_mom}
$$
%
%
%
Here, $\boldsymbol{v}$ is the fluid velocity field, $p$ is the
pressure field and $\mathsfbi{T}$ is the polymeric contribution to the
stress tensor, which in turn is given by the Oldroyd-B constitutive
relation \citep{larson1988constitutive} as follows
\begin{equation}
  W \left(\frac{\p \mathsfbi{T}}{\p t}+\left( \boldsymbol{v} \bcdot \bnabla \right)\mathsfbi{T}-\mathsfbi{T} \bcdot \left(\bnabla \boldsymbol{v}\right)  -\left(\bnabla \boldsymbol{v} \right)^{T} \bcdot\mathsfbi{T}\right) +\mathsfbi{T} = \frac{(1-\beta)}{\Rey} \{\bnabla\boldsymbol{v}+(\bnabla\boldsymbol{v})^T\}.
  \label{eqn:constitutive_relation}
\end{equation}
The solvent to solution viscosity ratio is denoted by
$\beta = \eta_s/\eta$, where the solution
viscosity is $\eta = \eta_p + \eta_s$, $\eta_s$ and $\eta_p$ being the
solvent and polymer viscosities
respectively; $\beta = 0$ and $1$ denote the UCM and Newtonian limits. 
For a fixed $\beta$, the dimensionless
%
 groups relevant to the stability of the Oldroyd-B
fluid above are the Reynolds number $\Rey=\rho U_{max} R/\eta $, the
Weissenberg number $W=\lambda U_{max} /R$ which is a ratio of the
polymer relaxation time $\lambda$ to the flow time scale.  The
Oldroyd-B model describes the stress in a dilute solution of polymer
chains modelled as non-interacting Hookean dumbbells
\citep{larson1988constitutive}, and is invariably the first model used
in the examination of elastic phenomena involving dilute polymer
solutions.  Consistent with the aforementioned microscopic picture,
the Oldroyd-B model assumes the relaxation time to be independent of
both the shear rate and the polymer concentration. Since the model
predicts a shear-rate-independent viscosity, the non-Newtonian
(elastic) effects in this model arise from an effective tension along
the streamlines (arising from flow-aligned dumbbells), which manifests
as a shear-rate-independent first normal stress different in viscometric flows.  This model has been extensively used,
and with considerable success, in earlier investigations of 
inertialess elastic instabilities in flows with curved streamlines
\citep{larson_etal_1990,shaqfeh1996,pakdel_McKinley_1996}. The
so-called Boger fluids constitute an experimental realization of this
constitutive model \citep{boger_nguyen_1978}.
As discussed later in the manuscript (in Section~\ref{sec:samanta_chandra}, where we use scaling
arguments in the context of the FENE-P model to assess the role of shear thinning), while shear-thinning can play an
important role especially in flow through microtubes
\citep{samanta_etal2013,chandra_etal_2018}, the Oldroyd-B model does
have the necessary ingredients to qualitatively predict the
instabilities observed in experiments.


\subsection{Base state}
The base-state velocity profile is the
classical Hagen-Poiseuille profile
because the  nonlinear terms in the upper-convected  derivative of 
derivative of the polymer shear stress $T_{rz}$ are identically zero. The
nondimensional base flow velocity vector is given by:
\begin{equation}
\setlength{\arraycolsep}{2pt}
\renewcommand{\arraystretch}{1.3}
\overline{\boldsymbol{v}} = \left[ \begin{array}{c} \bar{v}_{r}\\ 0\\ \bar{v}_{z}\\ 
\end{array}  \right]= \left[ \begin{array}{c} 0\\0\\U(r)\\ \end{array}  \right],
\label{eqn:basevel}
\end{equation}
where $U(r)=1-r^2$ for pipe Poiseuille flow. Here, and in what
follows, base state quantities are denoted by an overbar. The polymer
contribution to the stress tensor in the base state is given
by 
\begin{align}
\setlength{\arraycolsep}{4pt} 
\renewcommand{\arraystretch}{1.3} 
\overline{\mathsfbi{T}}=\left[ \begin{array}{ccc} 
\bar{\tau}_{rr} & 0 & \bar{\tau}_{rz}\\ 
0&\bar{\tau}_{\theta\theta}&0\\
\bar{\tau}_{zr}&0&\bar{\tau}_{zz}\\
\end{array}  \right]= \frac{1}{\Rey} \left[
  \setlength{\arraycolsep}{6pt} 
\begin{array}{ccc} 0 & 0 & U'\\ 
0&0&0\\ 
U'&0&2(1-\beta)WU'^2\\ 
\end{array}  \right], 
\end{align} 
where, $f'\equiv \mathrm{D}f \equiv
\frac{\mathrm{d}f}{\mathrm{d}r}$. 
Unlike the velocity profile, the base-state stress profile differs from that of a Newtonian fluid in having a tension along the streamlines proportional to the square of the velocity gradient.

\subsection{Linear stability analysis}
A temporal linear stability analysis is carried out wherein the base-state above is subjected to small amplitude axisymmetric perturbations. Because of the absence of a Squire-like theorem for
pipe flow even in the simpler Newtonian case, in general,
both axisymmetric and nonaxisymmetric disturbances need to be
considered for viscoelastic Oldroyd-B fluids. However, for the
parameter regime probed in this study, we find
axisymmetric disturbances alone to be unstable, and
this study is therefore restricted to axisymmetric disturbances. The total
velocity, pressure 
and stress are expressed in terms of their base-state values and
perturbations as
\begin{subeqnarray}
  \boldsymbol{v} &=& \boldsymbol{\overline{v}} + \boldsymbol{\hat{v}},\\
  p &=& \overline{p} + \hat{p},\\
  \mathsfbi{T} &=& \overline{\mathsfbi{T}} + \hat{\mathsfbi{T}}\, ,
  \label{eqn:perturb}
\end{subeqnarray}
with  $\hat{f}$ denoting the perturbation to the dynamical
quantity $f$. For axisymmetric disturbances, the perturbation
velocity and stress tensor are:
\begin{equation}
  \setlength{\arraycolsep}{4pt}
  \renewcommand{\arraystretch}{1.3}
  \hat{\boldsymbol{v}}= \left[ \begin{array}{c} \hat{v}_{r}\\ 0\\ 
\hat{v}_{z}\\ \end{array}  \right],\text{ and } \hat{\mathsfbi{T}}=
\left[ \begin{array}{ccc} \hat{\tau}_{rr} & 0 & \hat{\tau}_{rz}\\
         0&\hat{\tau}_{\theta\theta}&0\\
         \hat{\tau}_{zr}&0&\hat{\tau}_{zz}\\ \end{array}
     \right]. \label{eqn:stress_perturb} 
\end{equation}
Next, the perturbation quantities above are represented in the form of Fourier modes in the
axial ($z$) direction in the following manner: 
\begin{equation} 
\hat{f}(r,z;t) = \tilde{f}(r)\mathrm \exp\{\mathrm{i}k (z - c t)\},
\label{eqn:normal_modes} 
\end{equation} 
where $k$ is the axial wavenumber and $c = c_r + i
c_i$ is the complex wave speed. The flow is temporally unstable
(stable) if $c_i > 0$ $(< 0)$. Substituting Eq.~\ref{eqn:normal_modes}
in the linearized versions of
Eqs.~\ref{eqn:conti_and_mom}--\ref{eqn:constitutive_relation}, we
obtain the following set of linearized governing equations:
\begin{eqnarray} 
\label{eqn:linearized_eqns} 
&\left(\mathrm{D} + \frac{1}{r}\right)\tilde v_{r}+\mathrm{i} k\tilde
  v _{z}=0, \label{eqn:lin_conti}\\ 
&G \tilde{v}_r = -\mathrm{D}\tilde{p} +
  \left[\left(\mathrm{D}+\frac{1}{r}\right)\tilde{\tau}_{rr}+\mathrm{i}k\tilde{\tau}_{rz}-\frac{\tilde{\tau}_{\theta\theta}}{r}\right]
  + \frac{\beta}{\Rey}\mathrm{L}\tilde{v}_r,\label{eqn:lin_rmom}\\
&G \tilde{v}_z +U'\tilde{v}_r= -\mathrm{i}k\tilde{p} +
  \left[\left(\mathrm{D}+\frac{1}{r}\right)\tilde{\tau}_{rz}+\mathrm{i}k\tilde{\tau}_{zz}\right]
  +
  \frac{\beta}{\Rey}\left(\mathrm{L}+\frac{1}{r^2}\right)\tilde{v}_z,\label{eqn:lin_zmom}\\ 
&H\tilde{\tau}_{rr}=2\frac{(1-\beta)}{\Rey}\left(\mathrm{D}+W\mathrm{i}kU'\right)\tilde{v}_r,\label{eqn:lin_rrstress}\\ 
&H\tilde{\tau}_{rz}-WU'\tilde{\tau}_{rr} = \frac{(1-\beta)}{\Rey} [ \{ \mathrm{i}k-W\left(U''-U'\mathrm{D}-2\mathrm{i}kWU'^2\right) \} \tilde{v}_r \nonumber \\
                                                      & +\left(\mathrm{D}+W\mathrm{i}kU'\right)\tilde{v}_z ],\label{eqn:lin_rzstress}\\
&H\tilde{\tau}_{\theta\theta} = 2\frac{(1-\beta)}{\Rey} \frac{\tilde{v}_r}{r},\label{eqn:lin_ttstress}\\ 
&H\tilde{\tau}_{zz}-2WU'\tilde{\tau}_{rz} = 2\frac{(1-\beta)}{\Rey} [
  -2W^2U'U'' \tilde{v}_r
  +\{\mathrm{i}k+WU'\left(\mathrm{D}+2W\mathrm{i}kU'\right)\}\tilde{v}_z
  ] 
\label{eqn:lin_zzstress},
\newline 
\end{eqnarray} 
where $G = \mathrm{i}k(U-c)$, $H = 1 + WG$ and $\mathrm{L} =
\left(\mathrm{D}^2 + \frac{\mathrm{D}}{r}-\frac{1}{r^2}-k^2\right)$. 
The no-slip boundary conditions 
$
\tilde{v}_r = 0$ and $\tilde{v}_{z} = 0$ are applicable at $r = 1$,
while
at  $r = 0$, the conditions 
$
\tilde{v}_r = 0$ and $\tilde{v}_{z} = \text{finite}$,
corresponding to regularity of axisymmetric
disturbances in the vicinity of the centerline, are used 
\citep{batchelor_gill_1962,khorrami_etal_1989}.

\subsection{Numerical method}
We use two independent formulations to solve the viscoelastic eigenvalue problem for the wavespeed $c$. In the first, the governing equations for perturbation stresses (Eqs.~\ref{eqn:lin_rrstress}--\ref{eqn:lin_zzstress}) are substituted in
Eqs.~\ref{eqn:lin_rmom}--\ref{eqn:lin_zmom} to obtain two linearized
ordinary differential equations corresponding to the
momentum balances in $r$- and $z$-directions in addition to Eq.~\ref{eqn:lin_conti}, and the dependent variables in this formulation are $\tilde{v}_r$, $\tilde{v}_z$ and $\tilde{p}$.
 In the second formulation, we directly solve the system of linear equations (Eqs.~\ref{eqn:linearized_eqns}--\ref{eqn:lin_zzstress}), with $\tilde{v}_r/r$ as the dependent variable instead of $\tilde{v}_r$, with the other variables being $\tilde{v}_z$, $\tilde{p}$, $\tilde{\tau}_{\theta \theta}$, $\tilde{\tau}_{rr}$, $\tilde{\tau}_{rz}$, and $\tilde{\tau}_{zz}$. 
 The simplified
equations represent a homogeneous eigenvalue problem, and are solved
using the standard spectral collocation
numerical scheme based on Chebyshev polynomials \citep{boyd, TrefethenBook}. 
Results from the two different spectral approaches show excellent agreement. Further, the eigenvalues obtained from the spectral method were verified using a shooting method \citep{ho_denn_1977,lee_finlayson_1986} implemented for the first formulation, based on an adaptive
step size Runge-Kutta integrator and a Newton-Raphson procedure for determining the eigenvalue. 
The integration for the shooting method was carried out from a point near
the centerline $r = \epsilon$ (with $\epsilon \rightarrow 0$) to the
pipe wall at $r = 1$. The velocities at $r = \epsilon$
were obtained using a Frobenius series expansion
\citep{garg_rouleau_1972} about the regular singular point $r = 0$.  The shooting method
gives very accurate (based on our choice of tolerance, typically
$10^{-9}$) results when sufficiently close initial guesses are
provided, whereas the number of polynomials $N$ required for
convergence of eigenvalues in the spectral method depends mainly on
the nature of the eigensolutions and the parameter values. Typically, the $N$ required for convergence of eigenvalues
for finite $\beta$ is in the range $150$--$200$, while that for the UCM
limit ($\beta \rightarrow 0$), is in the range
$400$--$500$. There is no prior literature which reports the
eigenspectrum for pipe flow of an Oldroyd-B fluid,
and hence our numerical procedure was benchmarked in the Newtonian
limit (obtained by setting $W = 0$ or $\beta = 1$).  Results in this limit are available, for instance, in 
\cite{SchmidHennigson1994,SchmidHenningsonBook}.


\begin{figure}
  \centering
   \begin{subfigure}[htp]{0.48\textwidth}
    \includegraphics[width = 6cm]{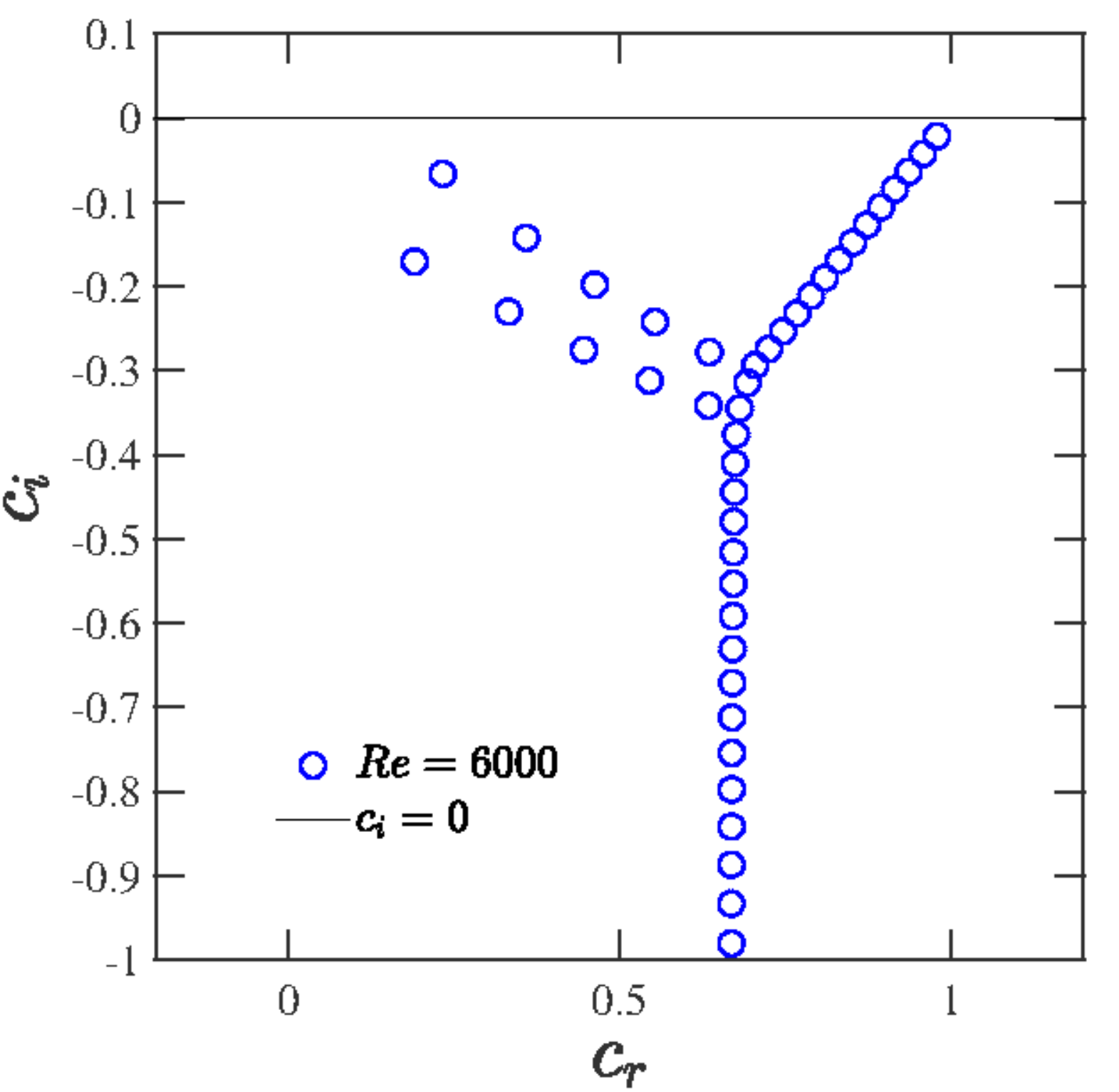}
    \caption{$Re = 6000$}
    \label{fig:NewtonianRe6000_k3}
  \end{subfigure}
   \begin{subfigure}[htp]{0.48\textwidth}
    \includegraphics[width = 6cm]{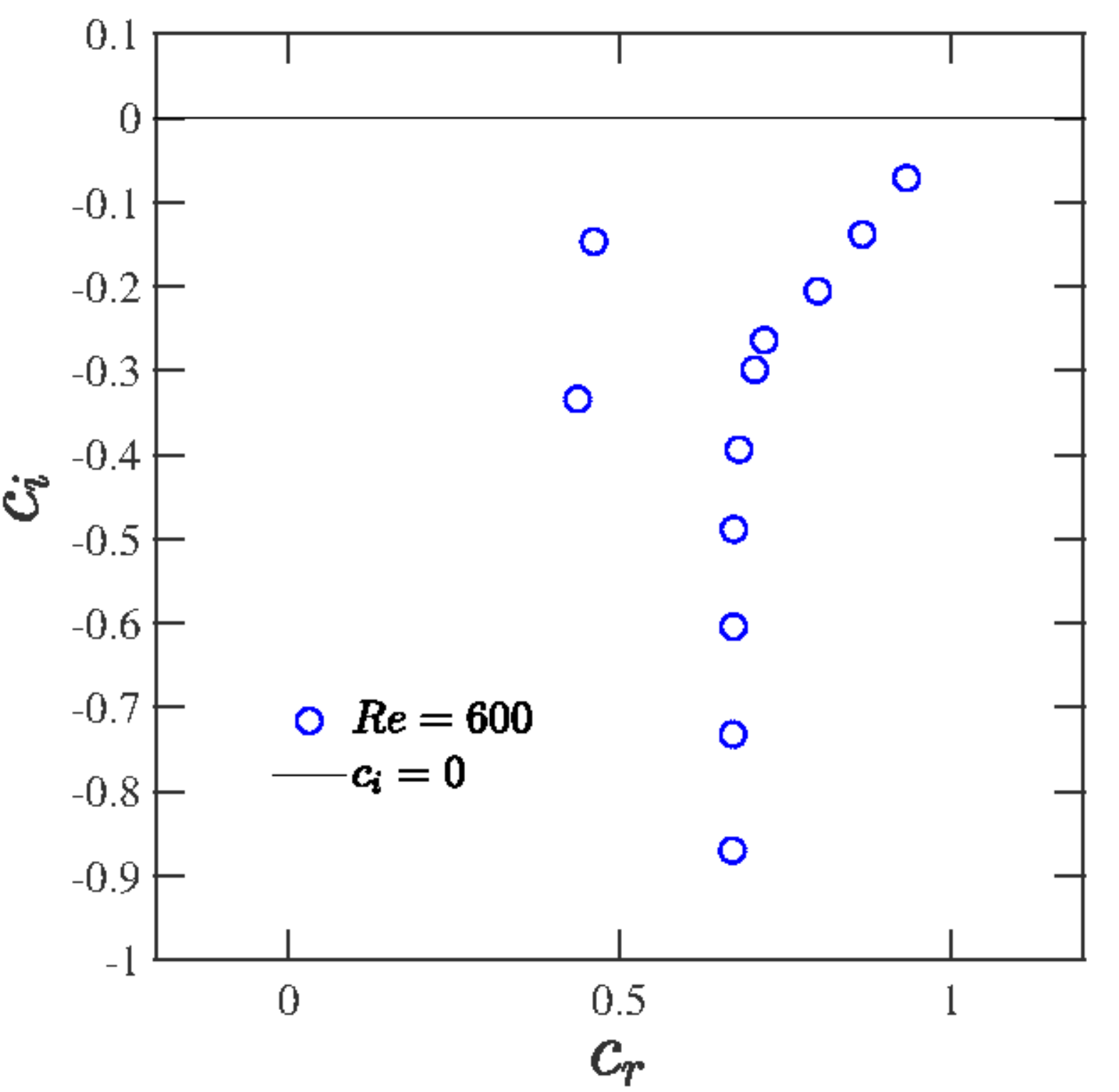}
    \caption{$Re = 600$}
    \label{fig:NewtonianRe600_k3}
  \end{subfigure}
  \caption{The `Y'-shaped eigenspectrum for Newtonian pipe flow subjected to axisymmetric 
  disturbances for $k = 3$, and for $Re = 6000$ and $600$.}
  \label{fig:Newtonianspectrum}
\end{figure}

\section{General features of the viscoelastic pipe flow eigenspectrum}
\label{sec:spectra_OB}

\begin{figure}
        \centering
        \begin{subfigure}[htp]{0.4\textwidth}
                \includegraphics[width=\textwidth]{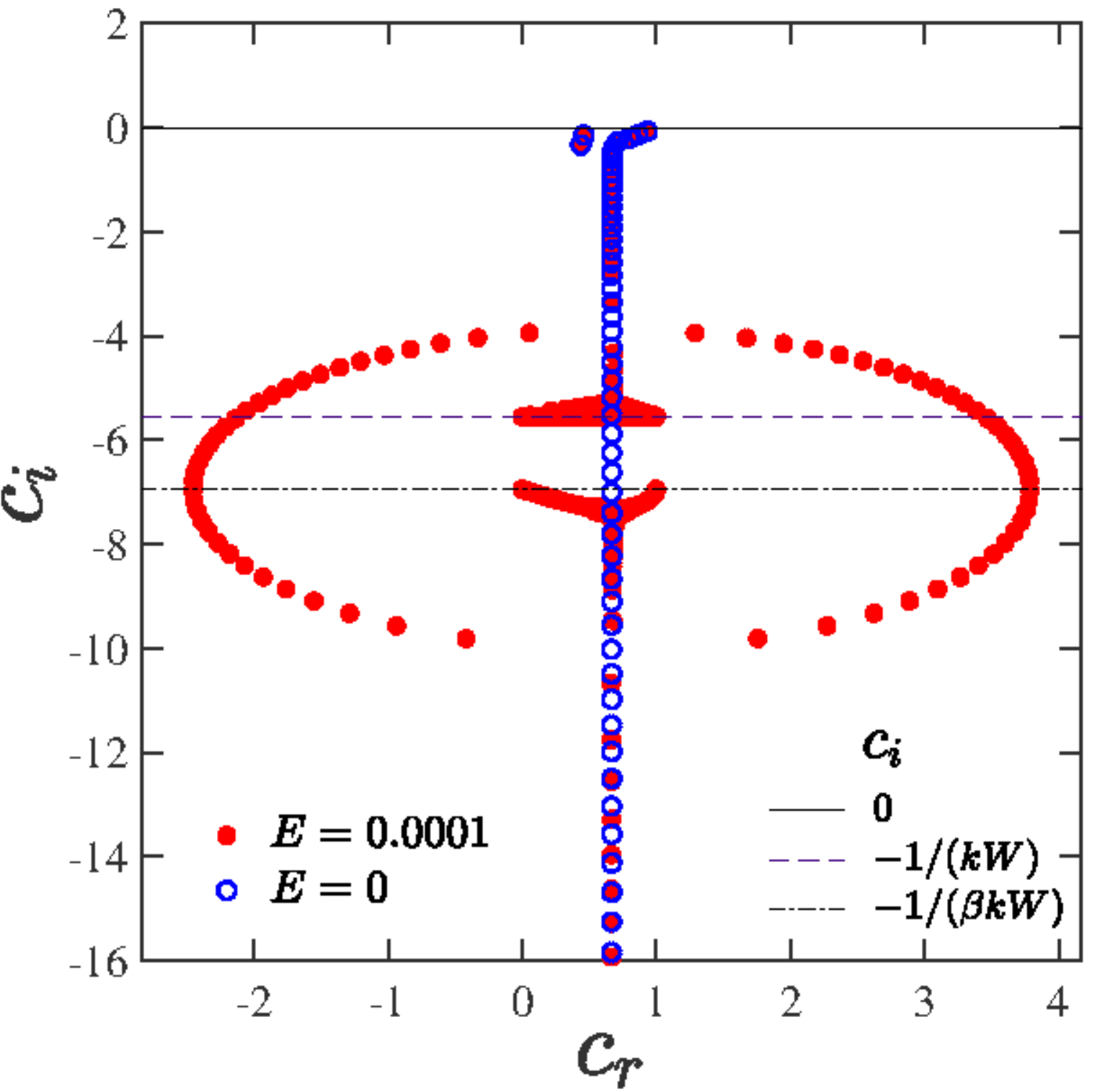}
                \caption{$E = 10^{-4}$}
                \label{fig:ring1e-4}
        \end{subfigure}
        \begin{subfigure}[htp]{0.4\textwidth}
                \includegraphics[width=\textwidth]{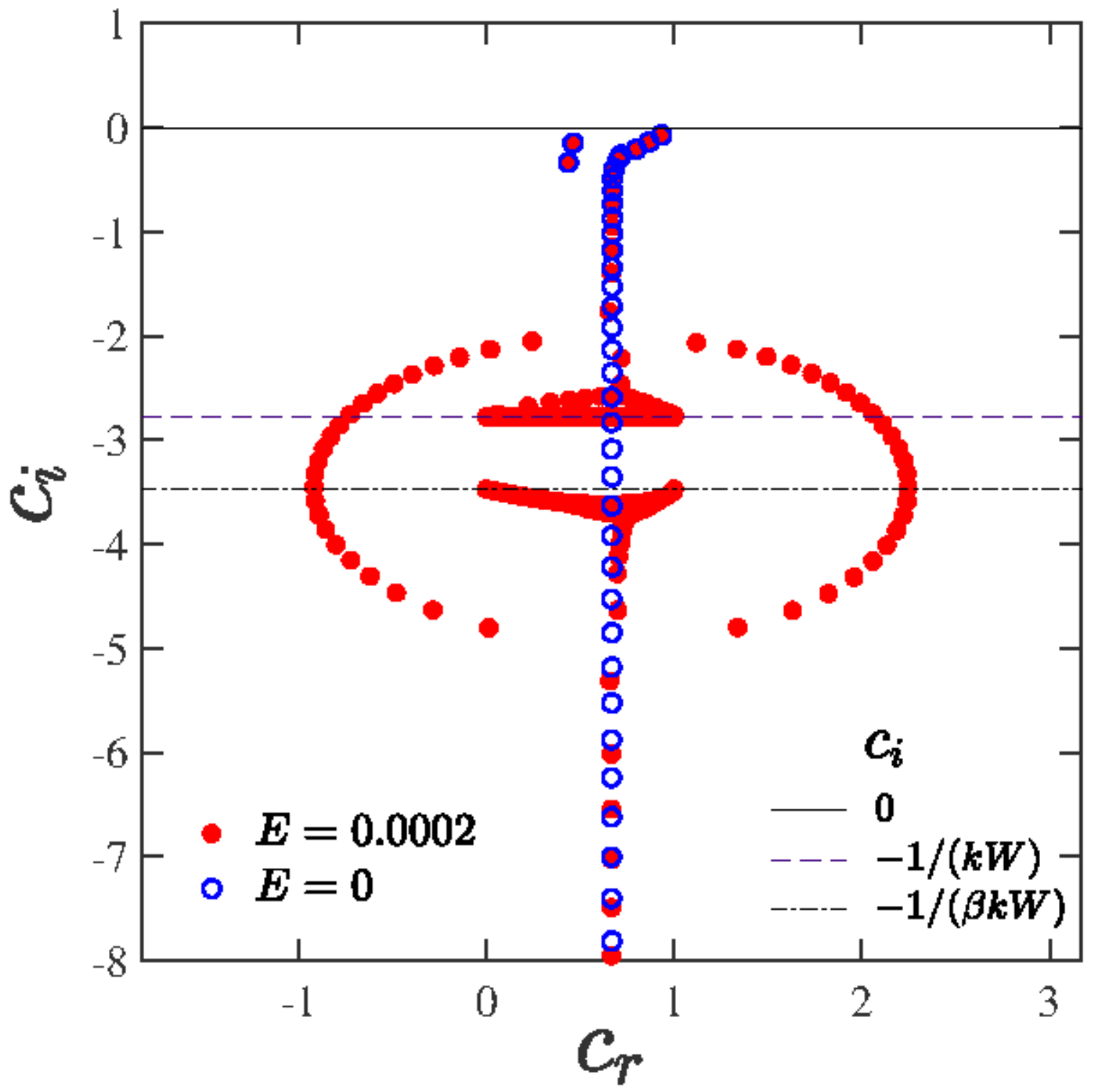}
                \caption{$E = 2\times 10^{-4}$}
                \label{fig:ring2e-4}
        \end{subfigure}\\
        \begin{subfigure}[htp]{0.4\textwidth}
                \includegraphics[width=\textwidth]{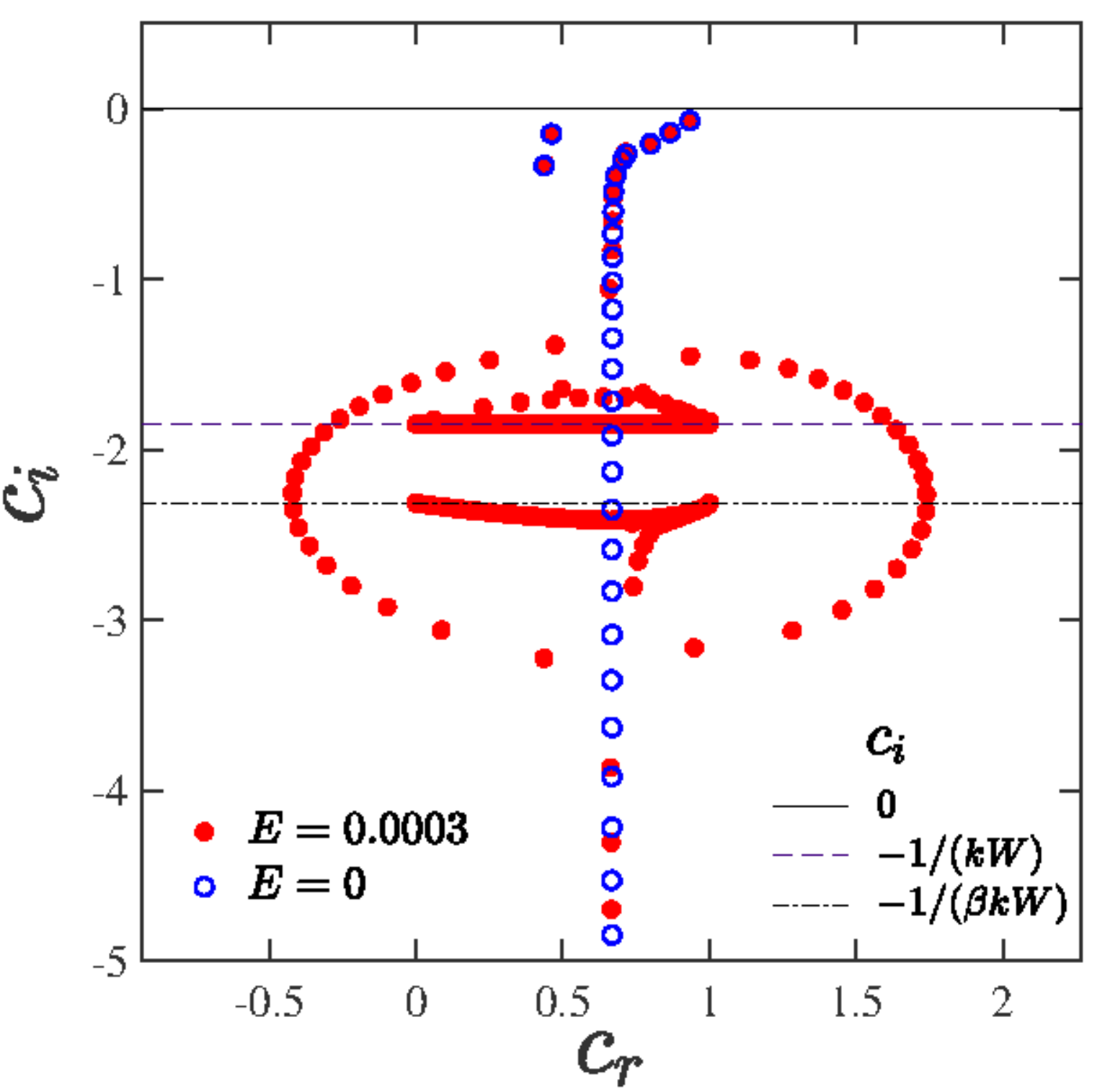}
                \caption{$E = 3\times 10^{-4}$}
                \label{fig:ring3e-4}
        \end{subfigure}
        \begin{subfigure}[htp]{0.4\textwidth}
                \includegraphics[width=\textwidth]{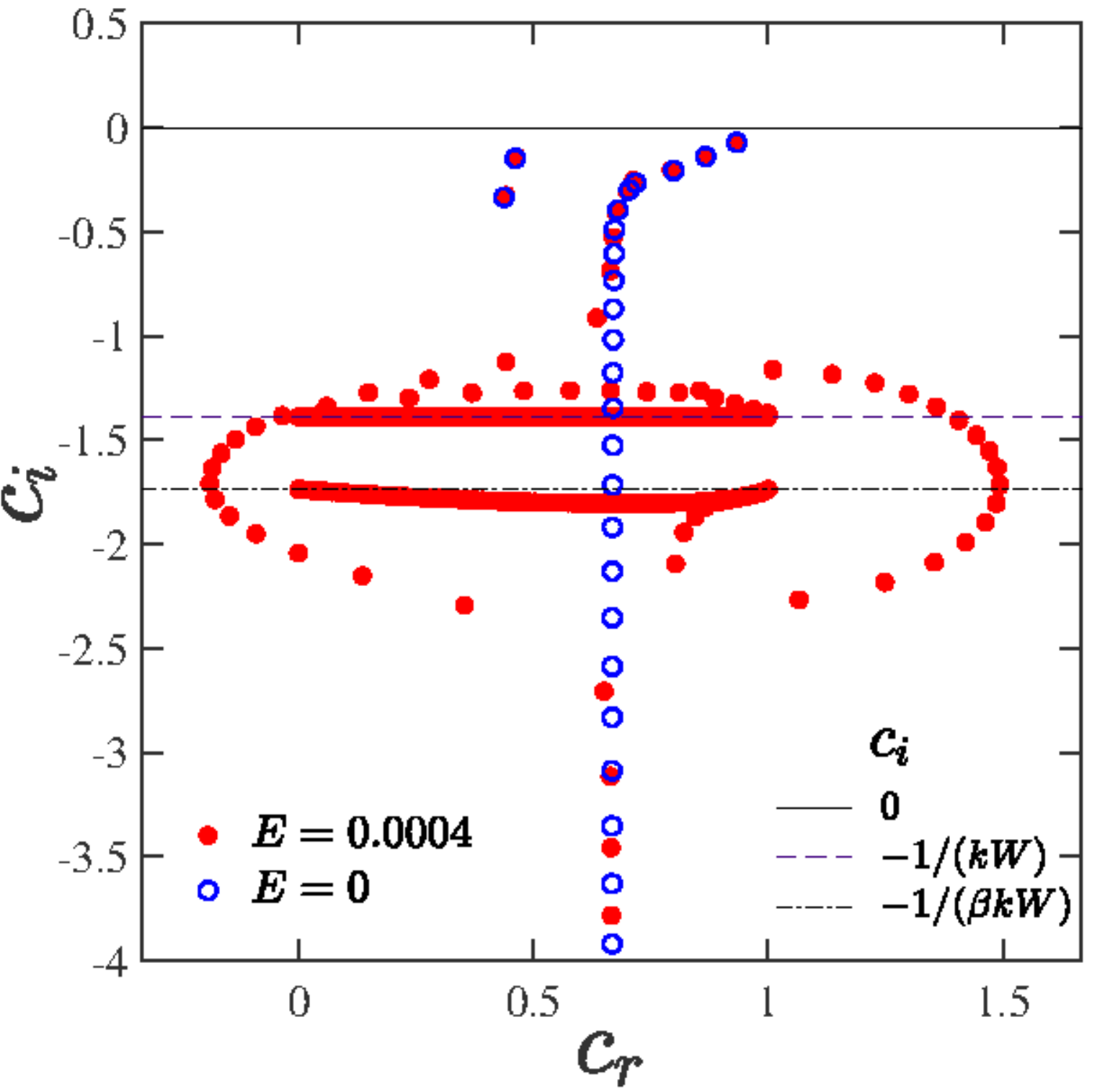}
                \caption{$E = 4\times 10^{-4}$}
                \label{fig:ring4e-4}
        \end{subfigure}\\
        \begin{subfigure}[htp]{0.4\textwidth}
                \includegraphics[width=\textwidth]{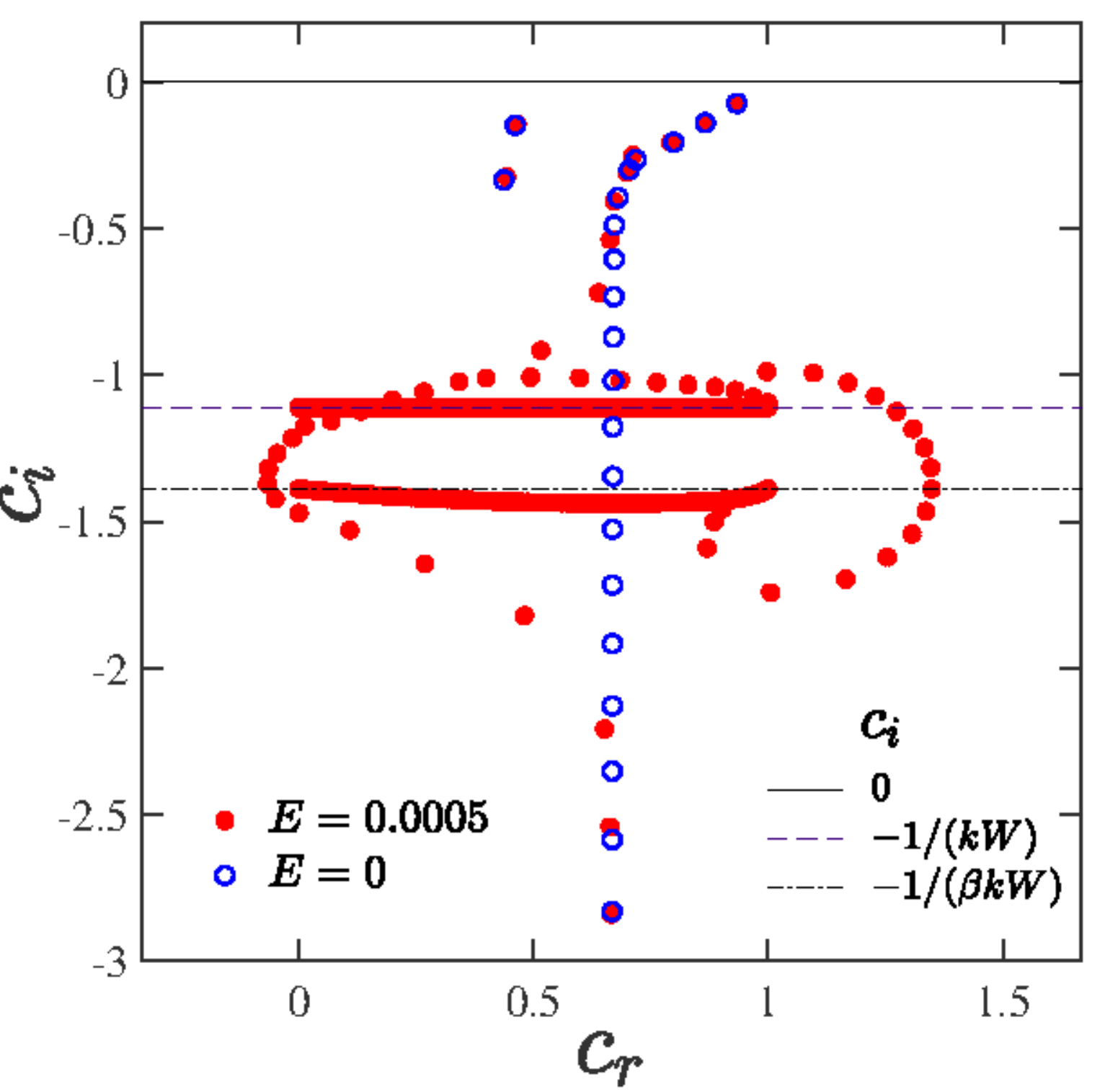}
                \caption{$E = 5\times 10^{-4}$}
                \label{fig:ring5e-4}
        \end{subfigure}
        \begin{subfigure}[htp]{0.4\textwidth}
                \includegraphics[width=\textwidth]{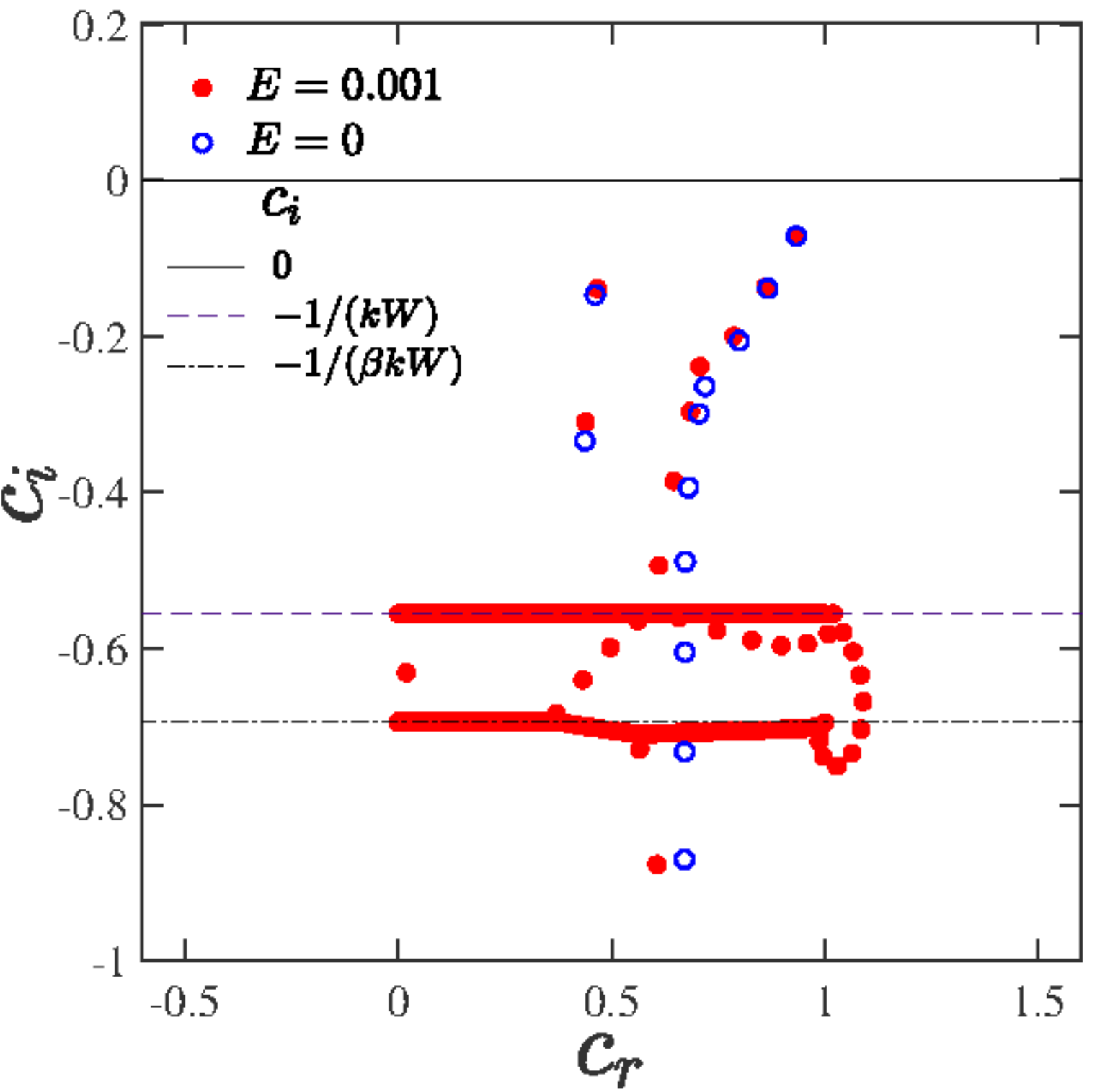}
                \caption{$E = 10^{-3}$}
                \label{fig:ring1e-3}
        \end{subfigure}
                        
        \caption{Eigenspectra for pipe flow of an Oldroyd-B fluid
          at $\beta = 0.8, \Rey = 600$ and $k = 3$, and for different
          $E$ in the range $5 \times
          10^{-4}$--$10^{-3}$. The eigenspectra are obtained for
          $N=200$, and there is excellent convergence of the spectra
          for $N=200$ and $250$ (not shown). An elliptical-ring structure is
          prominent at the lower $E$'s, but is absent beyond $E =
          10^{-3}$. The vertical locations of the CS1 and CS2 lines and the Newtonian spectrum for the same $\Rey$ and $k$ are  shown for reference. } 
\label{fig:ring}
\end{figure}

We first discuss results obtained for pipe Poiseuille flow of
Oldroyd-B fluids, with the extensive aid of eigenspectra, and demonstrate how the viscoelastic spectrum differs substantially 
from its Newtonian counterpart. Sections~\ref{subsec:varyingE} and \ref{subsec:varyingbeta}, 
 respectively, consider the variation in the eigenspectrum with increasing $E$ (from zero) at a fixed $\beta$, and with variation in $\beta$ at a fixed $E$. The focus is on the locations of the least stable modes, and as to how they change with changing $E$ and $\beta$.
Alongside, we also demonstrate (Secs.~\ref{sec:centermode} and \ref{subsec:originfixedEvaryingbeta})  how the continuous
spectra (henceforth abbreviated as `CS') play an important role in the emergence of the eigenmode (a center mode) that eventually becomes unstable. In
Sec.~\ref{sec:shekar}, we contrast the nature of the
least stable modes in viscoelastic pipe and channel flows, showing, in particular, that
for the parameters corresponding to viscoelastic channel flow where the wall (TS) mode is least stable \citep{shekar_etal_2019}, pipe flow has the center mode as its least stable mode.
The center mode instability is characterized further using neutral stability curves in the $Re$-$k$ plane at fixed $E$ and $\beta$
(Sec.~\ref{sec:neutral_curves}), which are shown to collapse when plotted using
suitable rescaled variables (Sec.~\ref{sec:collapses}).
The variation of the minima of the
$Re$-$k$ neutral curves (the critical Reynolds number $Re_c$) and the
corresponding critical wavenumber $k_c$ is explored (Sec.~\ref{sec:scalings})
for different $E$
and $\beta$, and scaling relationships are obtained in the limit $E \ll 1$ and
$E \ll 1$, $(1-\beta) \ll 1$.
It is then shown that the scaling
results inferred from the numerics are consistent with those obtained from a
boundary-layer analysis near the pipe centerline. We finally compare our theoretical
predictions with recent experimental and DNS studies in
Sec.~\ref{sec:recent_studies}.

%

\begin{figure}
        \centering
        \begin{subfigure}[htp]{0.4\textwidth}
                \includegraphics[width=\textwidth]{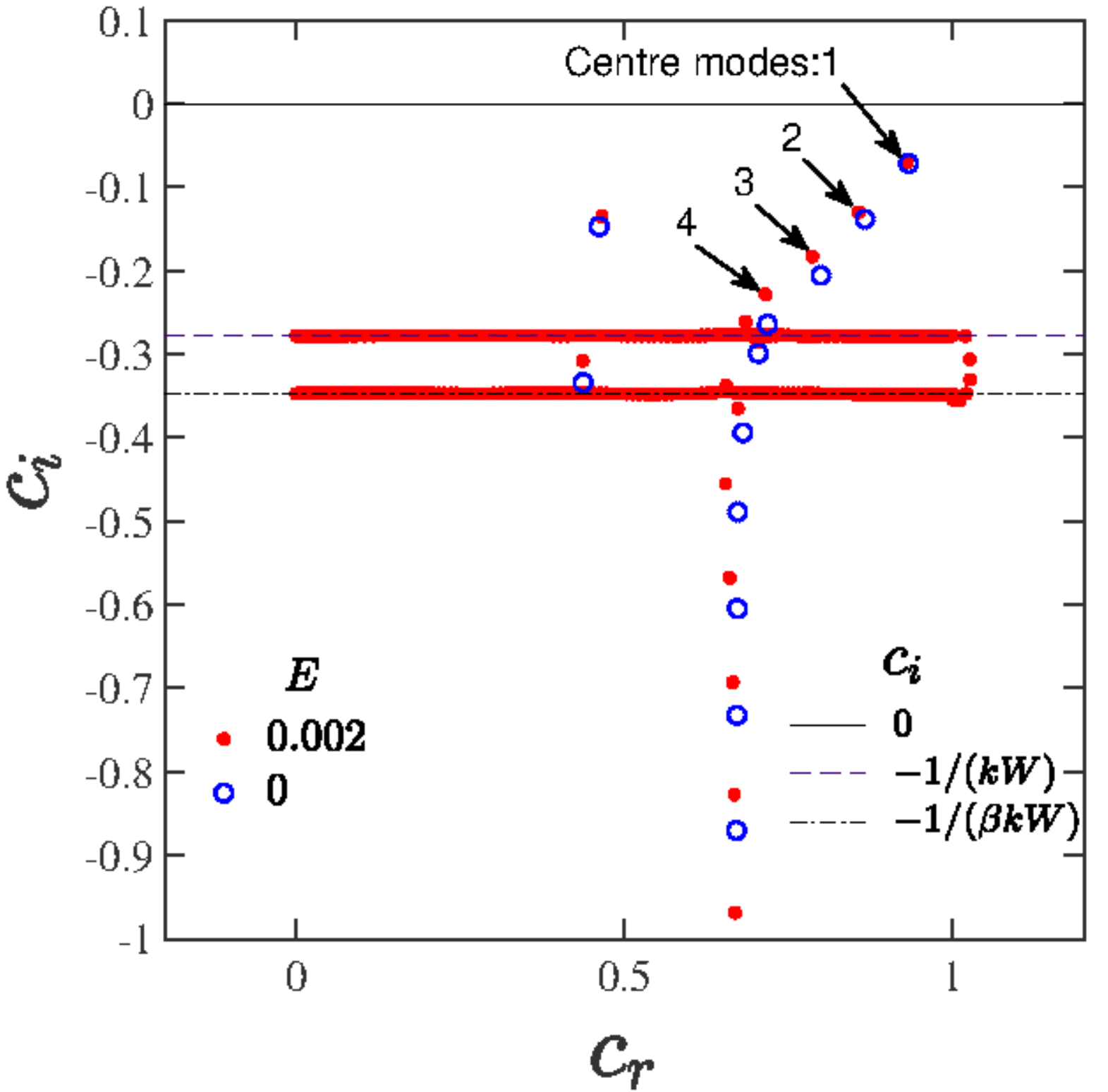}
                \caption{$E = 0.002$}
                \label{fig:ues_beta0pt8E0pt002Re600k3}
        \end{subfigure}
        \begin{subfigure}[htp]{0.4\textwidth}
                \includegraphics[width=\textwidth]{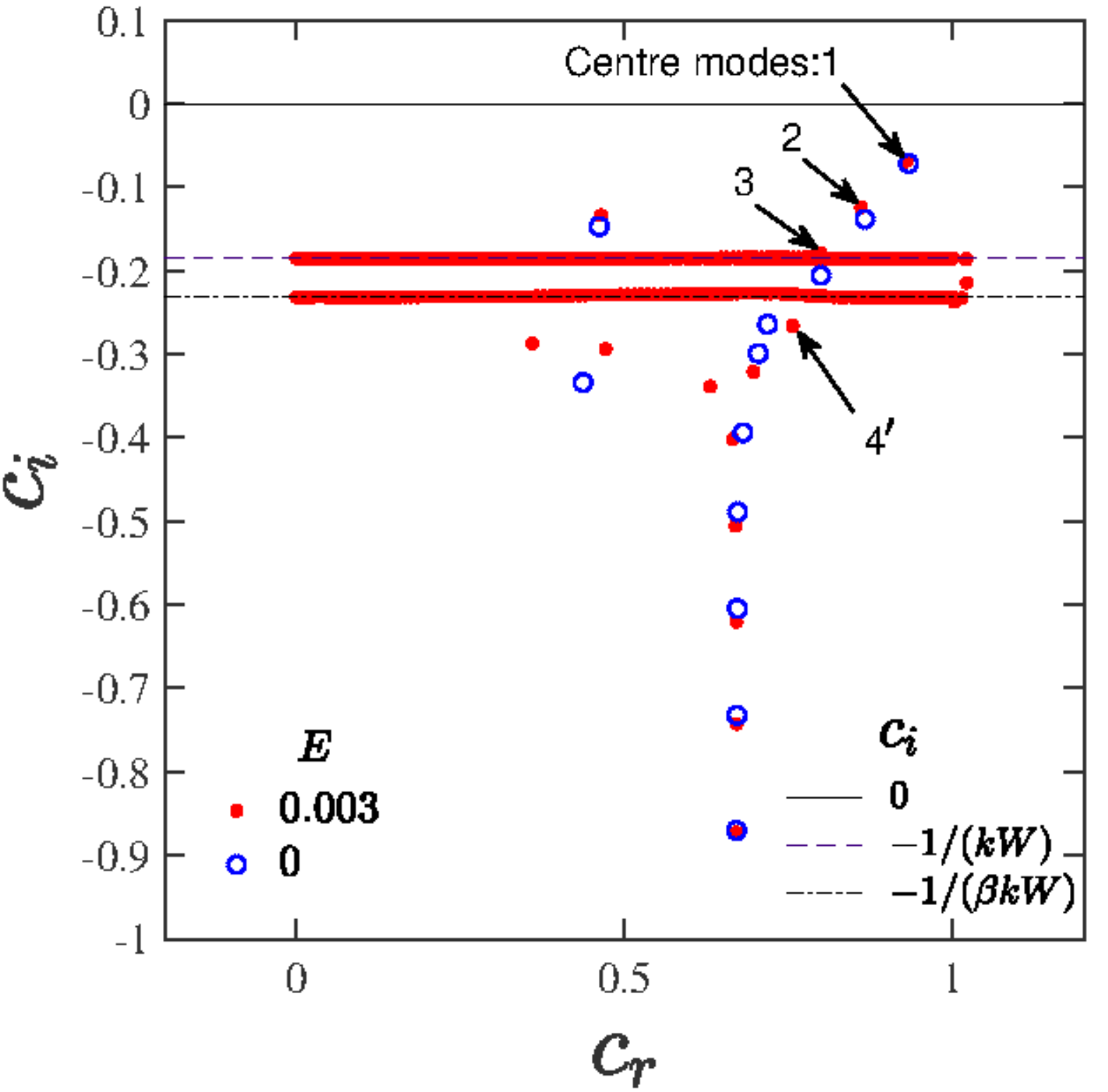}
                \caption{$E = 0.003$}
                \label{fig:ues_beta0pt8E0pt003Re600k3}
        \end{subfigure}
        \begin{subfigure}[htp]{0.4\textwidth}
                \includegraphics[width=\textwidth]{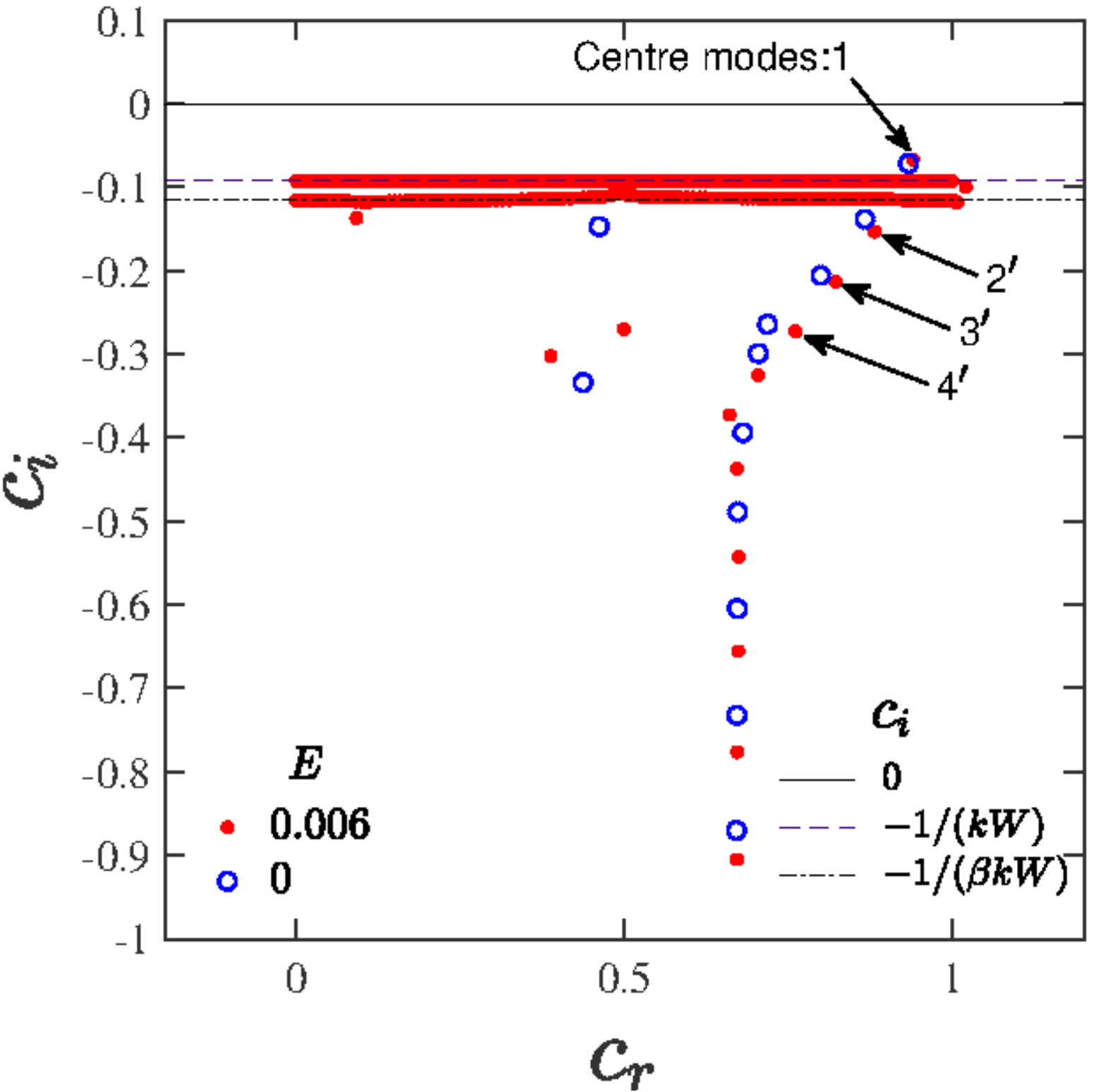}
                \caption{$E = 0.006$}
                \label{fig:ues_beta0pt8E0pt006Re600k3}
        \end{subfigure}
        \begin{subfigure}[htp]{0.4\textwidth}
                \includegraphics[width=\textwidth]{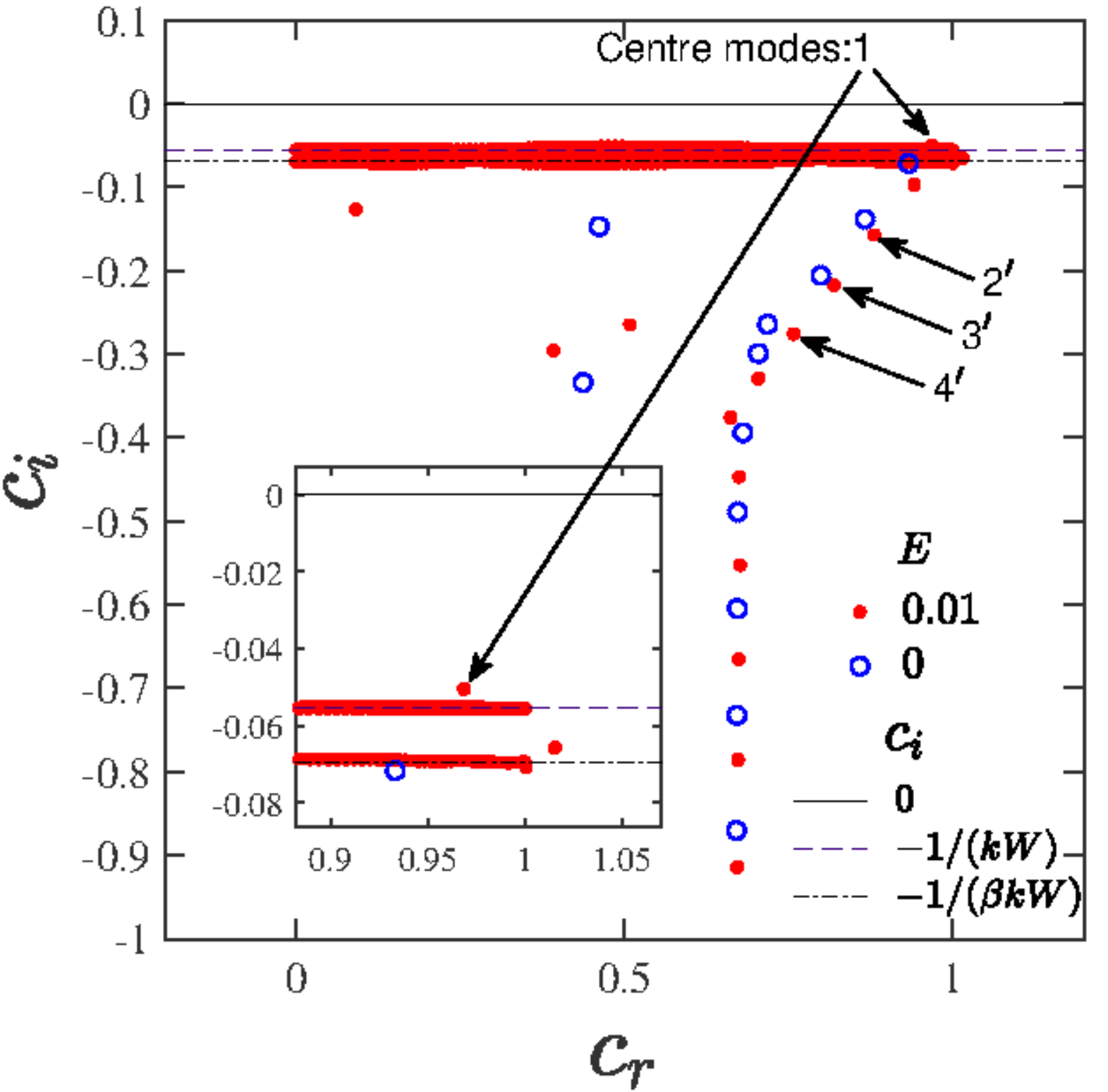}
                \caption{$E = 0.01$}
                \label{fig:ues_beta0pt8E0pt01Re600k3}
        \end{subfigure}
        \begin{subfigure}[htp]{0.4\textwidth}
                \includegraphics[width=\textwidth]{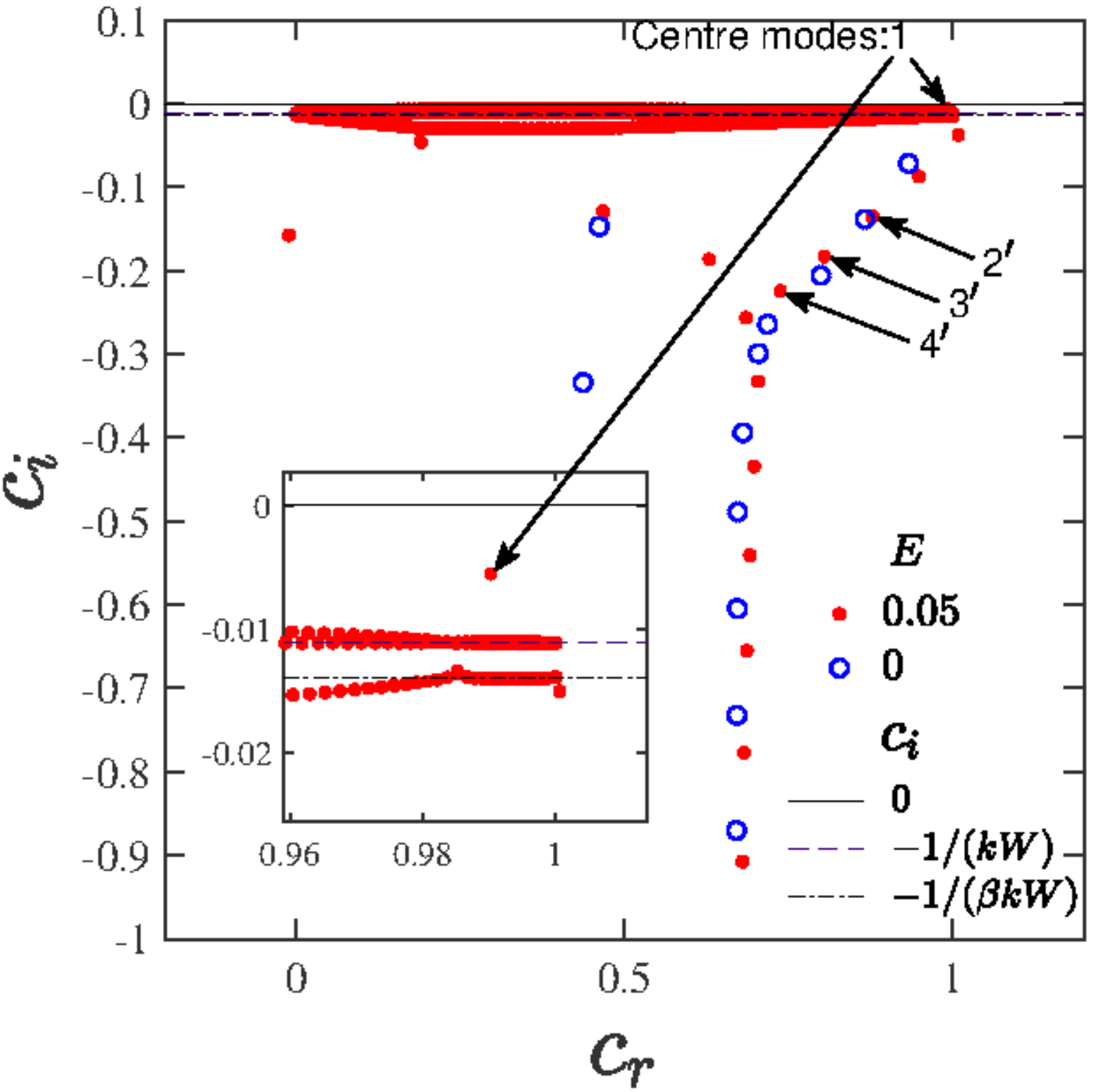}
                \caption{$E = 0.05$}
                \label{fig:ues_beta0pt8E0pt05Re600k3}
        \end{subfigure}
        \begin{subfigure}[htp]{0.4\textwidth}
                \includegraphics[width=\textwidth]{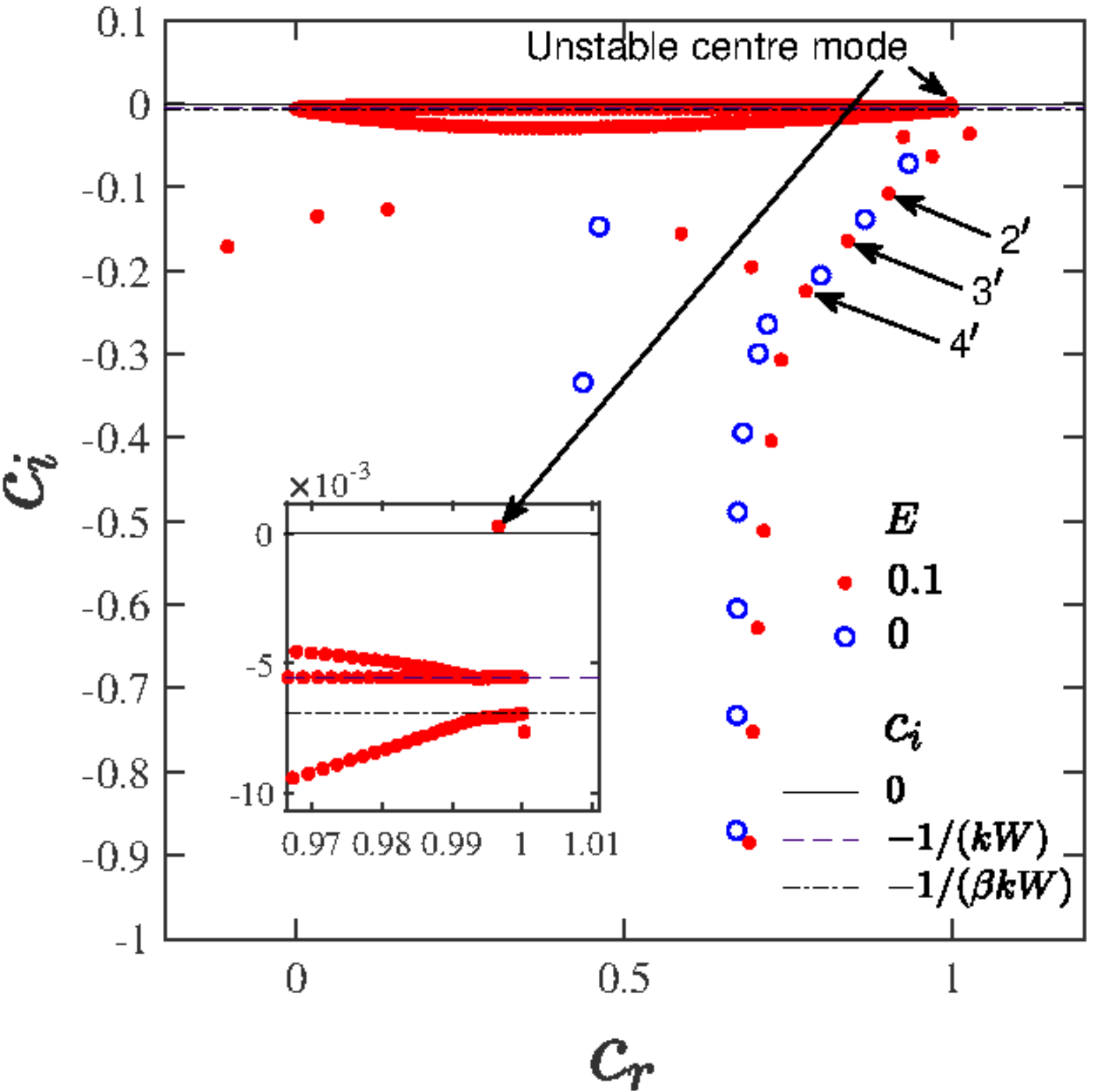}
                \caption{$E = 0.1$}
                \label{fig:ues_beta0pt8E0pt1Re600k3}
        \end{subfigure}
        \caption{Eigenspectra for the Oldroyd-B fluid for different $E$
in the range $0.002$--$0.1$, at $\beta = 0.8$, $\Rey = 600$, $N = 200$, and $k = 3$.  The spectra, shown for a narrower range of $c_r$ and $c_i$ compared to Fig.~\ref{fig:ring}, demonstrate how the discrete center modes (labelled $2$, $3$, and $4$) merge into and emerge out (labelled $2'$, $3'$, and $4'$) of the CS as $E$ is increased.    The least stable center mode (labelled 1) always stays above the CS, and eventually becomes unstable at $E = 0.1$. The vertical locations of the CS lines and the Newtonian spectrum at the same $\Rey$ and $k$ are shown for reference.}
\label{fig:ues_beta0pt8EERe600k3}
\end{figure}

\subsection{Spectra at fixed $\beta$ and different $E$}
\label{subsec:varyingE}

Figures~\ref{fig:NewtonianRe6000_k3} and  \ref{fig:NewtonianRe600_k3} show the
eigenspectra for Newtonian pipe flow at $Re = 6000$ and $Re = 600$ respectively. 
The spectrum
 has the well-known `Y'-shaped structure 
 \citep{SchmidHenningsonBook,mack_1976} at $Re = 6000$, but this is only beginning to form in
the spectrum at  $Re = 600$.  The Y-shaped structure is
  comprised of three branches: (i) the `A
branch' corresponding to `wall modes' with 
$c_r \rightarrow 0$ for $\Rey \gg 1$ on the top left; (ii) the `P
branch' which consists `center modes' with
$c_r \rightarrow 1$ for $\Rey \gg 1$ on the top right, and (iii) the
`S branch' which consists modes with $c_r \approx 2/3$ extending down to $c_i = -\infty$. For $Re \gg 1$, the decay rates of the
least stable center and wall modes vary as $|c_i| \sim Re^{-1/2}$ and
$|c_i| \sim Re^{-1/3}$ respectively \citep{meseguer_trefethen_2003}, implying that the center modes are the least stable at large $\Rey$. 
Although these scalings need not necessarily hold for 
the moderate $Re$ ($= 600$) considered in Figs.~\ref{fig:ring} and \ref{fig:ues_beta0pt8EERe600k3} below, the center mode is nevertheless found to be the least stable one. 
Consistent with previous studies
\citep{SchmidHennigson1994,SchmidHenningsonBook}, all
modes for Newtonian pipe flow are found to be stable. 
We discuss  the nature of the least stable mode in more detail below in Sec.~\ref{sec:shekar}. 

The spectra for pipe flow
of an Oldroyd-B fluid reduce to the Newtonian one either when
$E \rightarrow 0$ (at fixed $\beta$) or when $\beta \rightarrow 1$ (at
fixed $E$). We therefore first examine the effect of viscoelasticity as $E$ is increased from zero, with  $\beta$ fixed at $0.8$.
Figures~\ref{fig:ring} and
\ref{fig:ues_beta0pt8EERe600k3} show the viscoelastic eigenspectra for $Re =600$, with $E$ ranging from $5 \times 10^{-4}$ to $0.1$.  The  values of $\beta$ and $Re$ are chosen so they are close to the experimental conditions of \cite{samanta_etal2013} and our earlier theoretical work \citep{Garg2018}.
With increasing $E$, Figs.~\ref{fig:ring} and   \ref{fig:ues_beta0pt8EERe600k3} show 
the classical Y-shaped
structure of the Newtonian spectrum to be altered by elasticity in a singular manner.
There are  important differences between the two spectra even for the smallest $E$'s, 
the most prominent of these being the appearance of two continuous spectra for
the viscoelastic case, similar to viscoelastic plane
shear flows
\citep{renardy1986linear,sureshkumar1995linear,graham_1998,wilson1999,grillet2002,chokshi_kumaran_2009}.
It is now well understood \citep{graham_1998,chaudhary_etal_2019,roy_subramanian2018} that
the continuous spectra arise from the local nature of the constitutive
model for the polymeric stress, and disappear
when non-local diffusive effects are incorporated in the constitutive relation 
(see Sec.~\ref{subsec:stressdiffusion}).
The eigenvalues corresponding to the
CS are obtained by setting to zero the coefficient of the highest
order derivative in the differential equation governing the
stability. This coefficient turns out to be the product $[1 + \mathrm{i}kW(U-c)] [1 + \beta\mathrm{i}kW(U-c)]$, which leads to a pair of horizontal `lines' in the
$c_r$--$c_i$ plane with $c_i = -1/(kW)$ and $c_i = -1/(\beta k W)$, and with
$0 \leq c_r \leq 1$. Henceforth, these two continuous spectra are
respectively abbreviated as `CS1' and `CS2' respectively, with CS1 being present even in the limit of a UCM fluid, and CS2 being present only when there is a solvent contribution
($\beta \neq 0$), receding to $c_i = -\infty$ in the limit $\beta \rightarrow 0$.

Figure~\ref{fig:ring} explores the spectra for the smallest $E$'s  
(ranging from $10^{-4}$ to $10^{-3}$), the range of $c_r$ and $c_i$
being chosen so as to provide a larger view of the spectra.
Here, in addition to the modified Y-shaped structure
of the Newtonian spectra and the two CS lines, there exist
a class of modes which form a `ring' that
surrounds the continuous spectra at small  $E$'s of $O(\sim 10^{-4})$. Similar to CS1 and CS2 ,
all modes belonging to the ring structure are stable for
the range of $E$ explored. For small $E$'s, the modes on the
ring appear to be symmetrically distributed (Figs.~\ref{fig:ring1e-4}
to \ref{fig:ring3e-4}) about the S branch, forming an approximate
ellipse. 
As $E$ is increased, the modes move towards the
continuous spectra with the ring getting smaller in size. For $E = 4\times10^{-4}$, 
these modes move closer, intermingling with 
the other modes which emerge from the CS
(Fig.~\ref{fig:ring4e-4}), and the ring structure is now fully
distorted. At still higher $E\sim 10^{-3}$ (see Figs.~\ref{fig:ring5e-4} and \ref{fig:ring1e-3}), 
the modes originally on the ring
collapse, wrapping around the CS in an irregular manner.  To understand the origin of the ring structure, it
is relevant to recall a prominent feature of
the viscoelastic spectra  (at nonzero $Re$) in the  UCM limit ($\beta = 0$): 
an infinite sequence of discrete modes corresponding to damped shear waves 
in a viscoelastic fluid  (discussed below in 
Sec.~\ref{subsec:varyingbeta}), and are referred to as the high-frequency-Gorodtsov-Leonov (`HFGL') modes \citep{gl67,KumarShankar,chaudhary_etal_2019}. This sequence corresponds to $c_i = -1/(2kW)$, and extends to infinity in either direction parallel to the $c_r$ axis.
%
%
As we demonstrate below 
in Fig.~\ref{fig:hfgltoring}, at any finite $\beta$, the infinite-in-extent HFGL line curves downwards, eventually meeting the S branch, and thereby leading to the aforementioned ring structure for sufficiently small $E$. 


In Fig.~\ref{fig:ues_beta0pt8EERe600k3}, we explore the spectra for larger $E$, in the range $0.002$--$0.1$, with the ranges of $c_r$ and $c_i$ being chosen to provide a more magnified view of the spectra.
As $E$ is increased, the
vertical locations of the two CS move up towards $c_i = 0$, and in the
process, the discrete `elastic' center modes (labelled 2, 3, 4 in Fig.~\ref{fig:ues_beta0pt8E0pt002Re600k3}; and that lie above the CS) disappear into the
CS. As $E$ is  increased further, new discrete elastic center modes  (now shown with the labels $2'$, $3'$, and $4'$ in Figs.~\ref{fig:ues_beta0pt8E0pt006Re600k3} and \ref{fig:ues_beta0pt8E0pt01Re600k3}, which lie below the CS) emerge out of the CS. The labelling of the modes that emerge below the CS are for the purposes of reference only, and there is no connection between these modes (with primes) and the ones (without primes) that disappeared into the CS. This was ascertained from the absence of any resemblance between the eigenfunctions of the modes that disappear into and reappear from the CS.
A more detailed account of the evolution of the center modes as $E$ is varied is provided below in Fig.~\ref{fig:c_vs_E_modes1to4beta0pt8Re600k3}.
Importantly,  the least stable center mode (labelled 1) does not merge into the CS, and always stays above it.  As $E$ is increased to $0.1$, this center mode becomes unstable, and corresponds to the instability first reported in
\cite{Garg2018}.
This scenario of the unstable center mode being a smooth continuation of its stable Newtonian counterpart is, however,  sensitive to  $\beta$, and  we show below (and in more detail in Sec.~\ref{sec:centermode}) that 
there exist other parameter regimes where the center mode  that  eventually becomes unstable  emerges out the CS with increasing $E$, and there is no connection to the least stable Newtonian center mode. 
Other new stable center modes (with $c_r \rightarrow 1$; see Figs.~\ref{fig:ues_beta0pt8E0pt01Re600k3}--\ref{fig:ues_beta0pt8E0pt1Re600k3})
and wall modes (with $c_r \rightarrow 0$, and even negative; see Figs.~\ref{fig:ues_beta0pt8E0pt05Re600k3} and \ref{fig:ues_beta0pt8E0pt1Re600k3}), which have no Newtonian counterparts, 
 appear below the CS with increasing $E$.  Increase in $E$ has a stabilizing effect on these
modes. 
The aforementioned annihilation and creation  of discrete modes with increase in $E$ occurs because both the continuous spectra are branch cuts \citep{wilson1999}
for Poiseuille flow. Note that, for plane Couette flow, only CS2 is a branch cut.
It is well known that discrete eigenmodes can appear
or disappear out of the branch cut as parameters are varied, and this aspect is discussed further in Sec.~\ref{sec:centermode} in the specific context of the center mode.
\begin{figure}
  \centering
  \begin{subfigure}[htp]{0.48\textwidth}
    \includegraphics[width=\textwidth]{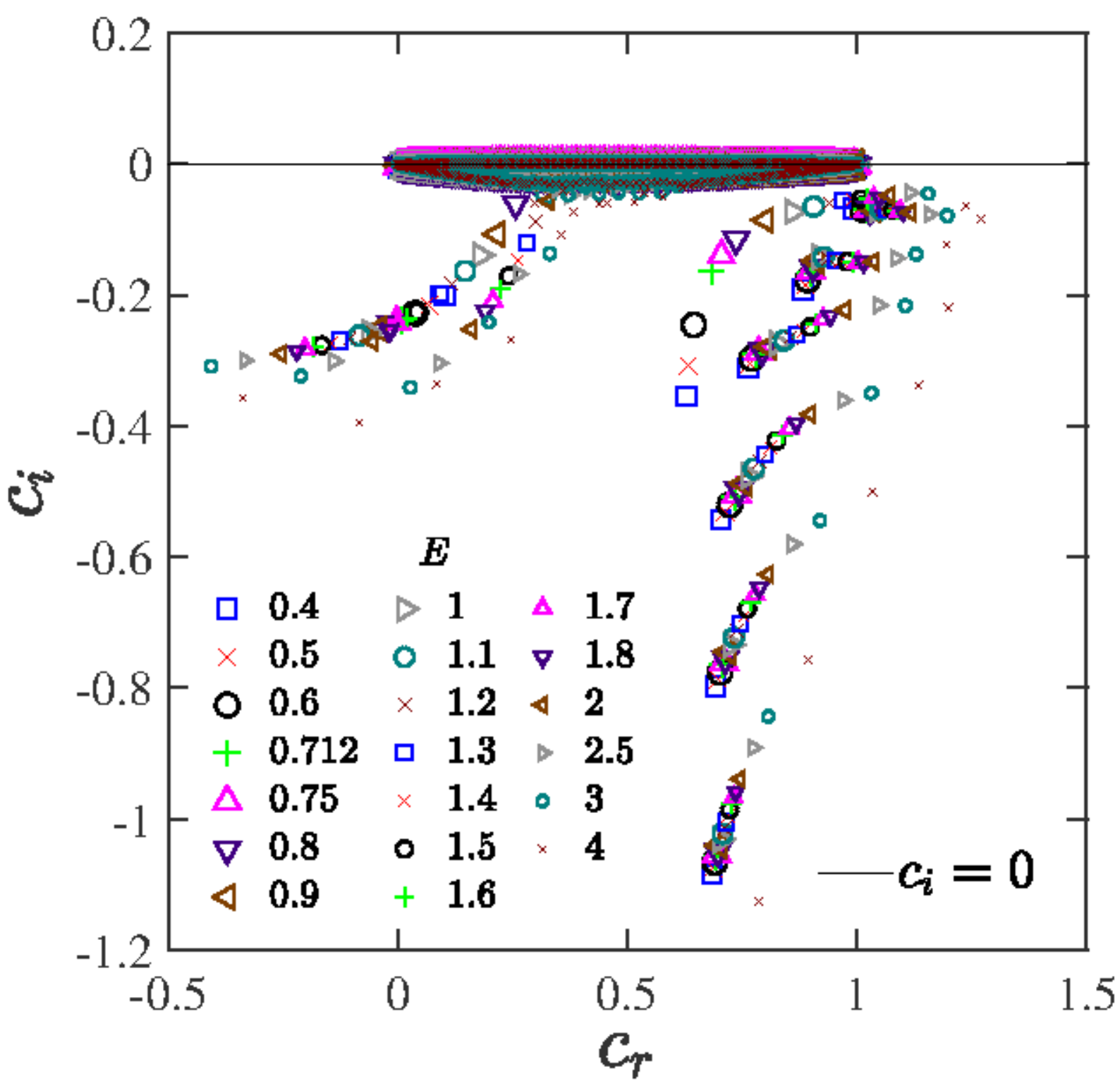}
    \caption{Unfiltered spectra}
    \label{fig:ues_beta_pt96_Re500_k1}
  \end{subfigure}
  \begin{subfigure}[htp]{0.48\textwidth}
    \includegraphics[width=\textwidth]{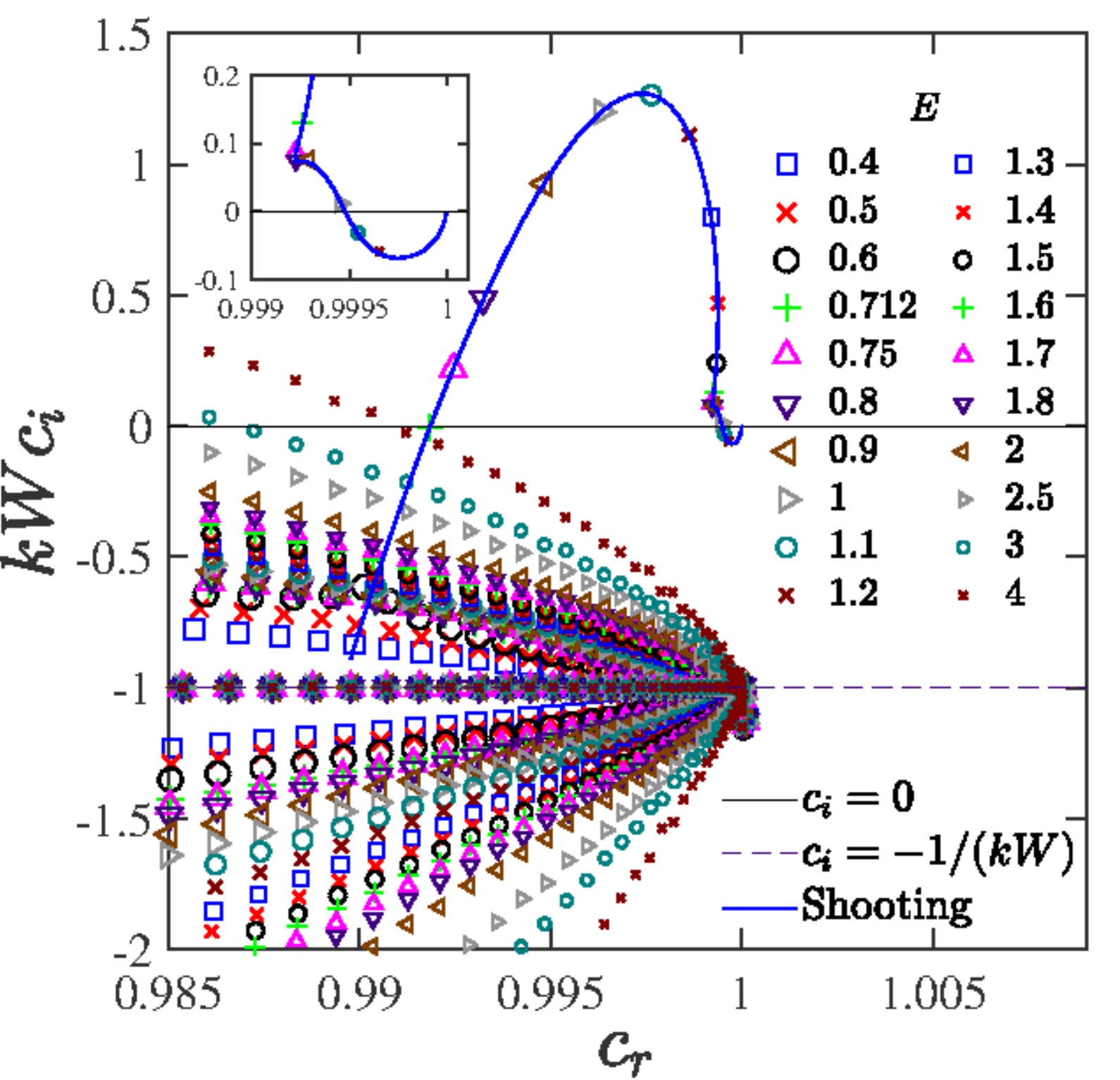}
    \caption{Magnified region near $c_r = 1$}
    \label{fig:ues_beta_pt96_Re500_k1_zoomed}
  \end{subfigure}
                
  \caption{(a) Eigenspectra for different values of $E$ for
    $\beta=0.96,\Rey=500, k=1$; (b) Enlarged version of region in
    panel (a) near the unstable  center mode. The scaled growth rate $kWc_i$ fixes
    the vertical location of both the CS (for $\beta = 0.96$, CS1 and CS2 lie very close to each other).
    The continuous line for the trajectory of the unstable center mode is obtained using the shooting method.}
  \label{fig:ues_beta_pt96}
\end{figure}

Figure~\ref{fig:ues_beta_pt96} shows the spectra in the near-Newtonian limit of $\beta = 0.96$
and for $E$ ranging over the interval $(0.4,4)$, overlaid in a single plot, in order to demonstrate the
variation of not just the (eventually) unstable center mode,
but also of the other stable modes.  
For the higher $E$'s considered in Fig.~\ref{fig:ues_beta_pt96}, the two CS's lie very close to $c_i = 0$ (and to each other for the chosen $\beta$), and the modes in the Newtonian P-branch have therefore already disappeared into the CS, with new modes emerging from below. Thus, the trajectories of the modes shown in
Fig.~\ref{fig:ues_beta_pt96_Re500_k1} are for the modes that start off below the CS.
The zoomed  version in Fig.~\ref{fig:ues_beta_pt96_Re500_k1_zoomed} shows the
spectra in terms of the scaled growth rate $kWc_i$, which fixes the vertical location of both the CS (for fixed $\beta$), and allows one to focus on the trajectory of the unstable center mode with varying $E$.
The continuous curve indicating the trajectory of the center mode, as $E$ is varied, is
obtained using the shooting method with much finer increments in $E$.
This figure shows that the center mode first emerges out of the CS, in the form of a bump in the 
continuous spectrum balloon, at $E \approx 0.6$, and becomes unstable as $E$ is increased to $0.712$. 
 The center mode remains unstable
for $0.712 \leq E \leq 2.5$, but 
becomes stable for $E > 2.5$, with $|c_i|$ eventually scaling as $1/E$ for large $E$. Thus,
Figs.~\ref{fig:ues_beta0pt8EERe600k3} and \ref{fig:ues_beta_pt96_Re500_k1_zoomed} show that there are two qualitatively different trajectories of the unstable center mode with increasing $E$. For the lower $\beta$ $(=0.8)$,  the 
center mode appears as a smooth continuation of the least stable Newtonian center mode, while for $\beta = 0.96$, it emerges from the continuous spectrum, with no obvious connection to the Newtonian spectrum. This aspect is discussed in more detail below in Sec.~\ref{sec:centermode}.

\begin{figure}
  \centering
  \begin{subfigure}[htp]{0.35\textwidth}
    \includegraphics[width=\textwidth]{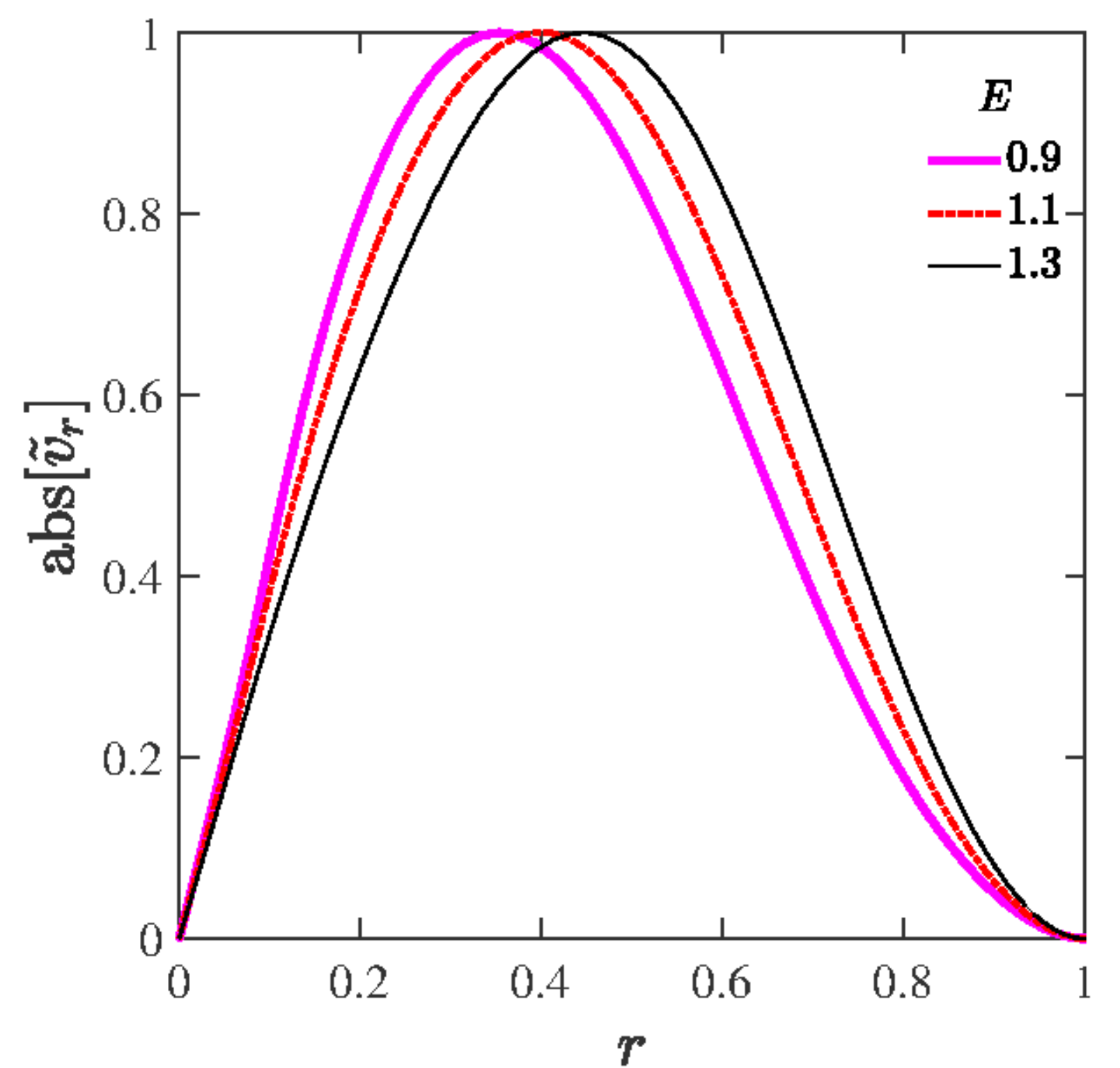}
    \caption{Radial velocity}
    \label{fig:ef_Vr_beta0pt8E0pt002Re600k3}
  \end{subfigure}
  \begin{subfigure}[htp]{0.35\textwidth}
    \includegraphics[width=\textwidth]{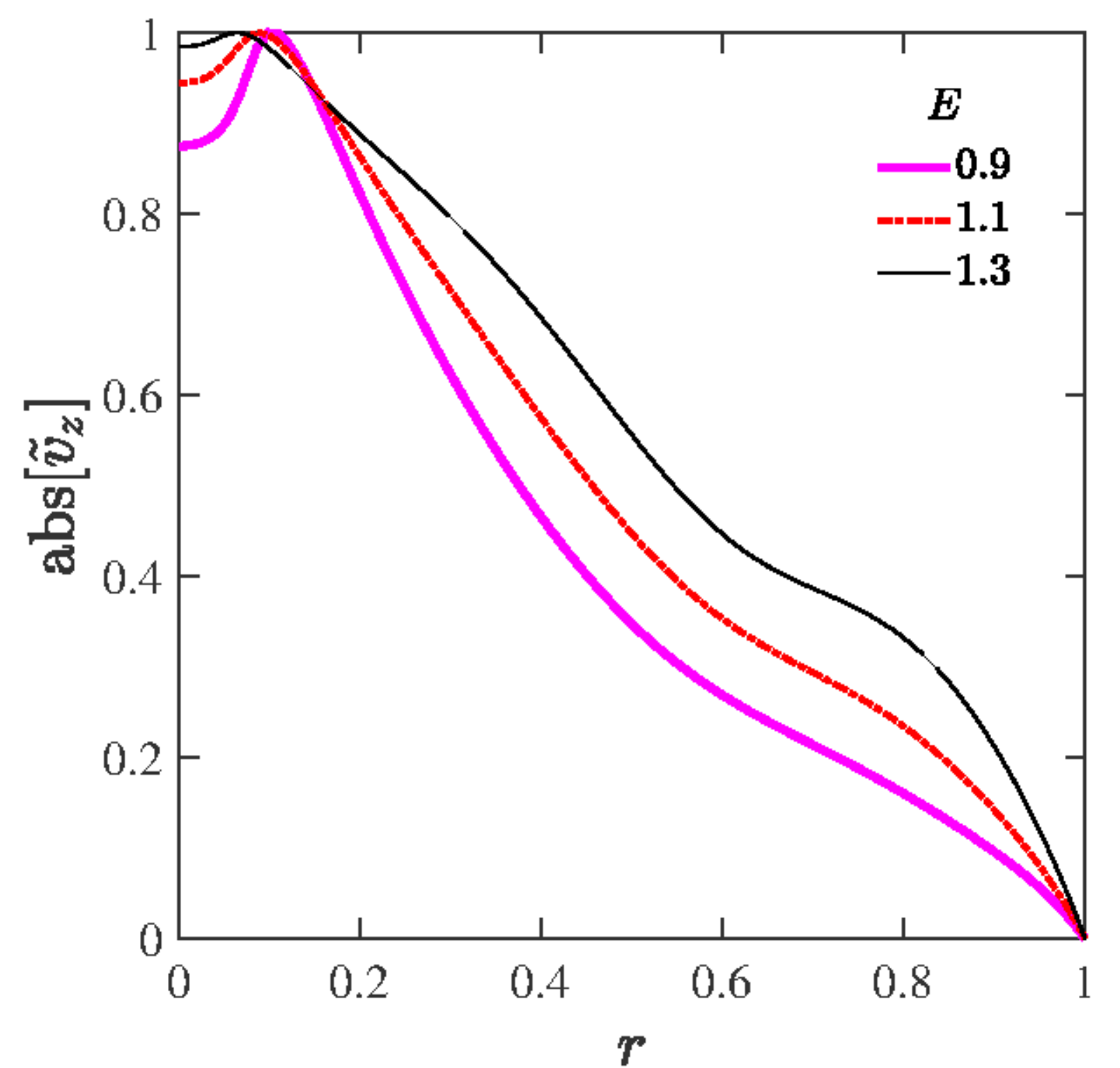}
    \caption{Axial velocity}
    \label{fig:ef_Vz_beta0pt8E0pt003Re600k3}
  \end{subfigure}
    \begin{subfigure}[htp]{0.35\textwidth}
    \includegraphics[width=\textwidth]{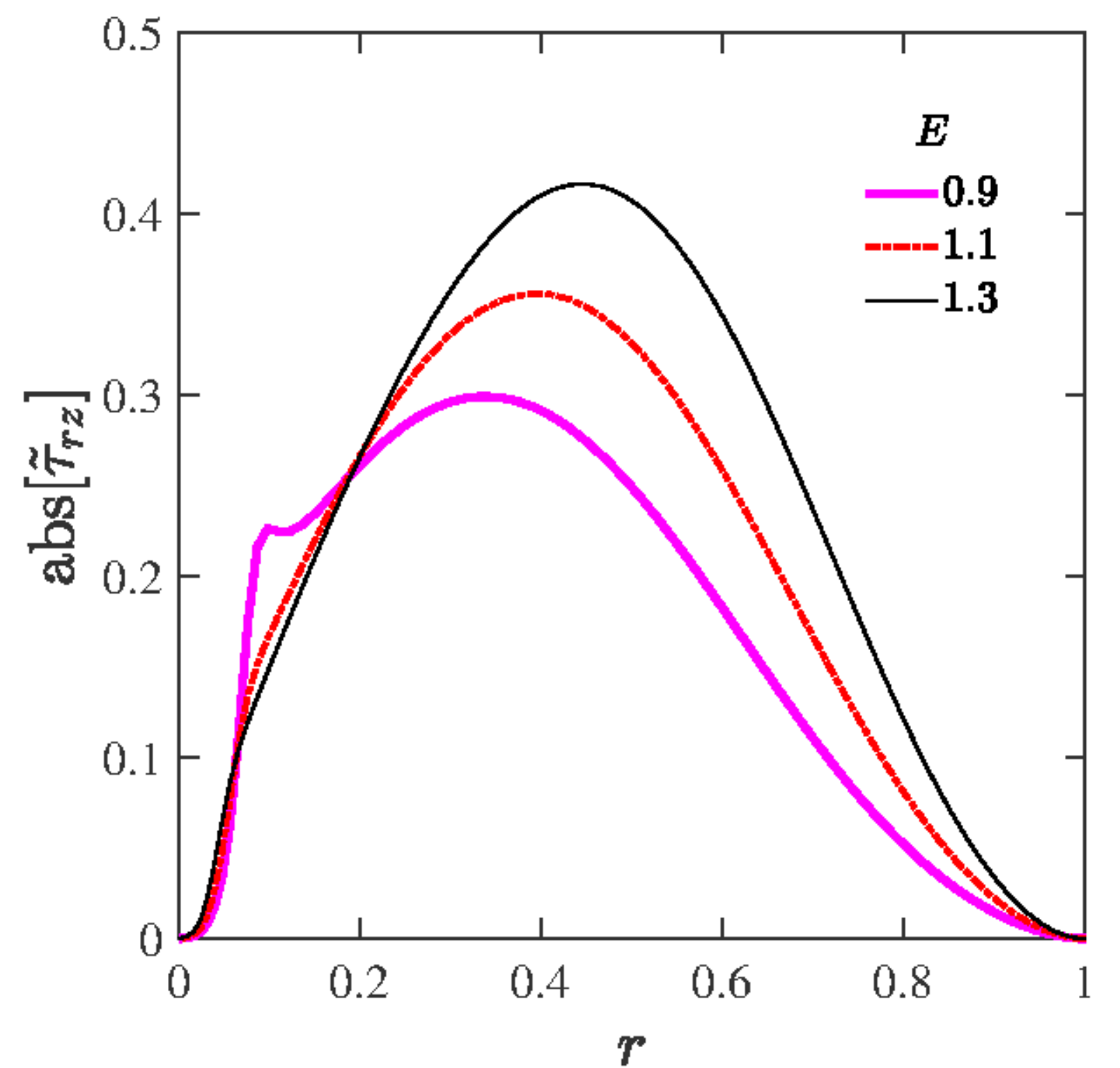}
    \caption{$rz$ polymer stress}
    \label{fig:ef_Taurz_beta0pt8E0pt003Re600k3}
  \end{subfigure}
   \begin{subfigure}[htp]{0.35\textwidth}
    \includegraphics[width=\textwidth]{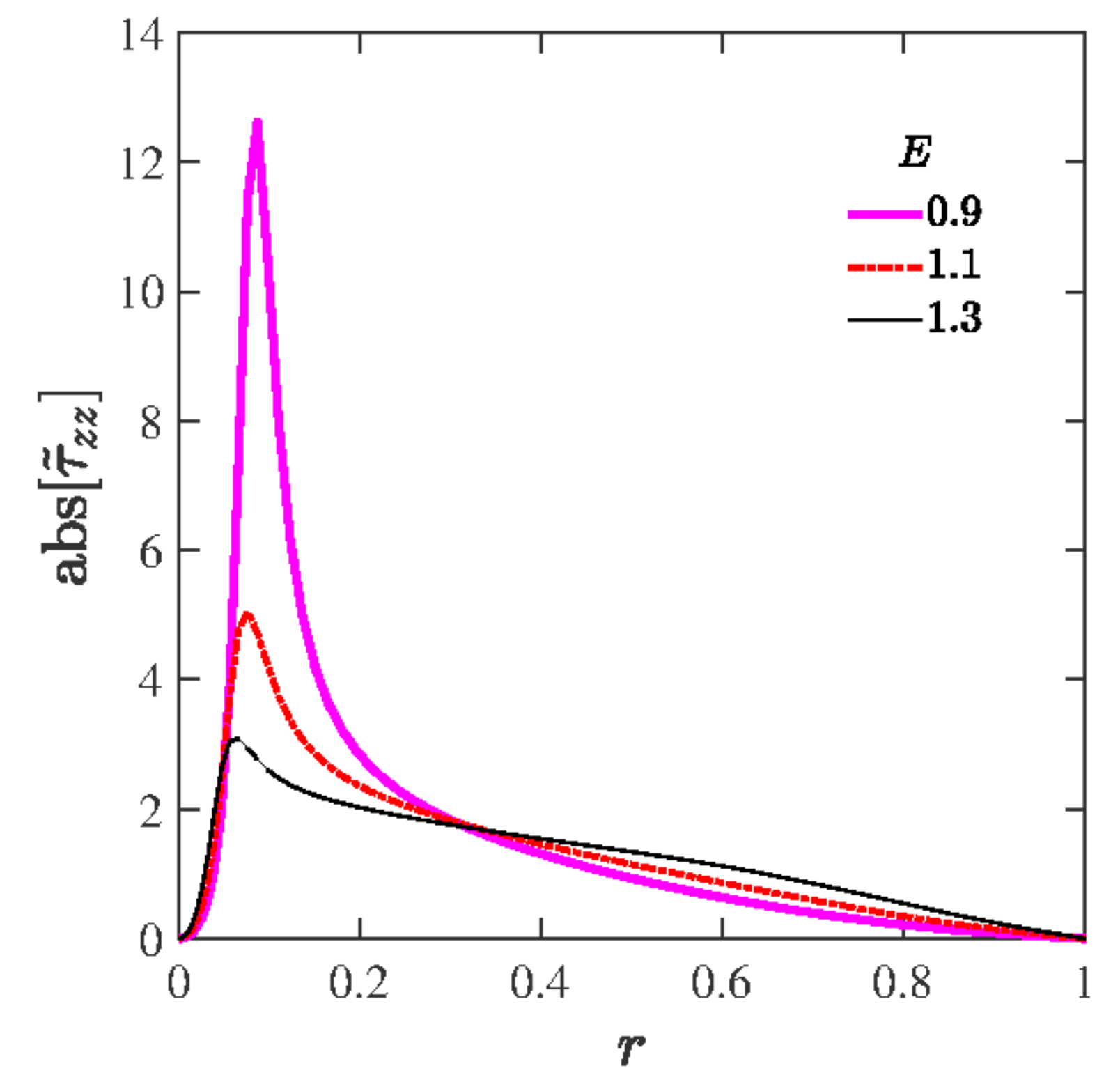}
    \caption{$zz$ polymer stress}
    \label{fig:ef_Tauzz_beta0pt8E0pt003Re600k3}
  \end{subfigure}
  \caption{Velocity and polymer stress eigenfunctions
    corresponding to the unstable centre modes 
    (in
    Fig.~\ref{fig:ues_beta_pt96_Re500_k1_zoomed})
    at different $E$ for
    $\beta = 0.96, \Rey = 500$ and $k=1$.}
  \label{fig:ef_beta0pt8EERe600k3}
\end{figure}

Figures~\ref{fig:ef_Vr_beta0pt8E0pt002Re600k3}--\ref{fig:ef_Tauzz_beta0pt8E0pt003Re600k3} show the
velocity ($v_r$ and $v_z)$  and stress ($T_{rz}$ and $T_{zz}$) 
eigenfunctions, for different $E$,  corresponding to few of the
unstable center modes shown in Fig.~\ref{fig:ues_beta_pt96_Re500_k1_zoomed}. The velocity and $T_{rz}$ eigenfunctions are 
largely insensitive to variations in $E$, but the $T_{zz}$ eigenfunction shows a distinct and sharp peak for the smaller $E$ ($ = 0.9$) near the radial location where the phase speed of the disturbances equals the local base flow velocity. 
While the amplitudes of
the axial velocity eigenfunctions in
Fig.~\ref{fig:ef_beta0pt8EERe600k3} are larger near the central core
region of the pipe,  the disturbance fields 
are nevertheless spread across the entire pipe cross-section for the parameters considered. As shown below
(Section~\ref{sec:scalings}), only for sufficiently large $Re$ ($> 1000$) does the localization of the velocity eigenfunctions near the center become prominent. It is worth emphasizing this feature here because recent studies \citep{shekar_etal_2019} have inaccurately characterized the center mode instability, analyzed in \cite{Garg2018} and the present work, as always being localized in the vicinity of the centerline regardless of $\Rey$.


\subsubsection{The origin of the center mode at fixed $\beta$ and varying $E$}
\label{sec:centermode}


\begin{figure}
  \centering
  \begin{subfigure}[htp]{0.48\textwidth}
    \includegraphics[width=\textwidth]{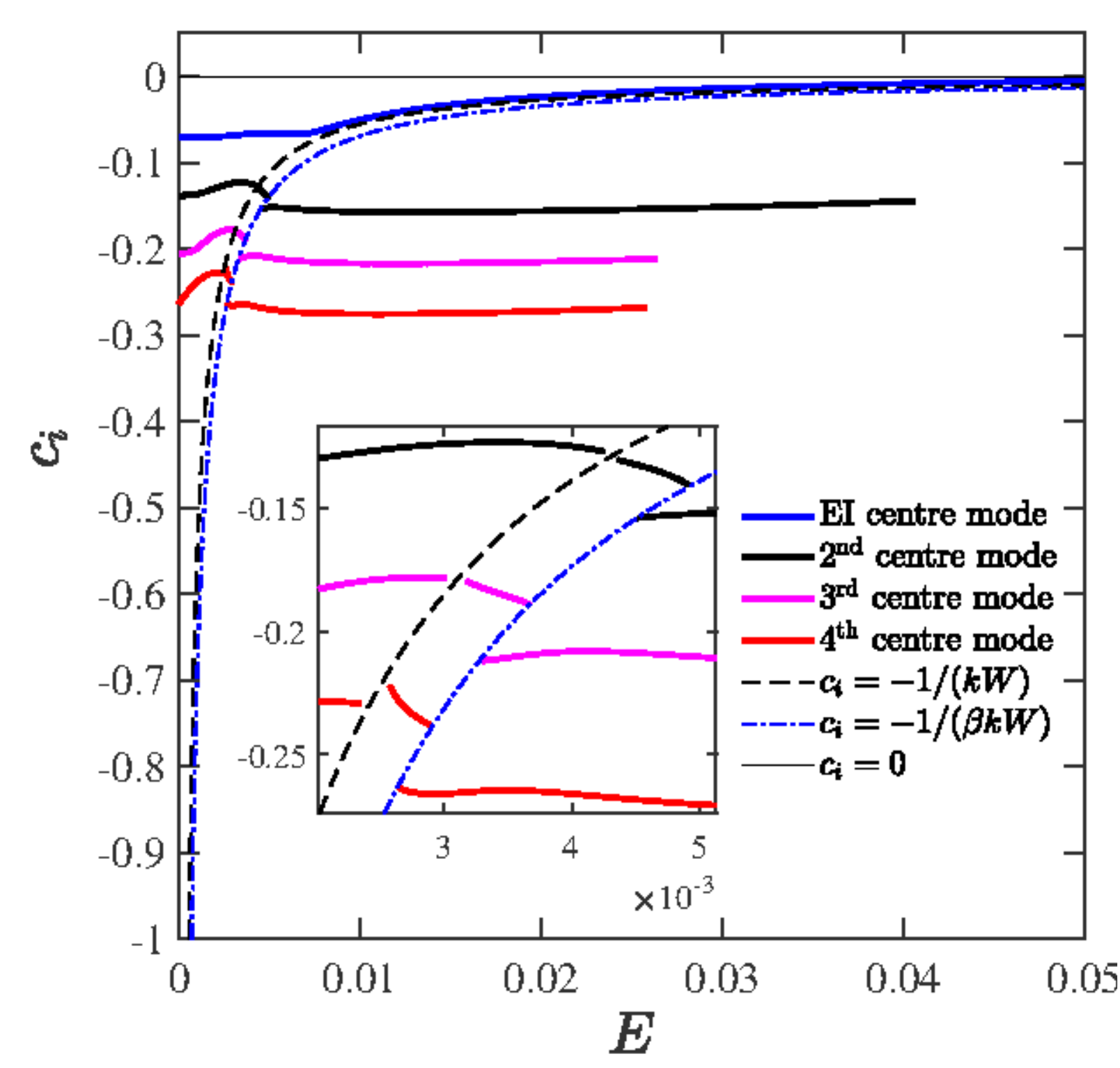}
    \caption{Growth rates}
    \label{fig:ci_vs_E_modes1to4beta0pt8Re600k3}
  \end{subfigure}
  \begin{subfigure}[htp]{0.48\textwidth}
    \includegraphics[width=\textwidth]{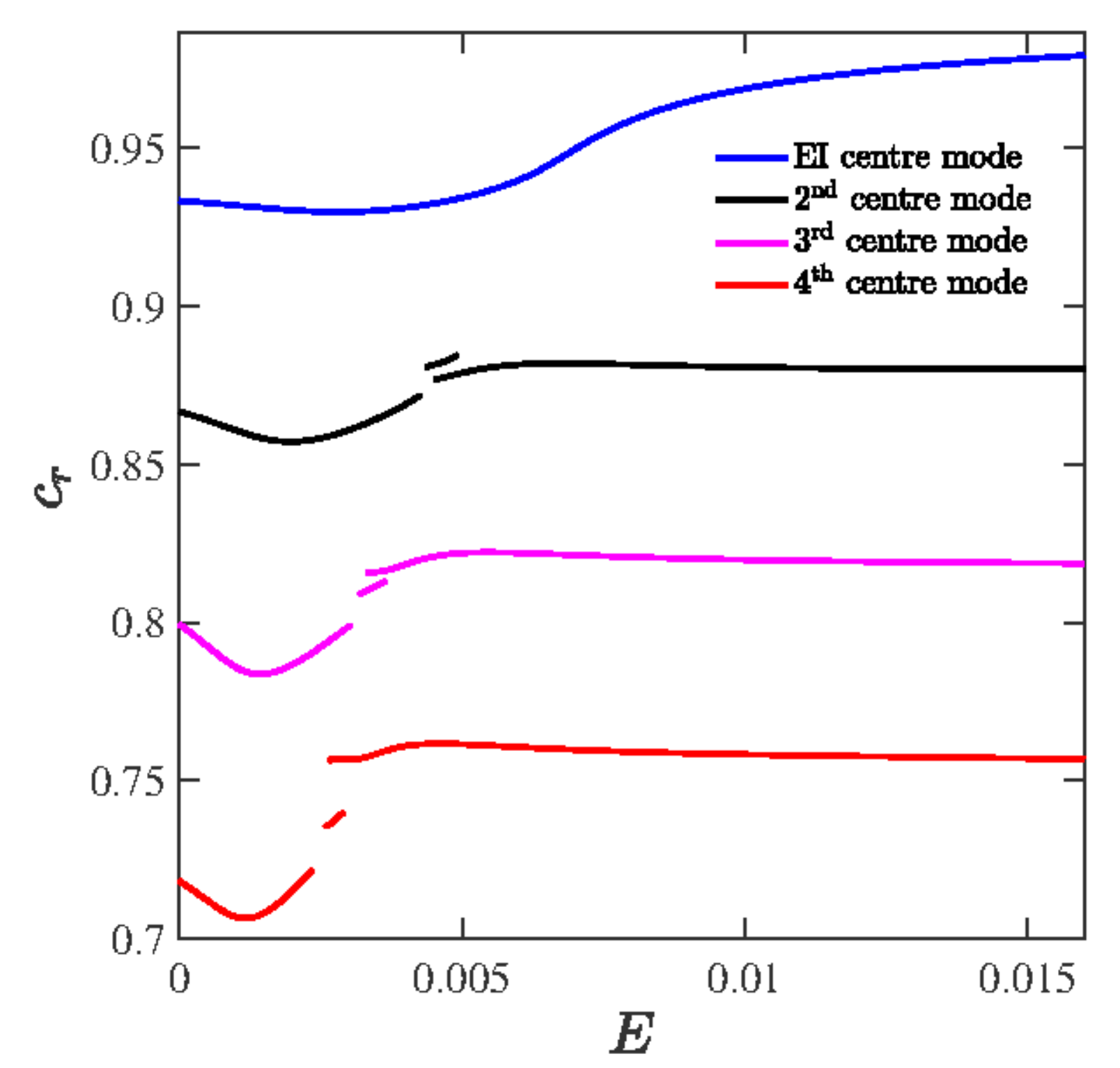}
    \caption{Phase speeds}
    \label{fig:cr_vs_E_modes1to4beta0pt8Re600k3}
  \end{subfigure}
  \caption{Effect of increasing $E$ on the first four
    least-stable center modes of
    Newtonian origin for $\beta = 0.8, \Rey = 600$ and
    $k = 3$. (a) Growth rate vs. $E$, with the inset presenting the
magnified view of the jumps suffered by the individual modes as they cross
CS1 ($c_i = -1/(kW)$) and CS2 ($c_i = -1/(\beta kW)$). 
    (b) Phase speeds
    corresponding to the modes shown in panel (a).}
  \label{fig:c_vs_E_modes1to4beta0pt8Re600k3}
\end{figure}
%

The origin of the center mode is more clearly demonstrated in  Fig.~\ref{fig:ci_vs_E_modes1to4beta0pt8Re600k3} through the variation of $c_i$ with $E$ for the
first four least-stable modes from the Newtonian P-branch, obtained using the shooting method. 
For $\beta = 0.8$, consistent with the spectra in Fig.~\ref{fig:ues_beta0pt8EERe600k3},
  the least stable
Newtonian center mode always lies above the CS (Fig.~\ref{fig:ci_vs_E_modes1to4beta0pt8Re600k3}),  smoothly
continuing with increasing $E$, eventually becoming unstable for
$E \approx 0.1$ (shown later in inset~(A) of Fig.~\ref{fig:ci_vs_E_bbRe600k3}). 
 However, the other (more) stable Newtonian center modes   (labelled $2$,$3$,$4$ in Fig.~\ref{fig:ues_beta0pt8E0pt002Re600k3}) vanish 
 into CS1 as $E$ is increased, and new modes appear
out of CS1 with further increase in $E$, subsequently suffering a second jump across the CS2 line.  The modes that emerge out of CS2 were the ones identified as and $2'$,$3'$ and $4'$ in the spectra in Fig.~\ref{fig:ues_beta0pt8EERe600k3}. 
This feature is also
evident in the variation of the phase speeds with $E$ in Fig.~\ref{fig:cr_vs_E_modes1to4beta0pt8Re600k3}. An analogous 
phenomenon was reported by \cite{chokshi_kumaran_2009} for the least
stable wall mode in plane Couette flow of an Oldroyd-B fluid.

\begin{figure}
  \centering
  \begin{subfigure}[htp]{0.48\textwidth}
   \includegraphics[width=\textwidth]{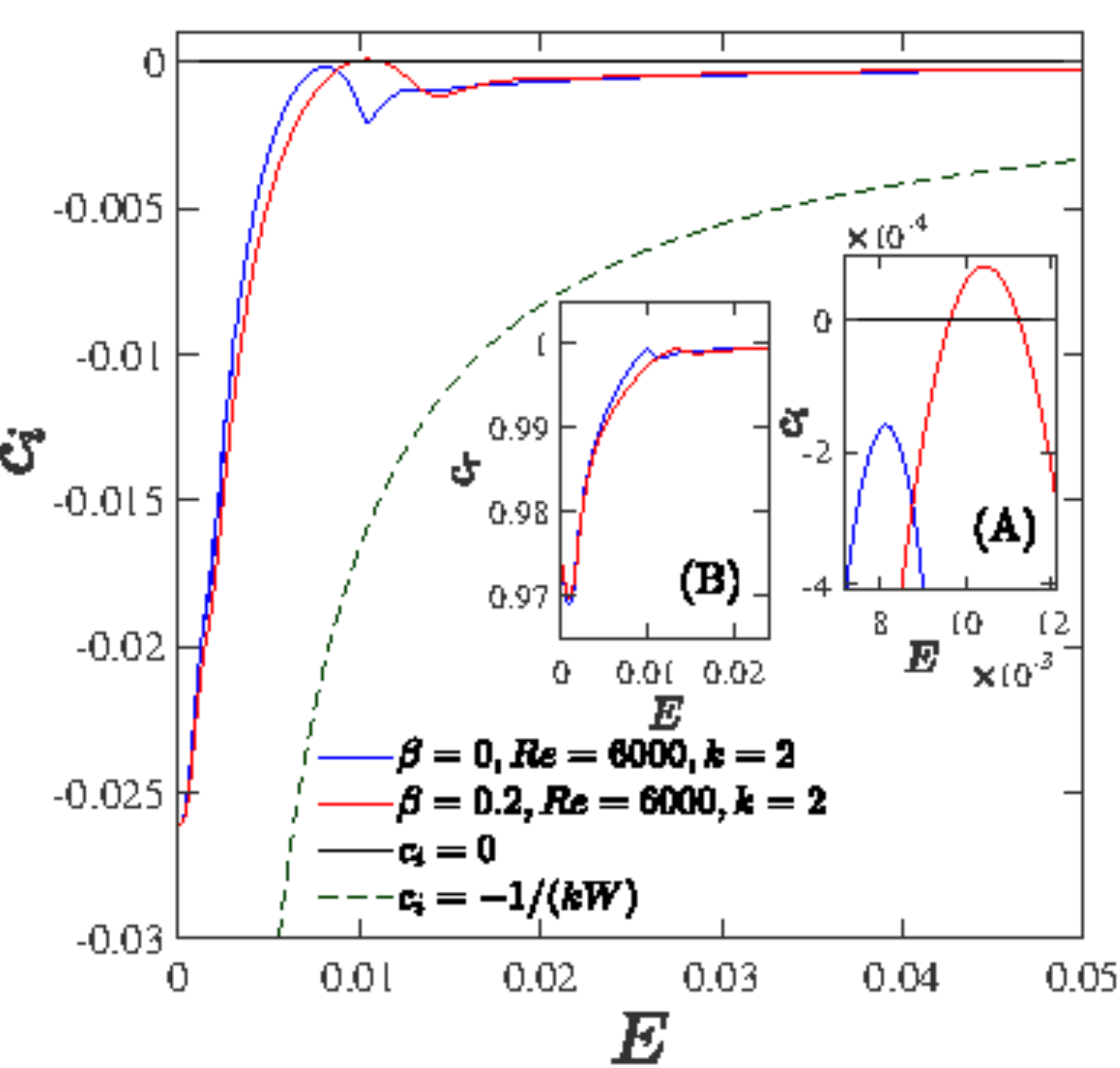}
    \caption{$\beta = 0$ and $0.2$}
    \label{fig:ci_vs_E_Re6000k2}
  \end{subfigure}
  \begin{subfigure}[htp]{0.48\textwidth}
    \includegraphics[width=\textwidth]{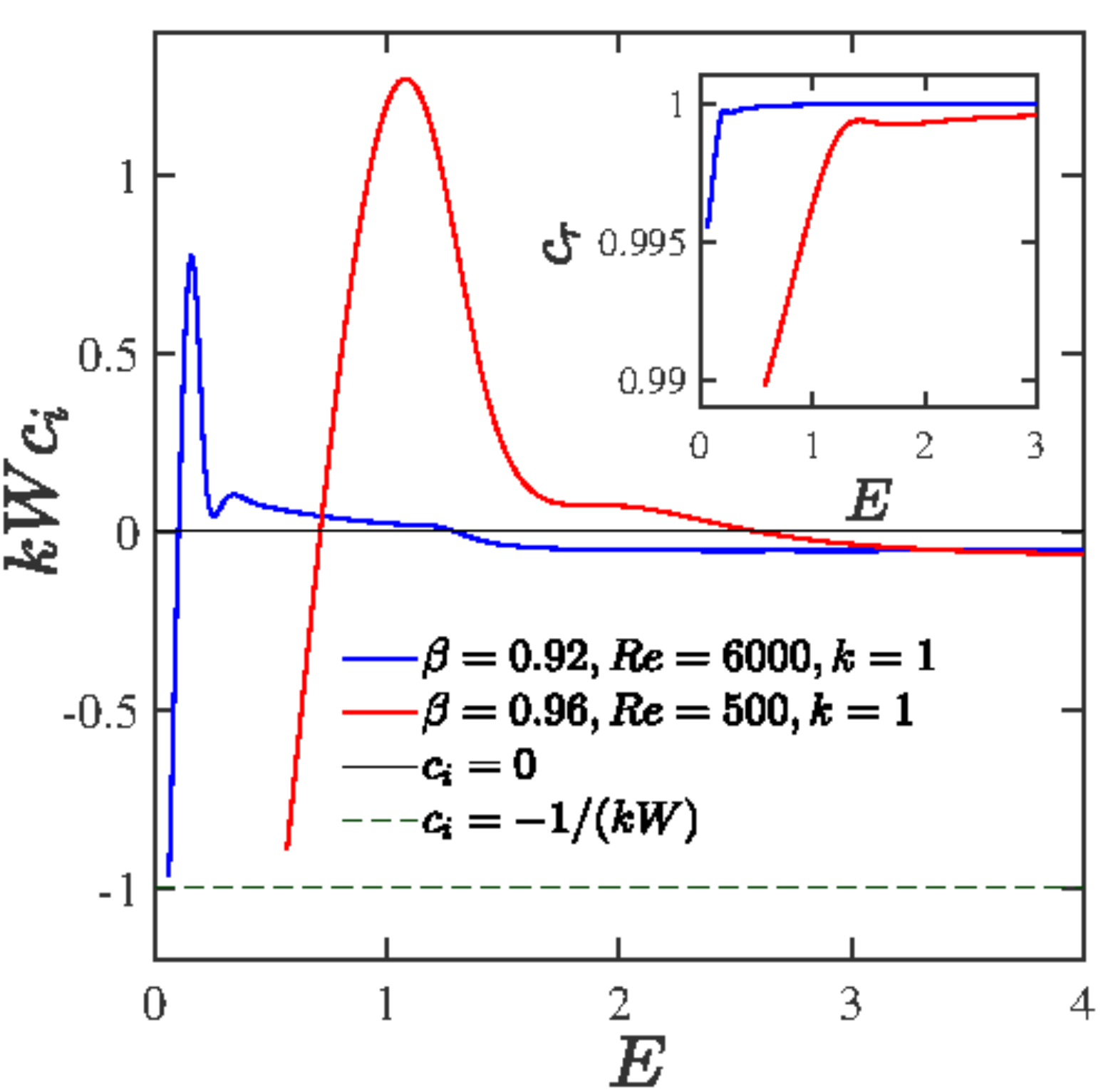}
    \caption{Scaled growth rate for $\beta = 0.92$ and $0.96$}
    \label{fig:ci_vs_E_Rk}
  \end{subfigure}
                
  \caption{Effect of increasing $E$ on the least stable Newtonian center mode for UCM and Oldroyd-B fluids: (a) $c_i$ for the least stable center mode at $Re = 6000$
    for $\beta = 0$ and $0.2$, and the inset~(A) shows the enlarged
    region near unstable range of the mode. Inset~(B) shows the corresponding phase speeds. (b) Scaled growth rate of  elasto-inertial modes at $Re = 6000$, $\beta = 0.92$ and $Re = 500$, $\beta = 0.96$, and the inset
    shows the corresponding phase speeds.}
  \label{fig:c_vs_E_Rk}
\end{figure}

In Fig.~\ref{fig:ci_vs_E_Re6000k2}, we examine the effect of
increasing $E$ in the UCM and near-UCM limits at fixed $\Rey$ and $k$ (note that, regardless of the value of $\beta$, $E = 0$ corresponds to the Newtonian limit).
For $\beta = 0$, the decay rate of the least stable Newtonian center mode decreases with increasing $E$, 
even to the point of reducing to $\sim 2 \times 10^{-4}$ at $E \approx 8 \times 10^{-3}$ (about 1/100$^{th}$ of the decay rate in the Newtonian limit), but the mode remains stable. 
Since elastic effects are responsible for the unstable center mode, 
it might be expected that this instability should persist even in the absence of solvent
contribution to the stress. 
The eigenspectra for pipe flow of a UCM fluid were computed for a
vast range of parameters $0.5 < k < 3$, $100<Re<20000$, and $0<E<1$. Unlike the spectrum for plane channel flow of a UCM fluid
\citep{sureshkumar1995linear,chaudhary_etal_2019}, only stable modes
were obtained for pipe flow of a UCM fluid subjected to axisymmetric disturbances.  Thus, as originally
stated in \cite{Garg2018}, the center
mode instability in viscoelastic pipe flow requires the combined
effects of both the polymer elasticity and solvent viscous effects, in
addition to fluid inertia.

For $\beta = 0.2$ (see Fig.~\ref{fig:ci_vs_E_Re6000k2}), however, the least stable Newtonian center
mode does become unstable for $0.01 < E < 0.011$.
For both $\beta$'s, the center mode trajectory is similar to that shown in 
Fig.~\ref{fig:c_vs_E_modes1to4beta0pt8Re600k3}, in that it remains above the CS over the range of $E$ examined. 
The corresponding phase speeds (inset~(B) of Fig.~\ref{fig:ci_vs_E_Re6000k2}), for both $\beta = 0$ and $\beta = 0.2$,  show a weak non-monotonic
behaviour with $E$, although $c_r \leq 1$ for all
$E$.  The contrasting behaviour for $\beta$ close to unity (representing dilute solutions) is shown in Fig.~\ref{fig:ci_vs_E_Rk}. The main figure shows the variation of the scaled growth rate $kWc_i$ with $E$  for two different sets of (near-unity) $\beta$ and $Re$. The instability occurs at significantly larger values of $E \sim O(1)$, in contrast to Fig.~\ref{fig:ci_vs_E_Re6000k2}, and the unstable range of $E$'s is also larger.
 In contrast to the trend for $\beta = 0.2$, the continuation of the Newtonian center modes for both $\beta = 0.92$ and $0.96$ collapses into the CS at smaller $E$'s (not shown).
 It is the trajectories of the new discrete modes, that emerge from the CS at slightly larger $E$'s, and that become unstable for $E \sim O(1)$, that are shown in Fig.~\ref{fig:ci_vs_E_Rk}.
 The corresponding
phase speeds for $\beta = 0.92$ and $0.96$ are shown in the inset of
Fig.~\ref{fig:ci_vs_E_Rk}.

\begin{figure}
  \centering
  \begin{subfigure}[htp]{0.48\textwidth}
    \includegraphics[width=\textwidth]{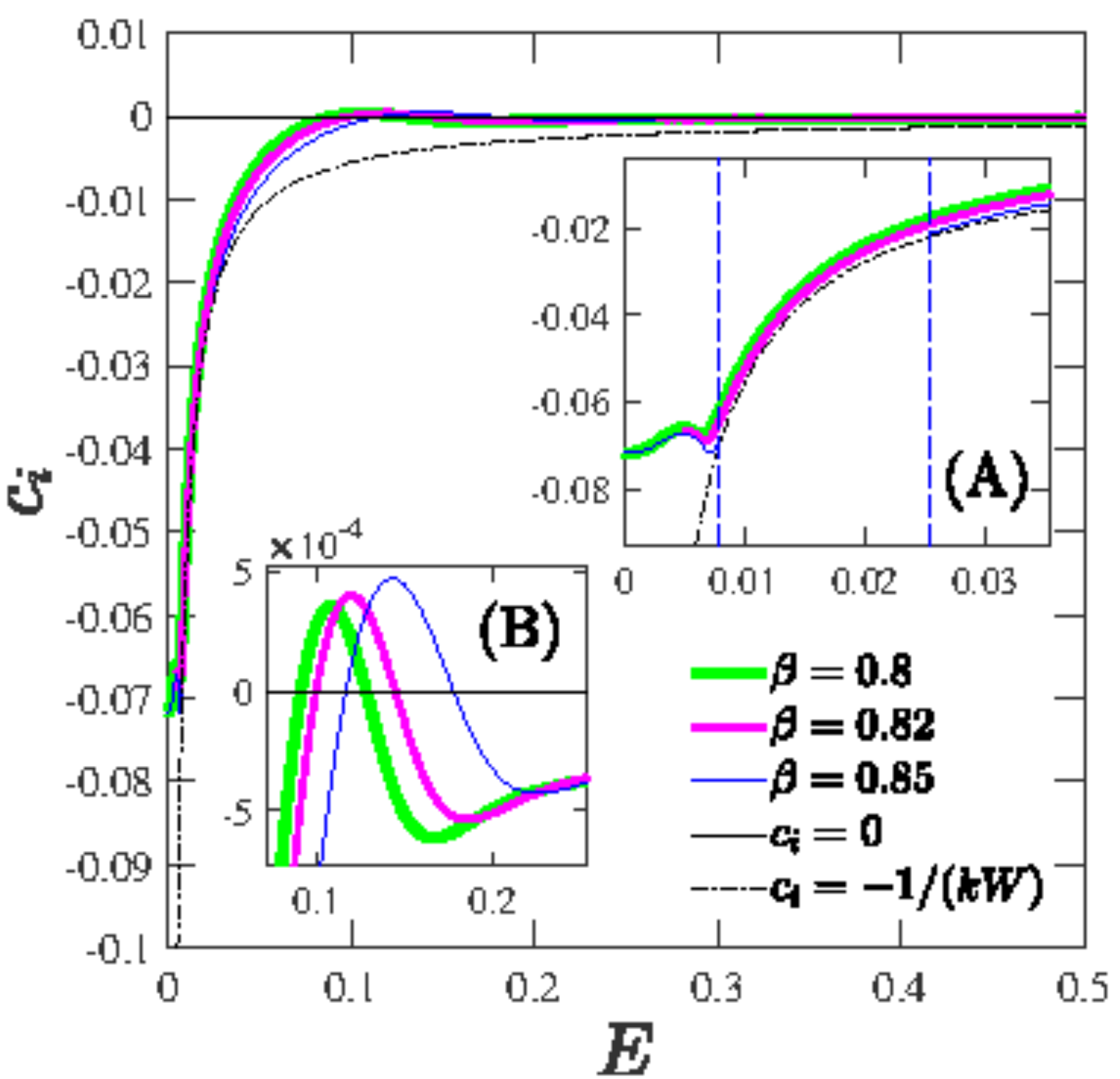}
    \caption{Growth rates}
    \label{fig:ci_vs_E_bbRe600k3}
  \end{subfigure}
  \begin{subfigure}[htp]{0.48\textwidth}
    \includegraphics[width=\textwidth]{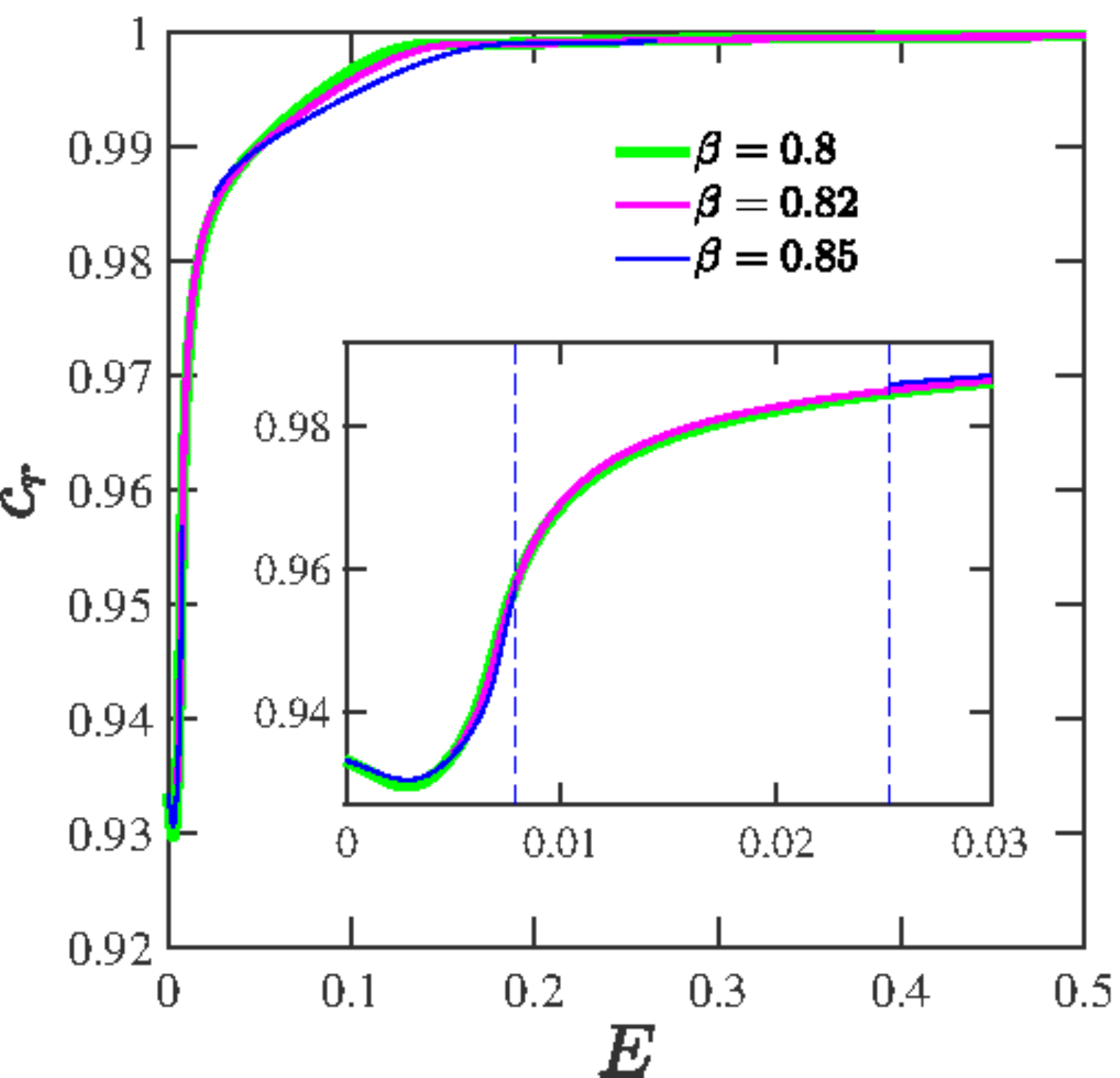}
    \caption{Phase speeds}
    \label{fig:cr_vs_E_bbRe600k3}
  \end{subfigure}
  \caption{Effect of increasing $E$ on the viscoelastic centre mode
     for fixed values of $\beta = 0.8, 0.82$ and
    $0.85, \Rey = 600$ and $k = 3$: (a) Growth rates, with inset (A) showing
    the enlarged region over the range of $E$ for which the
   centre mode is discontinuous (marked by vertical dotted lines) due to 
     CS1 with $c_i=-1/(kW)$.  
     The locations of CS2 (with $c_i = -1/(\beta k W)$) are not shown owing to the closely-separated values of $\beta$ used in this figure. Inset (B) shows the region where the
     centre mode is unstable for all the three values of
    $\beta = 0.8, 0.82$ and $0.85$. (b) Phase speeds corresponding to
    the modes shown in panel (a). Inset in panel~(b) shows the enlarged region near $E \rightarrow 0$.}
  \label{fig:c_vs_E_bbR600k3}
\end{figure}

In Figs.~\ref{fig:c_vs_E_modes1to4beta0pt8Re600k3} and \ref{fig:c_vs_E_Rk}, 
 we have seen two different trajectories for the center mode, as a function of $E$, depending on $\beta$. In order to clarify the change in the nature of the center mode trajectory - from a continuous variation of $c_i$ with increasing $E$ at smaller $\beta$, to a discontinuous variation for near-unity $\beta$ - 
Fig.~\ref{fig:ci_vs_E_bbRe600k3} shows the behaviour of the center mode for $\beta = 0.8$, $0.82$ and $0.85$. The center mode trajectory remains above CS1 until instability, for both $\beta = 0.8$ and $0.82$, while for $\beta = 0.85$, the center mode disappears into the CS at $E \approx 0.009$ (inset~(A) of Fig.~\ref{fig:ci_vs_E_bbRe600k3}). Thus, in this case, there exists a range $0.009\lessapprox E \lessapprox 0.024$ where the center mode does not
exist.  This range, which extends from the
 point of encounter of this
mode with CS1 to the point of emergence of the new mode from CS1 at higher $E$,  
varies with increasing $\beta$.
Evidently, the critical $\beta$, below which the center mode is a smooth continuation of the least stable Newtonian center mode, lies somewhere between  $0.82$ and $0.85$ (for $Re = 600$ and $k = 3$). Note that, despite the discontinuous transition in terms of the collapse into the CS's, the interval of instability in $E$ varies smoothly with increasing $\beta$ (the inset~(B) in Fig.~\ref{fig:ci_vs_E_bbRe600k3}). Figure~\ref{fig:cr_vs_E_bbRe600k3} shows the corresponding
phase speeds, and the enlarged region in the inset shows that the
trend for $c_r$ vs $E$ curves is more or less same for
$\beta = 0.8, 0.82$ and $0.85$, except for $\beta=0.85$,
where the absence of the mode in the interval $0.009\lessapprox E \lessapprox 0.024$ leads to a gap in the $c_r$ curve.

\begin{figure}
  \centering
  \includegraphics[width = 6cm]{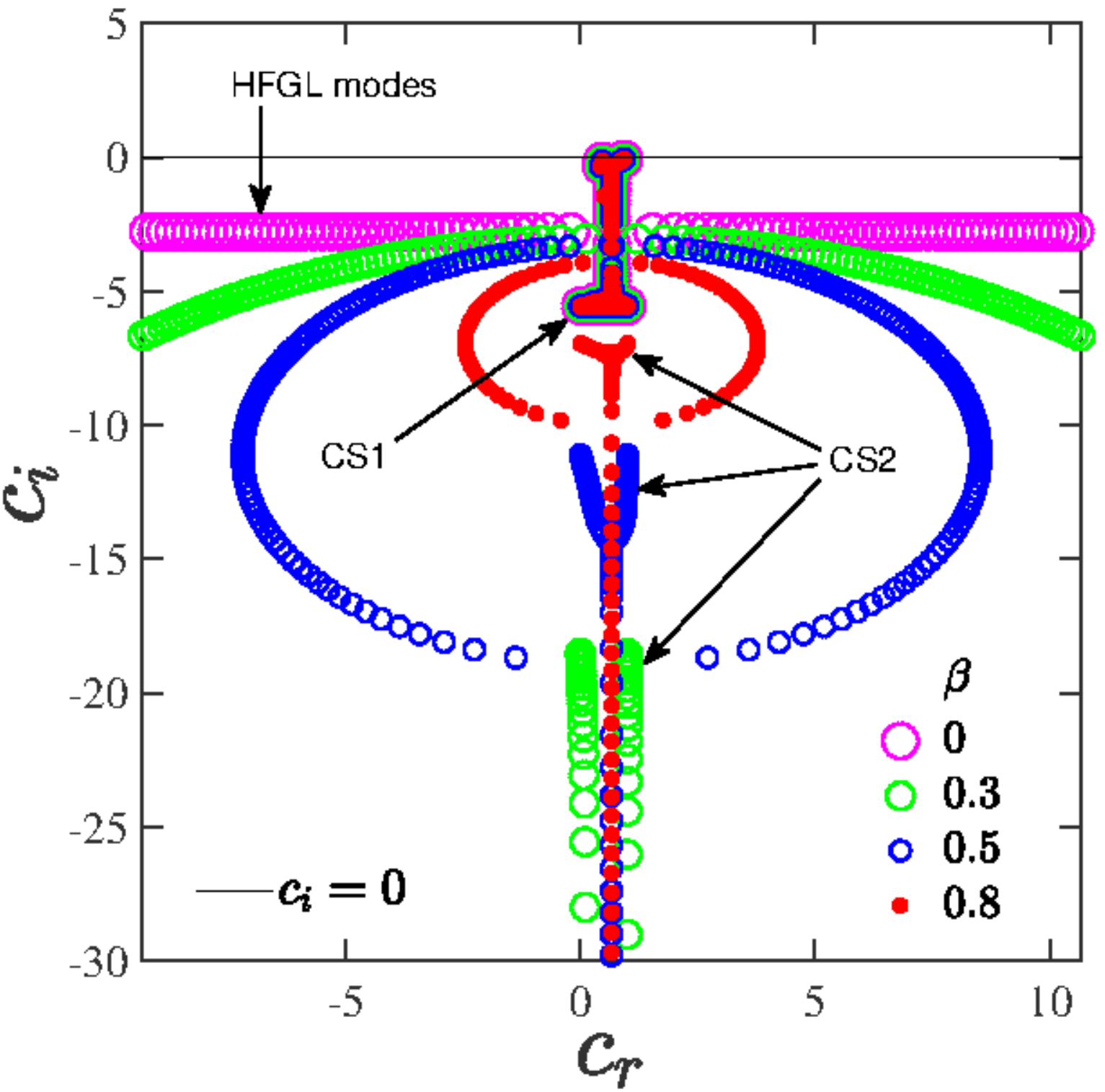}
  \caption{Eigenspectra for $E = 10^{-4}$, $Re = 600$, and $k = 3$ 
demonstrating the bending down of the HFGL modes with increasing $\beta$, leading to the appearance of a ring-like structure for the higher $\beta$'s.  The vertical location of CS1 is independent of $\beta$, while that of CS2 varies with $\beta$.}
  \label{fig:hfgltoring}
\end{figure}

%
%


\subsection{Spectra at fixed $E$ and different $\beta$}
\label{subsec:varyingbeta}
\begin{figure}
        \centering
        \begin{subfigure}[htp]{0.4\textwidth}
                \includegraphics[width=\textwidth]{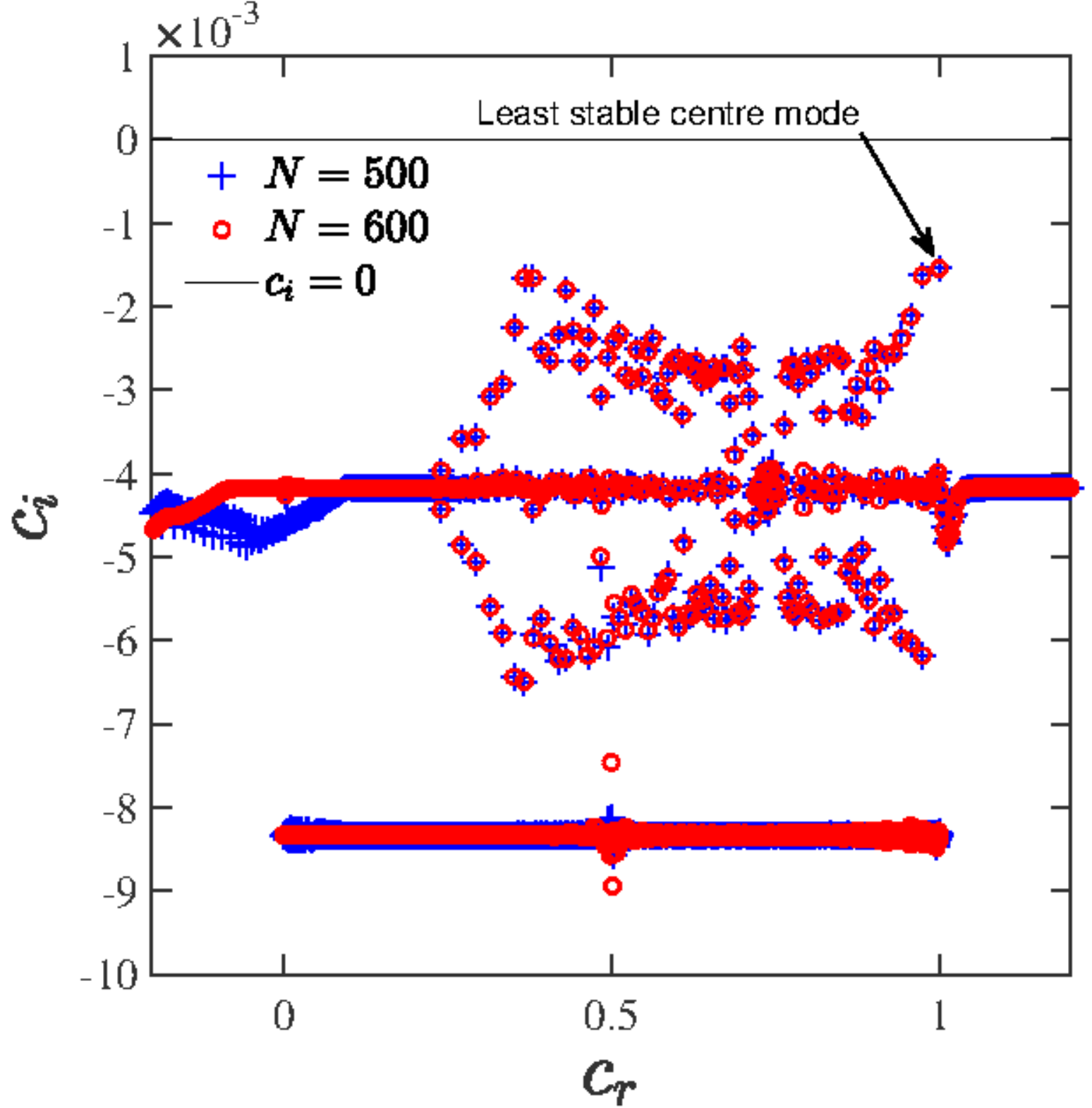}
                \caption{$\beta = 0$, UCM}
                \label{fig:es_beta0}
        \end{subfigure}
        \begin{subfigure}[htp]{0.4\textwidth}
                \includegraphics[width=\textwidth]{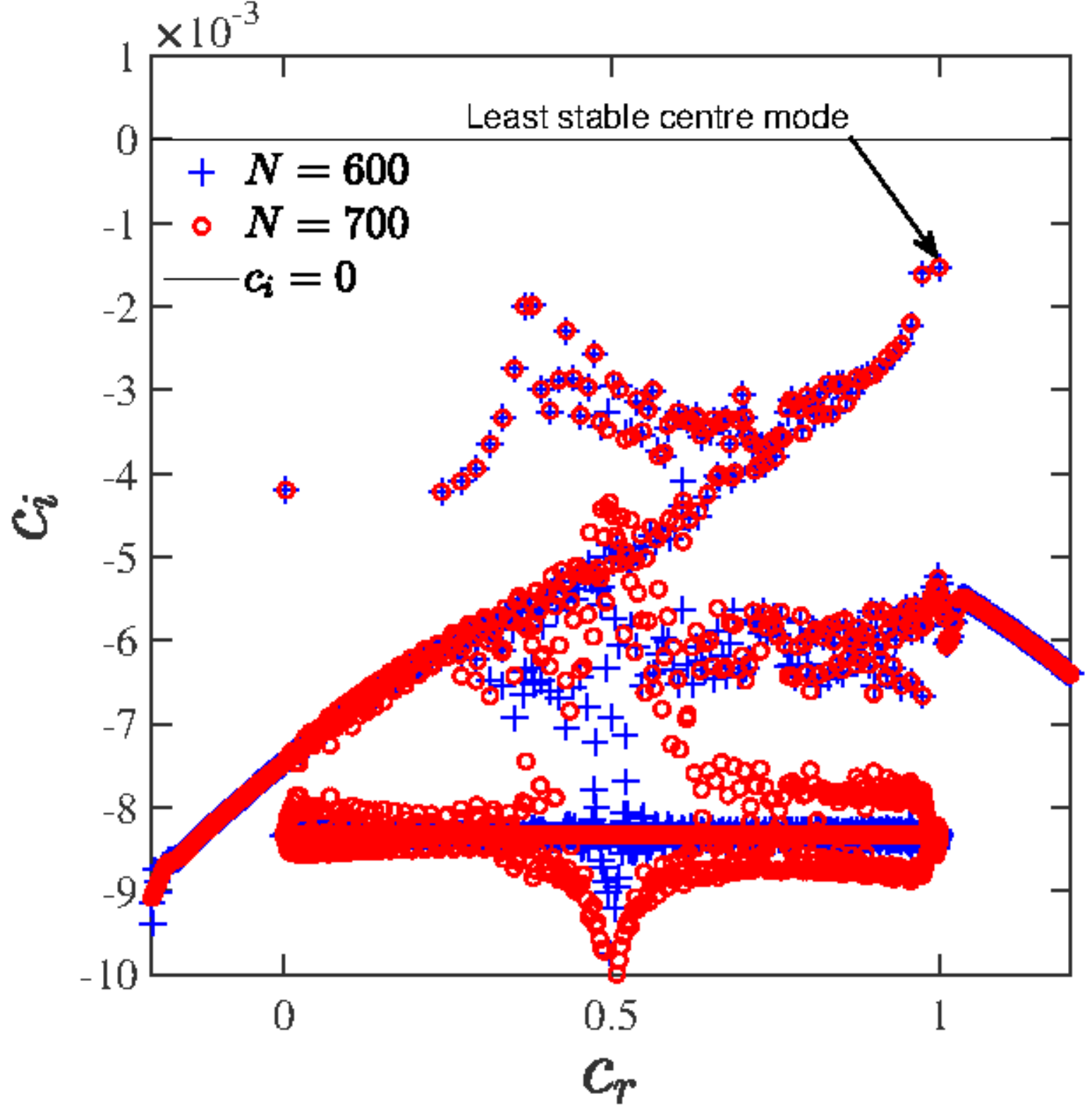}
                \caption{$\beta = 1\times 10^{-4}$}
                \label{fig:es_beta0pt0001}
        \end{subfigure}
        \begin{subfigure}[htp]{0.4\textwidth}
                \includegraphics[width=\textwidth]{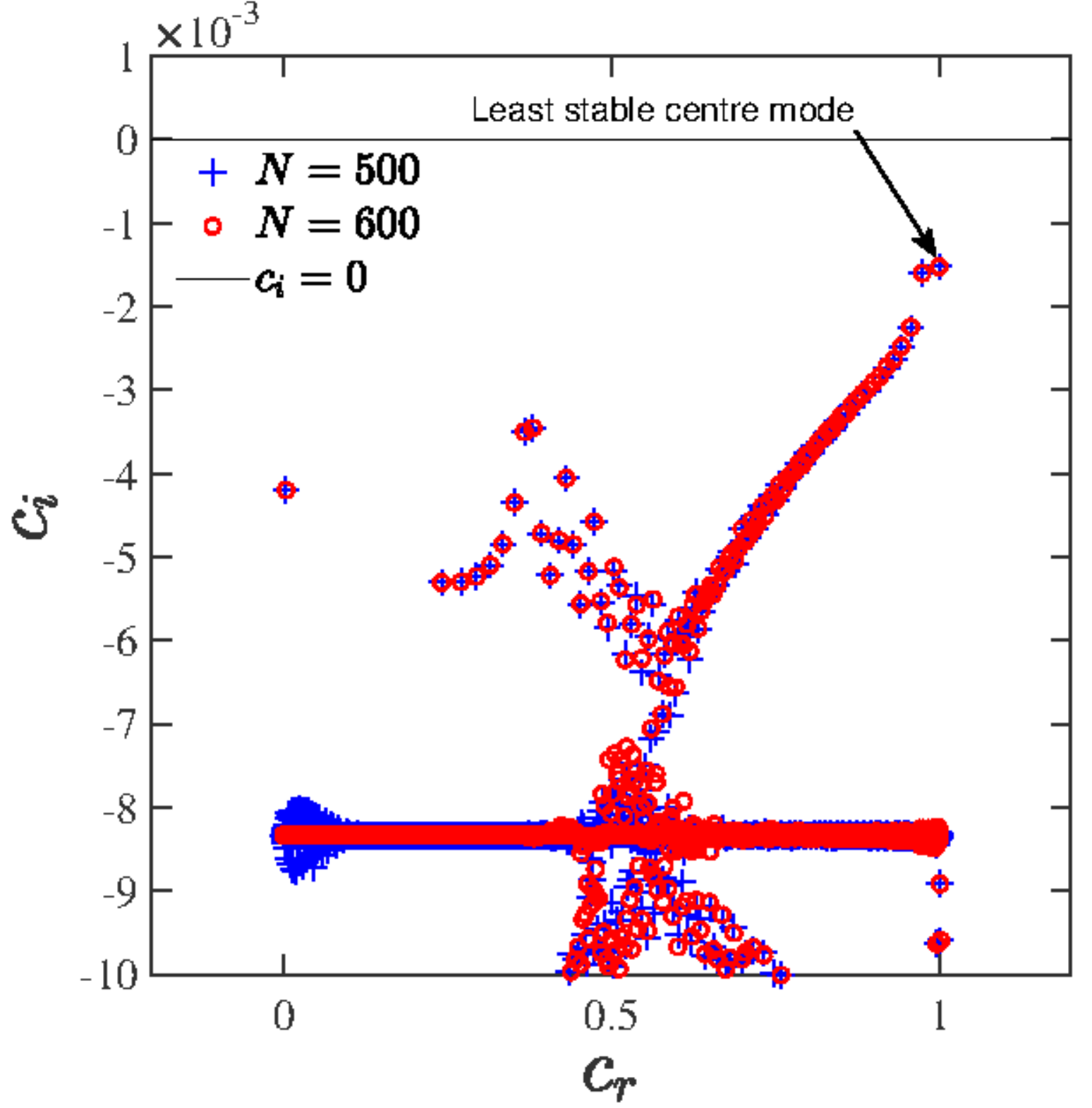}
                \caption{$\beta = 5\times 10^{-4} $}
                \label{fig:es_beta0pt0005}
        \end{subfigure}
        \begin{subfigure}[htp]{0.4\textwidth}
                \includegraphics[width=\textwidth]{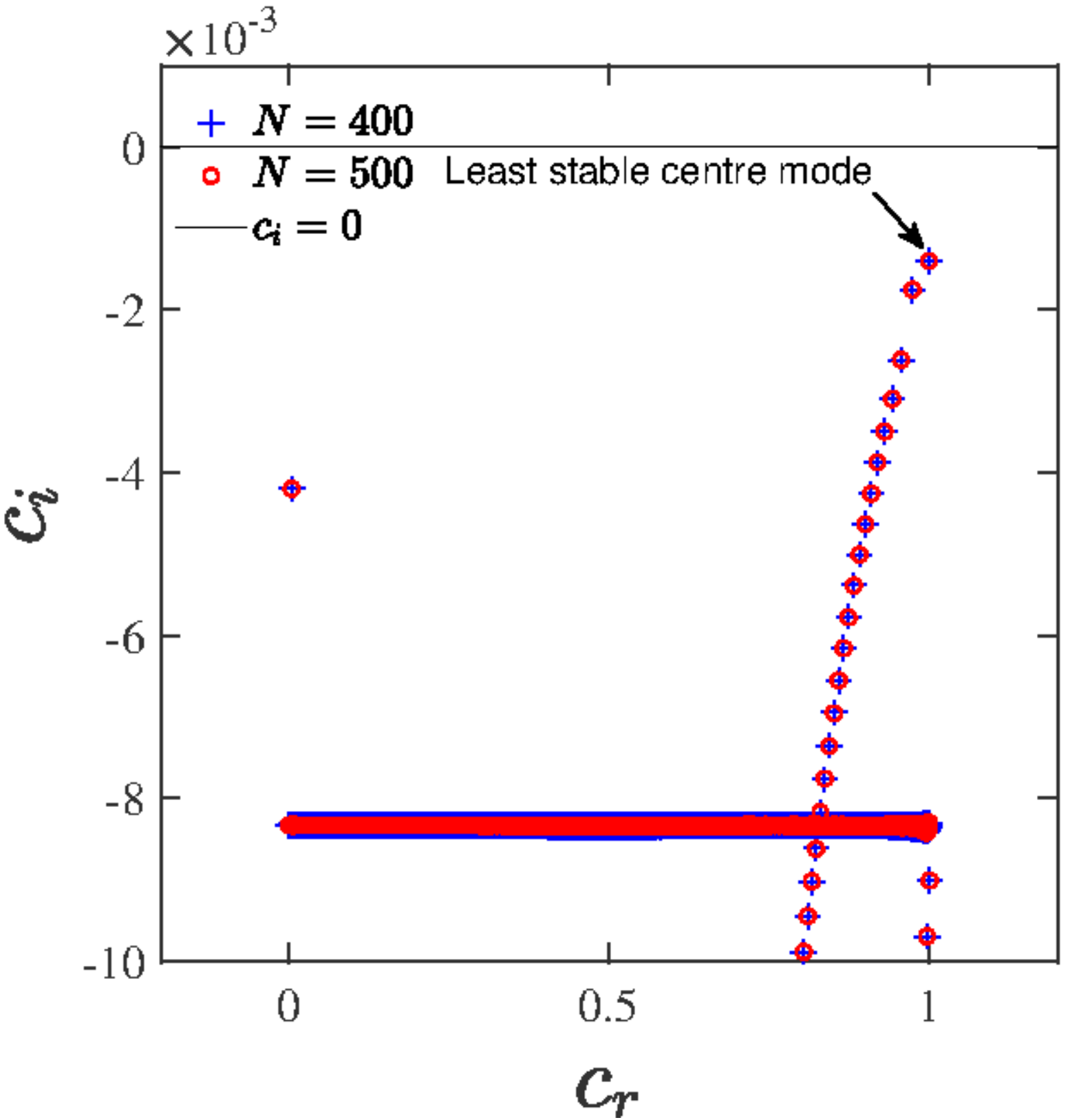}
                \caption{$\beta = 5\times 10^{-3} $}
                \label{fig:es_beta0pt005}
        \end{subfigure}
        \begin{subfigure}[htp]{0.4\textwidth}
                \includegraphics[width=\textwidth]{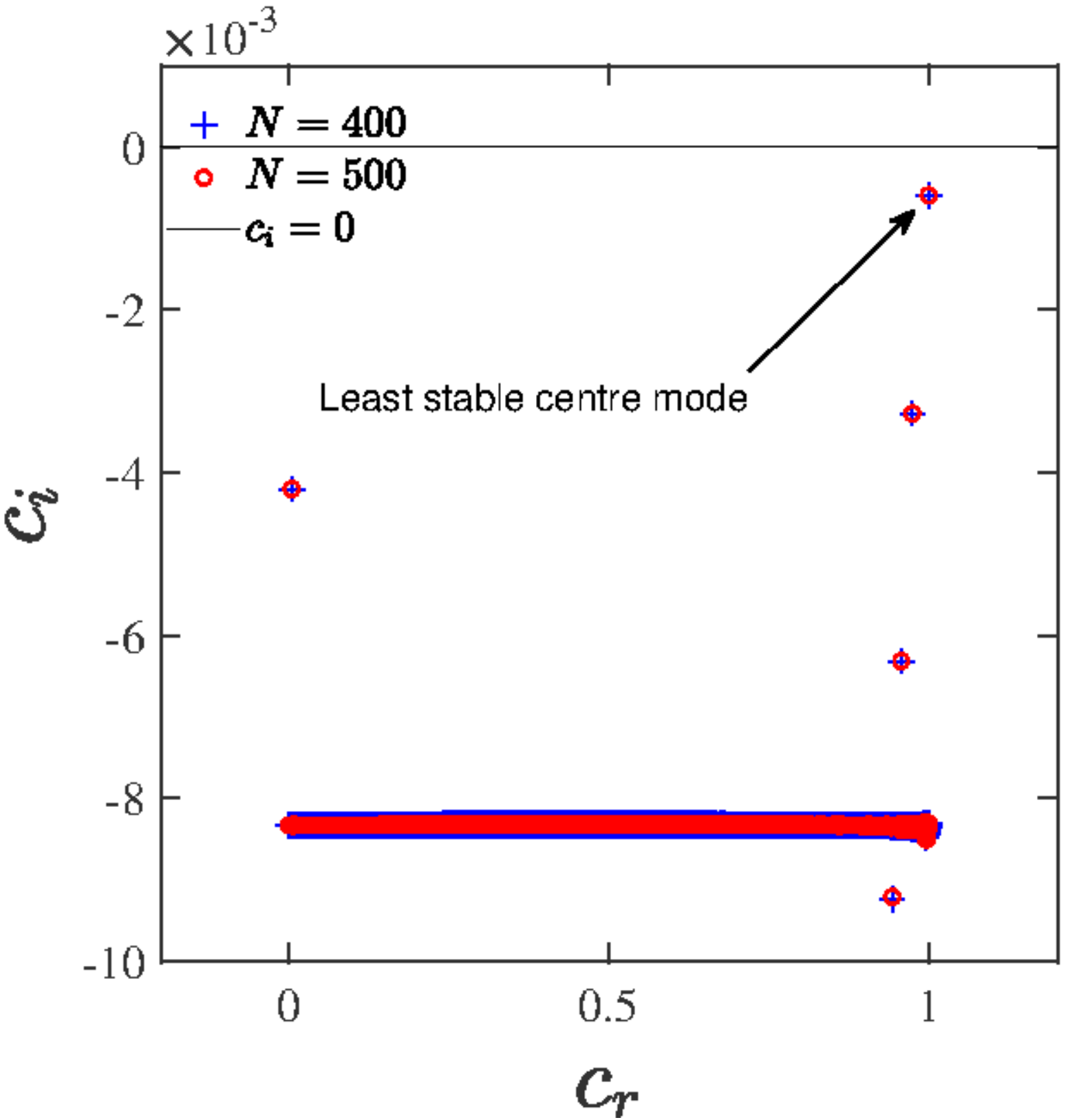}
                \caption{$ \beta = 5\times 10^{-2} $}
                \label{fig:es_beta0pt05}
        \end{subfigure}
        \begin{subfigure}[htp]{0.4\textwidth}
                \includegraphics[width=\textwidth]{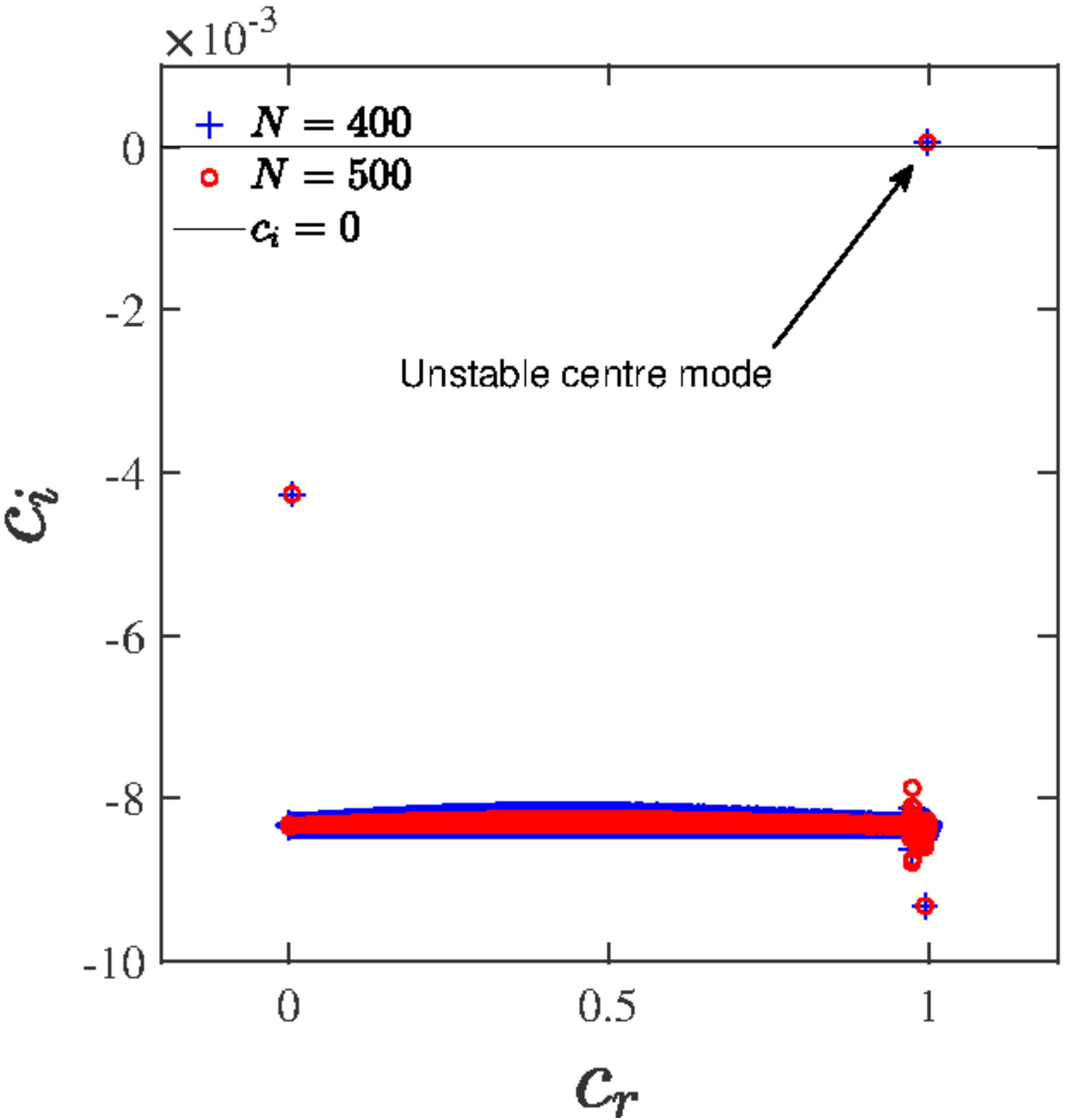}
                \caption{$\beta = 0.2$}
                \label{fig:es_beta0pt2}
        \end{subfigure}
                
        \caption{Unfiltered viscoelastic  eigenspectra for $E=0.01, \Rey=6000$, $k=2$ and different $\beta$. The decay rate of the least stable  centre mode in the UCM limit decreases with increase in $\beta$ and the center mode eventually becomes unstable at $\beta = 0.2$.}
                \label{fig:es_beta_variation_E0pt01k2}
\end{figure}

We next examine the viscoelastic eigenspectra as $\beta$ is  increased from zero, at fixed $E$.
We begin with Fig.~\ref{fig:hfgltoring} which illustrates the effect of increasing $\beta$, starting from the UCM spectrum, at $E = 10^{-4}$. A moderate $Re$ ($=600$) is chosen in order to keep the spectral features  relatively simple, requiring only a modest resolution (the number of collocation points $N$), and thereby allowing us to focus on the large-scale features. Figure~\ref{fig:hfgltoring} illustrates the singular feature of the bending down of the HFGL line for non-zero $\beta$. 
The bending down can be interpreted as a (very strong) stabilization of these modes due to the solvent viscosity.
For the larger $\beta$'s ($\beta = 0.5$ and $0.8$), the bending is `complete', leading to the ring-like structure within the range of $c_i$'s examined; this then clarifies the origin of the structure seen before in
Fig.~\ref{fig:ring}.  Figure~\ref{fig:es_beta_variation_E0pt01k2} shows spectra at a higher $Re$ ($=6000$), for different $\beta$, and with $E = 0.01$. The spectrum for $\beta = 0$ (Fig.~\ref{fig:es_beta0}) now has a more intricate structure, necessitating a zoomed view into the phase speed interval $(0,1)$. The features of the high-$Re$ UCM channel-flow spectrum were first explained in \cite{chaudhary_etal_2019}, and include 
CS1  (which appears as a balloon owing to the finite resolution), the  HFGL, and 
additional discrete modes with $c_r \in [0,1]$ which
lie on either side of the HFGL line , rougly along the contours of an `hourglass'. These  features of the UCM pipe-flow spectrum, in Fig.~\ref{fig:es_beta0}, are analogous to the channel flow case above.

Figures~\ref{fig:es_beta0pt0001}--\ref{fig:es_beta0pt005} show
the spectra for $\beta$ in the range $10^{-4}$--$5\times 10^{-3}$. 
Figure~\ref{fig:es_beta0pt0001} shows that even the smallest $\beta$ has a profound effect on the HFGL modes. In contrast to the UCM spectrum at $\Rey = 600$ (Fig.~\ref{fig:es_beta0}), where the bending of HFGL line became evident only for $c_r$'s well outside the base-state interval, the bending down of the HFGL modes is evident at $Re = 6000$ even for $c_r \in (0,1)$ - see Figs.~\ref{fig:es_beta0pt0001} and \ref{fig:es_beta0pt0005}. The bent HFGL line has all but disappeared as $\beta$ is increased to $10^{-3}$ (Fig.~\ref{fig:es_beta0pt0005}), again demonstrating that the HFGL modes are
rapidly damped by small amounts of solvent viscosity. Due to this
drastic stabilization even at rather small $\beta$, the HFGL modes in the original UCM spectrum become
irrelevant to the parametric regimes (corresponding to relatively
dilute solutions, with $\beta \sim 0.6$ and higher) explored later in
this study.  Further, the `density' of stable modes present in the 
hourglass structure in the UCM limit also decreases rapidly as $\beta$ is increased from zero, with the hourglass structure virtually absent for $\beta = 0.05$. 
Most importantly, while almost all other modes in the hourglass structure of the UCM spectrum are rapidly stabilized
with increasing $\beta$
(Figs.~ \ref{fig:es_beta0pt005}--\ref{fig:es_beta0pt2}), the least stable center mode (with
$c_r\approx 1$ and $c_i \rightarrow 0$) is rather unaffected
by the small increase in $\beta$. 
In fact, as shown in Figs~\ref{fig:es_beta0pt05} and \ref{fig:es_beta0pt2}, for the largest $\beta$ shown ($\beta = 0.2$), the center mode becomes unstable. Thus, as originally stated in Fig.~\ref{fig:ci_vs_E_Re6000k2}, it appears that all three effects, viz., elasticity, solvent viscous stresses and fluid inertia are important ingredients for the instability of the center mode in viscoelastic pipe flow.

\begin{figure}
  \centering
  \begin{subfigure}[htp]{0.48\textwidth}
    \includegraphics[width=\textwidth]{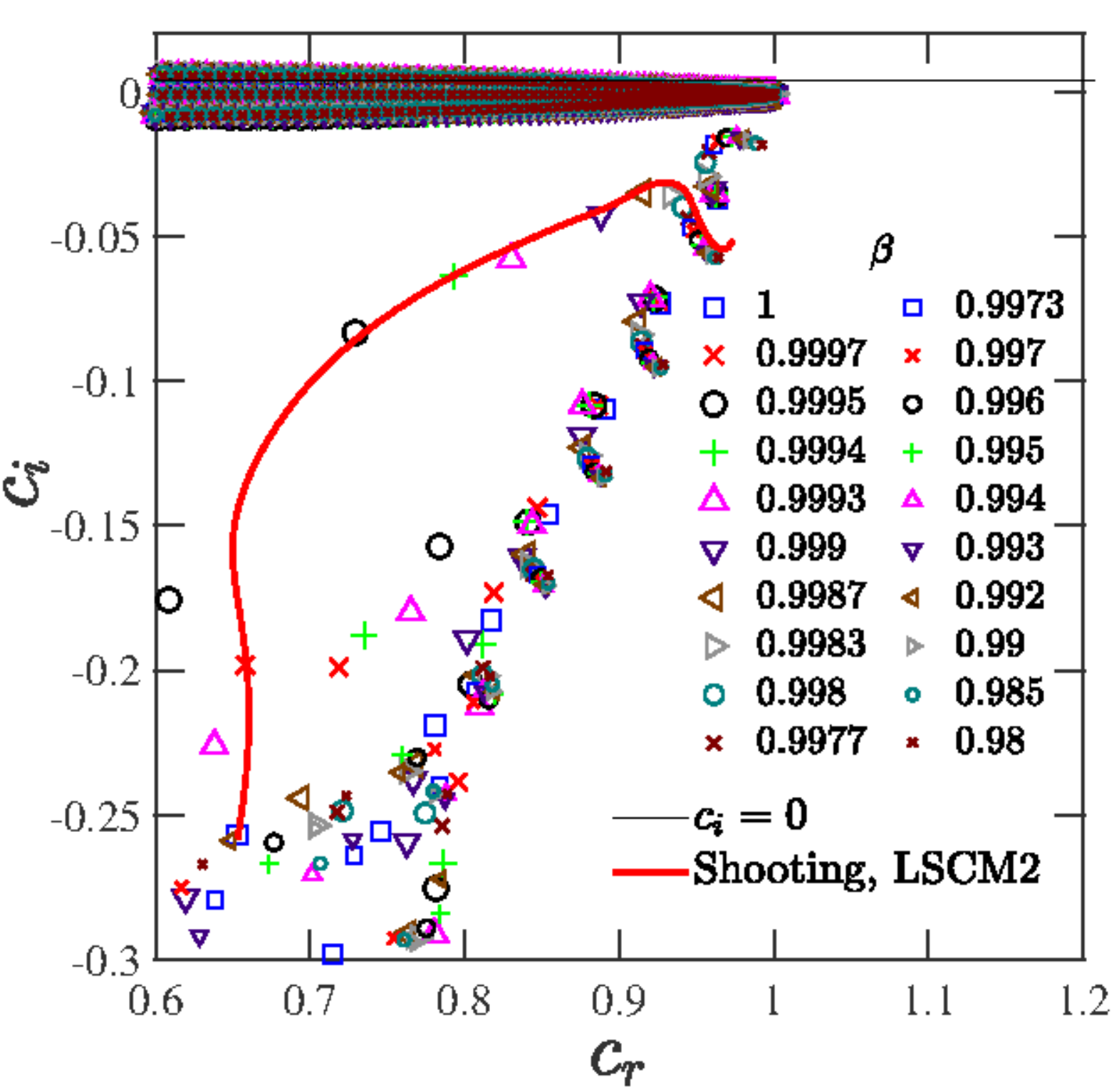}
    \caption{$(1-\beta) \in 0$ to $0.02$}
    \label{fig:es_E0pt15Re6000k1_beta0pt98to1}
  \end{subfigure}
  \begin{subfigure}[htp]{0.48\textwidth}
    \includegraphics[width=\textwidth]{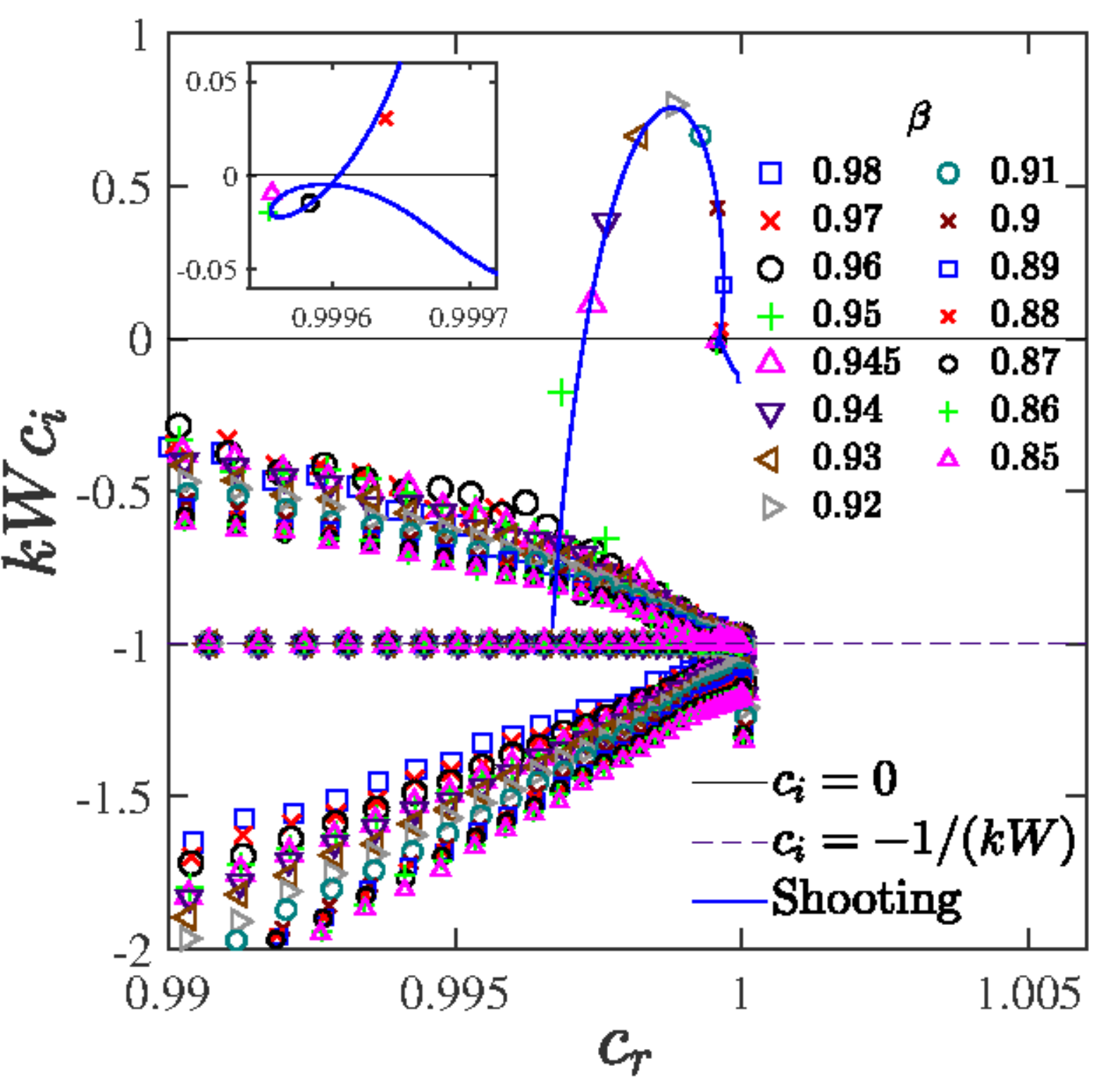}
    \caption{$(1-\beta) \in 0.02$ to $0.15$}
    \label{fig:es_E0pt15Re6000k1_beta0pt85to0pt98}
  \end{subfigure}
   
  \caption{Viscoealstic pipe flow eigenspectra for 
    $E=0.15, \Rey=6000, k=1$, and for varying $\beta$
    (a) $(1-\beta)=0$ to $0.02$, and (b) $(1-\beta)=0.02$ to
    $0.15$. Panel (a) shows the region in the vicinity of the $P$-branch.
    Panel (b) focuses on the trajectory of the center
    mode which becomes unstable for $\beta \in (0.88,0.945)$.}
  \label{fig:es_E0pt15Re6000k1_beta_near1}
\end{figure}

\begin{figure}
  \centering
  \begin{subfigure}[htp]{0.35\textwidth}
    \includegraphics[width=\textwidth]{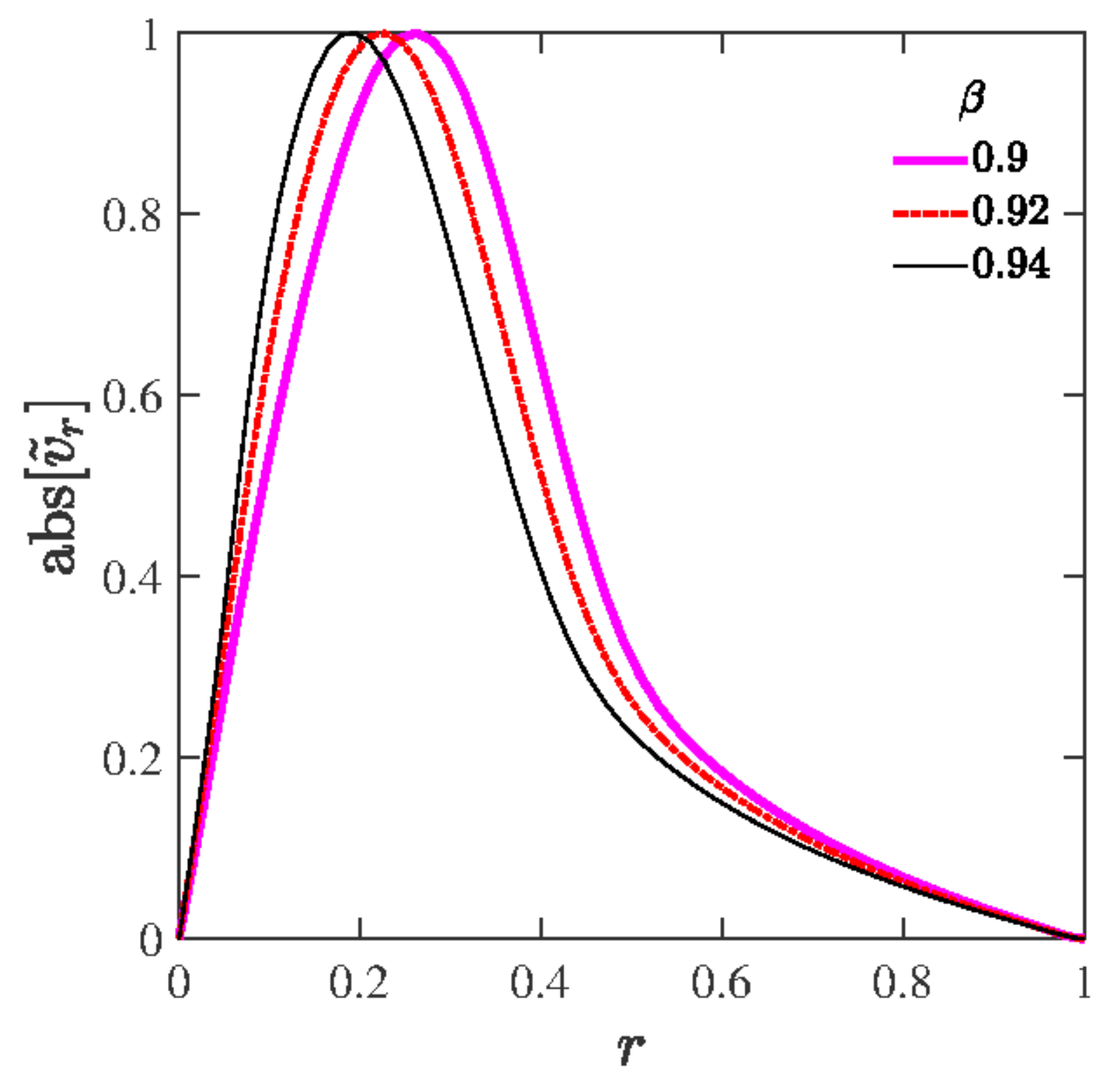}
    \caption{Radial velocity}
    \label{fig:ef_Vr_beta0pt8E0pt002Re6000k1}
  \end{subfigure}
  \begin{subfigure}[htp]{0.35\textwidth}
    \includegraphics[width=\textwidth]{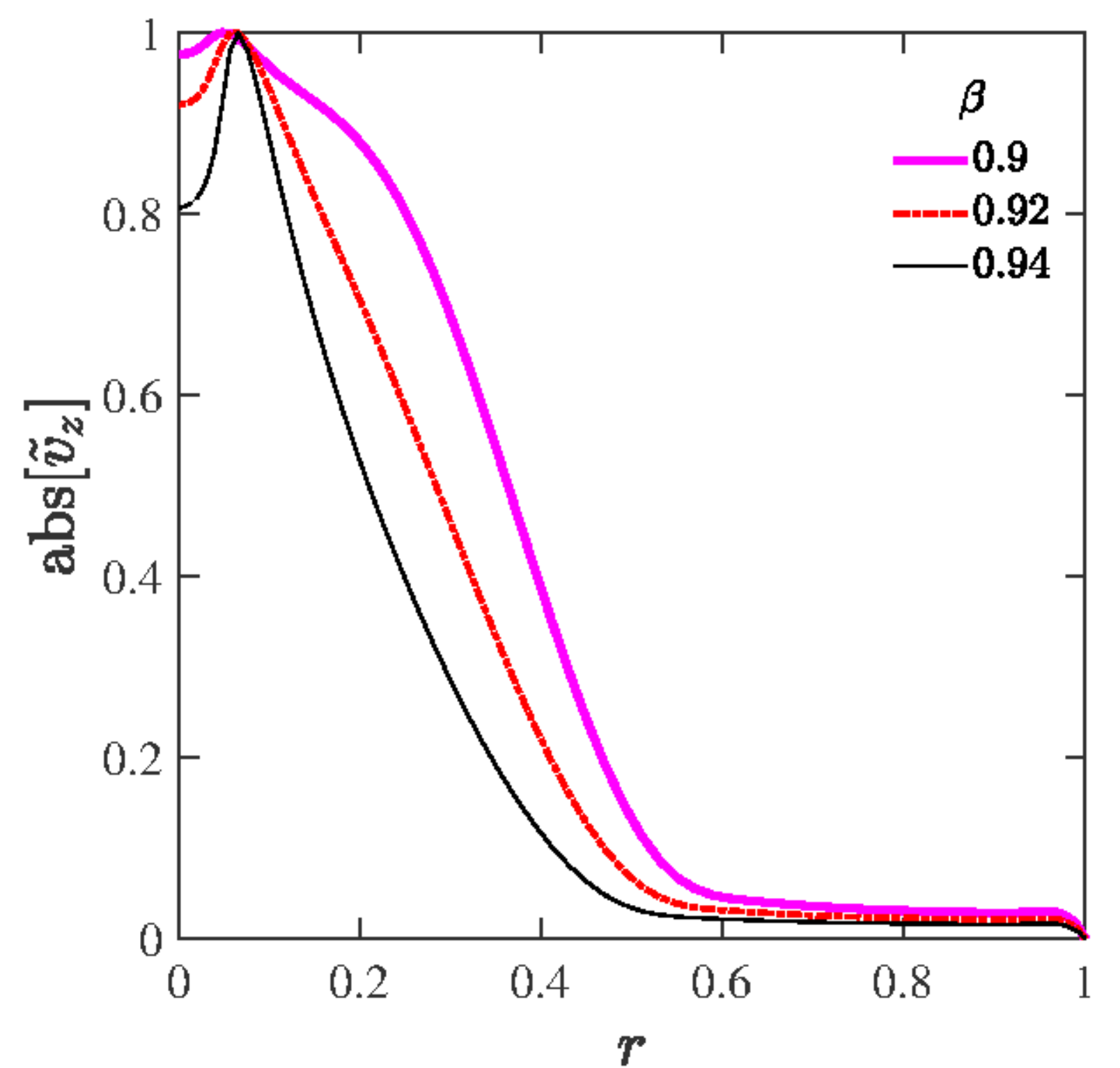}
    \caption{Axial velocity}
    \label{fig:ef_Vz_beta0pt8E0pt003Re6000k1}
  \end{subfigure}
 \begin{subfigure}[htp]{0.35\textwidth}
    \includegraphics[width=\textwidth]{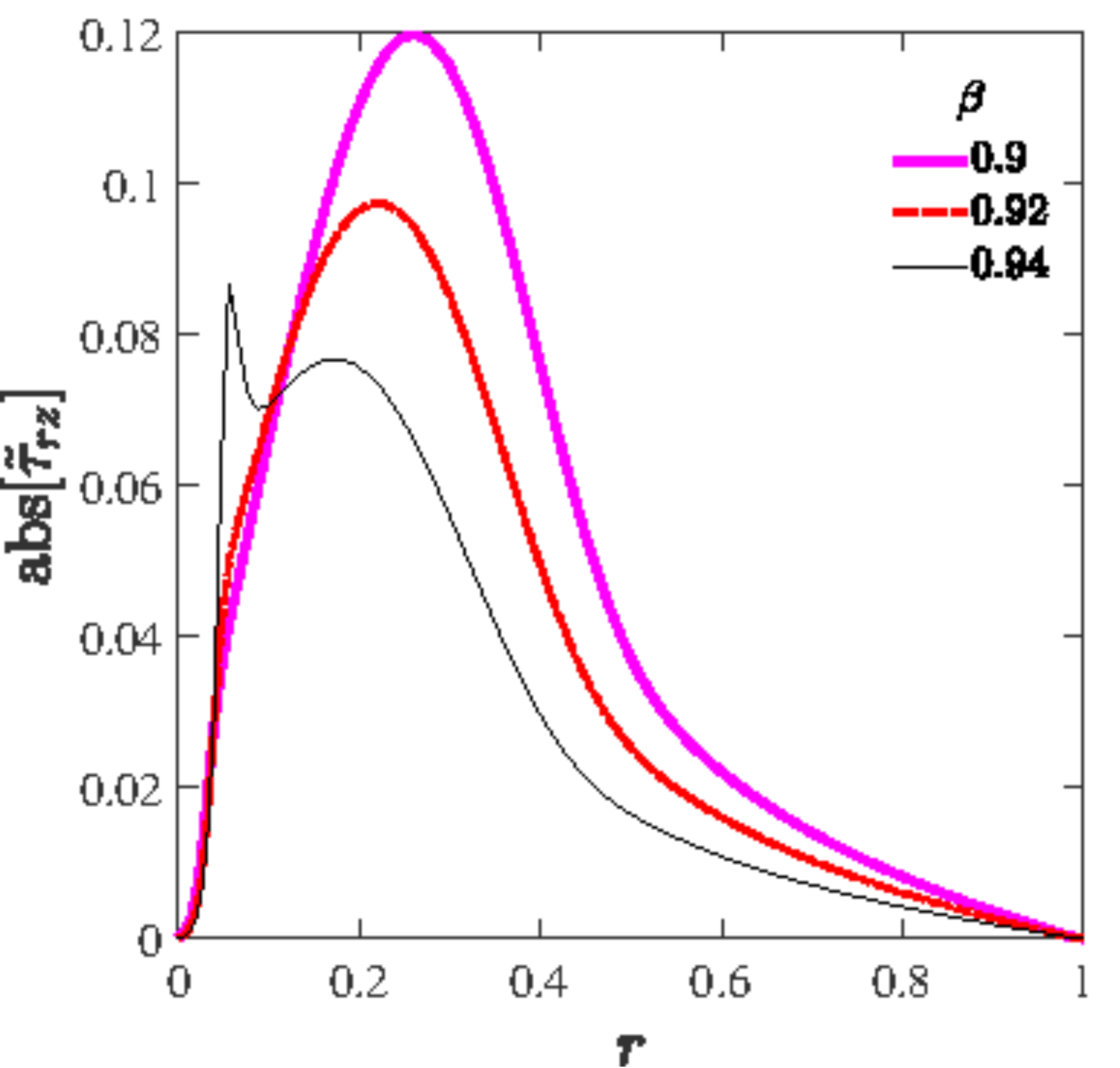}
    \caption{$T_{rz}$ stress}
    \label{fig:ef_Trz_beta0pt8E0pt002Re6000k1}
  \end{subfigure}
   \begin{subfigure}[htp]{0.35\textwidth}
    \includegraphics[width=\textwidth]{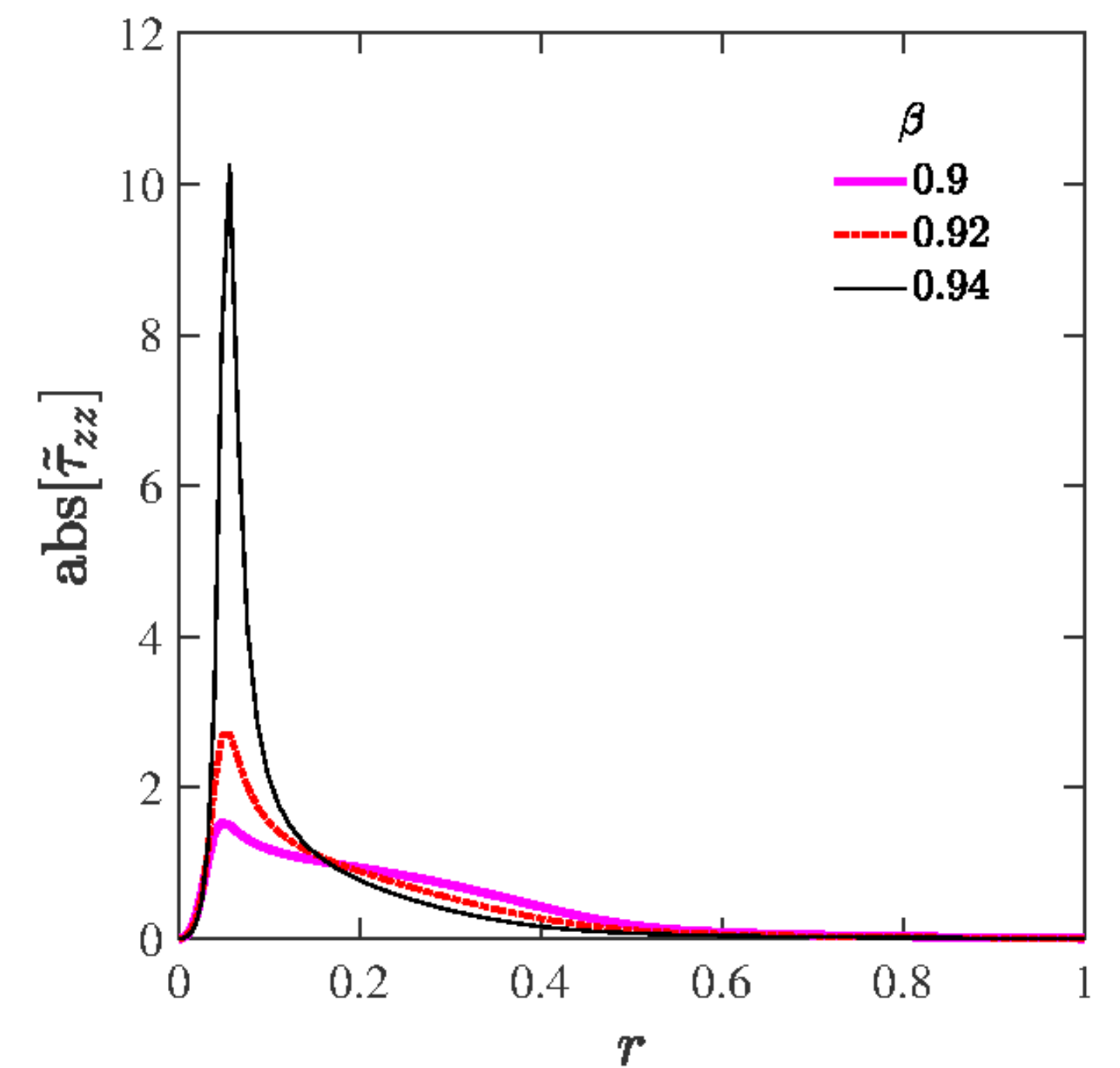}
    \caption{$T_{zz}$ stress}
    \label{fig:ef_Tzz_beta0pt8E0pt002Re6000k1}
  \end{subfigure}
  \caption{Velocity and polymer stress eigenfunctions
    corresponding to the unstable centre modes 
    (in
    Fig.~\ref{fig:es_E0pt15Re6000k1_beta0pt85to0pt98})
    at different $\beta$ for
    $E = 0.15, \Rey = 6000$ and $k=1$.}
  \label{fig:ef_beta0pt8EERe6000k1}
\end{figure}
In Fig.~\ref{fig:es_E0pt15Re6000k1_beta_near1}, we show the eigenspectra (overlaid) as
$\beta $ is reduced from unity, again at fixed $E$, $Re$ and $k$;
note that the $\beta$'s shown are all higher than the threshold value
for collapse into the CS (the analogue of that identified in
Fig.~\ref{fig:ci_vs_E_bbRe600k3}, but for $Re = 6000$). 
Figure~\ref{fig:es_E0pt15Re6000k1_beta0pt98to1} is for $\beta$'s close enough to unity that the center mode has not  emerged out of the CS yet (the other stable
modes, with $c_r \rightarrow 0$, are not shown).
 Thus, the trends in this figure pertain to all other (least stable) modes on the P branch.
Figure~\ref{fig:es_E0pt15Re6000k1_beta0pt98to1} shows no discernible trend in the
behaviour of the P branch modes with changing $\beta$. For instance, as $\beta$
is decreased, the least-stable Newtonian mode moves in the clockwise
sense in $(c_r,c_i)$-plane. In contrast, 
the mode LSCM2 smoothly continues from a Newtonian mode at the junction of the 'APS' structure present at $\beta = 0$. The remaining modes are, however, smooth  continuations of the modes of the 
Newtonian $P$ branch, but these move in the counter-clockwise sense with  decreasing $\beta$.
Eigenspectra  for smaller $\beta$ in the interval $0.85\leq\beta\leq 0.98$ are shown in
Fig.~\ref{fig:es_E0pt15Re6000k1_beta0pt85to0pt98}, the focus being on the center mode. The center mode first emerges at $\beta = 0.96$, and becomes unstable for $\beta \in [0.88,0.945]$. The smooth (blue) curve, passing through the spectral center mode eigenvalues, shows the trajectory of the center mode with decreasing $\beta$, obtained using the shooting method.
%
%
Thus, at a fixed $E$, the unstable center mode always emerges out of the CS as $\beta$ is decreased from unity.

The velocity and stress eigenfunctions corresponding to some of the unstable center modes 
in Fig.~\ref{fig:es_E0pt15Re6000k1_beta0pt85to0pt98} are shown in Fig.~\ref{fig:ef_beta0pt8EERe6000k1}. For the higher  $\Rey$ ($= 6000$), the axial velocity eigenfunctions are more localized near the center (as $\beta$ approaches unity),  compared to those for $\Rey = 600$ shown in Fig.~\ref{fig:ef_Vz_beta0pt8E0pt003Re600k3}.
The axial stress $T_{zz}$ (Fig.~\ref{fig:ef_Tzz_beta0pt8E0pt002Re6000k1})
shows a sharp peak as $\beta$ approaches unity at fixed $E$, similar to the feature that was seen earlier (Fig.~\ref{fig:ef_Tauzz_beta0pt8E0pt003Re600k3}), albeit with decreasing $E$ at a fixed $\beta$ for $Re = 600$. For the values of $\beta$ examined 
here, both the axial and radial eigenfunctions exhibit a rather smooth
variation with $r$, unlike the rapid, oscillatory variation (not shown) characteristic of
wall modes ($c_r \rightarrow 0$) for $\beta \rightarrow 0$. 
The latter are analogous to wall modes in viscoelastic channel flow whose structures was examined in detail by \cite{chaudhary_etal_2019} (see Fig.~20 therein); the overall similarity of the pipe and channel flow UCM spectra was already  discussed above.

\begin{figure}
  \centering
  \begin{subfigure}[htp]{0.48\textwidth}
    \includegraphics[width=\textwidth]{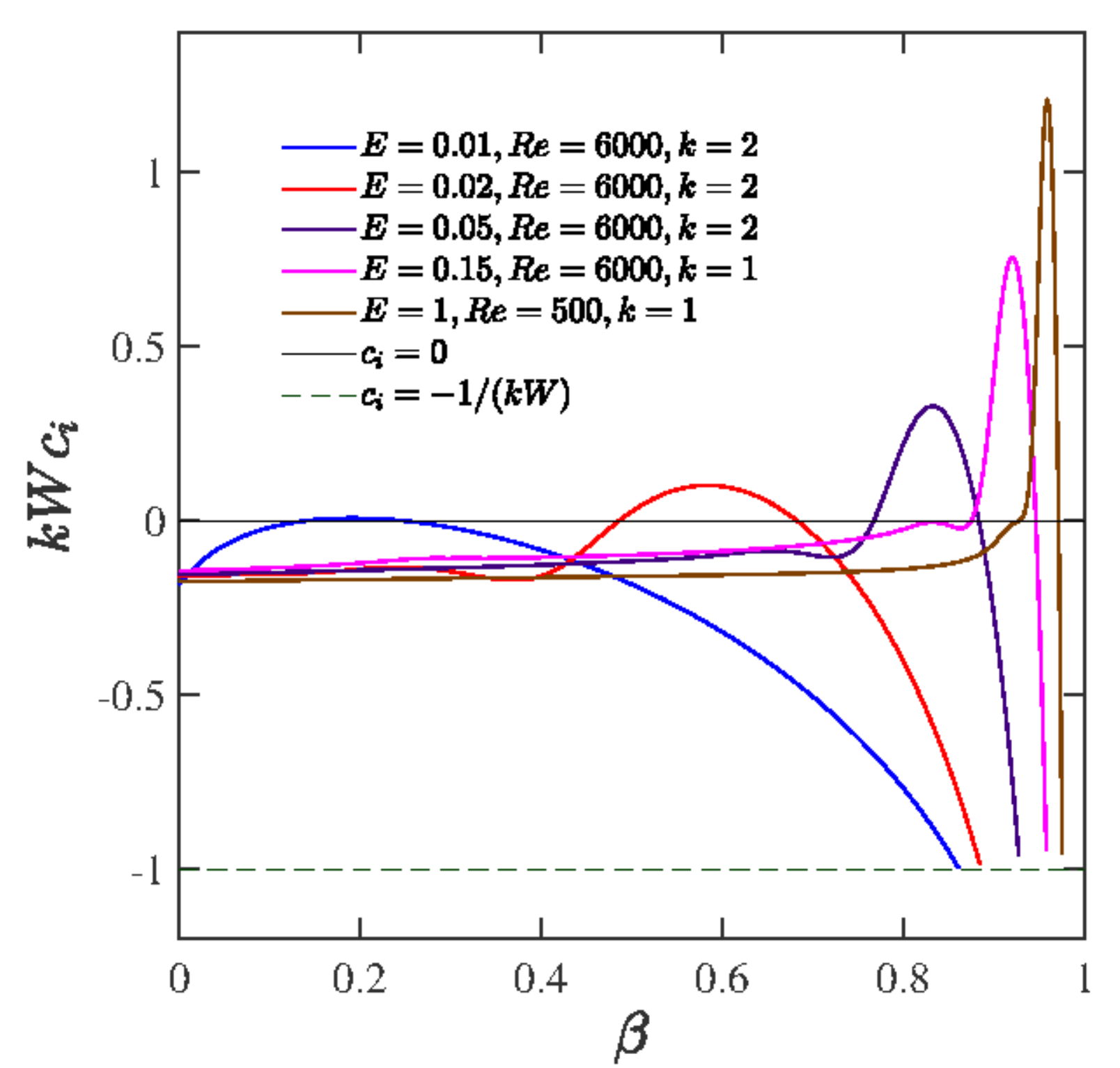}
    \caption{Growth rate}
    \label{fig:ci_vs_beta_ERk}
  \end{subfigure}
  \begin{subfigure}[htp]{0.48\textwidth}
    \includegraphics[width=\textwidth]{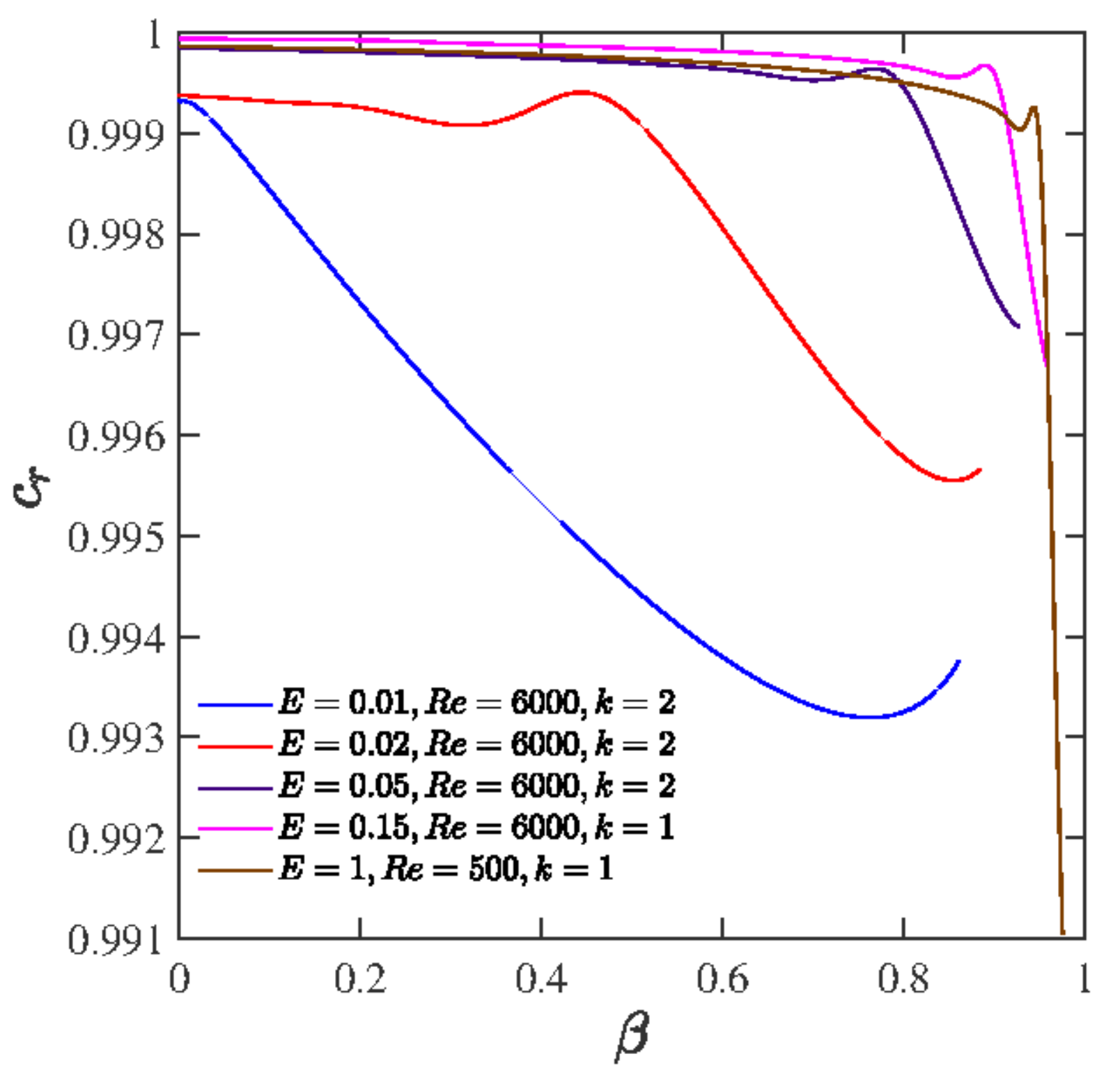}
    \caption{Phase speed}
    \label{fig:cr_vs_beta_ERk}
  \end{subfigure}
  \caption{(a) Scaled growth rate and (b) phase speed, for the
    center mode, with varying $\beta$, for
    different fixed sets of parameters
    $(E,\Rey, k) = (0.01, 6000, 2),(0.02, 6000, 2),(0.05, 6000,
    2),(0.15, 6000, 1)$ and $(1, 500, 1)$.}
  \label{fig:beta_variation}
\end{figure}


\subsubsection{The origin of the center mode at fixed $E$ and varying $\beta$}
\label{subsec:originfixedEvaryingbeta}


While the discussion pertaining to Fig.~\ref{fig:es_E0pt15Re6000k1_beta0pt85to0pt98} showed that center mode emerges out of the CS as $\beta$ is decreased from unity, in Fig.~\ref{fig:ci_vs_beta_ERk}, we address the question of what happens as $\beta$ is decreased down to zero (the UCM limit).  
This figure shows the variation of the scaled
growth rate, $kWc_i$, of the center mode with varying $\beta$ at
fixed $Re$, $E$ and $k$. 
 As $\beta$ is decreased from unity,
the center mode emerges out of  CS1 (when
$kWc_i = -1$) at a critical $\beta$, and becomes unstable 
as $\beta$ is decreased further. The critical $\beta$ corresponding to
the emergence of the center mode is closer to unity for higher
$E$. 
The range of unstable $\beta$ also approaches unity for larger $E$, while also narrowing down in extent, with a concomitant increase in the growth rate. A similar narrowing down occurs when 
$E$ approaches the lower threshold for the instability,
for the chosen $\beta$ (the blue curve in Fig.~\ref{fig:ci_vs_beta_ERk}). Figure~\ref{fig:cr_vs_beta_ERk} shows
that the corresponding $c_r$ remains close to (and less than) unity
for the entire range of $\beta$.

%
%


\begin{figure}
  \centering
  \begin{subfigure}[htp]{0.44\textwidth}
    \includegraphics[width=\textwidth]{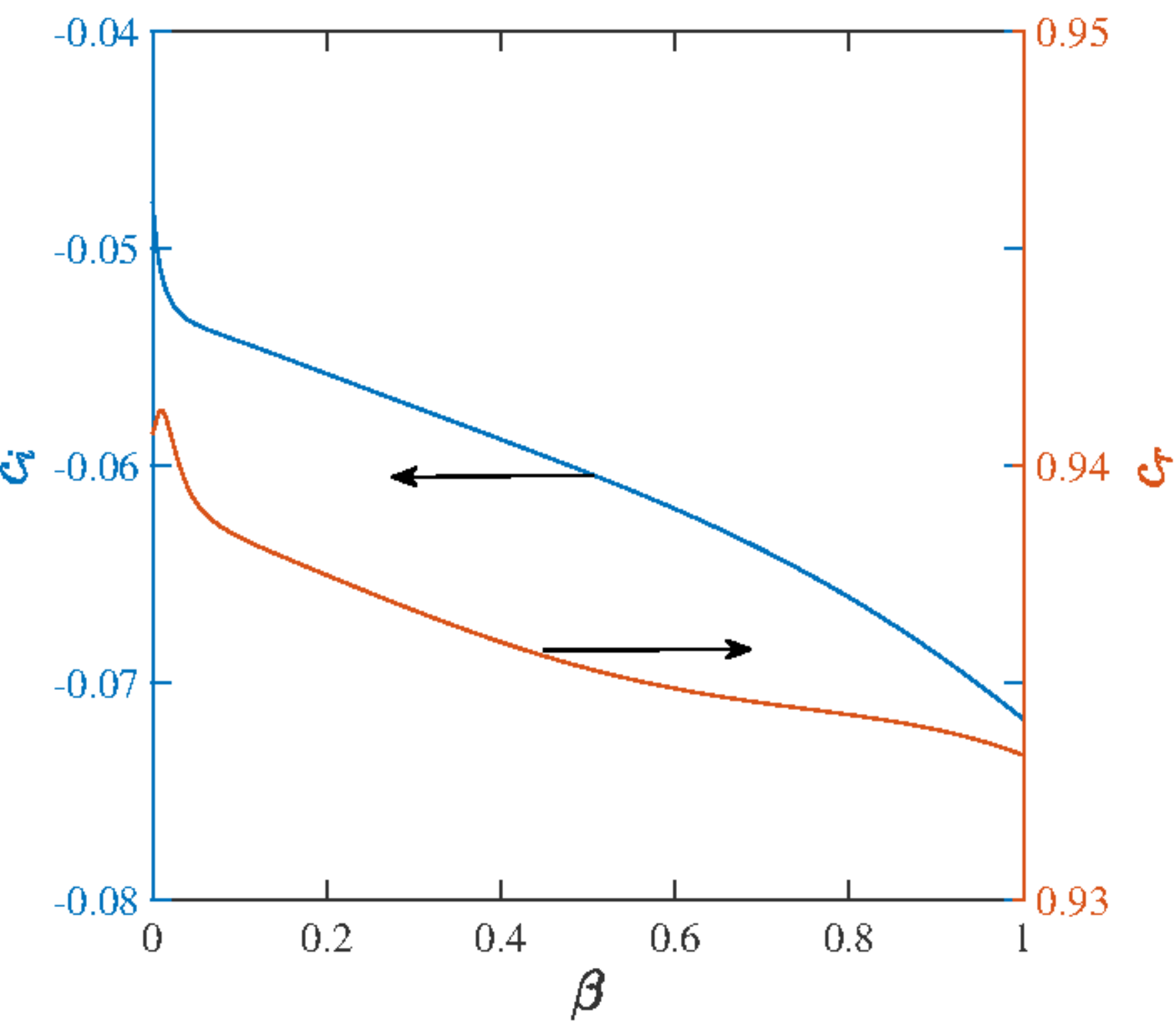}
    \caption{Growth rate and Phase speed}
    \label{fig:cicr_vs_beta_E0pt005Re600k3_Mode1}
  \end{subfigure}\\
  \begin{subfigure}[htp]{0.44\textwidth}
    \includegraphics[width=\textwidth]{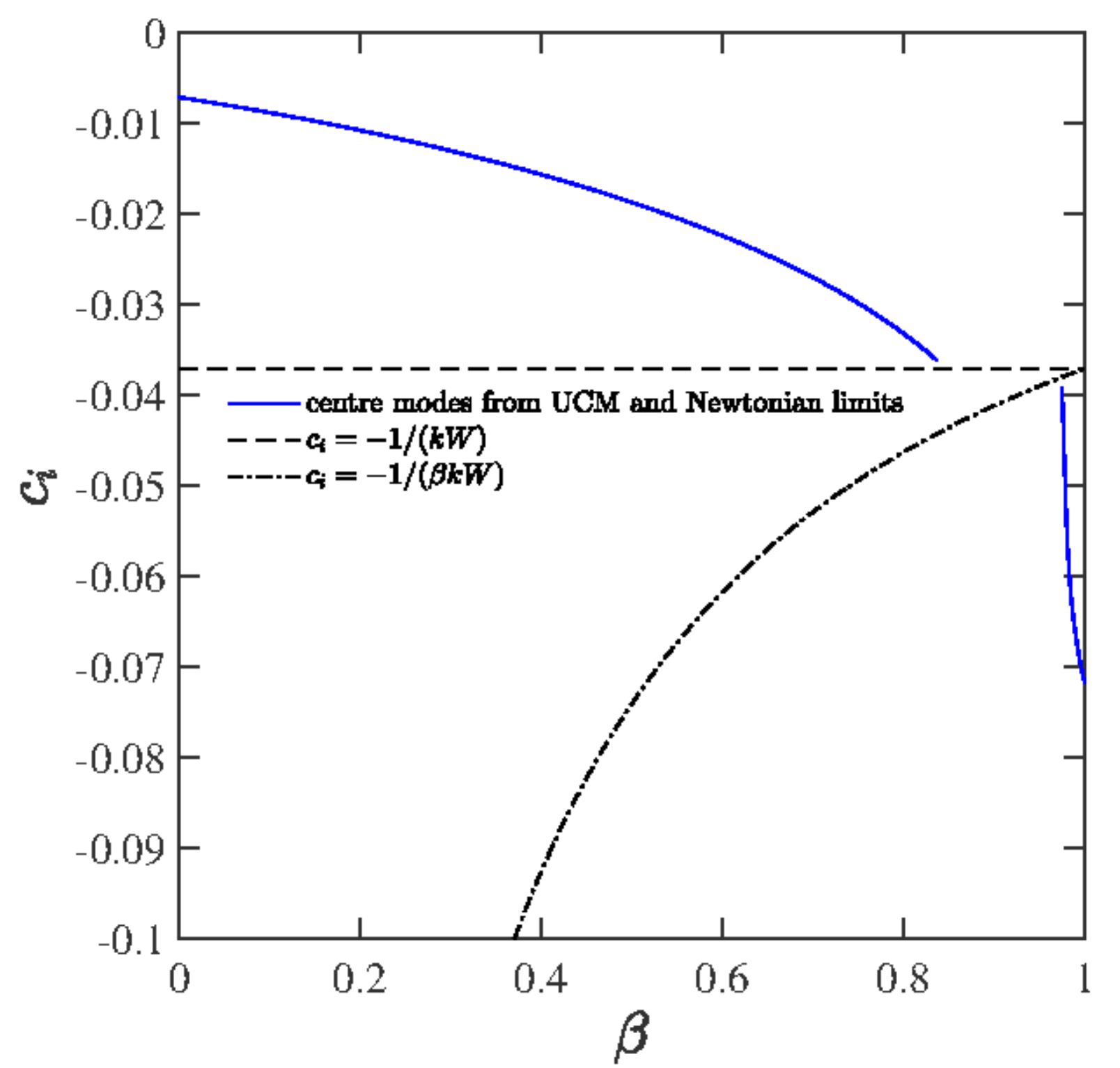}
    \caption{Growth rate}
    \label{fig:ci_vs_beta_E0pt015Re600k3_Mode1,1a}
  \end{subfigure}
  \begin{subfigure}[htp]{0.44\textwidth}
    \includegraphics[width=\textwidth]{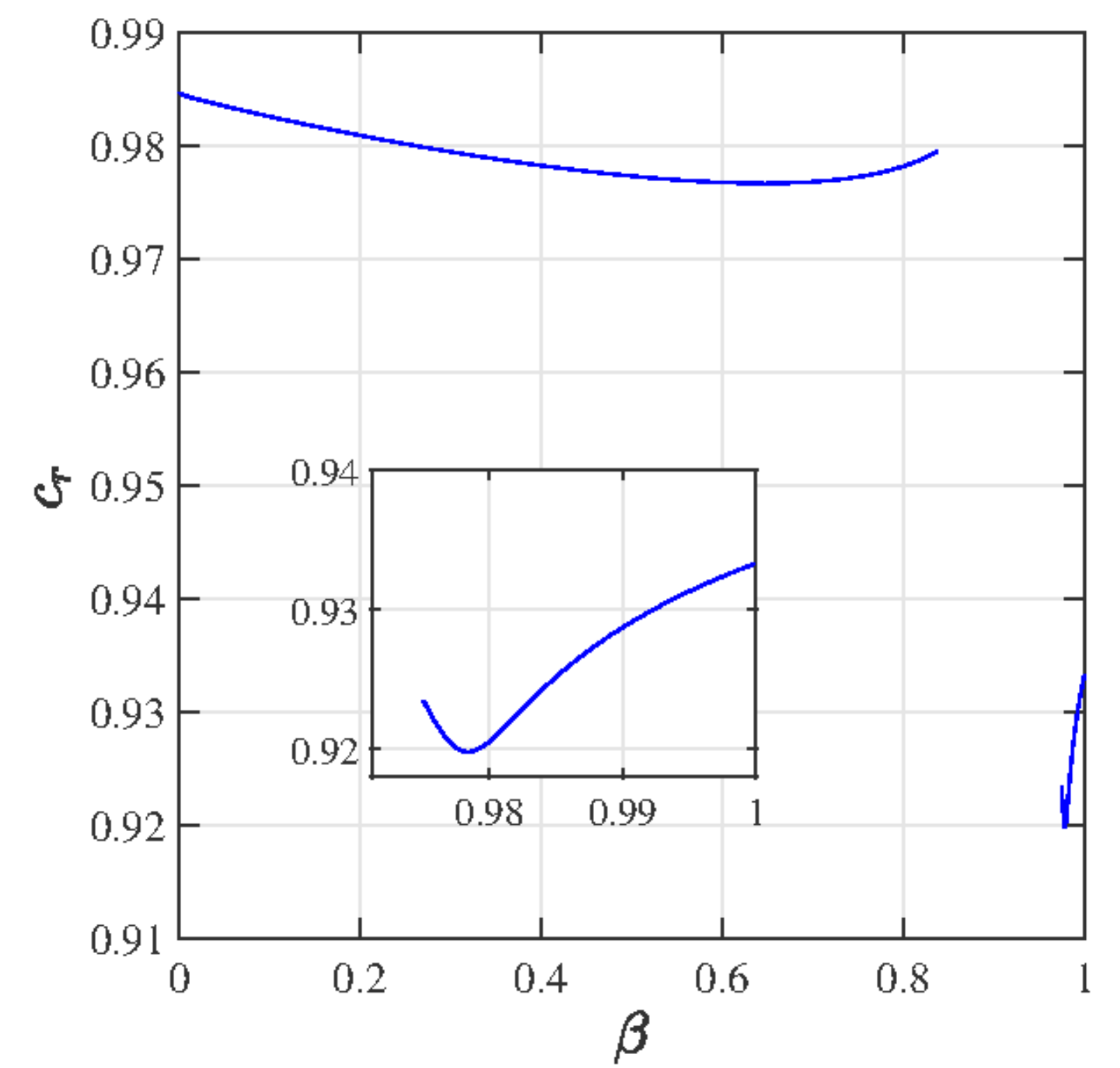}
    \caption{Phase speed}
    \label{fig:cr_vs_beta_E0pt015Re600k3_Mode1,1a}
  \end{subfigure}
  \begin{subfigure}[htp]{0.44\textwidth}
    \includegraphics[width=\textwidth]{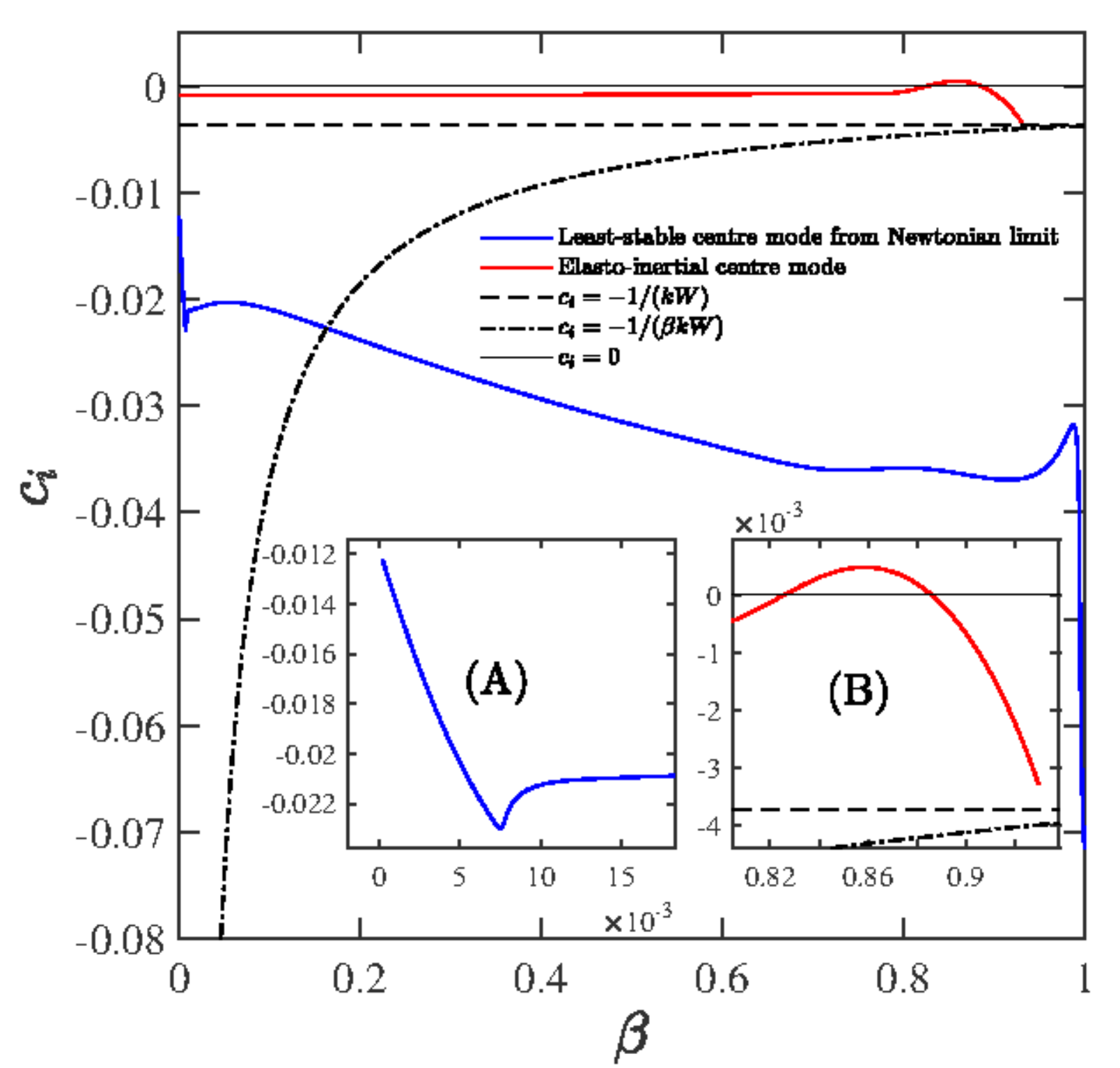}
    \caption{Growth rate}
    \label{fig:ci_vs_beta_E0pt15Re600k3_Mode1,1a}
  \end{subfigure}
  \begin{subfigure}[htp]{0.44\textwidth}
    \includegraphics[width=\textwidth]{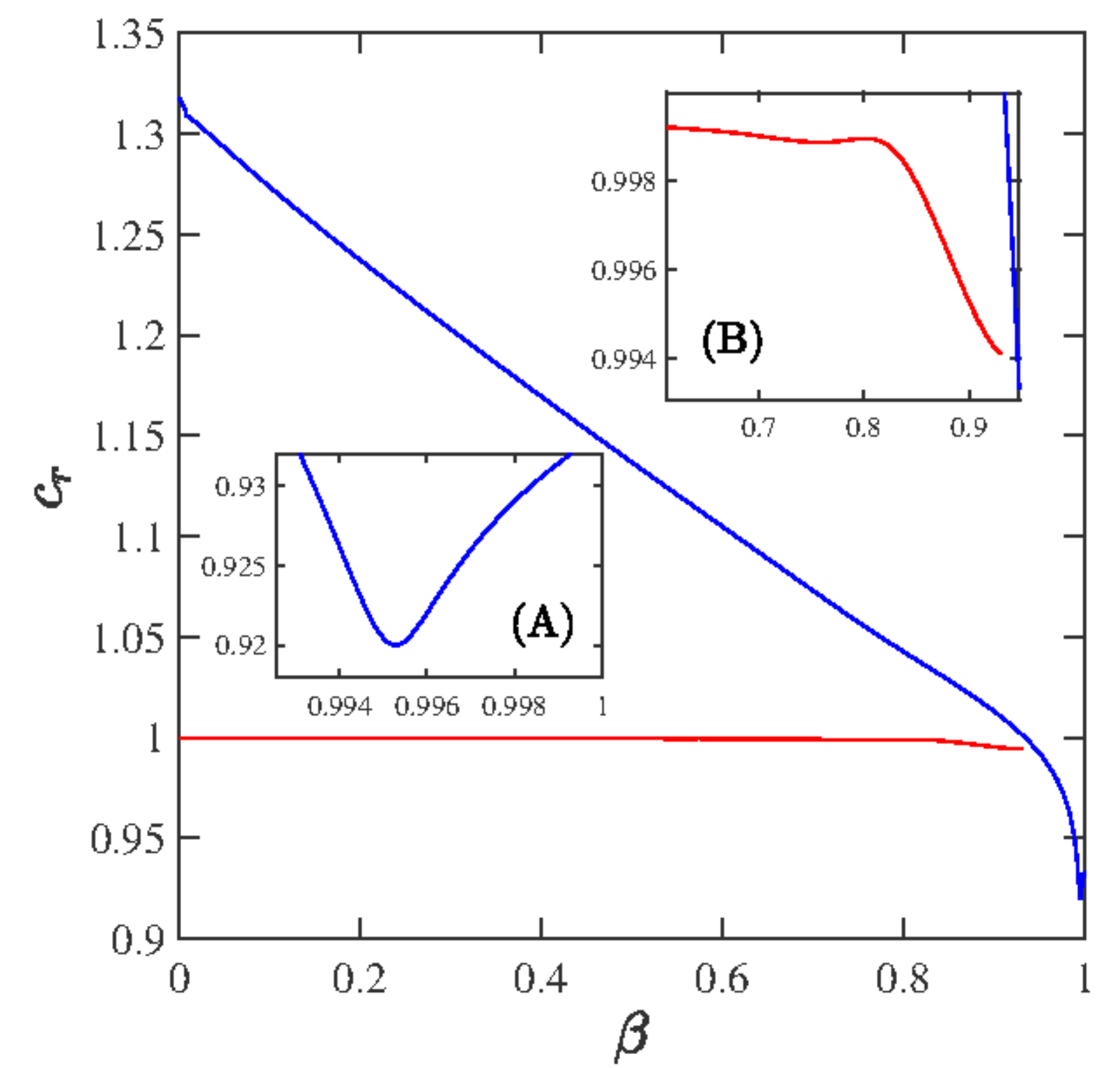}
    \caption{Phase speed}
    \label{fig:cr_vs_beta_E0pt15Re600k3_Mode1,1a}
  \end{subfigure}
                
  \caption{The three possible center mode trajectories   
   with variation in $\beta$ for
    $\Rey=600, k=3$ at various fixed values of the elasticity number: $E = 0.005$
    in panel (a), $E = 0.015$ in panels (b) and (c), and $E = 0.15$ in
    panels (d) and (e). Insets~$(A)$ in panels (d) and (e) show the
    enlarged regions near $\beta=0$ and $\beta = 1$,
    respectively. Inset~$(B)$ in panels (d) show the enlarged
    views of the unstable range of $\beta$s.}
  \label{fig:c_vs_beta_E0pt005to0pt15Re600k3_Mode1,1a}
\end{figure}

In Fig.~\ref{fig:c_vs_beta_E0pt005to0pt15Re600k3_Mode1,1a}, we show the three possible behaviours, within
the parameter regimes explored, for the trajectory of the least stable center mode as $\beta$ is varied from the UCM to the Newtonian limit.  For the smallest elasticities 
(e.g., $E = 0.005$ in Fig.~\ref{fig:cicr_vs_beta_E0pt005Re600k3_Mode1}), when the two CS are highly stable, and well outside the range of $c_i$'s shown,  the
center mode, while remaining stable,  smoothly continues all the way from the
UCM limit ($\beta = 0$) to the Newtonian ($\beta = 1$) limit without
suffering any discontinuities or abrupt endings. For moderate
elasticities (e.g., $E = 0.015$ shown in
Fig.~\ref{fig:ci_vs_beta_E0pt015Re600k3_Mode1,1a}), 
the
$c_i$ vs $\beta$ curve for the least stable center mode
starts from the Newtonian end ($\beta = 1$), but abruptly ends as it encounters CS2 from below. On the other hand, the least stable center mode in the UCM limit continues to finite $\beta$, abruptly ending at the location
of its encounter with CS1 from above. 
Corresponding phase speeds for $E = 0.015$ are shown in
Fig.~\ref{fig:cr_vs_beta_E0pt015Re600k3_Mode1,1a}, with the inset
showing an enlarged view near $\beta=1$,  where the variation of the phase speed
$c_r$ with $\beta$ is quite sharp.  
 For the chosen parameters, the center mode  still remains stable for all $\beta$. 
Finally, for 
higher elasticity (e.g., $E = 0.15$),  the $c_i$ vs $\beta$
curve for the least-stable mode from the Newtonian end 
continues all the way up to the
UCM limit without suffering discontinuities as shown in
Fig.~\ref{fig:ci_vs_beta_E0pt15Re600k3_Mode1,1a}, ending up as a center mode in the
UCM spectrum. Inset (A)
shows a  magnified view of the sharp variation of the $c_i$ curve near
$\beta=0$. The least stable center mode in the UCM limit behaves similar to the previous case of $E=0.015$, with an abrupt ending as it collapses onto CS1 from above, the only difference now being that the mode is unstable for a small
range of $\beta$ (due to the higher $E$); inset (B) provides the enlarged view of the
unstable range of $\beta$. 
The corresponding phase speeds for $E=0.15$
are shown in Fig.~\ref{fig:cr_vs_beta_E0pt15Re600k3_Mode1,1a} with
inset (A) showing the enlarged  view of the non-monotonic behaviour near the Newtonian limit.
Note that the Newtonian center mode does not suffer a jump despite crossing the CS2 curve ($c_i = -1/(\beta kW)$) in Fig.~\ref{fig:ci_vs_beta_E0pt15Re600k3_Mode1,1a}; this is only an apparent crossing since, as shown in Fig.~\ref{fig:cr_vs_beta_E0pt15Re600k3_Mode1,1a}, its phase speed exceeds unity, and it therefore `goes around' CS2 with decreasing $\beta$. In contrast, the discontinuities in the center mode trajectory, in Fig.~\ref{fig:ci_vs_beta_E0pt015Re600k3_Mode1,1a}, occur because $0 <  c_r < 1$.

%
%


\begin{figure}
        \centering
        \begin{subfigure}[htp]{0.45\textwidth}
                \includegraphics[width=\textwidth]{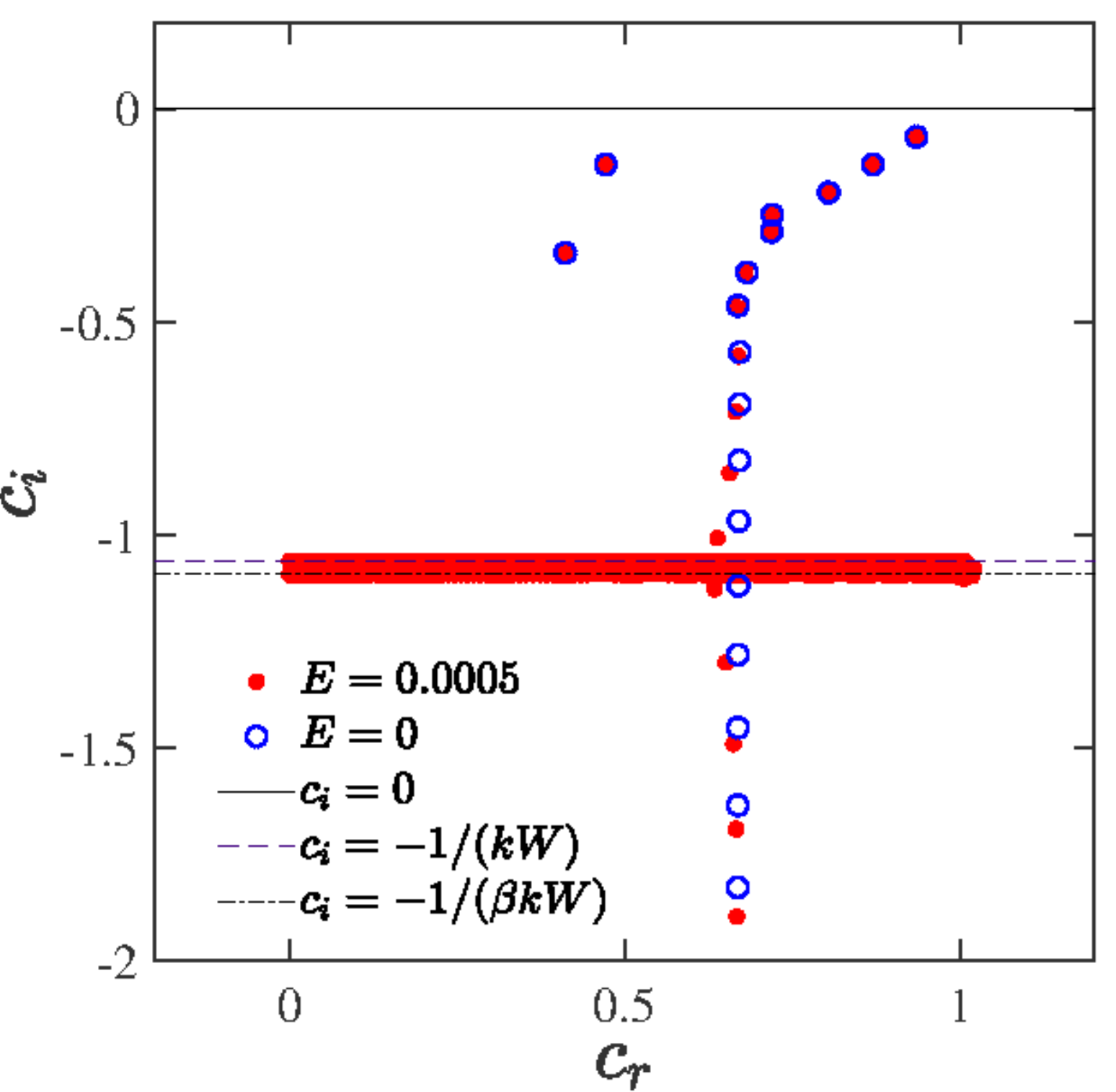}
                \caption{$E = 0.0005$}
                \label{fig:ues_beta0pt97E0pt0005Re1500k0pt4pi}
        \end{subfigure}
        \begin{subfigure}[htp]{0.45\textwidth}
                \includegraphics[width=\textwidth]{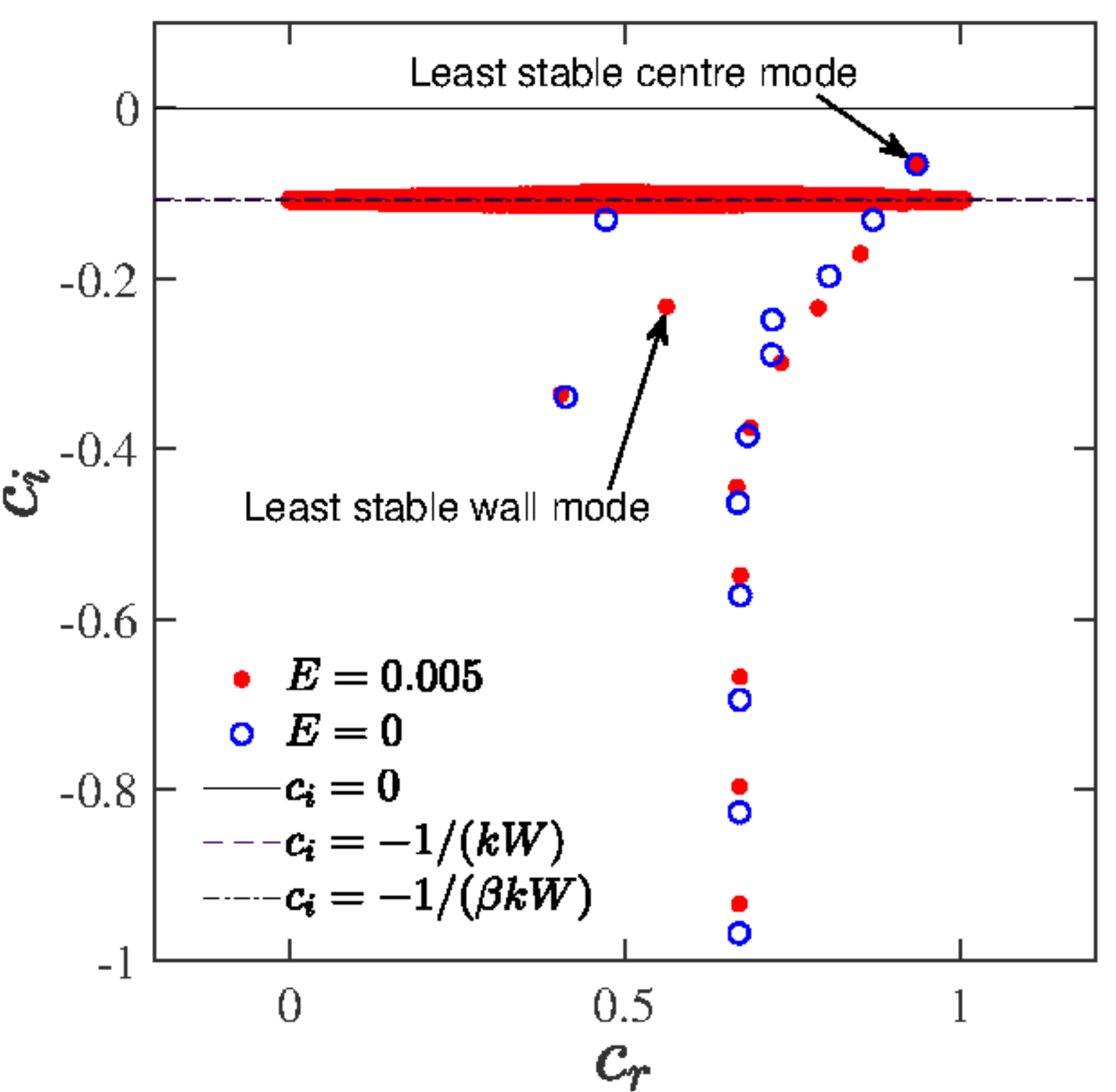}
                \caption{$E = 0.005$}
                \label{fig:ues_beta0pt97E0pt005Re1500k0pt4pi}
        \end{subfigure}
        \begin{subfigure}[htp]{0.45\textwidth}
                \includegraphics[width=\textwidth]{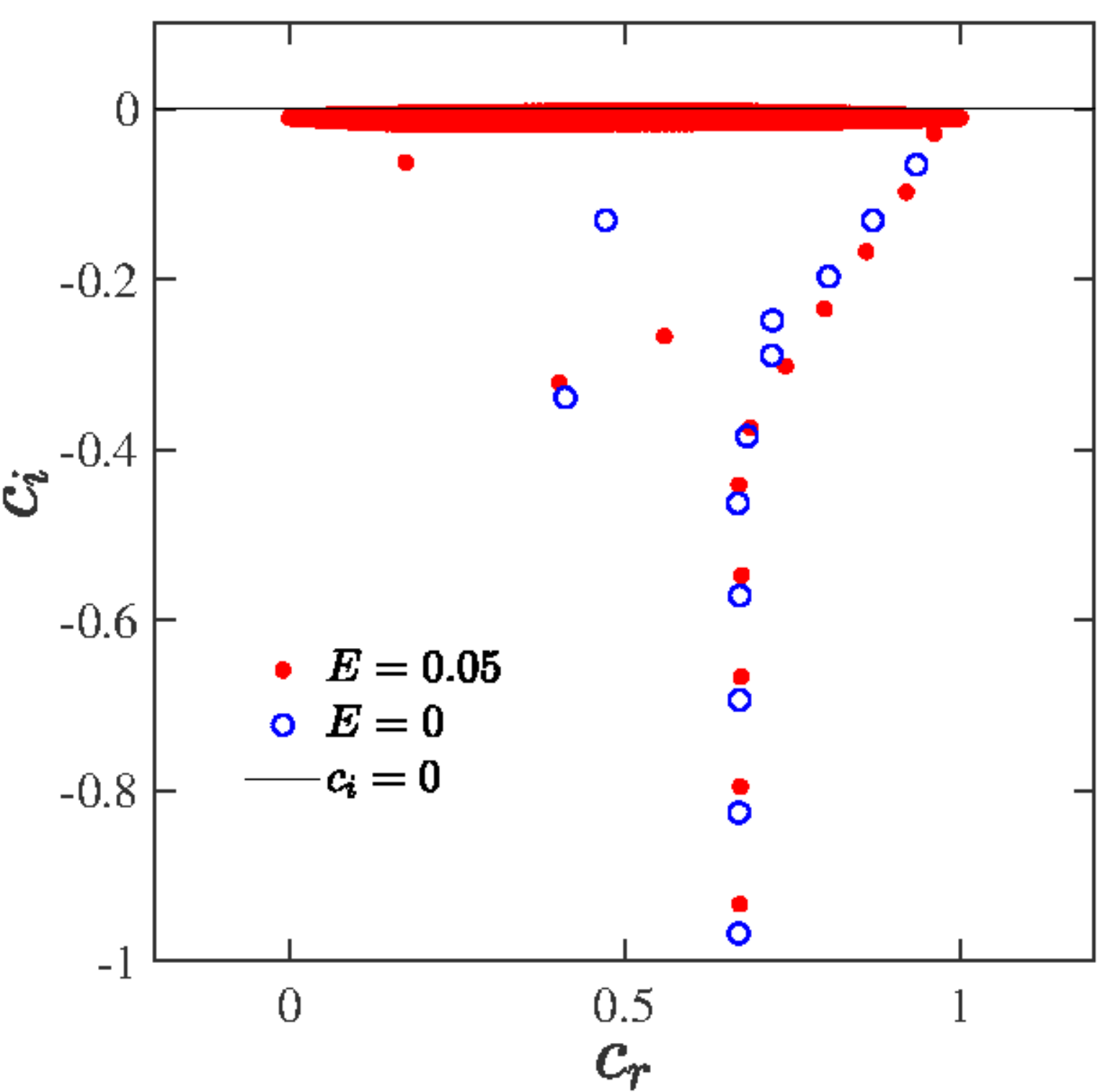}
                \caption{$E = 0.05$}
                \label{fig:ues_beta0pt97E0pt05Re1500k0pt4pi}
        \end{subfigure}
        \begin{subfigure}[htp]{0.45\textwidth}
                \includegraphics[width=\textwidth]{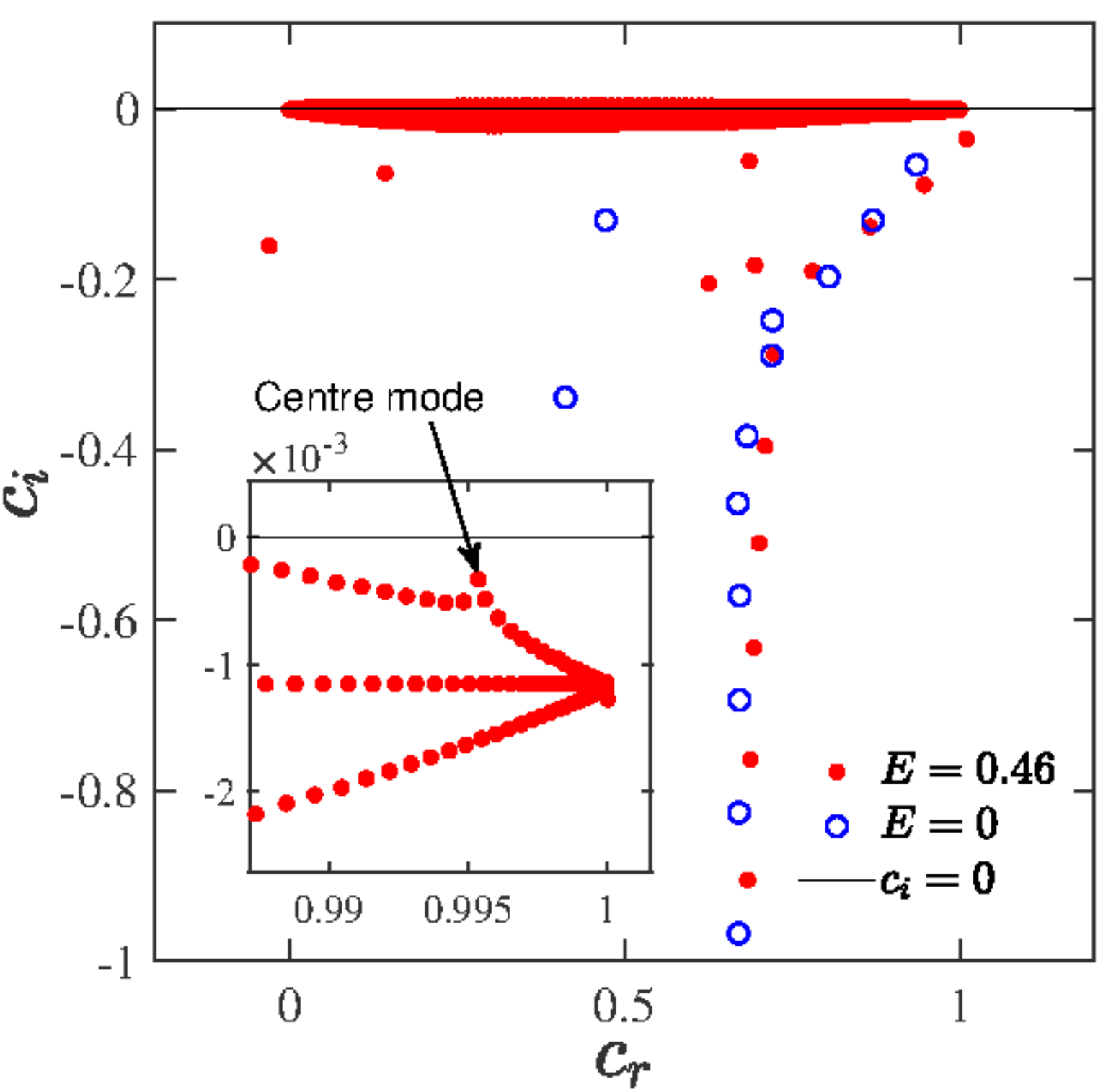}
                \caption{$E = 0.46$}
                \label{fig:ues_beta0pt97E0pt46Re1500k0pt4pi}
        \end{subfigure}
        \begin{subfigure}[htp]{0.45\textwidth}
                \includegraphics[width=\textwidth]{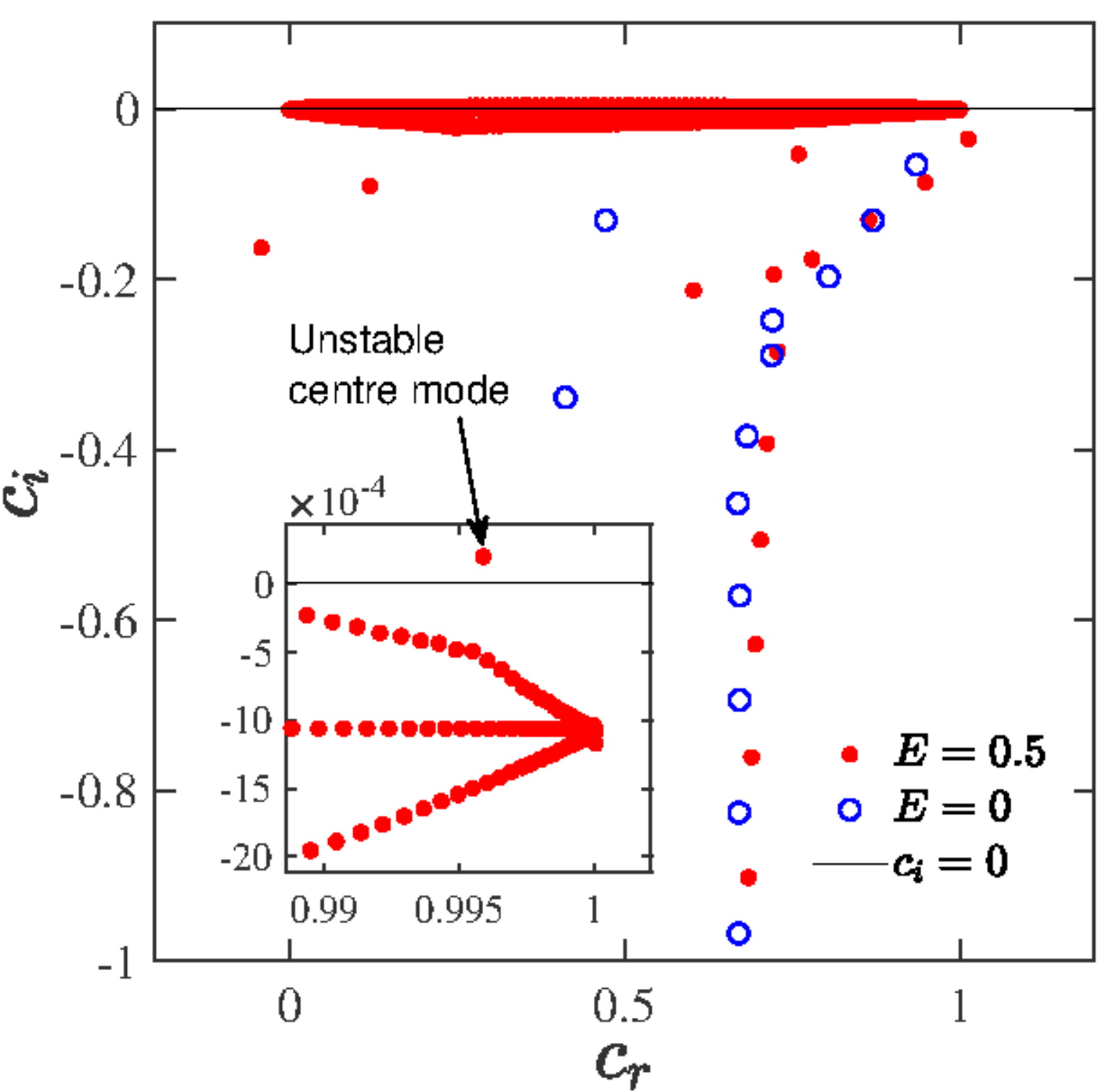}
                \caption{$E = 0.5$}
                \label{fig:ues_beta0pt97E0pt5Re1500k0pt4pi}
        \end{subfigure}
        \caption{Viscoelastic pipe-flow eigenspectra at $\beta = 0.97$ for different $E$: (a) $0.0005$, (b) $0.005$, (c) $0.05$, (d) $0.46$, and (e) $0.5$, compared with  the Newtonian eigenspectrum ($E =0$), for $\Rey = 1500$ and $k=0.4\pi$.}
				\label{fig:ues_beta0pt97EERe1500k0pt4pi}
\end{figure}

\begin{figure}
        \centering
        \begin{subfigure}[htp]{0.44\textwidth}
                \includegraphics[width=\textwidth]{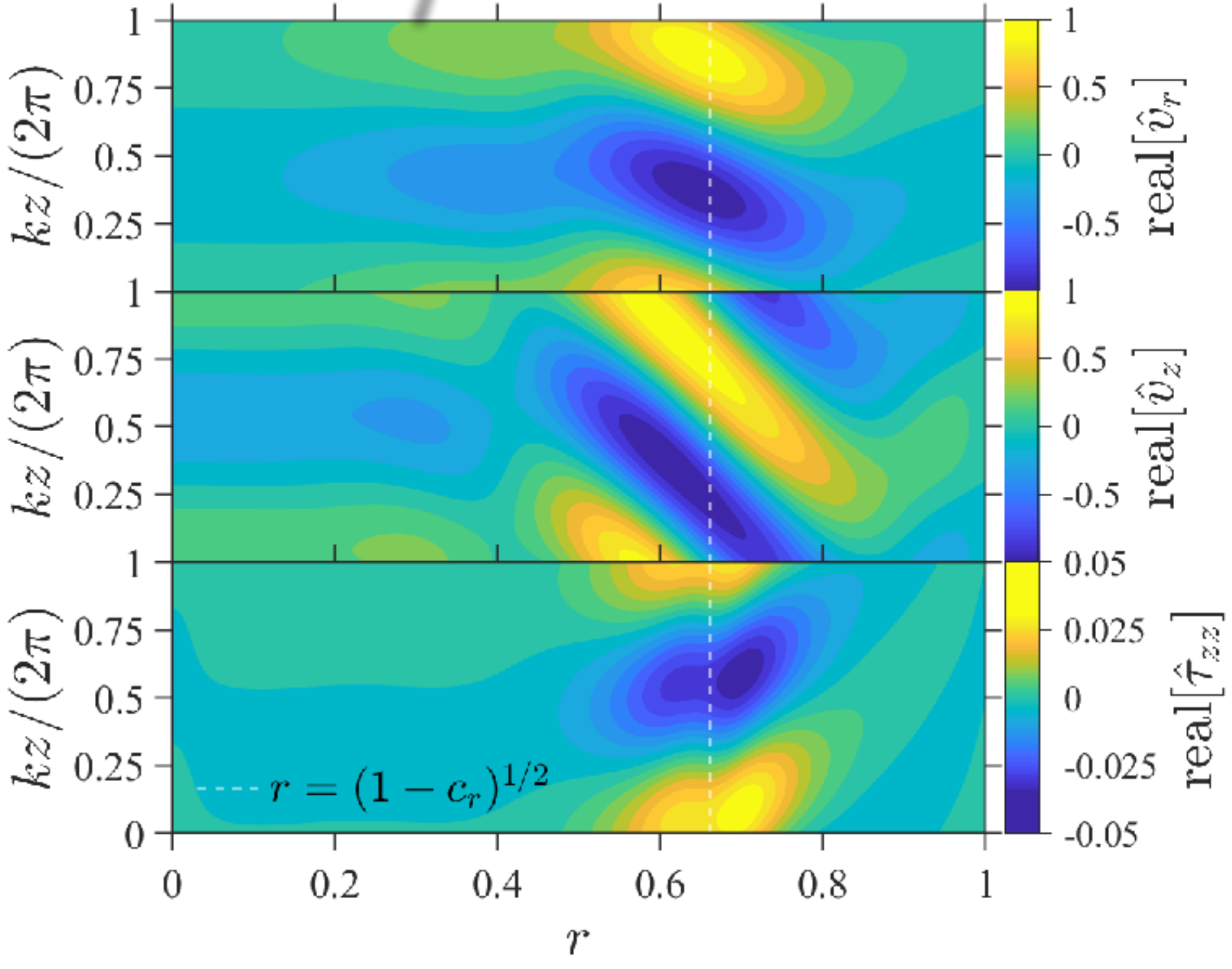}
                \caption{Least-stable wall mode}
                \label{fig:color_plots_wall_mode}
        \end{subfigure}\quad\quad
        \begin{subfigure}[htp]{0.44\textwidth}
                \includegraphics[width=\textwidth]{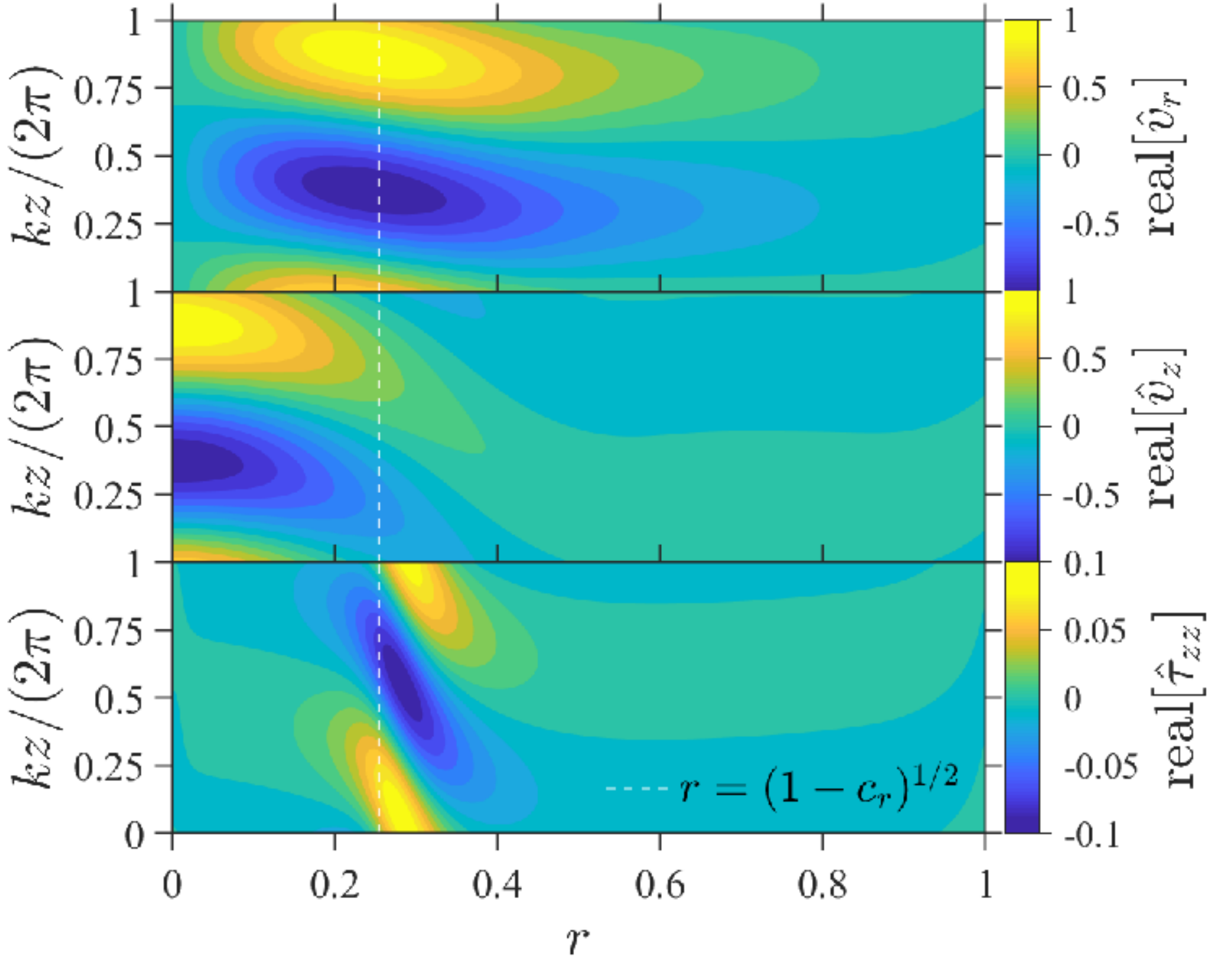}
                \caption{Least-stable centre mode}
                \label{fig:color_plots_centre_mode}
        \end{subfigure}
\caption{ Contours of the radial ($\hat{v}_r$), axial ($\hat{v}_x$) velocities and axial normal stress ($\hat{\tau}_{zz}$) in the $r$--$z$ plane  for the least stable wall and centre modes 
(marked in Fig.~\ref{fig:ues_beta0pt97E0pt005Re1500k0pt4pi}) in pipe flow
for $E = 0.005, \beta = 0.97, \Rey =1500$ and $k = 0.4\pi$:
(a) Least-stable wall mode $c = 0.561844626 - 0.233038878\mathrm{i}$ from the A-branch of the eigenspectrum, and (b) for the least-stable centre mode $c = 0.935239154- 0.064928230\mathrm{i}$ from the P-branch of the eigenspectrum. The location of the critical layer is shown using white dashed lines.}
\label{fig:color_plots}
\end{figure}

\subsection{Center vs. wall modes in viscoelastic pipe and channel flows} 
\label{sec:shekar}

In the results presented so far, we have characterized the behaviour of the elasto-inertial center mode as a function of $E$ and $\beta$. Although this  mode may either be directly related to a Newtonian center mode (for $\beta$'s below a threshold), or be disconnected from the Newtonian spectrum (for $\beta$'s above), the interpretation is nevertheless that the elasto-inertial turbulence observed in recent experiments \citep{samanta_etal2013,choueiri_etal_2018,chandra_etal_2018} is the outcome of a linear instability associated with this center mode. In sharp contrast to this picture, in a recent effort, \cite{shekar_etal_2019}  have argued based on DNS simulations  and a singular-value decomposition analysis that elasto-inertial turbulence in channel flow 
might instead be closely related to the elastically modified TS mode. As is well known, the TS mode is the least stable wall mode in the Newtonian limit, and this remains true for the range of elasticities considered by the authors. Thus, the premise of \cite{shekar_etal_2019} continues to be along the lines of a sub-critical bifurcation to EIT, similar in spirit to the earlier efforts of 
 \cite{meulenbroek_etal_2003,morozov_saarloos2005,morozov_saarloos2007} in the inertialess limit, and to the work of \cite{stone_graham2002,stone_graham2003,stone_graham2004,Li_Graham_2007} based on an elastic modification of 3D ECS structures. The main difference is that the bifurcation ascribed by \cite{shekar_etal_2019} is supposedly to a finite amplitude 2D mode, with EIT-like dynamics. 
The authors reported results for $\Rey = 1500$ (where the Newtonian flow is turbulent), $\beta = 0.97$, and for $0 < W < 50$. It is worth noting that, for these parameters, the elastically modified ECS's originally examined by Graham and co-workers \citep{Li_Graham_2007} also exist, although \cite{shekar_etal_2019} restrict themselves to two-dimensional initial conditions.

While the present study is
restricted to linear (modal) stability of pipe flow of an Oldroyd-B
fluid, it is nevertheless instructive to compare the viscoelastic pipe and channel flow spectra in order to assess the relative importance of center and wall modes in these geometries. 
Such an assessment would help set the template (in terms of the relevant linear modes, both discrete and continuous) for a nonlinear bifurcation analysis.
We show representative eigenspectra for
pipe flow (in Fig.~\ref{fig:ues_beta0pt97EERe1500k0pt4pi})
for a range of $E$ that subsumes the range ($0 < E < 0.013$) considered by \cite{shekar_etal_2019}, for
$\beta = 0.97, \Rey = 1500$ and $k=0.4\pi$. In each panel, the
corresponding Newtonian spectrum is also shown for comparison (as open blue circles). For $0.0005 <E < 0.05$, with increasing $E$, 
discrete modes collapse into the CS, and new ones emerge from below,
similar to what was shown in Fig.~\ref{fig:ring}.
%
The center mode emerges from CS1 
at $E = 0.46$, (see inset of Fig.~\ref{fig:ues_beta0pt97E0pt46Re1500k0pt4pi}), becoming unstable at $E \approx 0.5$ (Fig.~\ref{fig:ues_beta0pt97E0pt5Re1500k0pt4pi}). 
Importantly, for the parameters considered in Fig.~\ref{fig:ues_beta0pt97EERe1500k0pt4pi}, 
the center mode always remains the least stable or unstable mode.  This feature remains true even 
for other regimes investigated in this study ($\Rey \in 100$--$2000$).
This is unlike Newtonian channel flow, where there is a range of parameters where the wall mode (i.e., the TS mode) is the least stable (or unstable), and this remains  true for small but finite $E$. 



An important feature of Newtonian pipe flow is the absence of a critical-layer singularity \citep{Drazinreid,SchmidHenningsonBook} for axisymmetric disturbances, as a result of which there is no axisymmetric analogue of the two-dimensional TS instability. This difference between pipe and channel flows appears to persist even in the presence of elasticity. In Fig.~\ref{fig:color_plots}, we show, via contour plots,  the spatial structure of the least stable center and wall modes marked in Fig.~\ref{fig:ues_beta0pt97E0pt005Re1500k0pt4pi}. 
Further, and in sharp contrast to viscoelastic channel flow, where the elastically-modified TS mode was shown to have the  $T_{xx}$  (the stream-wise component of the normal stress) eigenfunction strongly localized in
the critical layer (see Fig.~2 of \cite{shekar_etal_2019}), neither the least stable center nor the wall mode in pipe flow exhibits a comparably strong localization of $T_{zz}$; in fact, the  extent of localization is more stronger for the center mode.  For these reasons, the connection between
the (stable) TS wall mode to the elasto-inertial structures suggested by
\cite{shekar_etal_2019} (in the context of viscoelastic channel flow)
is not applicable for viscoelastic pipe flow. This aspect will be discussed in
more detail in a future communication  \citep{khalid_channel}, where we show
that, even for viscoelastic channel flows, the parameter regime relevant to the proposed TS-mode-based subcritical mechanism of \cite{shekar_etal_2019} is somewhat restricted.  It is worth emphasizing that all of the experiments on viscoelastic transition \citep[with the exception of][]{srinivas_kumaran_2017} pertain to the pipe geometry.
Further, and importantly, recent simulations in both the channel \citep{samanta_etal2013,sid_etal_2018} and pipe \citep{lopez_choueiri_hof_2019} geometries have found analogous (span-wise oriented) coherent structures, suggesting a common underlying mechanism for elasto-inertial transition.



%
%
%

\begin{figure}
  \centering
  \begin{subfigure}[htp]{0.48\textwidth}
    \includegraphics[width=\textwidth]{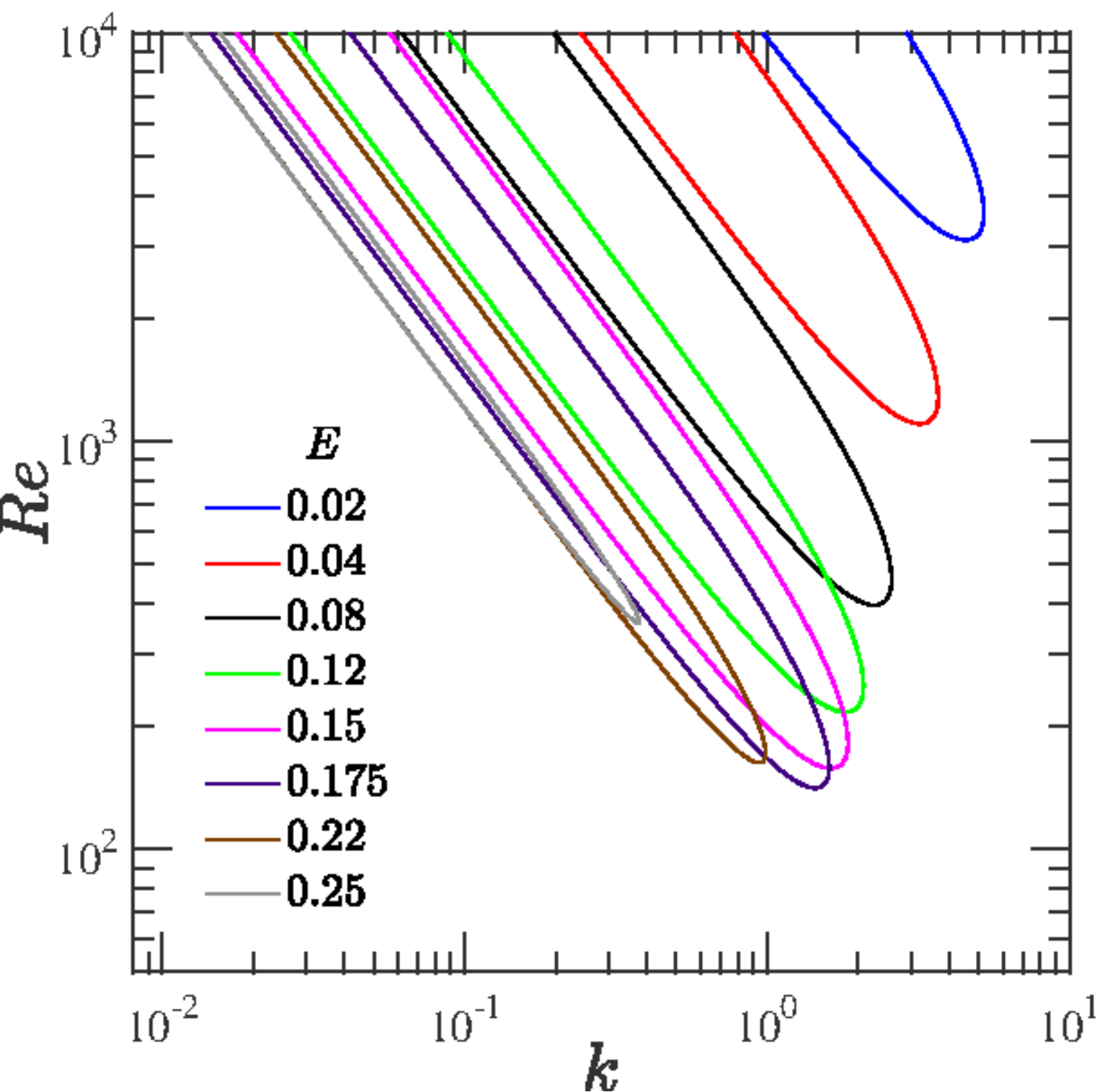}
    \caption{$\beta=0.65$}
    \label{fig:Re_vs_k_beta_0pt65}
  \end{subfigure}
  \begin{subfigure}[htp]{0.48\textwidth}
    \includegraphics[width=\textwidth]{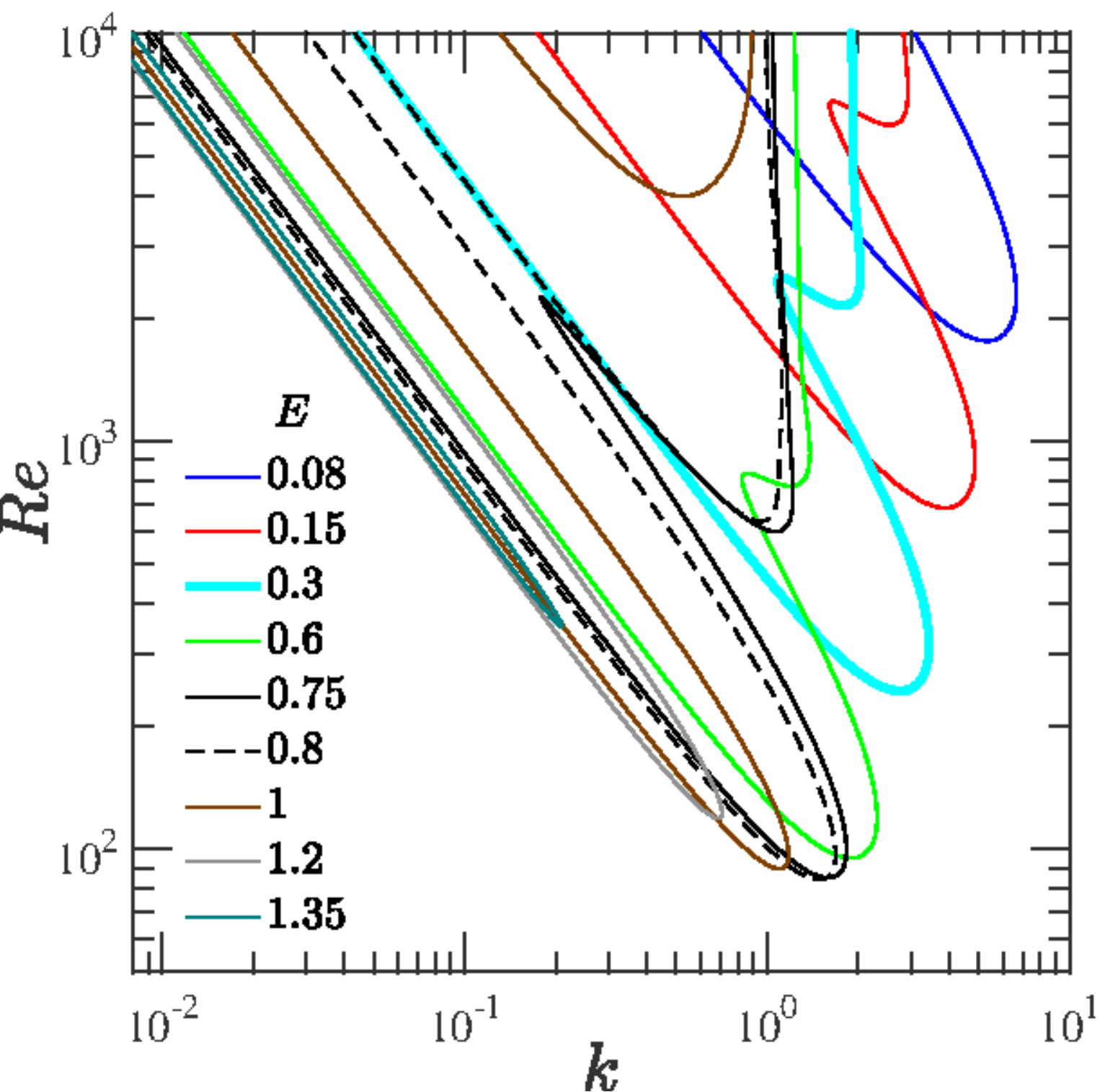}
    \caption{$\beta=0.9$}
    \label{fig:Re_vs_k_beta_0pt9}
  \end{subfigure}
                
  \caption{Neutral stability curves in the $Re$--$k$ plane for varying
    $E$ at: (a) $\beta=0.65$, and (b) $\beta=0.9$.}
  \label{fig:Re_vs_k_beta_fixed}
\end{figure}

\section{Neutral stability curves}
\label{sec:neutral_curves}
\begin{figure}
  \centering
  \begin{subfigure}[htp]{0.48\textwidth}
    \includegraphics[width=\textwidth]{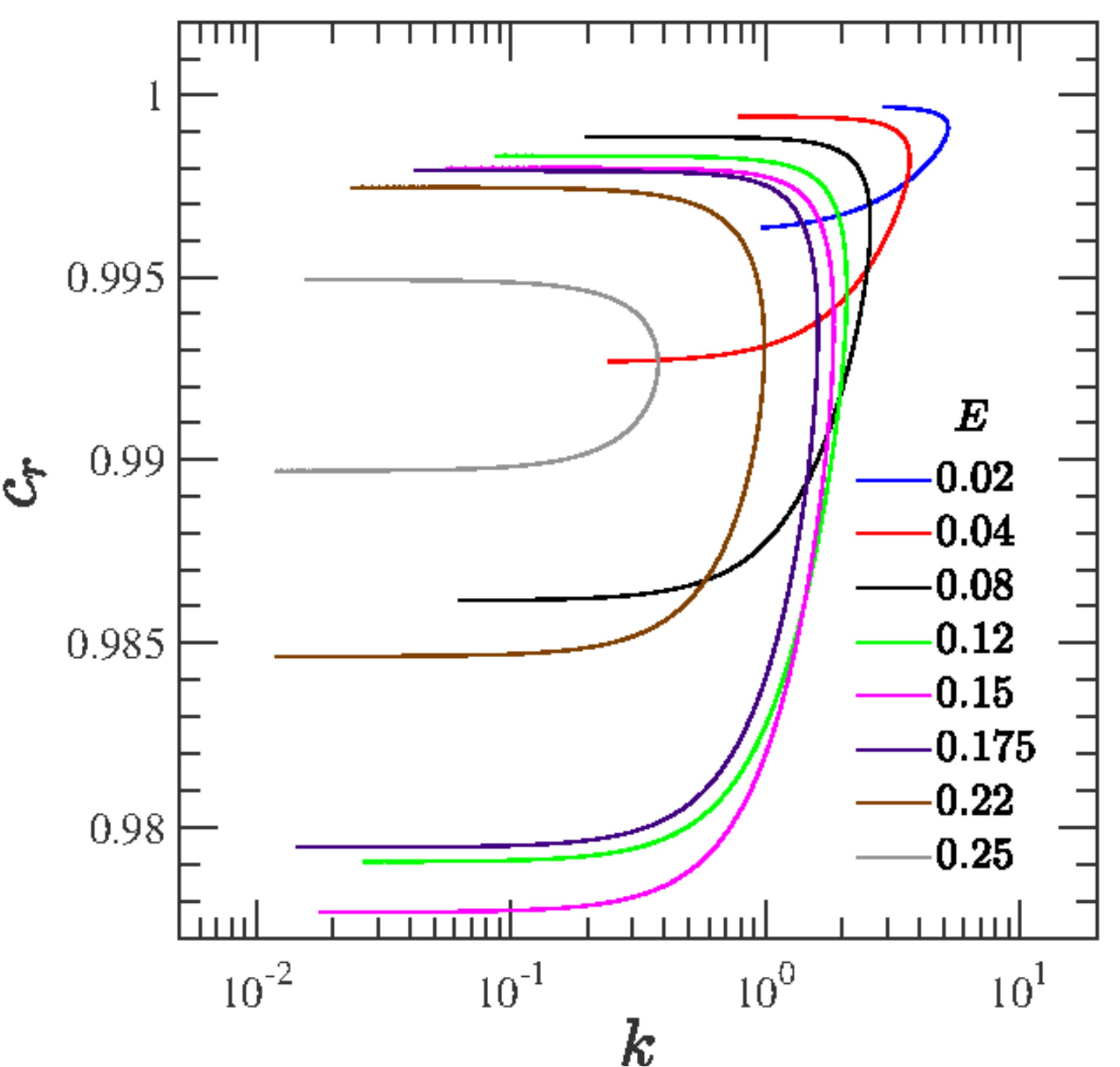}
    \caption{$\beta=0.65$}
    \label{fig:Cr_vs_k_beta_0pt65}
  \end{subfigure}
  \begin{subfigure}[htp]{0.48\textwidth}
    \includegraphics[width=\textwidth]{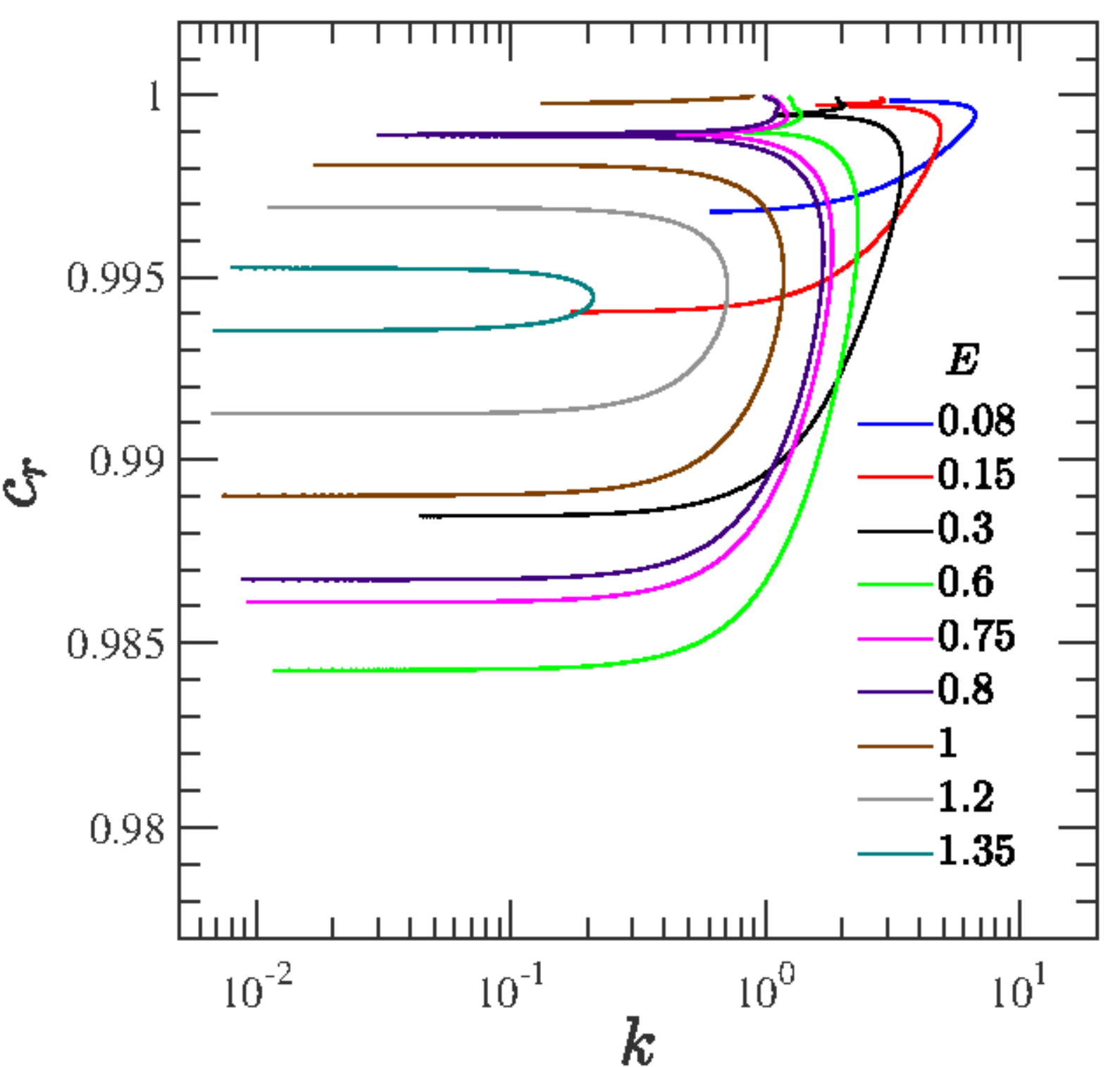}
    \caption{$\beta=0.9$}
    \label{fig:Cr_vs_k_beta_0pt9}
  \end{subfigure}
                
  \caption{The variation of the phase speed,  as a function of $k$, corresponding to the neutral curves for different 
    $E$ in Fig.~\ref{fig:Re_vs_k_beta_fixed} at two different values of $\beta$.}
  \label{fig:Cr_vs_k_beta_fixed}
\end{figure}

 Figures~\ref{fig:Re_vs_k_beta_0pt65} and \ref{fig:Re_vs_k_beta_0pt9} show
the neutral stability curves in the
$Re$-$k$ plane for fixed $\beta$ and $E$.  The curves are in the form of loops,
with the region inside the loop being unstable. While $Re \sim 1/k$
for $k \ll 1$ in the lower and upper branches of the loop for the smaller $\beta$ ($=0.6$), the upper
branch behaves in a different manner for $\beta = 0.9$.  In Fig.~\ref{fig:Re_vs_k_beta_0pt9}, the upper
branch has a non-monotonic behaviour as $E$ is increased, with a secondary minimum emerging
at a higher $\Rey$. This feature of multiple minima is reminiscent of a similar phenomenon observed, albeit for wall modes, in the UCM limit for plane channel flow \citep[see Fig.~15 of][]{chaudhary_etal_2019}.
The two minima move apart with increasing $E$, and for $E = 0.8$ and higher, the junction of the two distinct lobes in a given
neutral curve moves out of the range of $\Rey$ examined. Thus, the neutral curves  for $E \geq 0.8$ appear as a pair of disconnected envelopes. Both branches of the lower envelope exhibit the aforementioned $1/k$ scaling for small $k$. In contrast, only the lower branch of the upper envelope exhibits this scaling, with the upper branch being almost vertical (Fig.~\ref{fig:Re_vs_k_beta_0pt9}).
%
 The phase
speeds corresponding to the neutral curves shown in
Figs.~\ref{fig:Re_vs_k_beta_0pt65} and \ref{fig:Re_vs_k_beta_0pt9} are
shown in Figs.~\ref{fig:Cr_vs_k_beta_0pt65} and
\ref{fig:Cr_vs_k_beta_0pt9} respectively. Overall, the phase speeds
always remain close to, but less than, unity (the 
maximum  base-flow velocity). 
For the higher $\beta$, $c_r$ varies in a narrower range
close to unity, approaching it more closely at the higher $E$ (Fig.~\ref{fig:Cr_vs_k_beta_0pt9}), but never exceeding unity. Thus, the center mode  character of the
instability is preserved all along the neutral curves.
Similar to the two-lobed structure
of the neutral curves in the $Re$--$k$ plane for $\beta = 0.9$ (Fig.~\ref{fig:Re_vs_k_beta_0pt9}), a corresponding two-lobed structure is seen in the $c_r$--$k$ plane as well for $E \approx 0.15$ onwards.

For a given $E$ and
$\beta$, the  minimum of the neutral curve 
(the global one when there are multiple lobes) 
is the critical
Reynolds number ($\Rey_c$), the lowest Reynolds
number at which the flow is unstable. We
mainly focus on the lower curve only, because the critical Reynolds
number $\Rey_c$ lies on it. 
To begin with, an increase in $E$ shifts the neutral curves to lower $\Rey$ and $k$, but beyond a
certain critical $E$, the neutral curves again shift towards higher
$\Rey$.  Interestingly, the minima of the neutral curves are $O(100)$
for sufficiently high $E$ \citep[as first reported in our Letter;][]{Garg2018}, 
as opposed to a typical $\Rey$ of $O(2000)$ for the
Newtonian transition.
          \begin{figure}
            \center
            \includegraphics[width=0.5\textwidth]{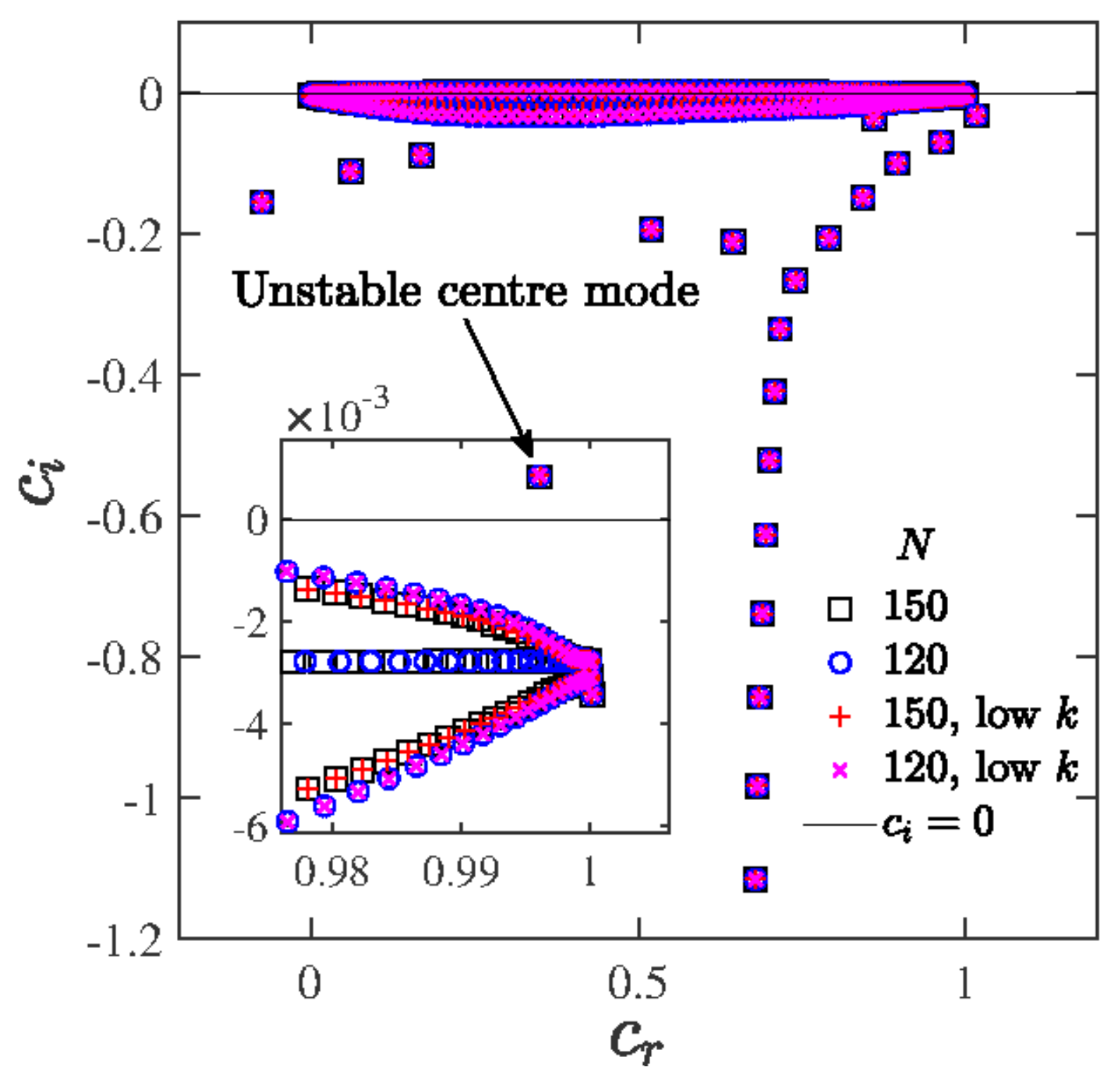}
            \caption{Comparison of the (unfiltered) asymptotic small-$k$ eigenspectrum with that
            obtained from the
           full problem for
              $\beta=0.9,E=0.15,k=0.3$ and $\Rey=8000$. Inset shows
              the zoomed region near the unstable mode.}
            \label{fig:es_lowk}
          \end{figure}
The $\Rey \propto k^{-1}$ scaling followed by the lower branches of the neutral curves
in Fig.~\ref{fig:Re_vs_k_beta_fixed} 
 suggests a regular perturbation analysis in the $k \ll 1$ limit wherein
Eqs. \ref{eqn:lin_conti}--\ref{eqn:lin_zzstress} can be 
simplified by systematically neglecting terms of $O(k)$ or higher. From the
neutral curves at fixed $E$, one obtains $\Rey = k^{-1}\tilde{\Rey}$,
$W = k^{-1}\tilde{W}$ for the $k$-scalings of the dimensionless parameters.
The radial velocity may be expanded as:
\begin{equation}
  \tilde{v}_r \equiv \tilde{v}_r^{(0)} + k \tilde{v}_r^{(1)} + k^2 \tilde{v}_r^{(2)} + ...\, ,
\end{equation}
which, when substituted in the continuity, $z$-momentum, $rr$-, $rz$- and
$zz$-stress equations, i.e., Eqs. \ref{eqn:lin_conti},
\ref{eqn:lin_zmom}--\ref{eqn:lin_rzstress} and \ref{eqn:lin_zzstress},
yields the following scalings at leading order:

\begin{equation}
  \left. \begin{array}{ll}  
           \displaystyle\tilde{v}_r \sim\tilde{v}_r^{(0)},\quad \tilde{v}_z \sim k^{-1}\tilde{v}_z^{(0)},\quad \tilde{p} \sim k^{-1}\tilde{p}^{(0)},\\[8pt]
           \displaystyle\tilde{\tau}_{rr} \sim k\tilde{\tau}_{rr}^{(0)},\quad\tilde{\tau}_{rz} \sim \tilde{\tau}_{rz}^{(0)},\quad\tilde{\tau}_{zz} \sim k^{-1} \tilde{\tau}_{zz}^{(0)},
         \end{array}\right\}
       \label{eqn:lowk_vrtozz}
     \end{equation}
     The above scalings are used in
     Eqs. \ref{eqn:lin_conti}--\ref{eqn:lin_zmom} to obtain the
     following simplified set of equations, to leading order in $k$:
     \begin{eqnarray}
       & (\mathrm{D}+\frac{1}{r}) \tilde{v}_r^{(0)} + \mathrm{i} \tilde{v}_z^{(0)} = 0,\label{eqn:lowk_conti}\\
       & \mathrm{D}\tilde{p}^{(0)} = 0,\label{eqn:lowkr}\\
       & -U'\tilde{v}_r^{(0)} + \{\frac{\beta}{\tilde{\Rey}} (\mathrm{D}^2 + \frac{\mathrm{D}}{r}) - \mathrm{i}(U-c)\}\tilde{v}_z^{(0)} - \mathrm{i}\tilde{p}^{(0)} + (\mathrm{D}+\frac{1}{r}) \tilde{\tau}_{rz}^{(0)} + \mathrm{i}\tilde{\tau}_{zz}^{(0)}=0.\label{eqn:lowkz} 
     \end{eqnarray}
     The boundary conditions become:
     \begin{equation}
       \left. \begin{array}{ll}  
                \displaystyle\tilde{v}_r^{(0)}=0=\tilde{v}_z^{(0)} \quad \mbox{at\ }\quad r=1,\\[8pt]
                \displaystyle  \tilde{v}_r^{(0)}=0, \tilde{v}_z^{(0)} = \text{ finite, }\tilde{p}^{(0)} = \text{ finite} \quad \mbox{at \ }\quad r=0,
              \end{array}\right\}
            \label{eqn:lowkbc}
          \end{equation}
          The simplified system comprising
          Eqs.~\ref{eqn:lowk_conti}--\ref{eqn:lowkz} was solved using
          a spectral method and the eigenspectrum obtained is
          compared with that for the full problem at $k = 0.3$ for the same
          parameters (Fig.~\ref{fig:es_lowk}); the inset zooms in on the unstable center mode. Both eigenspectra have a similar structure, and in particular, the center mode obtained from the low-$k$ analysis has the same phase speed and growth rate as that in the original problem.

\begin{figure}
  \centering
  \begin{subfigure}[htp]{0.45\textwidth}
    \includegraphics[width=\textwidth]{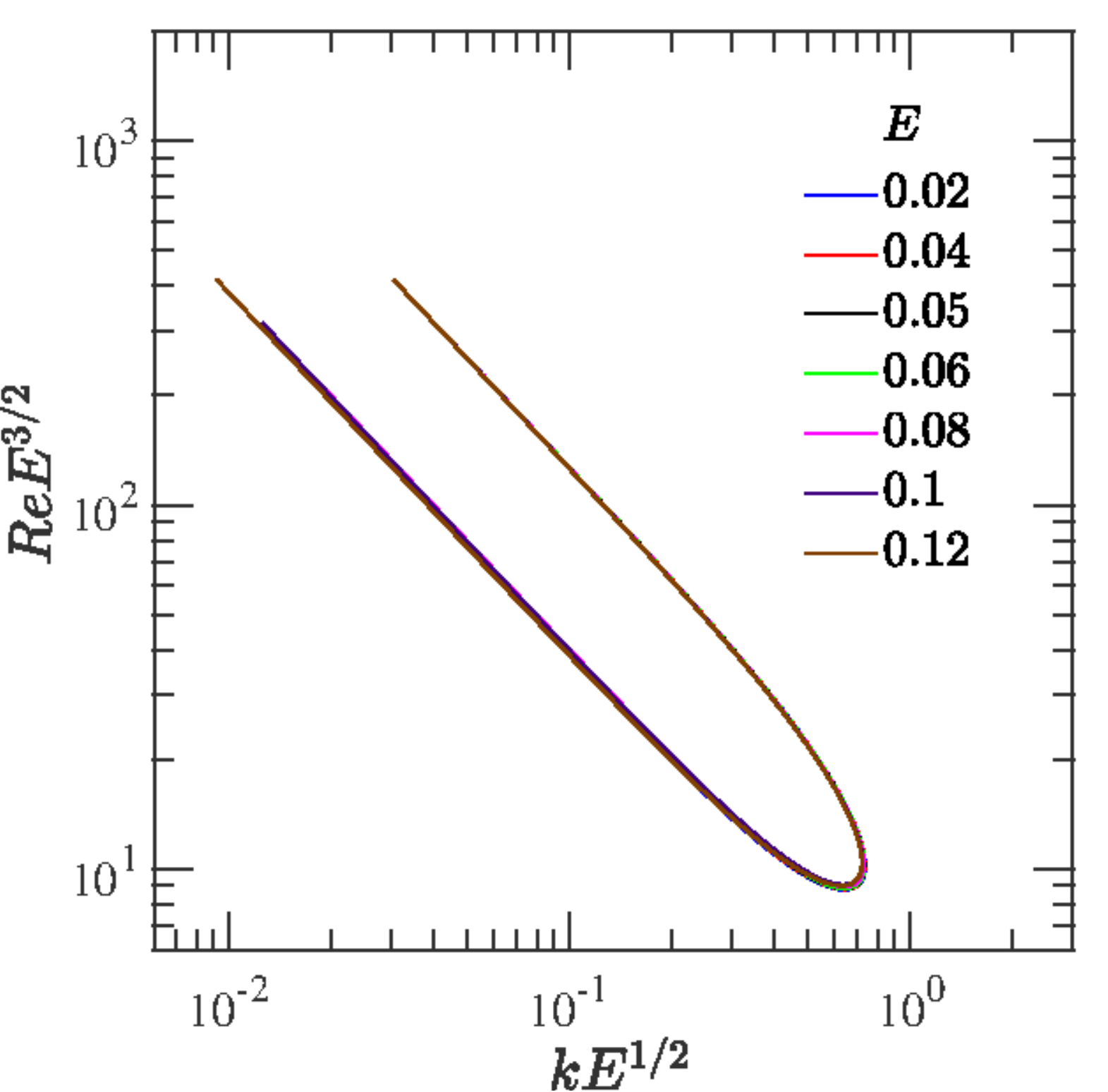}
    \caption{Rescaled neutral curves for $\beta = 0.65$}
    \label{fig:collapse_beta_0pt65}
  \end{subfigure}
  \begin{subfigure}[htp]{0.45\textwidth}
    \includegraphics[width=\textwidth]{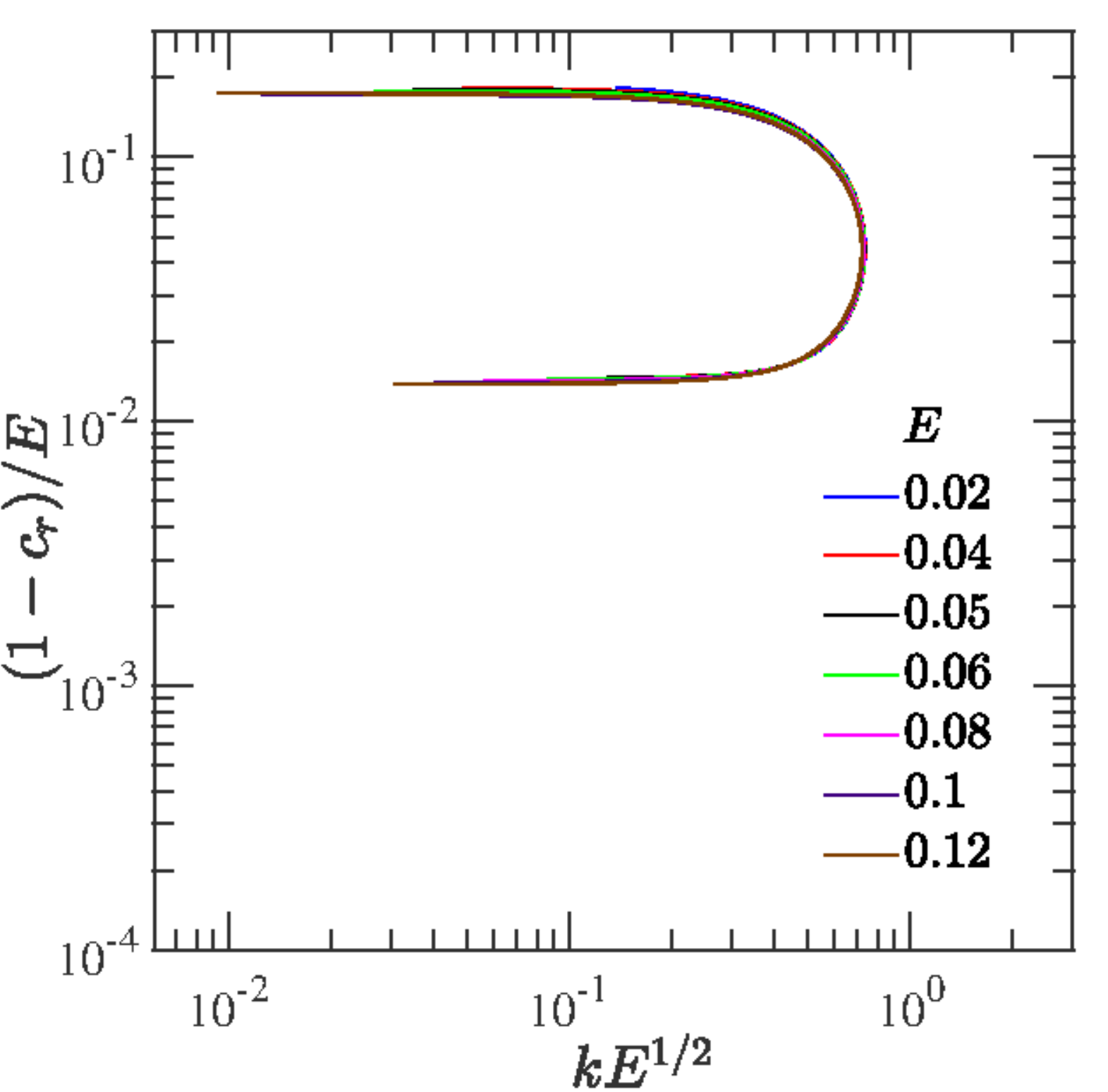}
    \caption{$(1-c_r)$ collapse for $\beta=0.65$}
    \label{fig:cr_vs_k_collapse_beta_0pt65}
  \end{subfigure}
 \begin{subfigure}[htp]{0.45\textwidth}
    \includegraphics[width=\textwidth]{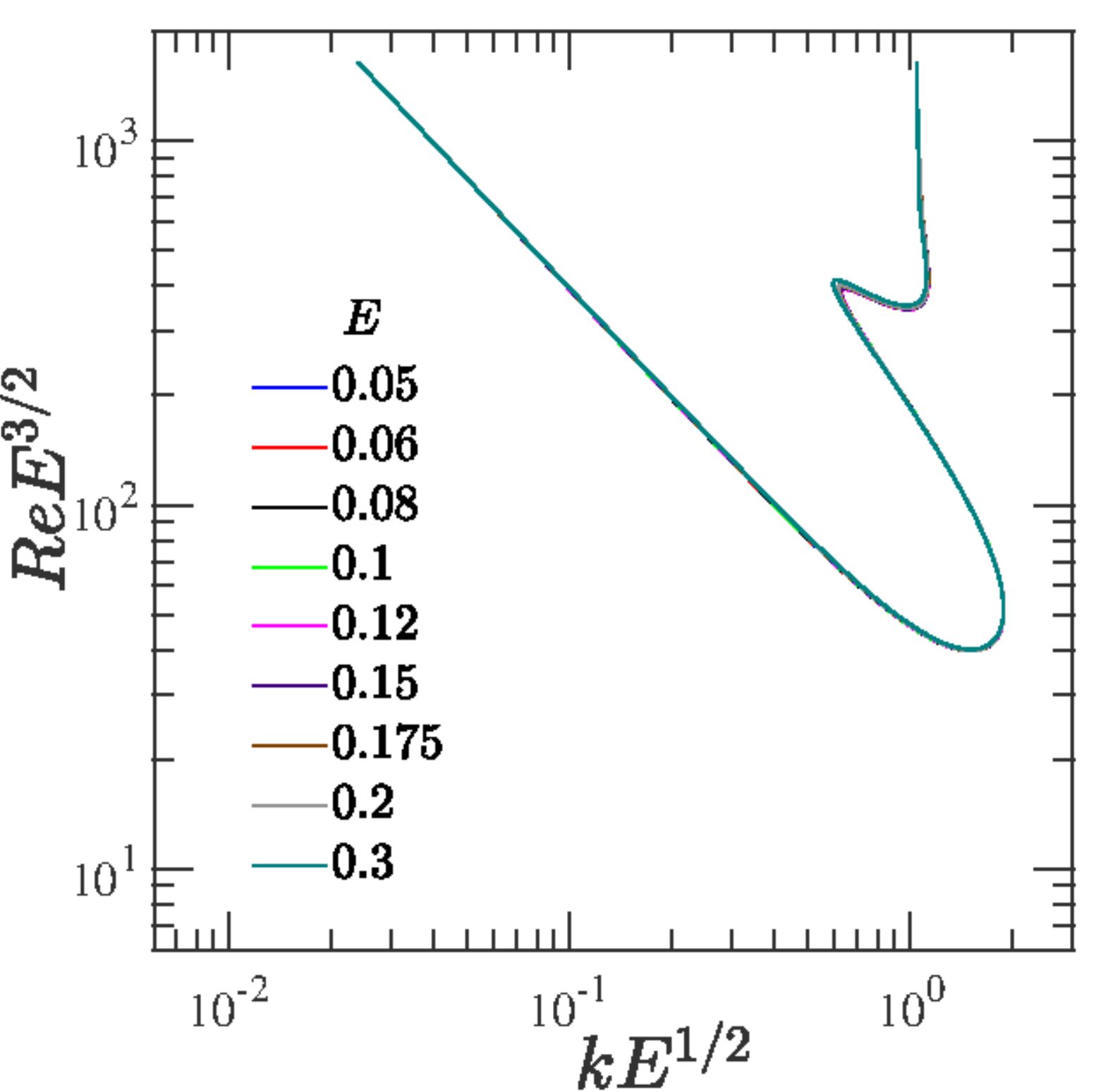}
    \caption{Rescaled neutral curves for $\beta = 0.9$}
    \label{fig:collapse_beta_0pt9}
  \end{subfigure}
  \begin{subfigure}[htp]{0.45\textwidth}
    \includegraphics[width=\textwidth]{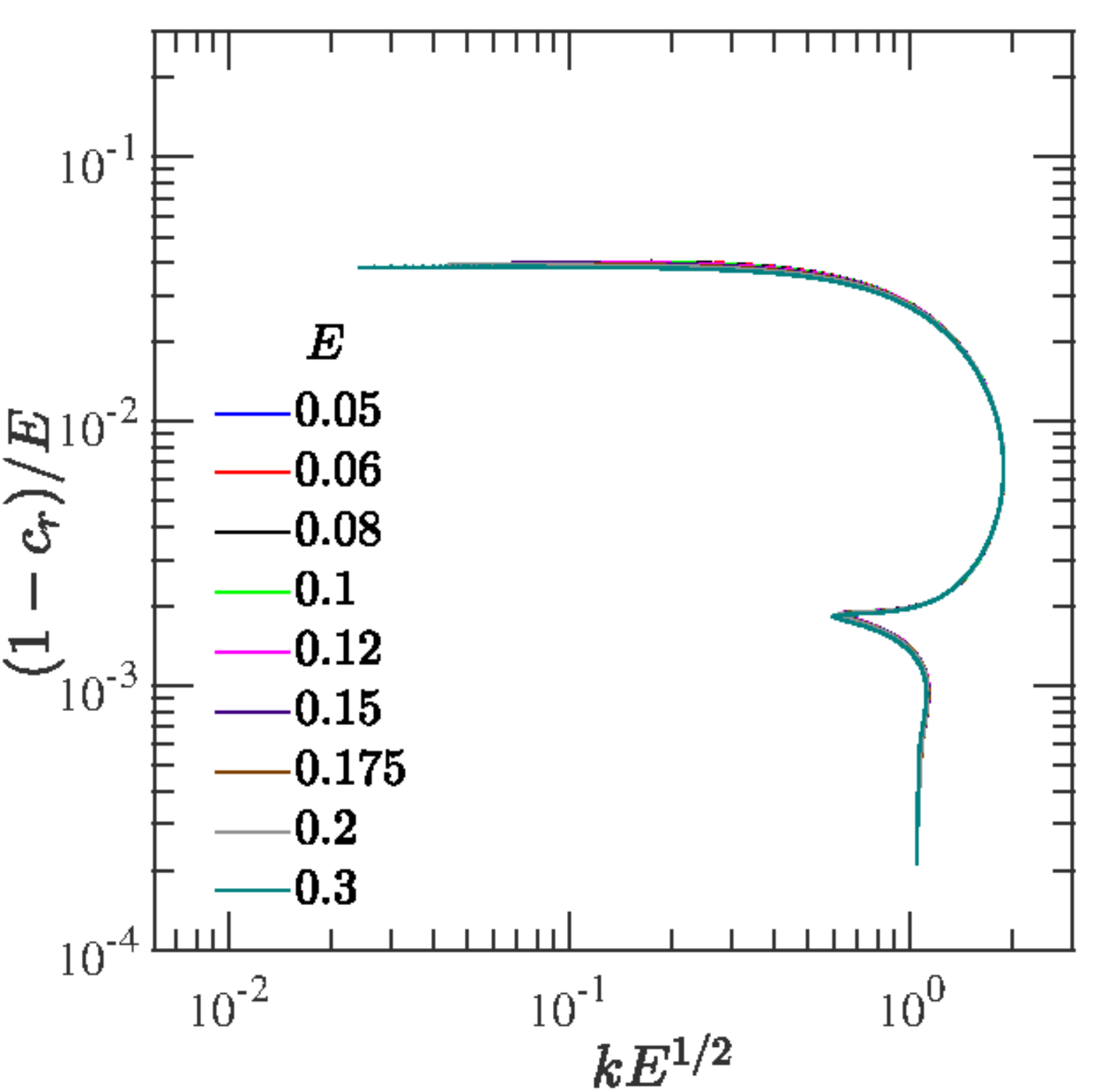}
    \caption{$(1-c_r)$ collapse for $\beta=0.9$}
    \label{fig:cr_vs_k_collapse_beta_0pt9}
  \end{subfigure}        
    \caption{Collapse of neutral curves for different $E$ in the $Re$-$k$ plane (panels (a) and (c)) and collapse of the corresponding phase speeds (panels (b) and (d)) for two different $\beta$.}
  \label{fig:collapses}
\end{figure}

\begin{figure}
  \centering
  \includegraphics[width=0.48\textwidth]{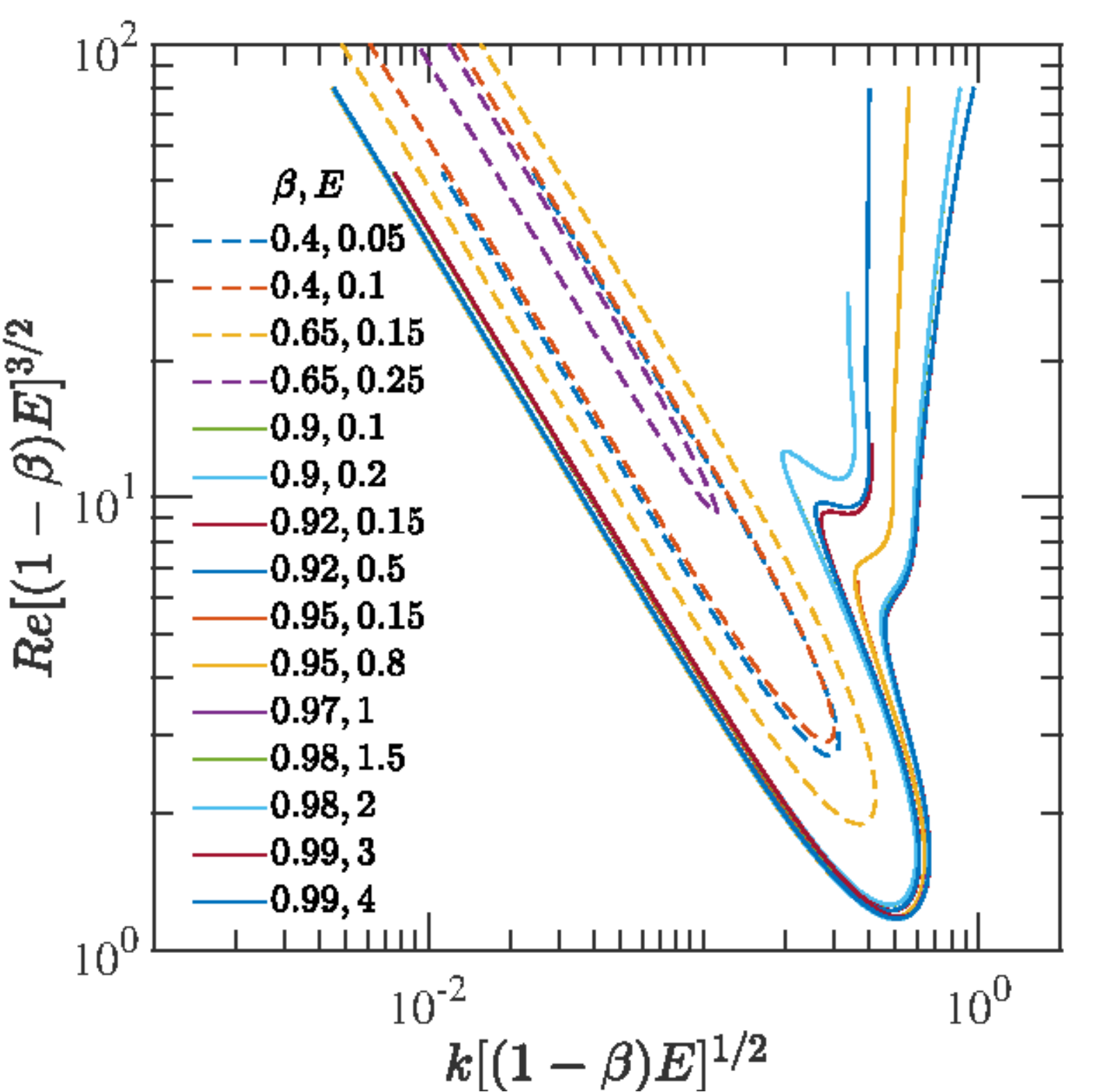}
  \caption{Neutral curves for different $\beta$ and $E$ plotted in terms of $\Rey[(1-\beta)E]^{3/2}$ vs. $k[(1-\beta)E]^{1/2}$:  The rescaled neutral curves collapse for $\beta \rightarrow 1$.}
  \label{fig:Re_vs_k_collapse_1mbE}
\end{figure}

\subsection{Collapse of neutral curves}
\label{sec:collapses}
The qualitatively similar character of the neutral curves at different $E$
in Fig.~\ref{fig:Re_vs_k_beta_fixed}
 is strongly suggestive of a
collapse upon suitable rescaling of both $\Rey$ and $k$ with the elasticity number
$E$. Figure~\ref{fig:collapses} shows that such a collapse is indeed
possible for sufficiently small $E$, when $Re$ is rescaled as
$\Rey E^{3/2}$ and $k$ as $k E^{1/2}$. 
%
%
These scalings are found to be
valid for fixed $\beta$, although the nature of the collapsed curve does depend 
on $\beta$ (as evident from Figs.~\ref{fig:collapse_beta_0pt65} and \ref{fig:collapse_beta_0pt9}).
Similarly, as shown in Figs.~\ref{fig:cr_vs_k_collapse_beta_0pt65} and
\ref{fig:cr_vs_k_collapse_beta_0pt9}, 
the curves for the rescaled phase
speed  $(1-c_r)/E$, plotted as a function of 
$kE^{1/2}$,  again exhibit a collapse, implying that $(1-c_r)$ is $O(E)$ for $E \ll 1$.

\begin{figure}
        \centering
        \begin{subfigure}[htp]{0.48\textwidth}
                \includegraphics[width=\textwidth]{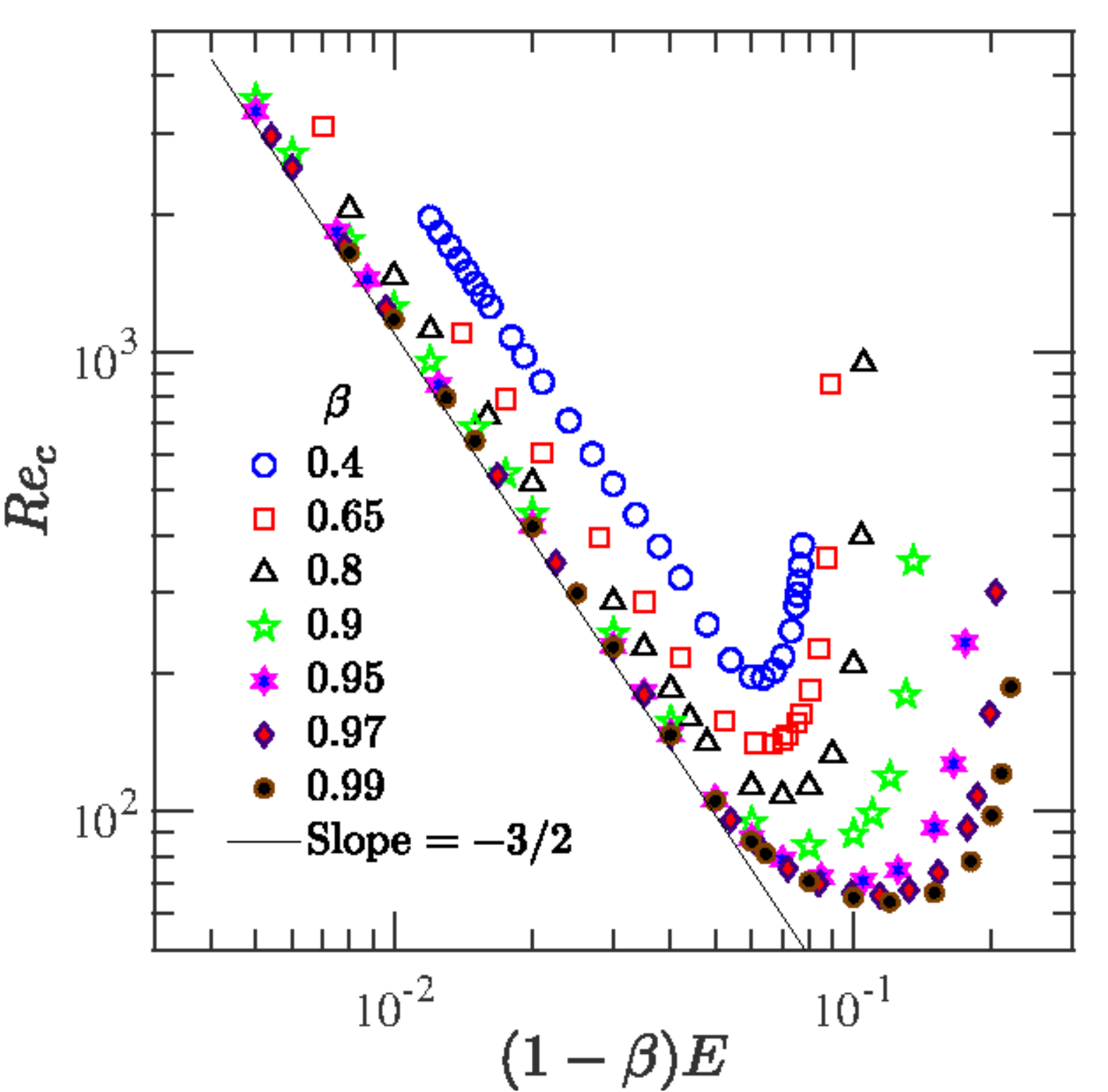}
                \caption{$Re_c$ vs. $E(1-\beta)$}
                \label{fig:Rec_vs_E_beta0pt4to0pt9}
        \end{subfigure}
         \begin{subfigure}[htp]{0.48\textwidth}
                \includegraphics[width=\textwidth]{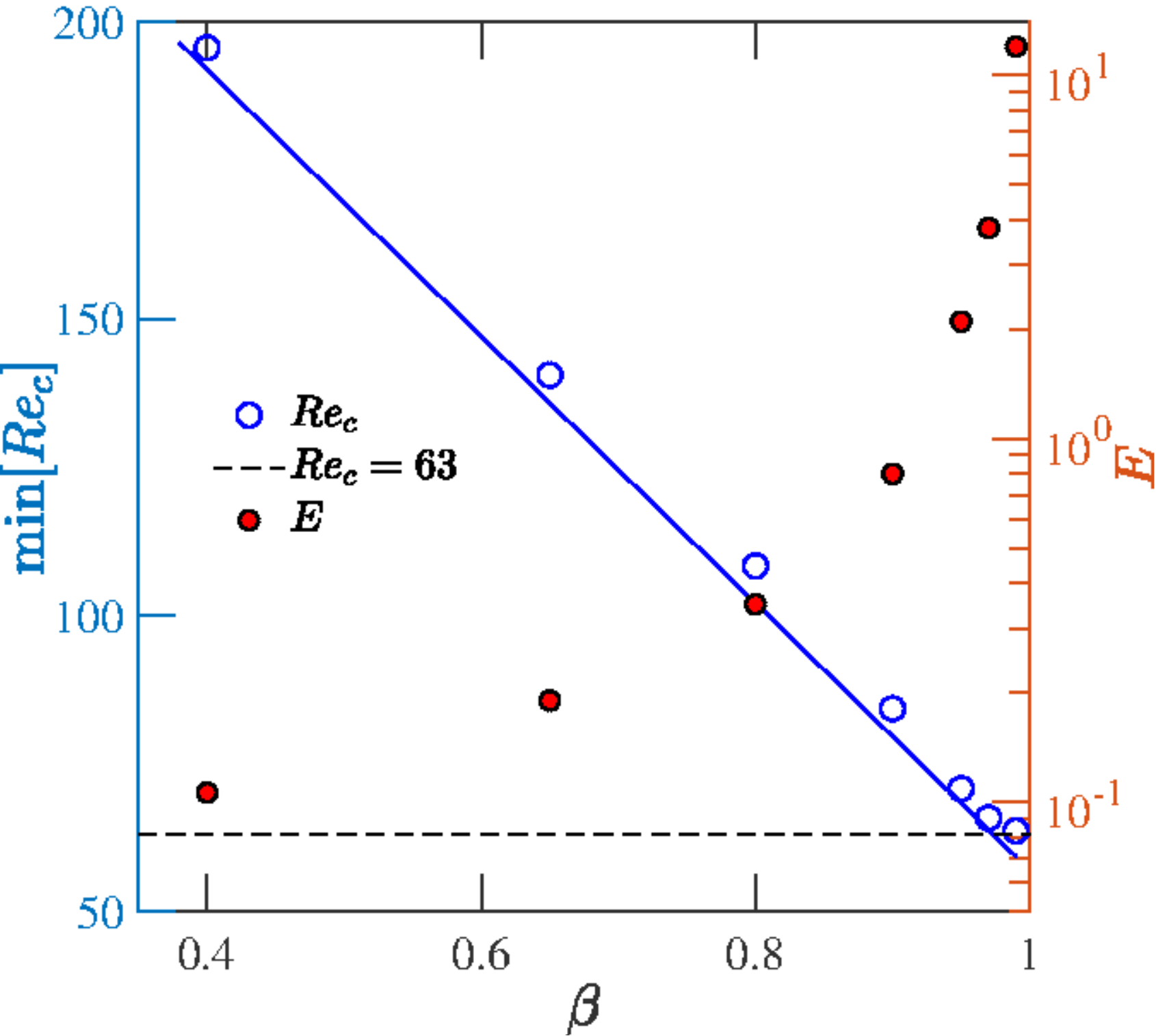}
                \caption{$\mathrm{min}[\Rey_c]$ vs $\beta$}
                \label{fig:minRe_c_vs_beta}
        \end{subfigure}
        \begin{subfigure}[htp]{0.48\textwidth}
                \includegraphics[width=\textwidth]{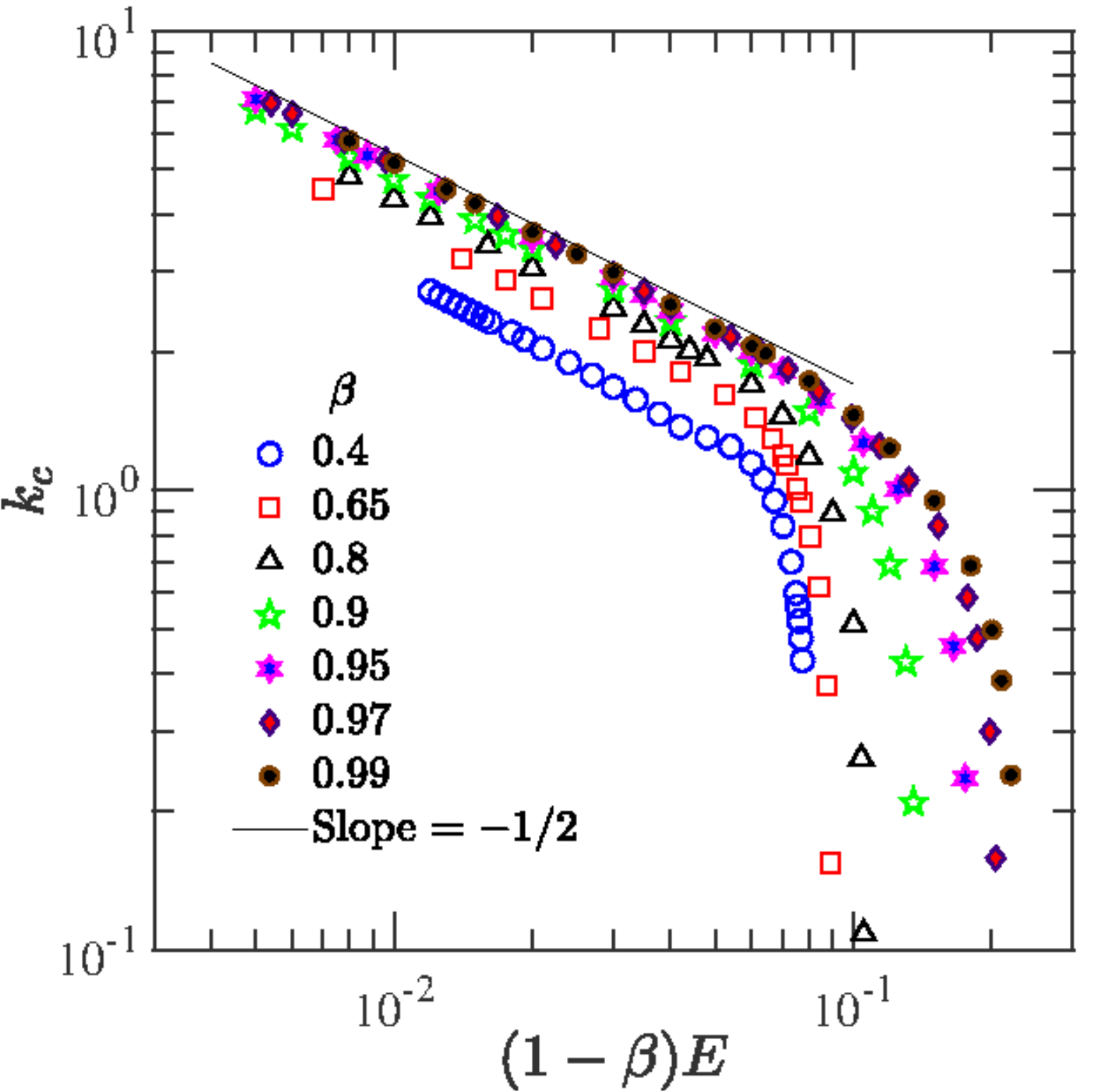}
                \caption{$k_c$ vs $E(1-\beta)$}
                \label{fig:kc_vs_E_beta0pt4to0pt9}
        \end{subfigure}
        \begin{subfigure}[htp]{0.48\textwidth}
                \includegraphics[width=\textwidth]{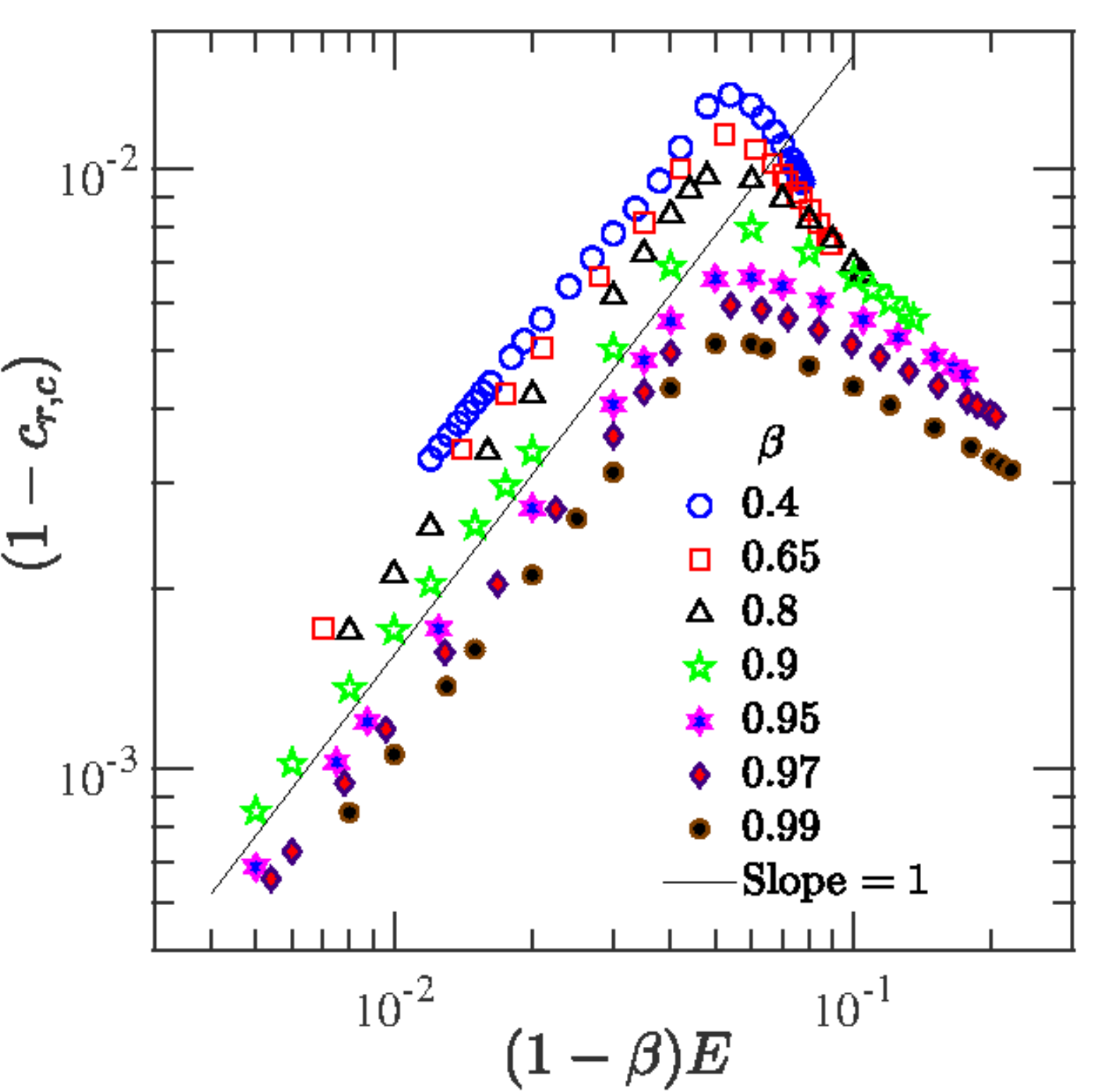}
                \caption{$(1-c_{r,c})$ vs $E(1-\beta)$}
                \label{fig:crc_vs_E_beta0pt4to0pt9}
        \end{subfigure}
       
        \caption{Variation of critical parameters with $E (1-\beta)$ for $\beta$ ranging from $0.4$ to $0.99$. (a) $Re_c$ vs $E(1-\beta)$, (b) The minima of $\Rey_c$ of panel (a) decreases approximately in a linear manner  with $\beta$, but appears to approach a finite value as $\beta \rightarrow 1$, while the corresponding $E$ diverges as $\beta \rightarrow 1$,
 (c) $k_c$ vs $E(1-\beta)$, and (d) $(1-c_{r,c})$ vs $E(1-\beta)$. $\Rey_c$ and $k_c$ follow the scalings $Re_c\propto [E(1-\beta)]^{-3/2}$ and $k_c\propto [E(1-\beta)]^{-1/2}$ respectively below a critical value of $E(1-\beta)$. }
                \label{fig:Rec_kc_crc_vs_E_beta0pt4to0pt9}
                
\end{figure}

%
%
While the collapse obtained above is for a fixed $\beta$ and for $E \ll 1$, 
a further collapse is obtained in the dual limit $E (1-\beta) \ll 1$,
 $(1-\beta) \ll 1$, when 
 the neutral curves are plotted in terms of $\Rey[(1-\beta)E]^{3/2}$
and $k [(1-\beta)E]^{1/2}$ as shown in
Fig.~\ref{fig:Re_vs_k_collapse_1mbE}, implying that the threshold
$\Rey$ and $k$ scale as $\Rey \propto [(1-\beta)E]^{-3/2}$ and
$k \propto [(1-\beta)E]^{-1/2}$ respectively, in this limit. 
The rescaled neutral curves in Fig.~\ref{fig:Re_vs_k_collapse_1mbE} begin to collapse onto a single one only for $\beta > 0.9$, the collapse being perfect for the lower branch, but less so for the upper ones.
  Thus,  the role of the solvent viscosity appears to be `universal' only as far as the lower branch is concerned.   
Importantly, however, since the critical $Re$ occurs on the lower branches of the neutral curves, the transition to the elasto-inertial turbulent state is governed by the combination $E(1-\beta)$ for 
$E (1-\beta) \ll 1$, $(1-\beta) \ll 1$.
It is worth noting that 
the nearly-vertical nature of the upper branch implies that the instability appears to exist in the limit of $\Rey \rightarrow \infty$, with $E$ fixed. 
An axisymmetric version of the `elastic Rayleigh' equation \citep[the elastic analogue of the classical Rayleigh equation; see][]{hinchrallison1995,roy_subramanian2018}, which also has $E(1-\beta)$ as the governing parameter, is known to govern the linearized dynamics of perturbations in this limit, and involves a balance of inertial and elastic forces in the fluid. There is, however,  no instability associated with 
the elastic Rayleigh equation for plane- \citep{kaffelrenardy2010} and pipe-Poiseuille 
\citep{chaudhary_inviscid} flows, and the lack of collapse of 
the (near-vertical) upper branches, and the implied instability for $Re \rightarrow \infty$, in Fig.~\ref{fig:Re_vs_k_collapse_1mbE}, betrays
therefore the singular nature of the inviscid elastic limit, with viscous effects playing
a likely role even as $Re \rightarrow \infty$.

          \begin{figure}
            \centering
            \begin{subfigure}[htp]{0.48\textwidth}
              \includegraphics[width=\textwidth]{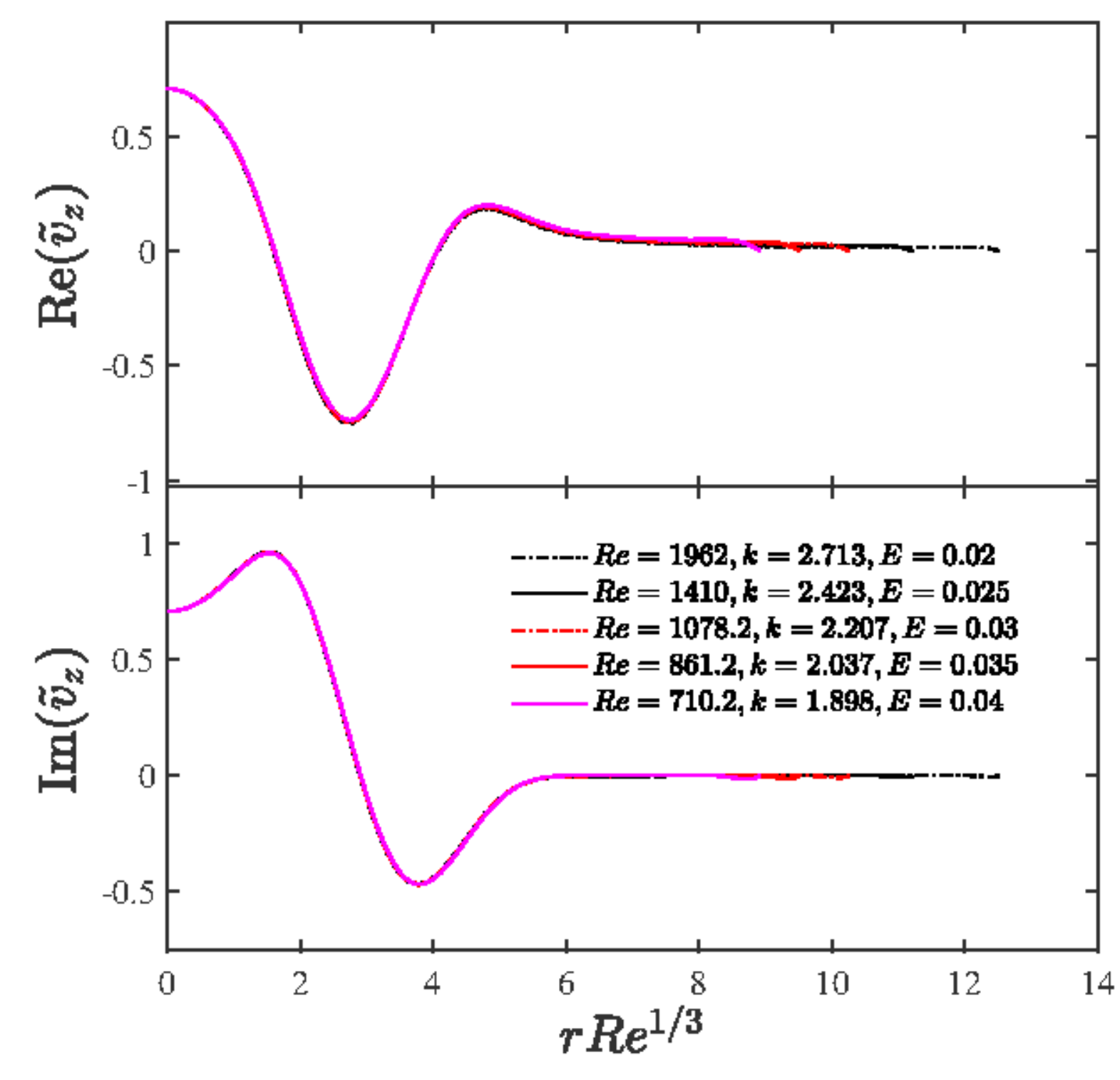}
              \caption{ $\tilde{v}_z$}
              \label{fig:Vz_collapse_1by3}
            \end{subfigure}
            \begin{subfigure}[htp]{0.48\textwidth}
              \includegraphics[width=\textwidth]{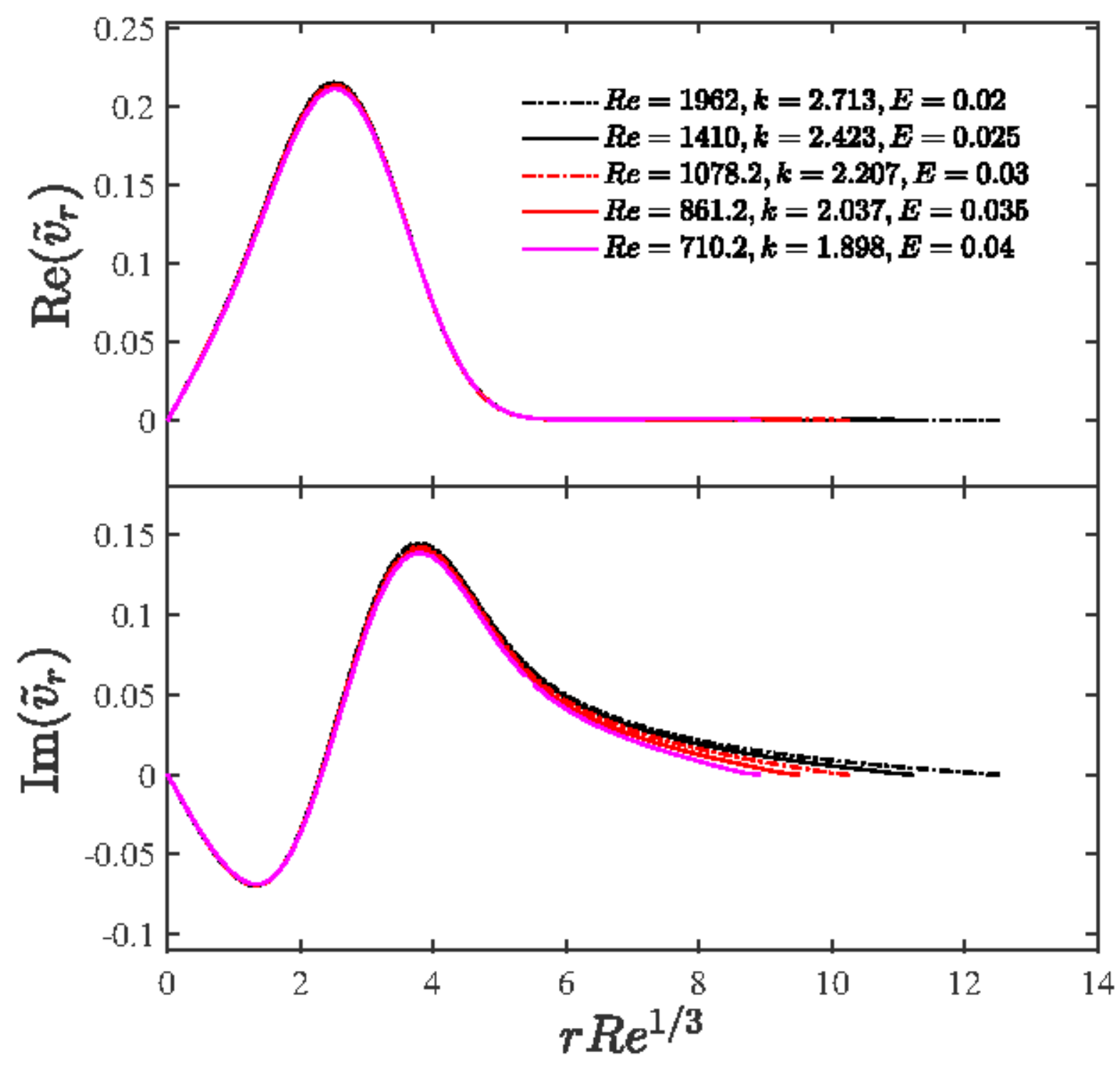}
              \caption{$\tilde{v}_r$}
              \label{fig:Vr_collapse_1by3}
            \end{subfigure}
                
            \caption{Collapse of eigenfunctions at different $\Rey$ and $k$ 
              along the lower branch of the neutral curve for $\beta = 0.4$.
               The eigenvalues (with $c_i = 0$) for which the rescaled eigenfunctions are shown are
              $c = 0.996707, 0.995912, 0.995131, 0.994367$ and
              $0.993621$ respectively.}
            \label{fig:eigfun_collapse_1by3}
          \end{figure}
          \subsection{Critical parameters and scalings}
          \label{sec:scalings}
          Figures~\ref{fig:Rec_vs_E_beta0pt4to0pt9},
          \ref{fig:kc_vs_E_beta0pt4to0pt9} and
          \ref{fig:crc_vs_E_beta0pt4to0pt9} show the variation of critical
          parameters $\Rey_c$, $k_c$ and $c_{r,c}$ with $E (1-\beta)$
          for different $\beta$. 
          Irrespective of $\beta$, the critical parameters conform to scaling laws for small $E(1-\beta)$; thus, $Re_c \propto (E(1-\beta))^{-3/2}$, $k_c \sim (E(1-\beta))^{-1/2}$, and 
$(1-c_{r,c}) \sim (1-\beta) E$. Further, the curves for $\beta > 0.9$ collapse onto a universal curve in this limit (as was expected from the findings of the previous section in the dual limit $E(1-\beta) \ll 1$, $(1-\beta) \ll 1$. The collapse breaks down for $E(1-\beta) > 0.05$, with the breakdown occurring
at the point where the original neutral curves in the $Re$--$k$ plane start shifting upwards (after becoming two-lobed), with the lower lobe shrinking in size with increasing $E$ (Fig.~\ref{fig:Re_vs_k_beta_fixed}).             
 As $E (1-\beta)$ is increased, $Re_c$ reaches a minimum value and beyond a threshold value of $E (1-\beta)$, it increases rather sharply
          indicating the flow to be stable beyond this threshold. However, this threshold shifts to higher $E(1-\beta)$ as $\beta \rightarrow 1$,  and $Re_c$ therefore continues to decrease for $\beta \rightarrow 1$, with the lowest $Re_c$ found being as small as $63$ (albeit for $E \sim 10$). The latter suggests that pipe flow of strongly elastic dilute polymer solutions
          can become unstable at an $\Rey$ much lower than that for their Newtonian counterparts.
%
%
%
%
%
%
           Figure~\ref{fig:minRe_c_vs_beta} shows that $Re_c$ in Fig.~\ref{fig:Rec_vs_E_beta0pt4to0pt9} decreases  approximately in a linear manner with $\beta$, although  there appears to be an eventual deviation from linearity for the highest $\beta = 0.99$ analyzed in this study. 
          The aforementioned deviation from linearity suggests the approach of $Re_c$ to a finite lower bound  regardless of $E$ or $\beta$, and that this lower bound is attained with $E(1-\beta)$ being finite. However, note that the corresponding $E$ diverges as $1/(1-\beta)$ for $\beta \rightarrow 1$, implying that the flow only becomes unstable for a very high $W$ in this limit.
   

          The scalings for the parameters
          ($\Rey$, $c_r$, $k$) with $E$ for $E\ll 1$, found above, may also be justified
          using a scaling analysis for the boundary layer near the centerline, as briefly outlined
          in \cite{Garg2018}.  In the
          limit $\Rey \gg 1$, $E\ll 1$, there is a `core' region
          around the centerline with (dimensionless) extent 
          $\delta \ll 1$, where inertial, elastic, and viscous stresses are equally important. 
          The scalings for $\Rey, k, \delta$ and $c_r$
          in terms of $E$ can be derived by rescaling 
          Eqs. \ref{eqn:lin_conti}--\ref{eqn:lin_zzstress} in the
          region near the centerline as follows. The radial coordinate
          $r$ near the centerline can be expressed as $ r=\delta\xi$, with $\xi \sim O(1)$.
          For $E\ll 1$, our numerical
          results show that $k_c$ becomes large for $ E \ll 1$, and so we set
          $k \sim \delta^{-1}$ in the analysis, as suggested by the continuity equation.
          The eigenvalue $c$ approaches unity in the said limit, and
          as $r \rightarrow 0$, $U\sim 1$, $(U-c) \sim \delta^2$
           and we therefore expand $c$ as
          $c=1+\delta^2c^{(1)}$. The derivatives near the centerline
          get rescaled as
          $  \frac{\mathrm{d}}{\mathrm{d}r}=\frac{1}{\delta}\frac{\mathrm{d}}{\mathrm{d}\xi} \equiv \delta^{-1} \mathrm{D_1}\, $.
          The base-flow profile becomes
          $U = 1-r^2 \equiv 1-\delta^2\xi^2$, and 
          Eqs.~\ref{eqn:lin_conti}--\ref{eqn:lin_zzstress} take the following
          forms near the centerline:
          \begin{eqnarray}
            &\delta^{-1}(\mathrm{D_1}+\xi^{-1})\tilde{v}_r+\mathrm{i}k\tilde{v}_z=0,\label{eqn:scaling_conti}\\
            &-\mathrm{i}k\delta^2(c^{(1)}+\xi^2)\tilde{v}_r=-\delta^{-1}\mathrm{D_1}\tilde{p}+\{\delta^{-1}
              (\mathrm{D_1}+\xi^{-1})\tilde{\tau}_{rr} \nonumber \\
            & +\mathrm{i}k\tilde{\tau}_{rz}-\delta^{-1}\xi^{-1}\tilde{\tau}_{\theta\theta}\} + \beta\Rey^{-
              1}\delta^{-2}\mathrm{L_1}\tilde{v}_r,\label{eqn:scaling_rmom}\\
            &-\mathrm{i}k\delta^2(c^{(1)}+\xi^2)\tilde{v}_z-2\delta\xi\tilde{v}_r=-\mathrm{i}k\tilde{p}+[\delta^
              {-1}(\mathrm{D_1}+\xi^{-1})\tilde{\tau}_{rz}  \nonumber \\
            &+\mathrm{i}k\tilde{\tau}_{zz}] +\beta\Rey^{-1}\delta^{-2}(\mathrm{L_1}+\xi^{-2})\tilde{v}_z,\label
              {eqn:scaling_zmom}\\
            &\{1- \mathrm{i}kW\delta^2(c^{(1)}+\xi^2)\} \tilde{\tau}_{rr} = 2(1-\beta)\Rey^{-1}(\delta^{-
              1}\mathrm{D_1}-2W\mathrm{i}k\delta\xi)\tilde{v}_r,\label{eqn:scaling_rrstress}
          \end{eqnarray}
          \begin{eqnarray}
            &\{1- \mathrm{i}kW\delta^2(c^{(1)}+\xi^2)\} \tilde{\tau}_{rz} +2W\delta\xi\tilde{\tau}_{rr} = (1-
              \beta)\Rey^{-1} [\{\mathrm{i}k\nonumber \\
            &-2W(1-\xi\mathrm{D_1}+4W\mathrm{i}k\delta^2\xi^2)\}\tilde{v}_r+(\delta^{-1}\mathrm{D_1}-2W\mathrm
              {i}k\delta\xi)\tilde{v}_z],\label{eqn:scaling_rzstress}\\
            &\{1- \mathrm{i}kW\delta^2(c^{(1)}+\xi^2)\} \tilde{\tau}_{\theta\theta} = 2(1-\beta)\Rey^{-1}\delta^
              {-1}\xi^{-1}\tilde{v}_r,\label{eqn:scaling_ttstress}\\
            &\{1- \mathrm{i}kW\delta^2(c^{(1)}+\xi^2)\} \tilde{\tau}_{zz} +4W\delta\xi\tilde{\tau}_{rz} = 2(1-
              \beta)\Rey^{-1} [-8W^2\delta\xi\tilde{v}_r\nonumber \\
            &+\{\mathrm{i}k-2W\delta\xi(\delta^{-1}\mathrm{D_1}-4W\mathrm{i}k\delta\xi)\}\tilde{v}_z],\label
              {eqn:scaling_zzstress}
          \end{eqnarray}
          where
          $\mathrm{L_1}=(\mathrm{D_1}^2+\xi^{-1}\mathrm{D_1}-\xi^{-2}-k^2\delta^{2})$. 
           In the $r$-momentum
          equation, a balance of inertial stresses and solvent viscous
          stresses gives $\delta \sim \Rey^{-1/3}$.
          The left-hand side of the linearized constitutive equations
          reveal that in order for the elastic  and viscous
          contributions to be of the same order, we require
          $W \sim \delta^{-1}$, or, after using $\delta \sim \Rey^{-1/3}$, $\Rey \sim E^{-3/2}$. 
          We thus obtain the following scaling relationships for $Re \gg 1$, $E \ll 1$,          
          along the neutral curve:
          \begin{equation}
            \delta \sim Re^{-1/3},\quad k\sim Re^{1/3},\quad \Rey\sim E^{-3/2},\text{ and } (1-c)\sim Re^{-2/3}\, .
            \label{eqn:scalings}
          \end{equation}  
          As shown 
          in Fig.~\ref{fig:eigfun_collapse_1by3},
          the eigenfunctions for different $Re$ and $k$ along the lower branch of the neutral curve do exhibit a collapse when plotted as a function of the boundary layer coordinate $\xi$.   
          \begin{figure}
            \centering
            \begin{subfigure}[htp]{0.48\textwidth}
              \includegraphics[width=\textwidth]{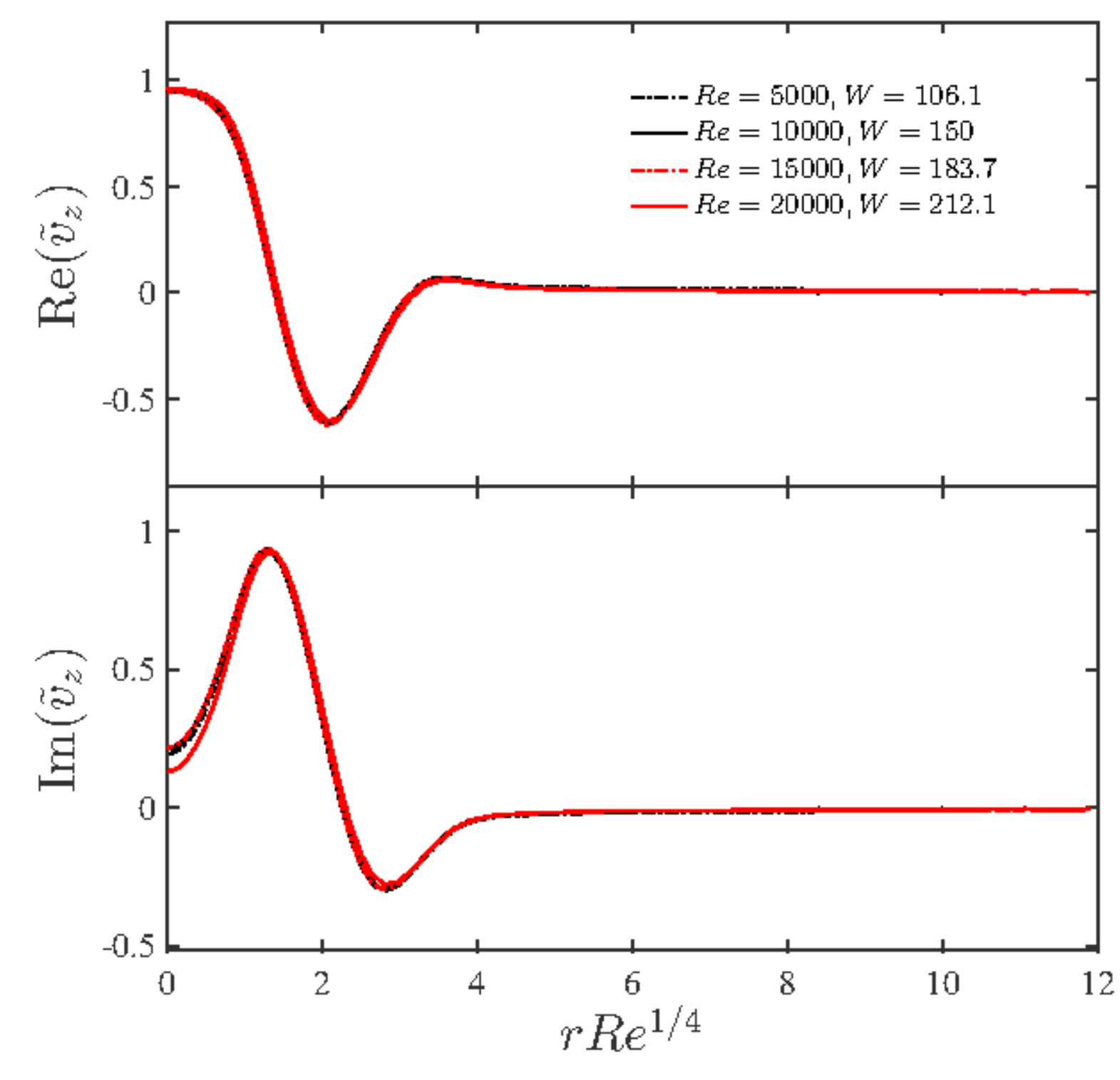}
              \caption{$\tilde{v}_z$}
              \label{fig:Vz_collapse_1by4}
            \end{subfigure}
            \begin{subfigure}[htp]{0.48\textwidth}
              \includegraphics[width=\textwidth]{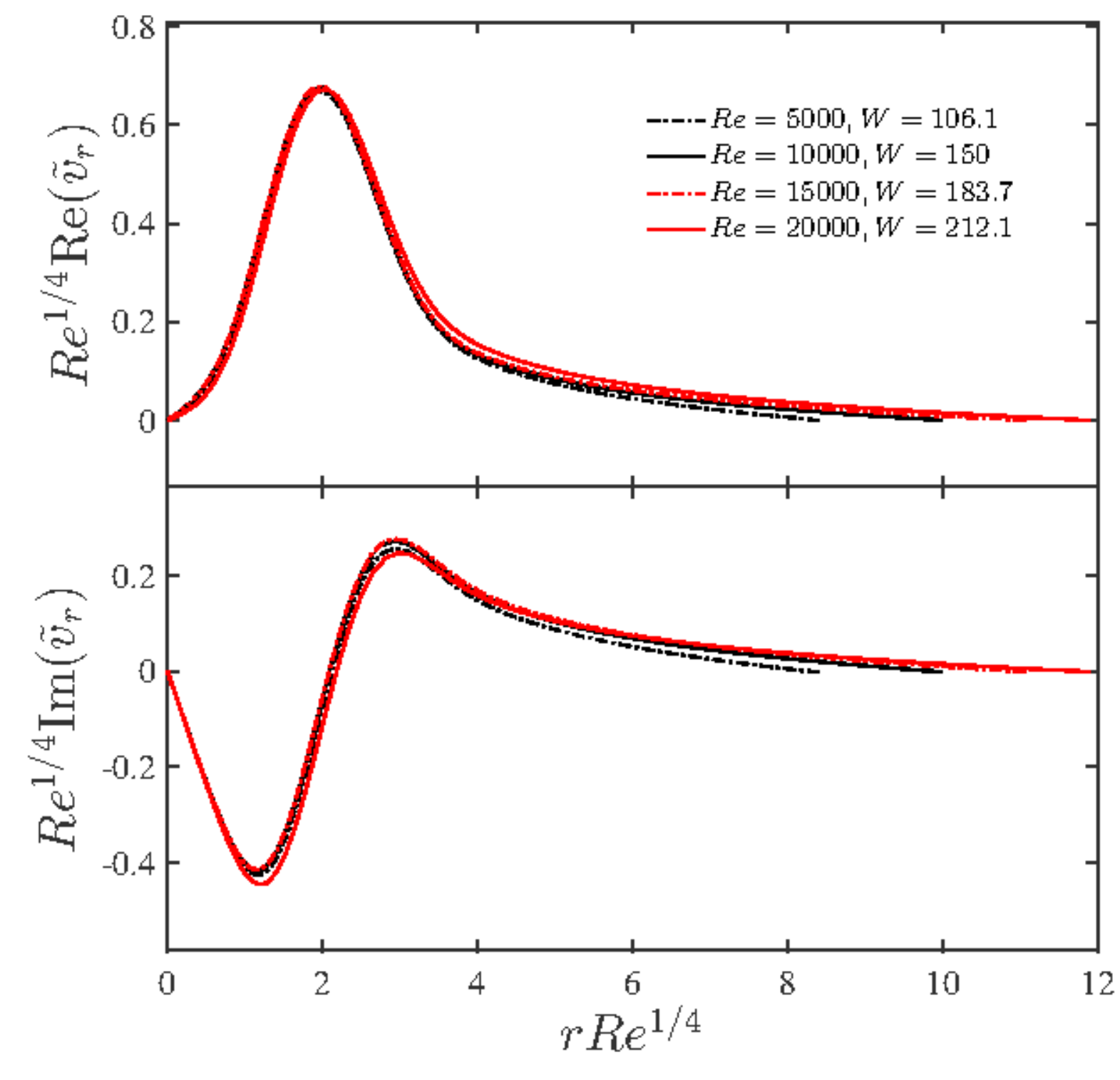}
              \caption{$Re^{1/4} \tilde{v}_r$}
              \label{fig:Vr_collapse_1by4}
            \end{subfigure}
                
            \caption{Collapse of eigenfunctions at different $Re$ and
              $E$, but for a fixed $k = 1$ for (a) axial velocity, and
              (b) scaled radial velocity, on scaled radial-axis in the limit
              $Re\rightarrow \infty$ and $W\rightarrow \infty$ for a
              fixed $W/Re^{1/2}$ and $\beta=0.5$, corresponding to the
              eigenvalues
              $c=0.994842+0.000112i,0.996321+0.000103i,0.996985+0.000093i$
              and $0.997383+0.000085i$ respectively.}
            \label{fig:eigfun_collapse_1by4}
          \end{figure}
		  It is also possible to obtain the following  scalings for a fixed $k$        
          by rescaling the Eqs. \ref{eqn:lin_conti}--\ref{eqn:lin_zzstress}:    
          \begin{equation}
            \delta\sim Re^{-1/4},\quad \Rey\sim E^{-2},\quad (1-c)\sim Re^{-1/2},\text{ and } \tilde{v}_z \sim \Rey^{1/4}\tilde{v}_r.
          \end{equation}
          These scalings are illustrated by the collapse of the
          $\tilde{v}_r$ and $\tilde{v}_z$ eigenfunctions, corresponding
          to the unstable mode,  as shown in
          Fig.~\ref{fig:eigfun_collapse_1by4} for a few selected pairs
           ($\Rey, W$) that are large enough to justify
          the limit $(\Rey, W)\rightarrow\infty$ such that $\Rey \sim E^{-2}$.  
               \begin{figure}
            \centering
            \begin{subfigure}[htp]{0.48\textwidth}
            \includegraphics[width =
            \textwidth]{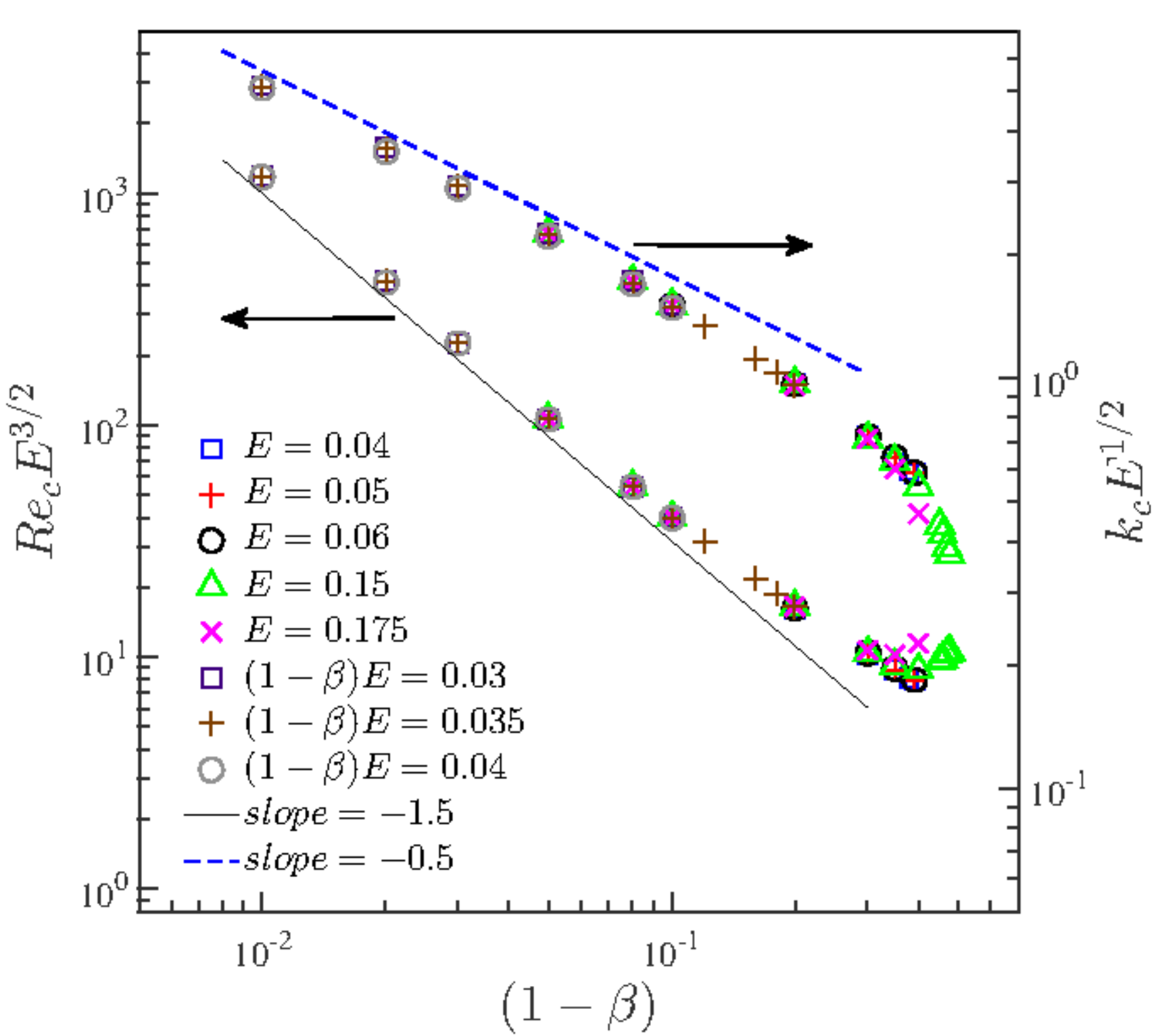}
            \caption{$Re_c E^{3/2}$ and $k_c E^{1/2}$ vs $(1-\beta)$}
            \label{fig:critical_Re_and_k_vs_1mbeta}
              \end{subfigure}
                \begin{subfigure}[htp]{0.48\textwidth}
            \centering
            \includegraphics[width =
            \textwidth]{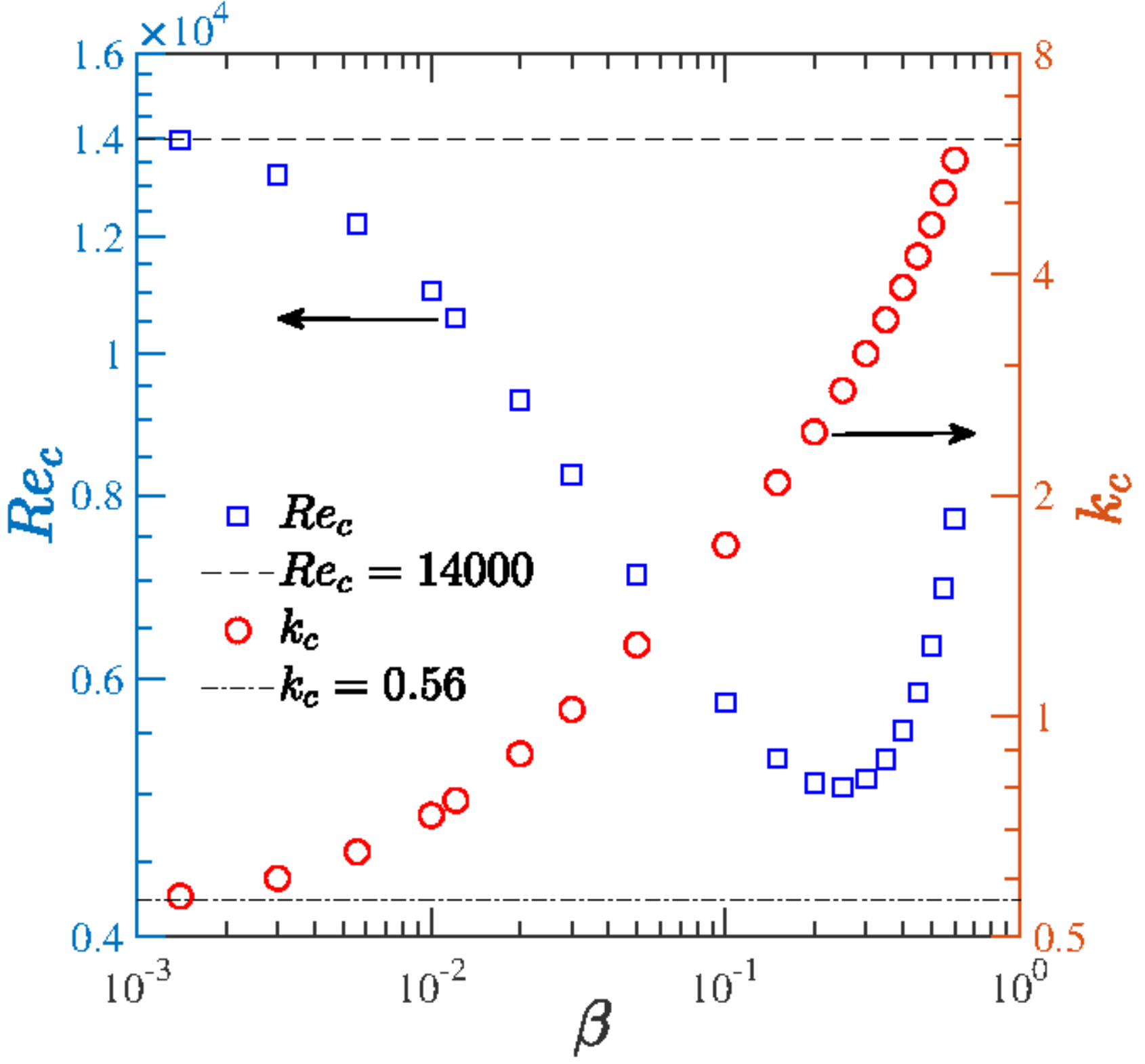}
            \caption{$Re_c$, $k_c$ vs. $\beta$ at $E = 0.01$}
            \label{fig:critical_Re_and_k_vs_beta_E0pt01}
          \end{subfigure}
          \caption{(a) Rescaled critical parameters $Re_c E^{3/2}$ and $k_c E^{1/2}$ vs. $(1-\beta)$
              fall on straight lines of slopes $-3/2$ and $-1/2$,
              respectively, for smaller $(1-\beta)$ implying
              $Re_c\propto [(1-\beta)E]^{-3/2}$ and
              $k_c\propto [(1-\beta)E]^{-1/2}$ in the limit
              $\beta \rightarrow 1$. (b) Variation of critical parameters $\Rey_c$ and
              $k_c$ with $\beta$ for
              $E=0.01$.}
            \label{fig:Reneutrals}
          \end{figure}

          Figure.~\ref{fig:critical_Re_and_k_vs_1mbeta} shows that the
          critical Reynolds number ($\Rey_c$) and critical wavenumbers
          ($k_c$) for different values of $E$ and $(1-\beta)E$
          follow the scalings
          $\Rey_c\propto [(1-\beta)E]^{-3/2}$ and
          $k_c\propto [(1-\beta)E]^{-1/2}$ for $E(1-\beta) \ll 1$.
          We had earlier reported \citep{Garg2018} that $Re_c$ diverges weakly as $\beta^{-1/4}$ for $\beta \rightarrow 0$, based on results extending down to a $\beta$ of $0.025$.  However, new results 
for lower values of $\beta$ (down to $10^{-3}$) in Fig.~\ref{fig:critical_Re_and_k_vs_beta_E0pt01} show that $Re_c$ does not diverge as $\beta^{-1/4}$, but appears instead to diverge more weakly, or perhaps even asymptote to a constant. 
Thus far, we have not found any unstable mode in the UCM limit for the corresponding $\Rey$ and $E$. However, note that the structure of the center mode changes qualitatively for the
 smallest $\beta$'s, characterized by the onset of small-scale oscillations \citep{chaudhary_etal_2019}, and our efforts thus far prevent us from discriminating between $\Rey_c$ approaching a constant vis-a-vis a weak divergence in the said limit. 
          
        

\subsection{Role of stress diffusion on the unstable center mode}
\label{subsec:stressdiffusion}
As discussed in the Introduction, artificial stress diffusion is often used for regularization in
DNS studies of viscoelastic flows \citep{sureshkumar1995linear,sureshkumar_etal_1997,lopez_choueiri_hof_2019}. Recently, it has been shown that this additional diffusivity can qualitatively impact the stress dynamics \citep{gupta2019}, even to the extent of suppressing signatures associated with elasto-inertial turbulence \citep{sid_etal_2018}. In this section, therefore, we briefly examine the effect of stress diffusion on the onset of the center mode instability. The constitutive equation for the polymeric stress, Eq.~\ref{eqn:constitutive_relation}, is now augmented with the stress diffusion term:
\begin{equation}
W \left( \frac{\partial \boldsymbol{T}}{\partial t} + (\boldsymbol{v} \boldsymbol{.} \boldsymbol{\nabla}) \boldsymbol{T} -\boldsymbol{T}\boldsymbol{.} ( \boldsymbol{\nabla} \boldsymbol{v} ) -  ( \boldsymbol{\nabla} \boldsymbol{v} )^{T} \boldsymbol{.} \boldsymbol{T}  \right) + \boldsymbol{T} + \frac{D \lambda}{R^2} \nabla^2 \boldsymbol{T} = \frac{1-\beta}{Re} \{ \boldsymbol{\nabla} \boldsymbol{v} +( \boldsymbol{\nabla} \boldsymbol{v})^{T} \},
\label{diff}
\end{equation} 
where $D$ is the stress diffusivity. \cite{leal1989} showed that the stress diffusion term owes its origin to the translational diffusion of the polymer molecules and estimated the diffusivity $D$ to be $O(10^{-12})$ m$^2$/s.  Using relaxation times $\lambda \sim 10^{-3}$s reported in \cite{chandra_etal_2018} for polymer concentrations $\sim 500$ppm, and for tube diameters $\sim 0.1$--$1$mm, the dimensionless diffusivity $D \lambda /R^2  \sim 10^{-9} - 10^{-7}$. It is useful to represent diffusive effects using a Schmidt number $Sc \equiv \nu/D = E /(D \lambda/R^2)$, with $Sc \rightarrow \infty$ representing the absence of diffusion.
The linearized equations for viscoelastic pipe flow using Eq.~\ref{diff} were solved using a spectral method. Additional boundary conditions are now required for the stress components. At the pipe wall ($r=1$), the stress equation is imposed without the diffusivity, while a regularity condition for the stress is imposed at the centerline \citep{beris1999,lopez_choueiri_hof_2019}. A finite stress diffusivity regularizes the continuous spectrum modes and leads to an additional family of stable diffusive  modes. The decay rate of this family increases with increasing $(D \lambda /R^2)$. However, the discrete modes existing in the absence of stress diffusion are only weakly perturbed for small values of the diffusivity $(D \lambda /R^2 \rightarrow 0)$. 

\begin{figure}
     \centering
          \includegraphics[scale=0.4]{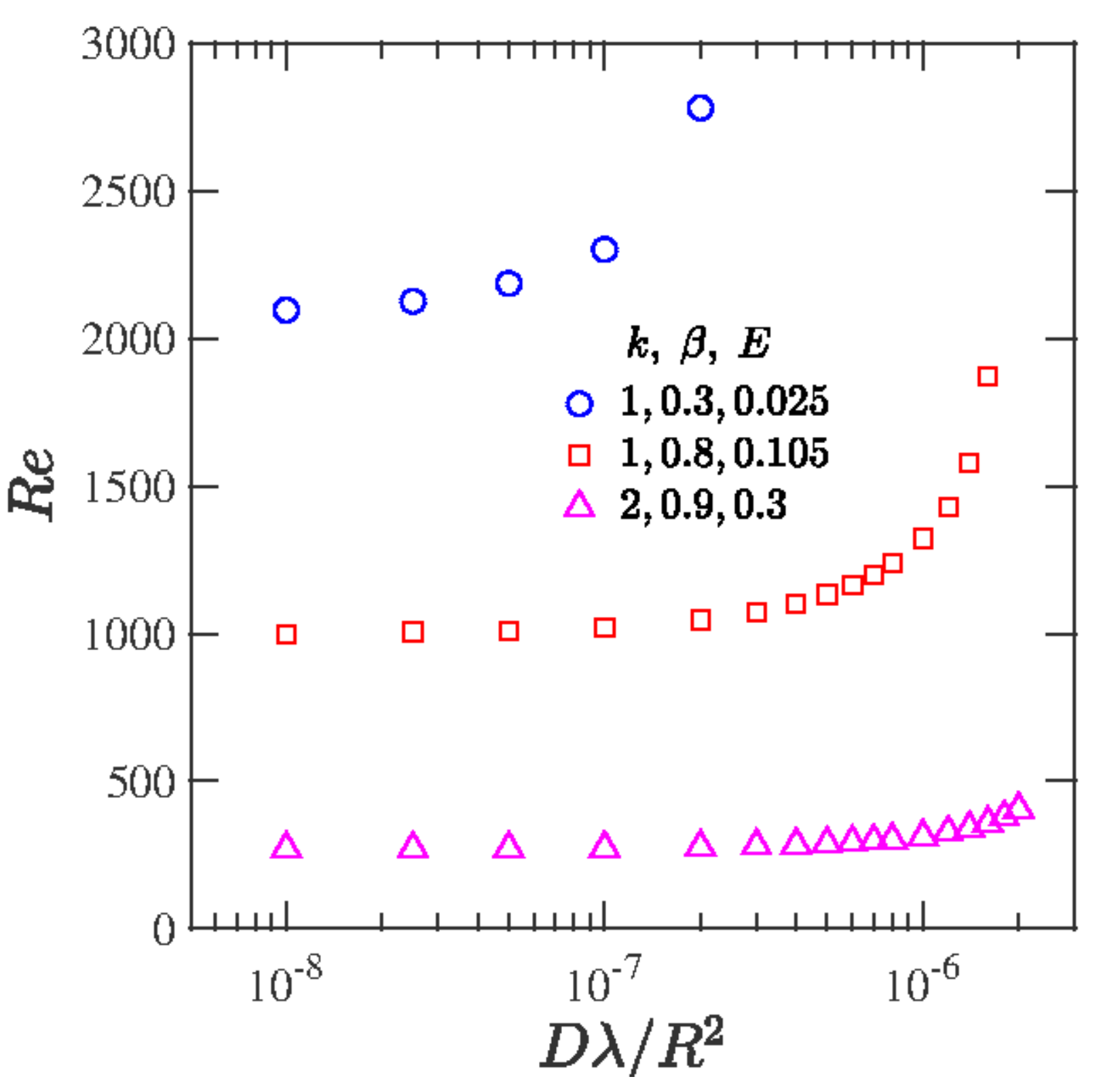}
          \caption{The effect of stress diffusion (characterized by $D \lambda /R^2$) on the threshold $Re$ required for onset of instability at different $E, \beta$ and $k$.}
          \label{fig:diff}
\end{figure} 

Figure~\ref{fig:diff} shows that the  $Re$ for onset of the center mode instability increases with increasing  ($D \lambda /R^2$), implying that the stress diffusivity has a stabilizing effect; for $D \lambda /R^2 \rightarrow 0$, the onset becomes independent of the diffusivity, approaching the values shown in
Fig.~\ref{fig:diff}.  The threshold diffusivity for stabilization is seen to depend on $E$ and $\beta$. Importantly, the instability continues to exist for the experimentally relevant values of $D \lambda/R^2 \sim$ $10^{-9}$--$10^{-7}$, but
would be suppressed at much larger values of $D \lambda /R^2 \sim 10^{-4}$--$10^{-2}$  
(or, equivalently, $Sc < 1000$, for $E \sim 0.1$)
used in earlier DNS studies \citep{sureshkumar1995linear,sureshkumar_etal_1997,lopez_choueiri_hof_2019}.
Thus, consistent with the results of \cite{sid_etal_2018}, the results of Fig.~\ref{fig:diff} reinforce the importance of using 
simulation techniques, which avoid an artificially enhanced diffusivity,  to access the axi-symmetric structures associated with the center-mode instability.


\subsection{Comparison with recent experimental and DNS studies}
          \label{sec:recent_studies}
          \subsubsection{Comparison with experiments}
          \label{sec:samanta_chandra}
\begin{figure}
            \centering
            \includegraphics[width =
            0.5\textwidth]{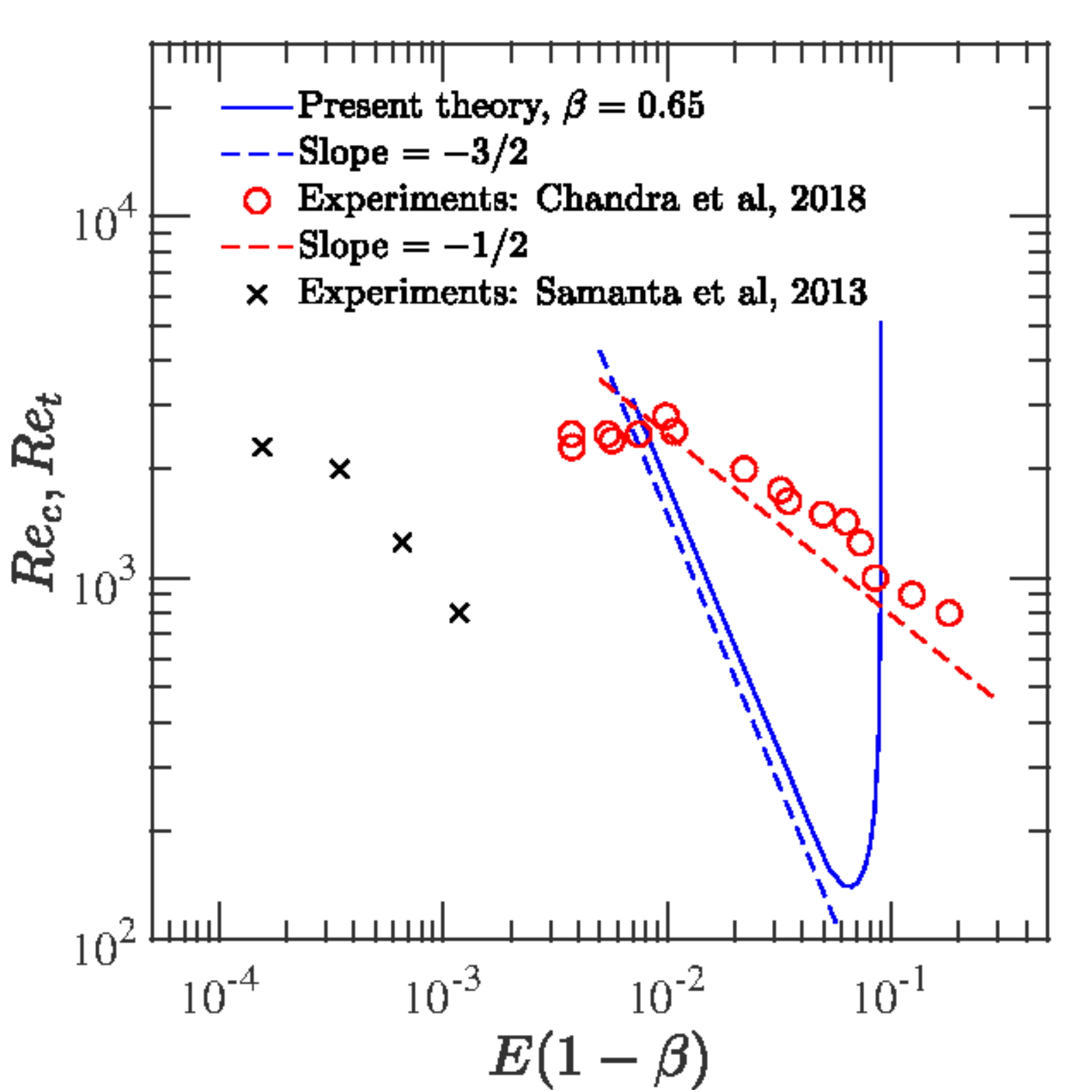}
            \caption{Comparison of the present theoretical predictions
              with experimental results of \cite{samanta_etal2013} and \cite{chandra_etal_2018}.}
            \label{fig:comparison_with_expts}
          \end{figure}

          We have replotted, in Fig.~\ref{fig:comparison_with_expts}, the results of \cite{samanta_etal2013} for the
          transition Reynolds number $\Rey_t$ as a function of $E(1-\beta)$, 
          based on the reported viscosities and relaxation times of the different polymer solutions used in the experiments. The present
          theoretical results yield similar critical Reynolds numbers
          $\Rey_c$ only at much higher values of $E(1-\beta)$.
          \cite{samanta_etal2013} estimated the  relaxation time using
          the CaBER technique \citep{shelley_mckinley2001}, in which the flow is extensional, and the polymer chains are highly stretched.  However, the CaBER procedure is known to have some disadvantages in the estimation of relaxation time for polymers in low-viscosity solvents due to the neglect of inertia in the filament thinning dynamics. The CaBER relaxation time also exhibits a significant concentration dependence even below the nominal overlap concentration \citep{clasen_etal_2006}.
          The data for
          $\Rey_t$ from the experiments of \cite{chandra_etal_2018},
          also plotted in Fig.~\ref{fig:comparison_with_expts}, shows good agreement 
 with the theoretical $Re_c$'s; in that, both threshold $Re$'s are of the same order of magnitude for comparable $E(1-\beta)$.   \cite{chandra_etal_2018}
          used small-amplitude oscillatory strain experiments
          to infer the relaxation times; in contrast to CaBER, the polymer chains are not greatly perturbed about their equilibrium conformations.
While the threshold $Re$'s from \cite{chandra_etal_2018} are comparable to theory, the latter predicts $\Rey_c \sim E^{-3/2}$  along the lower branch of the theoretical envelope, and  $\Rey_t \sim E^{-1/2}$ in  \cite{chandra_etal_2018}.        
           This difference in the
          scaling exponents could be due to shear thinning in
          the experiments, which can also
          significantly alter the parabolic nature of the base velocity profile. These  
          effects are not
          accounted for in the Oldroyd-B model used in this study. 
 The
          following scaling analysis examines the role of shear thinning on the scaling exponent characterizing the $Re_c$ vs. $E$ behaviour for small $E$. We begin by noting the limiting behaviour of viscosity and
		  relaxation time, for large $W$,
for the more realistic FENE-P model, where shear thinning arises on account of the chains being finitely extensible \citep{Bird1980}:
          \begin{equation}
            \eta_1 \sim \eta (\dot{\gamma} \lambda)^{-2/3}, \quad \text{and } \lambda_1 = \lambda (\dot{\gamma} \lambda)^{-4/3},
          \end{equation}
          where $\eta$ and $\lambda$ are viscosity and relaxation time
          at zero shear rate ($\dot{\gamma} = U/R$). The effective Reynolds number
          $\Rey_1$ and Weissenberg number $W_1$, evaluated using the
          shear-rate dependent viscosity and relaxation time, are
          given in terms of  those involving the corresponding zero-shear-rate quantities, as
          \begin{equation}
            \Rey_1 = \frac{\rho UR}{\eta_1} = E^{2/3}\Rey^{5/3}, \text{ and } W_1 = \frac{\lambda_1 U}{R},
            \label{eqn:Re_shear_thinning}
          \end{equation}
          and the effective elasticity number  $E_1$ becomes
          \begin{equation}
            E_1 = \frac{W_1}{\Rey_1}= \frac{\lambda_1 \eta_1}{\rho R^2} = E^{-1}\Rey^{-2}\, .
            \label{eqn:E_shear_thinning}
          \end{equation}
 We now postulate that the scaling for $\Rey_c$ determined
          above for an Oldroyd-B fluid, is valid for  a FENE-P fluid as
          well, but with $Re_1$ and $E_1$ replacing $Re$ and $E$ in order to account for the shear-rate
          dependence of viscosity and relaxation
          time.  Using  Eqs.~\ref{eqn:Re_shear_thinning}--\ref{eqn:E_shear_thinning}
          in the theoretical scaling
          $\Rey_{1,c} \propto E_{1}^{-3/2}$ gives $\Rey_c \propto E^{-5/8}$; the
           scaling exponent now being closer to that ($-1/2$)
          observed in experiments  \citep{chandra_etal_2018,chandra_shankar_das_2020}.
          A similar argument has been used earlier to successfully account the effect of shear thinning on the onset of inertialess elastic instability in Taylor-Couette flow \citep{larsonshearthinning1994}.

          \begin{figure}
            \centering
            \includegraphics[width =
            0.5\textwidth]{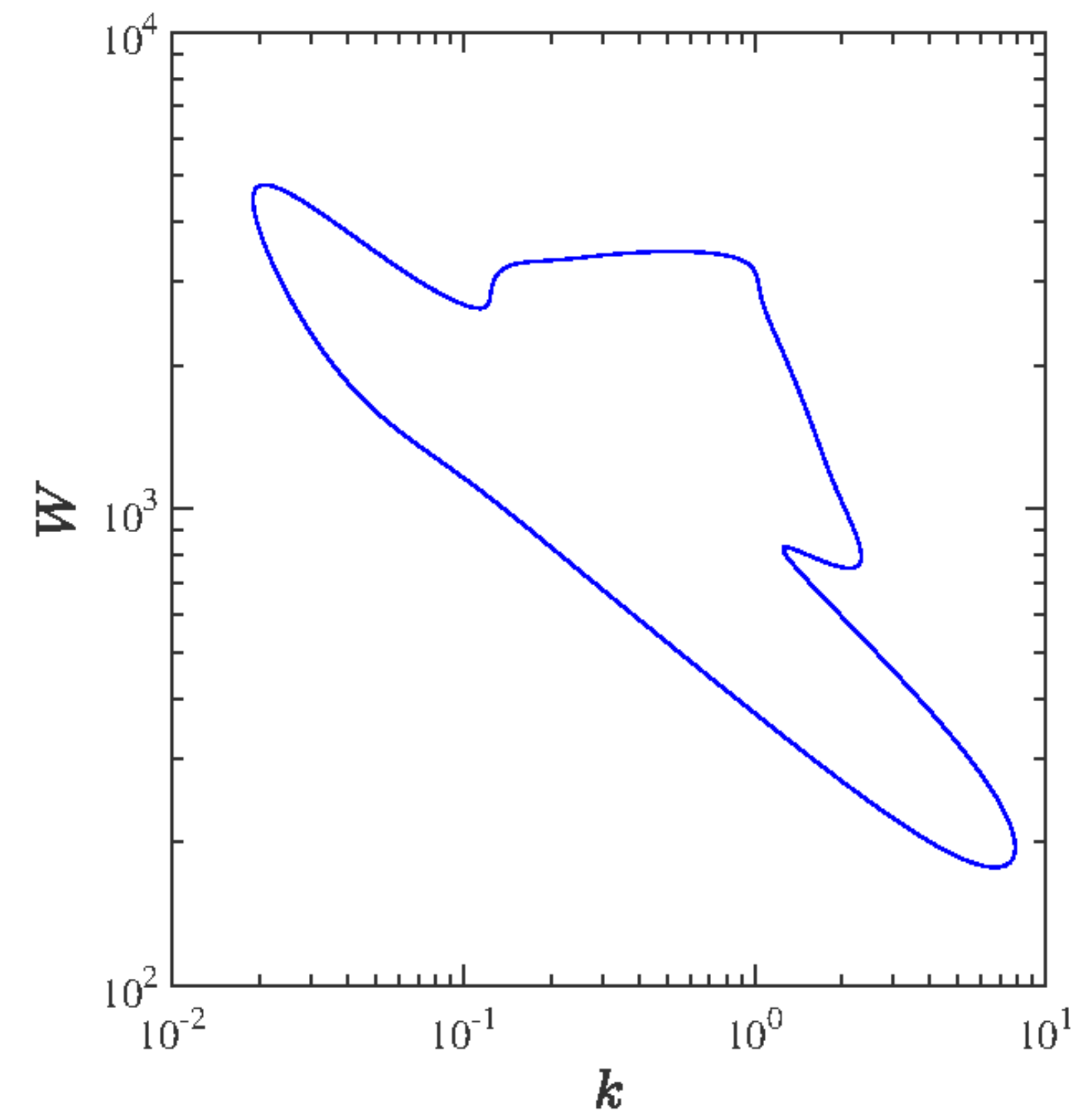}
            \caption{Neutral stability curve in the $W$--$k$ plane at fixed $\Rey = 3500$
              and $\beta = 0.9$; the region inside the loop is unstable.}
            \label{fig:W_vs_k_Re3500betapt9}
          \end{figure}
          \subsubsection{Comparison with DNS of
            \cite{lopez_choueiri_hof_2019}}
          \label{sec:lopez,choueri,hof}
      
          The recent DNS study by \cite{lopez_choueiri_hof_2019} on viscoelastic pipe flow 
          used the FENE-P model and showed that at
          a fixed $\Rey = 3500$,  the flow fully
relaminarizes as $W$ is increased, and at even larger $W$, the laminar
state again becomes unstable, with the post-instability friction factor approaching the MDR asymptote.  
It is worth noting that complete relaminarization was possible because of domains longer than those considered by Graham and co-workers who had observed the so-called hibernating state in the shorter domains \citep{xigrahamPRL2010,Xi_Graham_2012,graham_2014,Xi2019DRreview}. 
In Fig.~\ref{fig:W_vs_k_Re3500betapt9},
           we show the neutral stability curve 
          in the $W$--$k$ plane corresponding to the center mode
          instability for 
          $\Rey = 3500$ and $\beta = 0.9$ 
          (parameters corresponding to the DNS of \cite{lopez_choueiri_hof_2019}), according to which the flow is
           unstable in the range $176.9 <W<4783.6$. The closed loop in the $W$--$k$ plane, at a fixed $\Rey = 3500$, arises because the center mode instability is absent both in the low- and high-$W$ limits,  as can be inferred from the corresponding neutral curves in the $\Rey$--$k$ plane shown in Fig.~\ref{fig:Re_vs_k_beta_0pt9}.
          The range
          of $W$ corresponding to the $W$-$k$ loop where the linear instability of the center mode is found
          is significantly higher than the range ($16<W<80$)  over which EIT was observed in the
           simulations of \cite{lopez_choueiri_hof_2019}.
However, the Oldroyd-B model used here does not account for shear thinning effects inherent in the FENE-P model used in their simulations. 
          More work is thus needed to address these discrepancies between the
          theoretical predictions and experimental and/or DNS studies.

%

          \section{Conclusion and Outlook}
          \label{sec:conclusion}

          The present work builds on our earlier effort \citep{Garg2018} to
          provide the first comprehensive set of 
          results from a linear stability analysis of viscoelastic pipe flow using the  Oldroyd-B
          model. Contrary to the
          prevailing view, and in direct contrast to its Newtonian counterpart,
          pipe flow of an Oldroyd-B fluid is unstable to infinitesimal
          perturbations. The unstable eigenfunction is a center mode with phase
          speed close to the maximum of the base-state flow. We provide a detailed description of
          the emergence and nature of the unstable center mode, and its
          relation to the continuous spectra in the linearized
          spectrum.  
          Crucially,  despite the phase speed being close to unity (the rationale behind the `center mode' terminology),
          the eigenfunctions for the  unstable mode
          are not localized near the centerline in most of the parameter space, especially the region accessible
          to experiments. A bit surprisingly, perhaps, 
          the flow appears to be stable in the limit of an upper-convected Maxwell fluid, implying
          that the destabilizing mechanism involves a
          subtle interplay of fluid inertia, elasticity and solvent
          viscous effects. 
          In the asymptotic limit corresponding to dilute polymer solutions ($(1-\beta) \ll 1$ and 
          $E (1-\beta) \ll 1$), consistent with
          scaling arguments, the numerical results show that
          the critical Reynolds number scales as
          $Re_c \sim (E (1-\beta))^{-3/2}$, while the critical
          wavenumber scales as $k_c \sim (E (1-\beta))^{-1/2}$.  The radial lengthscale
           is now comparable to $k_c^{-1}$,  so that the unstable eigenfunction in this limit does become confined to a thin region in the vicinity of the pipe centerline.
          
%
%

          For $E$ and $\beta$ pertaining to the experiments of 
          \cite{samanta_etal2013} with polymer concentrations greater than $300$ppm, where the authors did
          observe the transition to be 
          supercritical, results from our linear stability
          theory yield much higher transition $Re$'s than the experiments.
           Equivalently, our results do predict a threshold $Re$ of $O(800)$, the one observed for the $500$ ppm solution in  \cite{samanta_etal2013}, but only at much higher $E$'s.
          This discrepancy  could perhaps be attributed to artifacts related to the CaBER procedure used by \cite{samanta_etal2013} to characterize the relaxation time. This procedure is known to lead to a spurious underestimation of the relaxation time (recall that the elasticity number $E$ is proportional to the polymer relaxation time) for solutions well below the nominal overlap concentration; there might be additional problems arising from use of low viscosity fluids \citep{clasen_etal_2006}.
%
          However, our
          theoretical predictions are broadly consistent with the
          observations of \cite{chandra_etal_2018}, who used small-amplitude oscillatory strain experiments
          to infer the relaxation time, wherein the polymer chains are not greatly perturbed about their equilibrium conformations. In their rheological characterization, the solvent viscosity was significantly enhanced to enable a measurable signal, while maintaining a fixed concentration in the dilute regime (unlike CaBER). For the range of $E$ and $\beta$ corresponding to the latter experiments,
          linear stability theory predicts $Re_c \sim 10^2$--$10^3$,
          while experiments report $Re_t \sim 800-1000$.  However,
          observations seem to
          satisfy the scaling relation
          $Re_t \sim (E (1-\beta))^{-1/2}$ in contrast to the $-3/2$
          exponent predicted by our theory (see Fig.~\ref{fig:Rec_vs_E_beta0pt4to0pt9}).  One aspect that could be
          relevant in experiments, but not accounted for in the Oldroyd-B model,
          is shear thinning. 
          Based on a scaling analysis for the FENE-P model that incorporated the asymptotic behavior of the relaxation time, and the resulting shear thinning, for large $W$, the aforementioned
          scaling exponent changes from 
          $-3/2$ to $-5/8$, the latter being closer to the
          experimental exponent of $-1/2$. Nevertheless, more work is
          needed to reconcile theoretical predictions and 
          observations, in terms of accurate 
          characterization of the polymer relaxation time (in the dilute regime), a careful
          detection of the onset of transition by multiple means (such as PIV and pressure-drop measurements), and by using
          realistic constitutive models (in the stability analysis) that extend across the overlap concentration, accounting for dynamics in both the dilute and semi-dilute regimes \citep{prabhakar2016}.

          Prior to the present work, the prevailing understanding of
          stability of viscoelastic flows and turbulent drag reduction
          was predicated on an elastic modification of the Newtonian
          transition scenario. The latter is known to be subcritical, wherein the actual transition is
          believed to be preceded by the appearance of
          (nonlinear) three-dimensional, exact coherent states (ECS). As $\Rey$ is increased, 
          the   
          laminar basin of attraction shrinks, with the concomitant appearance of more unstable ECS's.
          The turbulent trajectory is
           proposed to sample the phase space of such solutions in
          a chaotic manner \citep{budanur_etal_2017}; a sustained turbulent state, beyond a finite threshold $Re$, requires spatial-temporal dynamics that includes merging of localized ECS  solutions \citep{avila2011,chanty_kerswell,barkley2016}.  The work of Graham and co-workers 
\citep{stone_graham2002,stone_graham2003,stone_graham2004,Li_Graham_2007,graham_2014}          
          explored in detail the effect of viscoelasticity on the above scenario. 
          Specifically, \cite{Li_Graham_2007} have shown that
          viscoelasticity suppresses the appearance of the relatively simple (travelling waves) ECS in channel flow.  
An extrapolation, entailing an assumption that viscoelasticity has a similar effect  
 on the other ECS's, with a non-trivial time dependence, implies that the onset of the sub-critical transition is delayed by viscoelasticity.
As mentioned earlier in Sec.\ref{subsec:expts}, this conclusion has some
          indirect experimental support
          \citep{samanta_etal2013,chandra_etal_2018}.  
However, the transition scenario at higher $E$'s (equivalent to higher polymer concentrations), apart from being at lower $Re$'s, appears to have a fundamentally different character, being independent of external perturbations.  Our analysis,  consistent with these observations, shows that pipe flow is unstable to infinitesimal disturbances at sufficiently high $E$. Results from recent computations  \citep{sid_etal_2018} emphasize the importance of 2D (span-wise oriented) structures in the elasto-inertial turbulent state for channel flow, in marked contrast to the 3D Newtonian scenario, reinforcing the notion of an underlying instability to axisymmetric perturbations examined here. Importantly, the nature of
the elasto-inertial coherent structures identified in DNS studies of both viscoelastic channel and pipe-flow  \citep{sid_etal_2018,lopez_choueiri_hof_2019} are quite similar, pointing to a generic mechanism operative in both these geometries,
again consistent with our own finding of an analogous center mode instability in channel flow \citep{Garg2018,khalid_channel}.
          
%

          Our study is a clear call for a reassessment of the
          current understanding of turbulent drag reduction by
          polymers, in particular, of the nature of the
          maximum-drag-reduced (MDR) regime and its relation to both the laminar
          and 
          (Newtonian) turbulent states. Recent experimental results of
          \cite{choueiri_etal_2018} explicitly demonstrate the link
          between elasto-inertial turbulence and the
          maximum-drag-reduction (MDR) regime, by showing that the
          same physical mechanisms underlie the two states (at least for the moderate $Re$'s accessed in
          the experiments).  
         Both experiments and DNS studies 
          \citep{choueiri_etal_2018,shekar_etal_2019,lopez_choueiri_hof_2019} have also shown
          that  MDR can
          also be reached via a direct pathway from the laminar state
          of a polymer solution, without entering the Newtonian
          turbulent regime.  Thus, the
          terminology of `drag reduction' is somewhat ambiguous: the
          MDR state was traditionally viewed as a drag-reduced state
          from Newtonian turbulence upon addition of polymers.
          Based on the above picture, and the linear instability identified in this work, 
          we conjecture that the aforementioned direct pathway to MDR
          could be achieved via a nonlinear saturation of the
          elasto-inertial center mode instability of viscoelastic pipe flow, with a concomitant mild drag enhancement relative to the laminar state. The
          eigenfunctions corresponding to the unstable mode identified
          in this study should form a template for future nonlinear
          studies aimed at identifying novel nonlinear elasto-inertial
          structures that might play an important role in
          understanding the nature of the MDR state at large elasticity numbers.
          As a first step in this direction, results from a weakly-nonlinear
		  analysis \citep[along the lines of][]{stuart_1960,watson_1960}, which are largely 
		  consistent with the conclusions of the linear stability analysis presented here, will
		 be reported in a future communication. 

\begin{figure}
  \centering
  \includegraphics[width = 10cm]{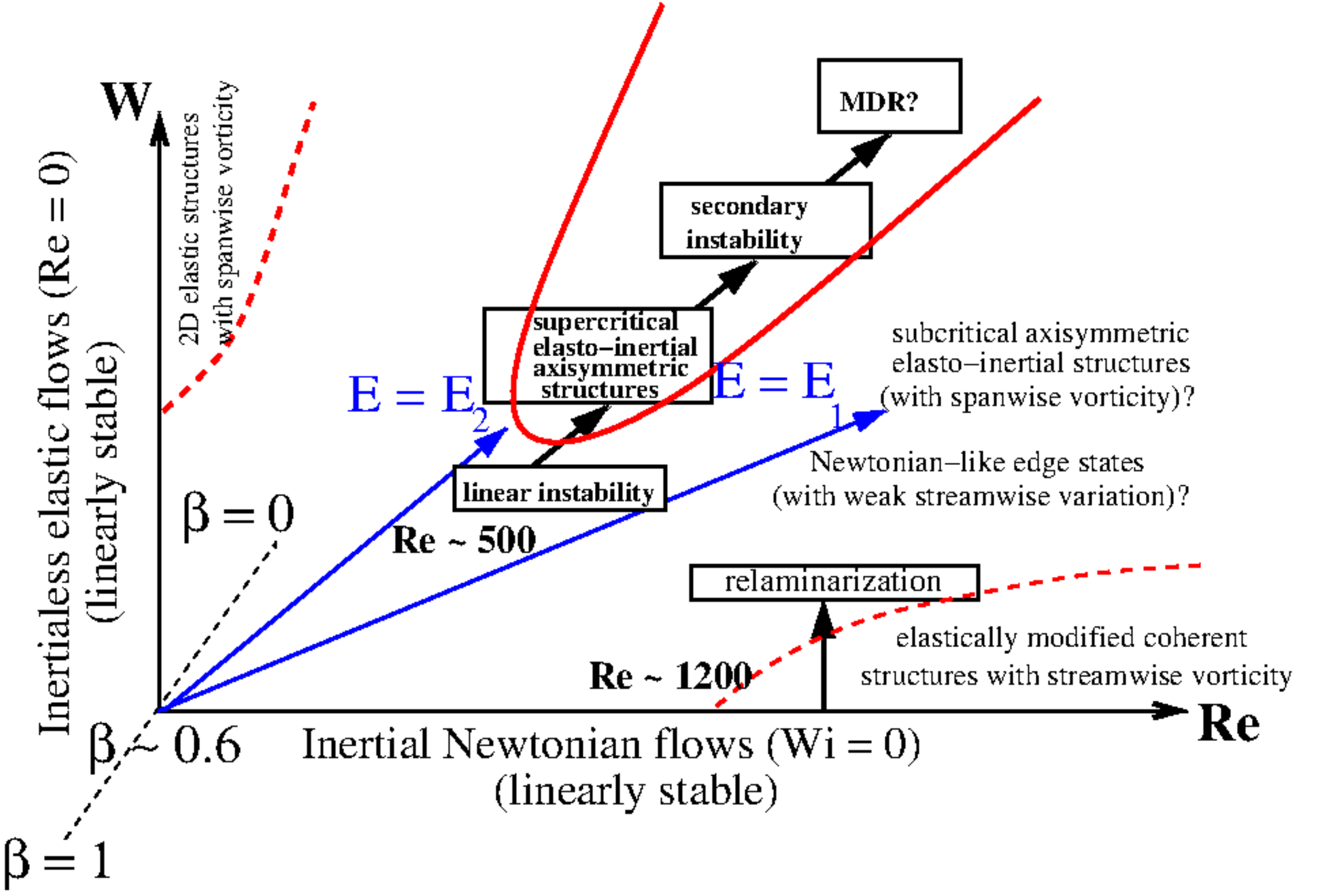}
  \caption{A schematic representation of various transition scenarios in
    viscoelastic pipe flow in the $\Rey$--$W$ (for a fixed $\beta$)
    plane. The boundaries shown using dotted red lines represent subcritical
    bifurcations, while the unstable boundary due to the center mode instability is shown using a continuous red line. The oblique blue lines indicate experimental paths in which flow rate is increased (in a pipe of a given diameter and for a given polymer solution) with $E = W/Re$ being constant; $E_2 > E_1$.}
  \label{fig:bigpicture}
\end{figure}


Based on the above discussion, it is useful to organize our
understanding of the various possible transition scenarios in
viscoelastic pipe flows in the form of a `phase diagram' in 
$\Rey$--$W$ (and $\beta$) space. This is shown as a schematic in Fig.~\ref{fig:bigpicture} for a fixed $\beta \sim 0.5$ and higher; for $\beta \rightarrow 0$, $Re_c \sim O(10^4)$ (see Fig.~\ref{fig:critical_Re_and_k_vs_beta_E0pt01}), and it is not clear if the linear instability (at such high $Re$) would continue to be practically relevant. Note that an experimental pathway representing an increase in flow rate (for a given pipe diameter and polymer solution) will appear as an oblique line (with slope $ E = W/Re$) in the $Re$-$W$ plane \citep{graham_2014,Xi2019DRreview}.
It helps to consider two limiting sequences in this plane. The first corresponds to increasing $Re$ at $W = 0$ - the Newtonian transition. The sub-critical nature of this transition and the underlying role of the ECS's is now relatively well established. The effect of viscoelasticity on this picture has been discussed in the earlier paragraph, the main idea being that the suppression of the ECS's by elasticity, at $Re$'s greater than the Newtonian threshold ($Re \sim 2000$), has been interpreted as the reason for a delayed transition; the regime of existence of the ECS's for $W = 0$, and the postponement of this regime with increasing $W$ is marked with a dashed red line near the $Re$-axis in  Fig.~\ref{fig:bigpicture}. 
It is worth noting that despite the sharp contrast in the linear (modal) eigenfunctions for pipe and channel flow (the absence of the TS wall mode in pipe flow being an example), the Newtonian ECS's have a similar character across all of the canonical shearing flows, consisting of counter-rotating vortices and stream-wise streaks in all cases. Thus, the extrapolation of the effects of viscoelasticity to the pipe geometry, based on the domain of existence of finite-$W$ ECS's in the channel geometry, is reasonable. Nevertheless, there is a need to examine the nature of elastically-modified pipe flow ECS's, as a function of $W$, in order to render the arguments quantitative. As already indicated above, such ECS-based arguments are no longer valid at higher $E$ when the ECS's are absent.

The understanding of the MDR regime attained at higher $E$'s (or higher $W$'s with $Re$ fixed), shown schematically by the path $E_1$ in Fig.~\ref{fig:bigpicture}, was until recently based on a series of minimal (channel) flow unit simulations carried out by Graham and co-workers \citep{xigrahamPRL2010,Xi_Graham_2012,graham_2014,Xi2019DRreview}. The hypothesis advanced was that of the dynamics in the MDR state corresponding to that of a largely unaltered Newtonian `edge state', a marginal state whose stable manifold forms the boundary separating the laminar fixed point and the turbulent attractor. This edge state manifests as prolonged periods of so-called hibernating turbulence, characterized by subdued fluctuatons and an associated weak stream-wise variation. The primarily Newtonian character of this edge-state was proposed as an explanation for the independence of the MDR regime with respect to polymer characteristics. In effect, the originally unstable Newtonian edge state is apparently stabilized at higher $E$'s. The connection between the disappearance of the ECS's in earlier work by Graham's group, and the subsequent appearance of a stabilized edge state at higher $E$'s, has not been clarified from a dynamical systems view point in terms of an appropriate viscoelastic state space (it is worth noting that, for the Newtonian case, most of the lower-branch travelling-wave solutions are known to lie on the aforementioned laminar-turbulent boundary, and are therefore edge states, albeit unstable). Importantly, however, the veracity of the above edge-state-based interpretation has been recently challenged by simulations in longer domains \citep{lopez_choueiri_hof_2019} where the hibernating state is found to give way to spatio-temporal intermittency, and subsequent relaminarization.

At sufficiently high $E$'s (shown schematically by the path $E_2$ in Fig.~\ref{fig:bigpicture}), there is the possibility of the center-mode instability (region in Fig.~\ref{fig:bigpicture} demarcated by a continuous red curve) leading to a direct pathway from the laminar state to a non-linear state characterized by essentially axisymmetric elasto-inertial structures that presumably arise from a saturation of the growing center mode. These structures might then form the backbone of EIT dynamics.
The identification of this pathway confirms the speculation of a linear instability at high $E$  \citep[see Fig.~4 of][]{graham_2014}, thereby augmenting the various possible transition scenarios in the $Re$-$W$ plane. Based on the $E$-intervals identified here, for the center-mode instability, there does appear to be a region in the $Re$-$W$ plane where the elastically modified ECS's are absent and pipe flow is still linearly stable. In this regime, one might either expect dynamics corresponding to the Newtonian edge state, proposed by Graham and coworkers (described above), or an entirely new set of subcritical elasto-inertial structures. In the latter regard, it is tempting to postulate
the 2D, nonlinear mechanism of \cite{shekar_etal_2019}, with signatures similar to
the least stable Newtonian TS mode, to play a role. In contrast to 
the qualitative similarities in the nature of Newtonian pipe and channel ECS's, however, and as pointed out in Sec.~\ref{sec:shekar}, differences in the axisymmetric pipe and two-dimensional channel flow viscoelastic spectra render this wall-mode based mechanism untenable for pipe flow. This is because the  center mode in Newtonian pipe flow, while being stable, still has a decay rate smaller than that of the wall mode. Further, there is a significant regime in the $Re$--$W$ plane where the continuous spectra are the least stable. Thus, unlike the proposal of \cite{shekar_etal_2019} for channel flow,  a novel sub-critical elasto-inertial dynamics, in pipe flow, would seem to have to account for the dynamics of the continuous spectrum at leading order \citep{balmforth2013pattern}. Further, any additional dynamics related to the discrete modes would still appear to be dominated by the  center mode on account of its lower decay rate. Based on this qualitative picture, we have indicated (in Fig.~\ref{fig:bigpicture}) the two possible mechanisms in the region that separates the regimes corresponding to the center-mode instability, and the subcritical ECS. Note that any pathway leading upto the EIT regime, at a fixed $Re$, involves a transition from coherent structures with stream-wise vorticity to those with span-wise vorticity.

Finally, at the highest $E$'s, one approaches the second limiting sequence in Fig.~\ref{fig:bigpicture}, which is that of increasing $W$ at $Re = 0$, and therefore, concerns the inertia-less transition to elastic turbulence \citep{Groisman2000},  which is believed to follow the traditional nonlinear route 
 \citep{stuart_1960,watson_1960}. van Saarloos and co-workers
  \citep{meulenbroek_etal_2003,meulenbroek_sarloos2004,morozov_saarloos2005,morozov_saarloos2007}, based on a viscoelastic analogue of the original Stuart-Landau expansion, have shown that inertia-less pipe flow undergoes a sub-critical bifurcation to a nonlinear two-dimensional state (represented using a dotted red line near $Re = 0$ in Fig.~\ref{fig:bigpicture}); the same is true for plane Couette and Poiseuille flows. This scenario has found some support in experiments \citep{pan_etal_2013}. These weakly nonlinear analyses are based on the existence of an unstable/weakly stable discrete mode  well separated from the remainder of the spectrum. This is indeed true for inertia-less viscoelastic flows where
the spectrum consists of a small number of discrete modes \citep{renardy1986linear,wilson1999}, in addition to the continuous spectra. However, the spectrum becomes far more complicated with increasing $Re$,
with there being no clear separation in the above sense \citep[see Fig.~\ref{fig:es_E0pt15Re6000k1_beta0pt98to1} and those in][]{chaudhary_etal_2019}. In fact, for moderate $Re$ and for small but finite $E$, as pointed out above, there 
exist scenarios wherein no discrete modes are present above the continuous spectrum (e.g., for $E < 0.6$ and $Re = 500$, $\beta = 0.96$ in Fig.~\ref{fig:ues_beta_pt96_Re500_k1}), thereby necessitating the 
 consideration of the CS at leading order in the nonlinear analysis.
Clearly, therefore, the nonlinear mechanisms proposed by 
van Saarloos and co-workers are restricted to modest $Re$, and cannot serve as an explanation for transition to EIT.

In summary, our finding of a linear instability in viscoelastic pipe
flow marks a possible paradigm shift from both classical and modern theoretical
work on Newtonian fluids, by providing a natural explanation
for the connection between the laminar state and the elasto-inertial state 
underlying the so-called maximum-drag-reduced regime.

\noindent{{\bf Equal contribution of authorship}}

I.~C and P.~G contributed equally to this work. 

\noindent{{\bf Declaration of interests}}

The authors report no conflict of interest.

\bibliographystyle{jfm} \bibliography{BiblioPipe}

\end{document}